박 사 학 위 논 문
Ph. D. Dissertation

# 인간형 로봇을 위한 머리전달함수 정위 단서 기반의 두 귀의 소리 사건 정위 및 감지 신경망

Binaural Sound Event Localization and Detection Neural Network based on HRTF Localization Cues for Humanoid Robots

2024

이 경 태 (李 炅 泰 Lee, Gyeong-Tae)

한 국 과 학 기 술 원

Korea Advanced Institute of Science and Technology

박 사 학 위 논 문

인간형 로봇을 위한 머리전달함수 정위 단서 기반의 두 귀의 소리 사건 정위 및 감지 신경망

2024

이 경 태

한 국 과 학 기 술 원

기계공학과

# 인간형 로봇을 위한 머리전달함수 정위 단서 기반의 두 귀의 소리 사건 정위 및 감지 신경망

이 경 태

위 논문은 한국과학기술원 박사학위논문으로
학위논문 심사위원회의 심사를 통과하였음

2023년 12월 14일

심사위원장  박 용 화 (인)
심 사 위 원  전 원 주 (인)
심 사 위 원  윤 용 진 (인)
심 사 위 원  이 승 철 (인)
심 사 위 원  최 정 우 (인)

# Binaural Sound Event Localization and Detection Neural Network based on HRTF Localization Cues for Humanoid Robots

Gyeong-Tae Lee

Advisor: Yong-Hwa Park

A dissertation submitted to the faculty of
Korea Advanced Institute of Science and Technology in
partial fulfillment of the requirements for the degree of
Doctor of Philosophy in Mechanical Engineering

Daejeon, Korea
December 14, 2023

Approved by

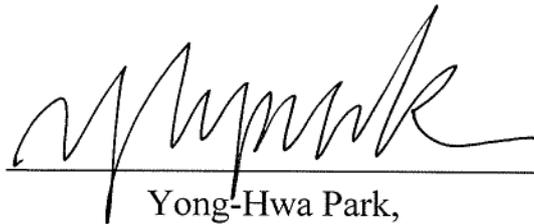

Yong-Hwa Park,
Professor of Mechanical Engineering

The study was conducted in accordance with Code of Research Ethics[1].

---

1) Declaration of Ethical Conduct in Research: I, as a graduate student of Korea Advanced Institute of Science and Technology, hereby declare that I have not committed any act that may damage the credibility of my research. This includes, but is not limited to, falsification, thesis written by someone else, distortion of research findings, and plagiarism. I confirm that my dissertation contains honest conclusions based on my own careful research under the guidance of my advisor.

DME 이 경 태. 인간형 로봇을 위한 머리전달함수 정위 단서 기반의 두 귀의 소리 사건 정위 및 감지 신경망. 기계공학과. 2024년. 180+xii 쪽. 지도교수: 박용화. (영문 논문)

Lee, Gyeong-Tae. Binaural Sound Event Localization and Detection Neural Network based on HRTF Localization Cues for Humanoid Robots. Department of Mechanical Engineering. 2024. 180+xii pages. Advisor: Park, Yong-Hwa. (Text in English)

초 록

인간형 로봇이 소리를 통해 상황을 인식하기 위해서는 주변 소리 사건의 종류 및 방향을 동시에 추정해야 한다. 또한 보청기나 원격현장감과 같은 인간-로봇 상호작용 기술에 적용하려면 인간과 같은 2채널 입력으로 구현할 필요가 있다. 그러나 수평 배열의 2채널 입력으로는 고도각 추정이 어려우며, 방위각 추정시 앞뒤 혼동이 발생한다. 이를 해결하기 위해, 두 귀 입력 특징에서 소리 사건의 시간-주파수 패턴 및 머리전달함수 정위 단서를 학습하여, 각 소리 사건의 종류 및 방향을 동시에 추정할 수 있는 두 귀의 소리 사건 정위 및 감지(binaural sound event localization and detection, BiSELD) 신경망을 제안한다. 학습을 위해 원점전달함수 측정 및 비인과성 보상에 관한 명확한 기준을 세워 머리전달함수를 측정하였고, 이를 수집한 소리 사건 데이터베이스와 합성하여 두 귀의 데이터셋을 구축하였다. 특히, 머리전달함수 정위 단서의 분석을 기반으로 BiSELDnet의 입력 특징으로 두 귀의 시간-주파수 특징(binaural time-frequency feature, BTFF)을 제안하였다. BTFF는 좌우 mel-spectrogram, 각 주파수 성분의 시간 변화율을 나타내는 좌우 V-map, 이간 시차를 추정하는 ITD-map, 앞뒤 혼동 해결의 단서로서 이간 음량차를 나타내는 ILD-map, 그리고 고도각 추정을 위한 스펙트럼 단서를 제공하는 좌우 SC-map 등 총 8채널의 특징맵으로 구성된다. 전방향, 수평면 그리고 정중면의 소리 사건에 대한 BTFF의 감지 및 정위 성능을 평가한 결과 그 효과를 확인하였다. 입력된 BTFF를 학습하여 각 소리 사건의 종류 별로 방향 벡터의 시계열을 출력하는 다양한 BiSELDnet을 구현하였으며, 이중 파라미터 수가 작고 성능이 가장 우수한 삼일체 모듈에 기반한 BiSELDnet을 선정하였다. 삼일체 모듈은 채널 간 상관관계가 낮은 BTFF의 학습에 적합한 깊이별 분리 합성곱을 기반으로 3×3, 5×5, 7×7 크기의 세 커널을 각각 3×3 크기의 커널로 인수분해하여 연결한 모듈로서 다양한 크기의 특징맵을 동시에 추출할 수 있는 장점이 있다. 또한, BiSELDnet이 학습한 것을 시각화하고, 입력 특징의 어느 부분이 감지 및 정위의 최종 결정에 기여하는지 확인하기 위해 벡터 활성화 맵 시각화를 제안하였다. 이를 통해 BiSELDnet이 소리 사건의 고도각 추정을 위해 N1 홈 주파수에 주목하는 것을 확인하였다. 마지막으로, 다양한 신호 대 잡음비의 도시 배경소음조건에서 수평면 또는 정중면의 소리 사건에 대해 BiSELD 모델과 최신 SELD 모델의 감지 및 정위 성능을 비교하였다. 비교 결과, 두 귀의 입력 조건에서 제안된 BiSELD 모델이 기존의 최신 SELD 모델 보다 성능이 우수함을 확인하였다.

핵 심 낱 말: 인간형 로봇, 두 귀의 소리 사건 정위 및 감지, 머리전달함수, 두 귀의 시간-주파수 특징, 삼일체 모듈, 깊이별 분리 합성곱, 벡터 활성화 맵.

Abstract

In order for a humanoid robot to recognize the situation through sound, it must simultaneously estimate the type and direction of surrounding sound events. Also, to be applied to hearing aids or human-robot interaction technologies such as telepresence, it is necessary to be implemented with two-channel input like a human. However, with horizontal two-channel input, it is difficult to estimate the elevation of sound event, and front-back confusion occurs when estimating the azimuth. To solve this problem, binaural sound event localization and detection (BiSELD) neural network is proposed, which can simultaneously estimate the class and direction of each sound event by learning the time-frequency pattern and head-related transfer function (HRTF) localization cues of sound event from a binaural input feature. For learning, HRTFs were measured by establishing clear standards for origin transfer function measurement and non-causality compensation, and binaural dataset was constructed by synthesizing the measured HRTFs with collected sound event databases. In particular, based on the analysis of HRTF localization cues, binaural time-frequency feature (BTFF) was proposed as the input feature for BiSELDnet. A BTFF consists of eight-channel feature maps: left and right mel-spectrograms; left and right V-maps showing the time change rate of each frequency component; ITD-map estimating interaural time difference (ITD) below 1.5 kHz; ILD-map representing interaural level difference (ILD) above 5 kHz with front-back asymmetry as a clue to solve the front-back confusion; and left and right SC-maps providing spectral cue (SC) above 5 kHz for the elevation estimation of sound event. The effectiveness of BTFF was confirmed by evaluating its detection and localization performance for sound events coming from omnidirectional, horizontal, and median planes. Using BTFF as input feature, a variety of BiSELDnets were implemented that output a time series of direction vectors for each sound event class. The magnitude and direction of each vector represent the activity and direction of the corresponding sound event class, allowing simultaneous detection and localization of sound events. Among them, BiSELDnet based on Trinity module, which has the best performance with a small number of parameters, was selected. Based on depthwise separable convolution, which is suitable for BTFF with low cross-channel correlation, Trinity module is implemented by factorizing each of the three concatenated kernels of size 3×3, 5×5, and 7×7 into kernels of size 3×3. It has the advantage of simultaneously extracting feature maps of various sizes from its input feature with a small number of parameters. In addition, vector activation map (VAM) visualization was proposed to visualize what BiSELDnet learned and check which parts of input feature contribute to the final decision of detection and localization. Through VAM visualization, it is confirmed that BiSELDnet focuses on the N1 notch frequency for the elevation estimation of sound event. Finally, the detection and localization performances of BiSELD model and state-of-the-art (SOTA) SELD model were compared for sound events in the horizontal or median plane under urban background noise conditions with various signal-to-noise ratios. The comparison results demonstrate that the proposed BiSELD model performs better than the existing SOTA SELD model under binaural input conditions.

# Contents

# List of Tables

# List of Figures

# Chapter 1. Introduction

## 1.1 Background

### 1.1.1 Humanoid Robot

Since most of the human infrastructure is adapted to the human body type, humanoid robots have long attracted attention in the field of human-machine interaction. A humanoid robot is a highly intelligent robot that requires the ultimate development of robotics technology because it must not only have a human form and movement ability, but also have human-like recognition [1–3]. Research on humanoid robots is largely divided into two fields. The first is research on mechanisms for bipedal locomotion, and the second is research on sensor-based artificial intelligence (AI) such as vision and voice recognition [4]. In the first few decades, most of the research has been focused on the development of humanoid bipedal robots. WABIAN robots [5], Honda humanoid robots [6], JOHNNIE [7], HRP [8], ASIMO [9], and HUBO [10] are well known humanoid robots. Recently, there has been a tendency to focus more on detection, recognition, and emotion research based on human-robot interaction rather than on bipedal humanoid research [11,12]. To interact with humans, a humanoid robot needs to be aware of its surroundings just like a human. Since vision is the most effective way to obtain information from the environment, early humanoid robots processed visual information primarily through a pair of actuated cameras [13]. However, vision is not the only sensing mechanism contributing to human interaction tasks [14]. Humans can determine the location of a sound source just by hearing, and can detect odors and obtain information about taste through the sense of smell. Accordingly, humanoid robots have been developed to include various sensing systems such as senses of hearing [15], smell [16], and taste [17]. Above all, audition seems to have become a standard sensing system for humanoid robots, playing a key role in tasks such as speech recognition, environmental awareness, and social interaction [12]. The first humanoid robot to integrate vision and audition was the WABOT-2 in 1984 [15]. Recently, some humanoid robots perform simple conversations with humans through speech recognition and synthesis [12]. However, to respond appropriately to a given situation, a humanoid robot should recognize not only speech but also various sound events occurring around it. Sound event detection and localization is essential for humanoid robots to effectively navigate and interact in its environment. This allows humanoid robots to be more aware of their surroundings. By identifying and pinpointing sound sources, humanoid robots can gather information about objects or events that may not be visible through cameras or other sensors. This situational awareness is critical for making informed decisions and responding appropriately to dynamic environments. In particular, it can contribute to the safety of both the humanoid robot and nearby humans. For instance, humanoid robots can identify emergency alarms, sirens, or other auditory warnings, and respond accordingly, such as navigating away from potential hazards or alerting human operators. Also, in scenarios where humanoid robots interact with humans, the ability to detect and localize sound events is crucial. For example, in a rescue mission, a crying baby may become trapped under a collapsed structure outside of the humanoid robot's field of view, as shown in Fig. 1.1. In such a situation, to detect and localize the baby, the humanoid robot needs to be able to compensate for the missing visual information by using auditory information based on the mounted microphones.

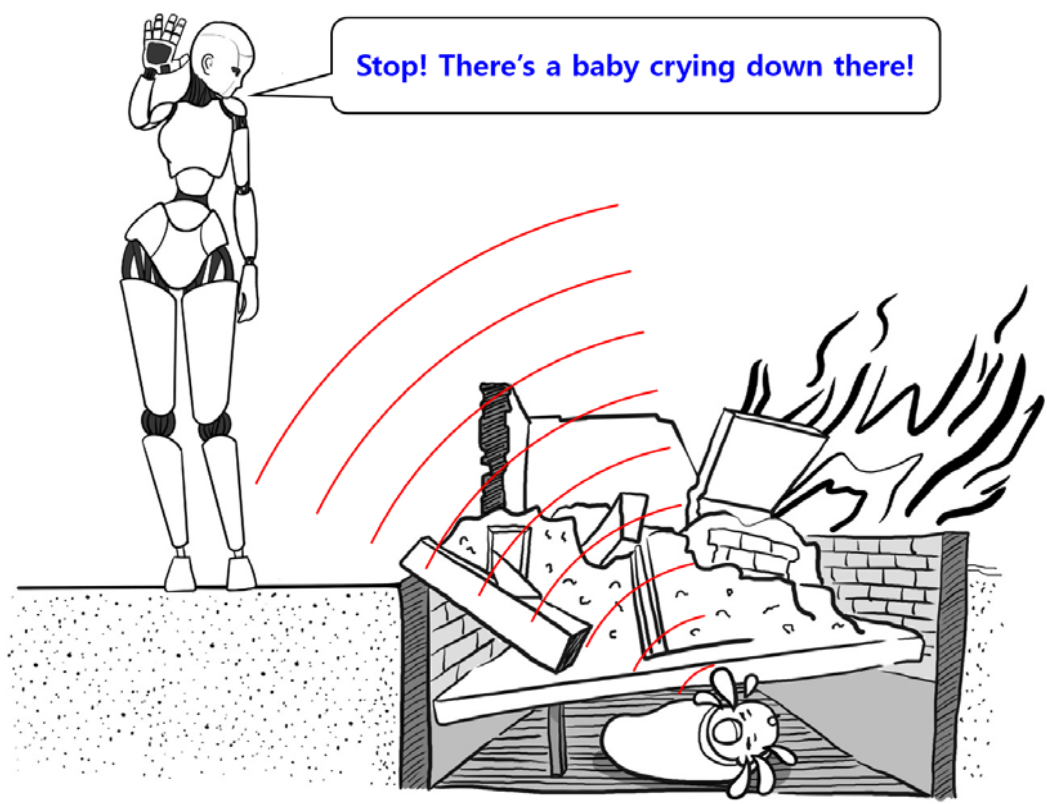

**Figure 1.1.** Out of sight scenario of a disaster site.

1.1.2 Computational Auditory Scene Analysis (CASA)

We receive a variety of sounds from all directions. The task of the auditory system is to parse complex audio mixtures to determine possible sound sources from incoming signals [18]. Bregman named this process auditory scene analysis (ASA) [19]. Although incoming sounds do not fully express their sources, our auditory perception is rarely confusing or misleading. However, little is known about the neural mechanisms by which the human brain accomplishes this feat. Over the past 40 years, several theories have been hypothesized to explain the processing of complex auditory scenes and perceptual phenomena. Most of these theories have been implemented in the form of computational models such as computational auditory scene analysis (CASA) [20]. Bregman divided the ASA into two steps. In the first step, input sounds are grouped in parallel by various algorithms that implement the Gestalt perception principle [21]. In the second step, these groups compete with each other and the winner is recognized. The result is a coherent sequence of sounds called an auditory stream or a perceptual object [19]. Finally, the perceptual objects can generate predictions for later sounds generated from the same source [22]. Although Bregman presented a theoretical framework for ASA through experiments, some of the ASA processes are not detailed. Therefore, based on the ASA theory, various CASA models have been proposed and they can be classified into three categories: Bayesian models, neural models, and temporal coherence models.

The Bayesian models use a predictive mechanism based on Bayesian inference, and the acoustic environment is described as state vectors estimated from the input. Competitions are coordinated through the adjustment of priors. That is, the current decomposition can affect the a priori probabilities of other objects. The model of Barniv and Nelken [23] assigns each input sound event to a class, which is determined by a Bayesian algorithm. In this model, the number of classes is not fixed, and a new class is defined when the class probability of an incoming event is low. The model implements competition between alternatives by dynamically adjusting the probability of occurrence of each class. The output is a discrete time series describing whether the model assigns all sounds to a single class or sorts them into two classes. In Nix and Hohmann's model [24], the acoustic environment is represented by a state vector, and features are extracted from the state vector. The goal of the model is to determine the posterior probability of the state vector. It requires prior knowledge of sound source and determines the most probable state vector distribution for a new input. The model defines *particles* consisting of a state vector and weights, and each particle predicts the next state vector. The weight of each particle is updated based on the prediction value, and competes with each other for prediction success.

The neural models are inspired by the neural networks of the brain. However, none of the models claim that their networks correspond to real networks located in the brain. Wang and Chang [25] faithfully reproduced the three areas (integration, separation, ambiguity) discovered by van Noorden [26] in the audio streaming paradigm. Each model object is represented by a synchronized oscillator, which is arranged in a two-dimensional network in terms of frequency and time. Output states are unified when all active oscillators are synchronized, and separated when unsynchronized. The separate output states compete with each other. The model of Pichevar and Rouat [27] is implemented in the form of a two-layer neural network, and separates voice and environmental sounds. The first layer of the model consists of oscillators arranged in two dimensions. The first dimension is frequency, and the second dimension is the cochlear channel. The activity of the two-dimensional oscillator is regulated by local excitatory and global inhibitory dynamics. In the second layer, the binding between oscillators is determined in such a way that neurons belonging to the same object are synchronized and neurons belonging to different objects are desynchronized. The model output is a binary mask for each object. The number of objects (the number of masks) is predefined. The model of Mill et al. [28] simulates the perceptual dynamics of auditory stream separation by focusing on multiple stable perception. The first step in the model is to continuously search for repeating patterns contained in the sequence and form a representation according to the predictions. Pattern links are formed probabilistically based on event similarity, and patterns made up of similar sounds are more likely to be discovered. If a repeating pattern is found, it is considered representative of a candidate object. In the second stage of competition, activation of the target expression is determined. The resulting activation of the model is divided into high-state and low-state objects, with high-state objects becoming the target of perception. The model of Rankin et al. [29] simulates perceptual multi-stability in the auditory streaming paradigm, and is implemented based on a tonotopic space with three neural units receiving input from the primary auditory cortex (A1). Competition is implemented as an inhibitory interaction between the three units, and model parameters are calibrated to match behavioral data. The output is either integrated or separated, and the perception is considered integrated if the integrated activity is greater than the average of the separated activities.

The temporal coherence models are based on the concept that sound emitted from the same source repeats its features, so if grouped by temporal coherence, they can be grouped together as an actual auditory object representation. Therefore, there is no need to implement competition between objects. The feature extraction is followed by grouping and clustering, and the outcome appears in perception with possible modulation by selective attention. The goal of this model is to simulate complex sound separation using computations of hypothesized cortical mechanisms. Inputs to the model of Krishnan et al. [30] are two-dimensional pitch representations (time and pitch) and four-dimensional feature representations (time, frequency, scale, and rate). These two input streams are first analyzed in parallel and then merged later. To group temporally coherent features, a sliding time window is used to compute a coincidence matrix. The coincidence matrix is used to determine which channels are temporally coherent, i.e. strongly correlated. The computed coincidence matrices are linked together by an extended correlation matrix at each time step. The columns of the extended correlation matrix are provided as input to a non-linear principal component analysis (nPCA), which is responsible for feature grouping. Two masks are output and used to reconstruct the separated sound sources. The model of Ma [31] is also based on the temporal coherence principle. The input representation includes frequency, scale, pitch, and location features. Grouping after feature extraction is performed similarly to the model of Krishnan et al. The output is a set of masks corresponding to the detected objects, the number of which is pre-defined. Target and non-target masks are classified through supervised prior training on mask classification. For this, a support vector machine (SVM) is implemented that classifies the data into two clusters. Training is performed on mixed utterances and used to set up a segregating hyperplane. Once calibrated, the SVM returns the mask's label (target or non-target) and the distance to the hyperplane. The model of Elhilali and Shamma [32] is similar to that of Krishnan et al., but clustering is based on the prediction of the next input. The prediction is based on an auto-regressive moving average (ARMA) model, which is a stochastic process with additive noise. The input is a four-dimensional feature representation (time, frequency, pitch, and scale). Since the model assumes two sound sources, the task of the grouping step is to assign the input to one of the two clusters at each time step. Predictions are used to determine classification and are implemented through stochastic latent variable. Using the object's previous state and recent input and output decisions, the current state of the latent variable is recursively estimated and the next input is predicted. At each step the input is assigned to the object that gives the closest prediction.

As reviewed above, conventional CASA models mainly implement algorithms that address specific aspects of ASA. Some models are inspired by general oscillatory mechanisms thought to operate within the brain, while others are supported by neural evidence obtained by large-scale brain imaging methods, while others are primarily based on behavioral evidence [18]. This is due to the lack of solid evidence for neural mechanisms involved in sound separation or auditory perception. This opens the possibility of integrating the fragmentary CASA models into a more comprehensive ASA theory. However, it is still too early to apply the existing CASA models for the sound event detection and localization in humanoid robots.

1.1.3 Sound Event Localization and Detection (SELD)

Sound event localization and detection (SELD) involves detecting the temporal activities of surrounding sound events and localizing their spatial locations when they are active [33,34]. The SELD task can be used to help robots navigate their surroundings and interact naturally with their environment [35–37]. It has played an essential role in many applications, such as audio surveillance [38,39], bio-diversity monitoring [40], and context-aware devices [41]. Since it was first included in the detection and classification of acoustic scenes and events (DCASE) challenge in 2019, significant progress has been made in the field of SELD research [42]. The SELD task can be divided into two sub-tasks, sound event detection (SED) and sound source localization (SSL).

The SED task is to detect the onset and offset of each sound event and attach a label to each detected sound event. It has often been approached using a variety of supervised classification methods that predict the frame-by-frame activity of each class of sound events. The classifiers mainly used in the methods are Gaussian mixture model (GMM) - hidden Markov model (HMM) [43], deep neural network (DNN) [44], recurrent neural network (RNN) [45–47], and convolutional neural network (CNN) [48,49]. Recently, a classifier that successively stacks CNN, RNN, and DNN has become mainstream, and it is called convolutional recurrent neural network (CRNN) [50,51]. Real-world sound events often overlap with other sound events, and the task of detecting all overlapping sound events is called polyphonic SED. Recently, several SED methods based on multi-channel audio input have been proposed to improve detection performance for overlapping sound events [50–54]. Comparing the performance of various microphone types for overlapping sound events, it was found that SED performance improved as spatial sampling increased, and first-order ambisonics (FOA) microphones achieved the best performance [50].

The SSL is the task of estimating the relative position of each sound event with respect to the microphone used for recording. In this study, SSL is performed by estimating the direction-of-arrival (DOA) of each sound event and is represented in the 2D spherical coordinate space of azimuth and elevation angles. Methods for the DOA estimation can be broadly classified into parametric-based approaches and DNN-based approaches. The spatiotemporal information of a sound field can be concisely described by several parameters, such as the locations of sound sources, the intensities of sound sources, and the shape of wavefronts [55]. The parametric-based approach searches for these essential parameters from the input acoustic signals. The well-known parametric methods include time-difference-of-arrival (TDOA) [56], steered-response-power (SRP) [57], multiple signal classification (MUSIC) [58], and estimation of signal parameters via rotational invariance technique (ESPRIT) [59]. Subspace methods such as MUSIC can be applied to a variety of array types and can estimate the DOA at high resolution for multiple sound sources. However, the methods require an estimate of the number of sound sources and are sensitive to reverberation and low signal-to-noise (SNR) conditions [59]. Recently, DNN-based methods are used to robustly estimate the DOA of sound under conditions unfavorable to parametric methods such as reverberation and low SNR. Methods [60–62] detected the DOA of overlapping sound events by estimating the number of sound sources from input data. Whereas most previous methods estimated the DOA within a fixed set of angles based on a classification approach, the methods [63,64] used a regression approach to generate continuous DOA estimates. Regarding the input format, methods [61,63,65] used circular and distributed microphone arrays to estimate the full azimuth, while methods [60,64,66] used linear arrays to estimate only the 180° range of azimuth. While most previous methods estimated only the azimuth angle [35–37,60,61,65,66], some estimated both the azimuth and elevation angles together [62,67]. Notably, the method [62] jointly estimated the azimuth and elevation angles using FOA microphones.

The SELD task is a relatively new research topic that unifies the tasks of SED and SSL (or DOA estimation). In 2019, Adavanne et al. [33] proposed a CRNN-based SELDnet, which made a new breakthrough in this field. In the SELDnet, the class and direction of each sound event are directly estimated using a series of convolutional layers, bidirectional gated recurrent unit (GRU) layers, and fully connected layers. The SELDnet became the baseline model for Task 3 of the DCASE Challenge in 2019. Therefore, it has inspired many other SELD models, and many DCASE Challenge candidate models have been built based on it with various modifications and improvements. Lu [68] integrated additional convolutional layers, and replaced bidirectional GRU (BiGRU) layers with bidirectional long-short term memory (LSTM) layers. Maruri et al. [69] used generalized cross-correlation with phase transform (GCC-PHAT) as input feature of the SELDnet. Cao et al. [70] connected log-mel spectrogram and GCC-PHAT and used them as input features of two separate CRNNs for SED and DOA estimation. Cao et al. [71] used log-mel spectrogram for SED, and used intensity vector and GCC features for DOA estimation. Ronchini et al. [72] investigated adding 1D convolutional filters to exploit information along the feature axis. Sampathkumar and Kowerko [73] augmented the baseline SELD model by providing more input features, such as log-mel spectrograms, GCC-PHAT, and intensity vector. Li et al. [74] used non-square convolutional filters and a unidirectional LSTM layer in their CRNN model. Bohlender et al. [75] proposed an extension of the model of Chakrabarty and Habets [76], in which LSTMs and temporal convolutional networks (TCNs) replaced the last fully connected layer of the former model. Nguyen et al. [77] introduced a sequence matching network using BiGRU layers to match SED and DOA output sequences. Cao et al. [78] proposed a two-branch network that used soft parameter sharing between the SED and DOA branches, and used multi-head self-attention to output predictions. Shimada et al. [79] proposed a unified output representation called activity-coupled cartesian direction of arrival (ACCDOA) to combine the loss of SED and the loss of DOA into a single optimization objective. In addition, Shimada et al. [80] extended ACCDOA to multi-ACCDOA and proposed auxiliary duplicating permutation invariant training to detect and localize sound events of the same class. Since 2021, the ACCDOA representation has been adopted by many researchers, allowing joint SED and SSL processes up to the last model layer [81]. Notably, Nguyen et al. [82] proposed the input feature called spatial cue-augmented log-spectrogram (SALSA) with exact time-frequency mapping between signal power and source directional cue, supporting both FOA and generic microphone array (MIC) formats. However, SALSA is computationally expensive because it requires computation of principal eigenvectors. Therefore, a computationally cheaper version of SALSA, SALSA-Lite, was proposed for MIC format [83].

As outlined above, due to the need for source localization, SELD task typically requires multi-channel audio inputs from a microphone array, such as FOA and MIC formats. Conventionally, SELD tasks use four audio input channels according to the input format of DCASE challenge. While a four-channel microphone array can offer advantages in capturing and processing audio signals, it also has some disadvantages. Implementing a four-channel microphone array can be more complex and expensive than a simpler, two-channel setup. It requires additional microphones, signal processing hardware, and calibration to ensure accurate spatial information. This complexity can increase the overall cost of the system. Four-channel audio generates a larger amount of data compared to two-channel systems. Storing, transmitting, or processing this increased data volume may require more storage capacity and higher bandwidth, which can be impractical in certain situations. Maintaining the accuracy and performance of a four-channel microphone array can be more demanding. Regular calibration and maintenance are essential to ensure that all channels are functioning correctly and that the spatial information is accurate. The physical size and form factor of a four-channel microphone array can be larger than simpler microphone setups. This can be a concern in applications where compactness is important.

### 1.1.4 Binaural Sound Event Localization and Detection

Recently, binaural input formats are frequently employed in deep learning-based SED, SSL, and SELD tasks due to their ability to replicate the natural auditory cues perceived by humans. By using two microphones positioned similarly to human ears, binaural recordings capture interaural time difference (ITD), interaural phase difference (IPD), interaural level difference (ILD), and spectral cues, providing essential information for humanoid robots to determine the direction of sound events in three-dimensional space. In scenarios where humanoid robots need to navigate, communicate, or collaborate with humans, binaural SED, SSL, and SELD become instrumental in enabling them to identify and react to specific auditory cues, contributing to their overall effectiveness and adaptability in diverse real-world applications. A summary of the most recent deep learning based binaural SED, SSL, and SELD methods is presented in Table 1.1.

For binaural SED, there is a recent trend to use multi-channel inputs, such as binaural inputs, instead of mono-channel input to improve detection performance for overlapping sound events. Based on CRNN architecture, Adavanne and Virtanen [84] performed SED on six sound event classes using binaural mel-spectrograms and binaural magnitude + phase spectrograms, respectively. Furthermore, Krause and Mesaros [85] tested several combinations of features such as mel-spectrogram, ILD, phase spectrogram, IPD, sine and cosine of IPD, and GCC-PHAT to investigate the impact of spatial information on the performance of SED and acoustic scene classification tasks, and found that using binaural features resulted in better model performance.

Regarding binaural SSL, many studies have used binaural features such as ITD, ILD, and IPD and monaural features such as magnitude and phase spectrograms as input to DNN, CNN, or CRNN. Youssef et al. [86] used ITD and ILD which were fed into input branches of an artificial neural network for the azimuth estimation of sound source. Roden et al. [87] compared the performance of a DNN with two hidden layers and different input types, such as ITD, ILD, magnitude + phase spectrograms, and real + imaginary spectrograms. Zermini et al. [88] proposed a method based on DNN and time-frequency masking for binaural sound source separation. In this method, the DNN is used to predict the DOA of sound sources which is then used to generate soft time-frequency masks for the recovery and estimation of the individual sound sources. Ma et al. [89] presented a system that uses DNNs and head movements for binaural localization of multiple sources in reverberant environments. The DNNs are used to learn the relationship between the source azimuth and binaural cues, consisting of ILDs and cross-correlation function. Yiwere and Rhee [90] concatenated the ILD and the cross-correlation of two binaural channels and fed it to a DNN to predict the azimuth and distance of sound sources. To estimate both the azimuth and elevation, Nguyen et al. [91] used a relatively small CNN in regression mode, with binaural input features such as ILD and IPD. Pang et al. [92] also used a CNN to estimate the azimuth and elevation of sound sources by processing ILD and IPD features in the time-frequency domain. In addition, Y. Yang et al. [93] exploited log-magnitudes and IPD as input features of CNN to predict the azimuth and elevation of speech sources. For CRNN-based binaural SSL, B. Yang et al. [94] proposed a network for robust binaural sound source localization. The network consists of a branched CNN module to separately extract the inter-channel magnitude and phase patterns, and a CRNN module for joint feature learning. Recently, Dwivedi et al. [95] developed a CNN to map linear prediction residual coefficients along with spectral notches to the corresponding elevation angles to build a framework for binaural localization in the median plane. In addition, García-Barrios et al. [96] studied CRNN-based binaural sound source localization, under the influence of head rotation in reverberant conditions. They presented various CRNN models to improve the performance of binaural SSL under head rotation, and the best performance was obtained when using quaternions as inputs for a CNN branch whose outputs were concatenated with spectral features. Notably, van der Heijden and Mehrkanoon [97] proposed neurobiological-inspired CNN models of human sound localization. The CNN models spatialized with human binaural hearing characteristics accurately predicted sound location in the horizontal plane.

The only related publication covering deep learning based binaural SELD is Wilkins et al. [98]. They presented a comparative analysis of the DCASE 2022 SELD baseline model across FOA, binaural, and stereo audio input formats. To fairly compare the three formats, they looked at the problem of sound event localization on the horizontal plane by removing the elevation component. As a result of comparison, it was shown that while detection and localization performance decreases given less informative audio inputs, binaural and stereo-based SELD models are still able to localize lateral sound events relatively well.

**Table 1.1.** Summary of deep learning based binaural SED, SSL, and SELD methods in the literature.

| | Author | Architecture | Input feature | Detection (classes) | Azimuth (range) | Elevation (range) |
|---|---|---|---|---|---|---|
| S E D | Adavanne and Virtanen [84] | CRNN | Mel-spectrogram, Magnitude + Phase | 6 | × | × |
| | Krause and Mesaros [85] | CRNN | Mel-spectrogram, ILD, Phase, IPD, sin(IPD) + cos(IPD), GCC-PHAT | 62 | × | × |
| S S L | Youssef et al. [86] | ANN | ITD, ILD | × | −45° ~ +45° | × |
| | Roden et al. [87] | DNN | ITD, ILD, Magnitude + Phase, Real + Imaginary | × | −30° ~ +30° | −10° ~ +50° |
| | Zermini et al. [88] | DNN | Mixing vector + ILD + IPD | × | −90° ~ +90° | × |
| | Ma et al. [89] | DNN | ILD, Cross-correlation | × | 0° ~ 360° | × |
| | Yiwere and Rhee [90] | DNN | ILD, Cross-correlation | × | 0° ~ 60° | × |
| | Nguyen et al. [91] | CNN | ILD + IPD | × | −45° ~ +45° | −30° ~ +30° |
| | Pang et al. [92] | CNN | ILD + IPD | × | −80° ~ +80° | −45° ~ +230.625° |
| | Yang et al. [93] | CNN | IPD, Log-magnitude | × | −80° ~ +80° | −45° ~ +230.625° |
| | Yang et al. [94] | CRNN | Log-magnitude, Phase | × | −80° ~ +80° | × |
| | Dwivedi et al. [95] | CNN | Linear prediction residual coefficients | × | × | −30° ~ +210° |
| | García-Barrios et al. [96] | CRNN | Mean magnitude + sin(IPD) + cos(IPD) + ILD, Quaternions | × | −180° ~ +180° | −35° ~ +35° |
| | van der Heijden and Mehrkanoon [97] | CNN | Bilateral auditory nerve representation | × | −180° ~ +180° | × |
| S E L D | Wilkins et al. (state of the art) [98] | CRNN | Mel-spectrogram + GCC | 13 | −180° ~ +180° | × |
| | Lee (proposed) | CRNN (depthwise separable) | Mel-spectrogram + V-map + ITD-map + ILD-map + SC-map | 12 | −180° ~ +180° | −30° ~ +60° |

Two-channel microphone array has advantages such as cost-effectiveness, reduced data volume, ease of maintenance, and small form factor. As an application example, hearing aids equipped with a two-channel SELD system can provide wearers with valuable spatial information about their acoustic environment. This awareness allows them to better locate and engage with sound sources, which is crucial for safety and overall situational awareness. Human brains are adept at processing binaural audio cues to determine the location of sound sources. A two-channel SELD system can replicate this natural ability, allowing hearing aid users to accurately locate sounds, such as approaching vehicles or voices in conversations. Moreover, when applying SELD system to a humanoid robot or telepresence technology, it is reasonable to implement it with a two-channel input that is close to human modality. The FOA format, the conventional input format of SELD system, not only requires a larger number of microphones and processing hardware compared to two-channel input, but also requires an additional microphone array structure. Due to its simplicity and compact form factor, the two-channel input format is suitable for the SELD system of humanoid robots. In telepresence applications, binaural audio input closely mimics the way humans perceive surrounding sounds in the real world as shown in Fig. 1.2. The information about the timbre and location of surrounding sounds is transmitted from the teleoperator to the operator. In the operator site, the incoming sound information is recombined, using dynamic binaural synthesis of spatial sound, to create an immersive sound experience through headphones [99]. The operator perceives a three-dimensional sound sensation with the exact direction of sound sources at the teleoperator site. This makes us feel like we are actually in the remote teleoperator site. Taking all of these into consideration, it is advantageous to implement a SELD system based on human-like two-channel input.

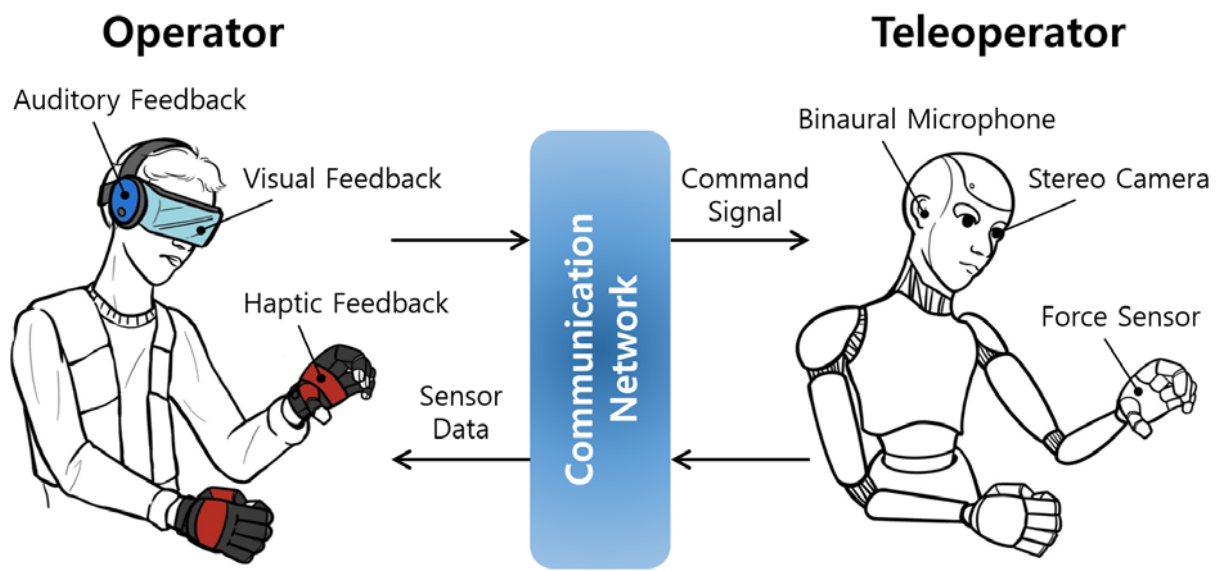

**Figure 1.2.** Multi-modal telepresence system using vision, audition, and haptics.

## 1.2 Research Objectives

The main objective of this research is to develop a two-channel SELD model for applications in the human-machine interaction field, especially humanoid robots. However, with a two-channel audio input arranged horizontally, it is difficult to estimate the elevation of sound source, and front-back confusion arises when estimating the azimuth. I was inspired by the fact that humans rely on head-related transfer function (HRTF) cues to solve these problems. Assuming an android robot, which is a humanoid robot with exact human form and modality, the binaural signals from the robot ear contain information about HRTF. Thus, the robot has to extract the HRTF information inherent in the binaural input signals to estimate DOA of each sound source. An ideal android robot will learn its HRTF information like humans, and simultaneously detect and localize surrounding sound events. This is the basic idea of my two-channel SELD approach for humanoid robots. As an incarnation of this idea, I propose a binaural sound event localization and detection (BiSELD) approach that simultaneously estimate the class and direction of each sound event from binaural inputs. This novel approach is based on deep learning of unique acoustic patterns and HRTF localization cues of each sound event, represented by a binaural time-frequency feature (BTFF) extracted from binaural audio inputs. The abstract concept of BiSELD is illustrated in Fig. 1.3.

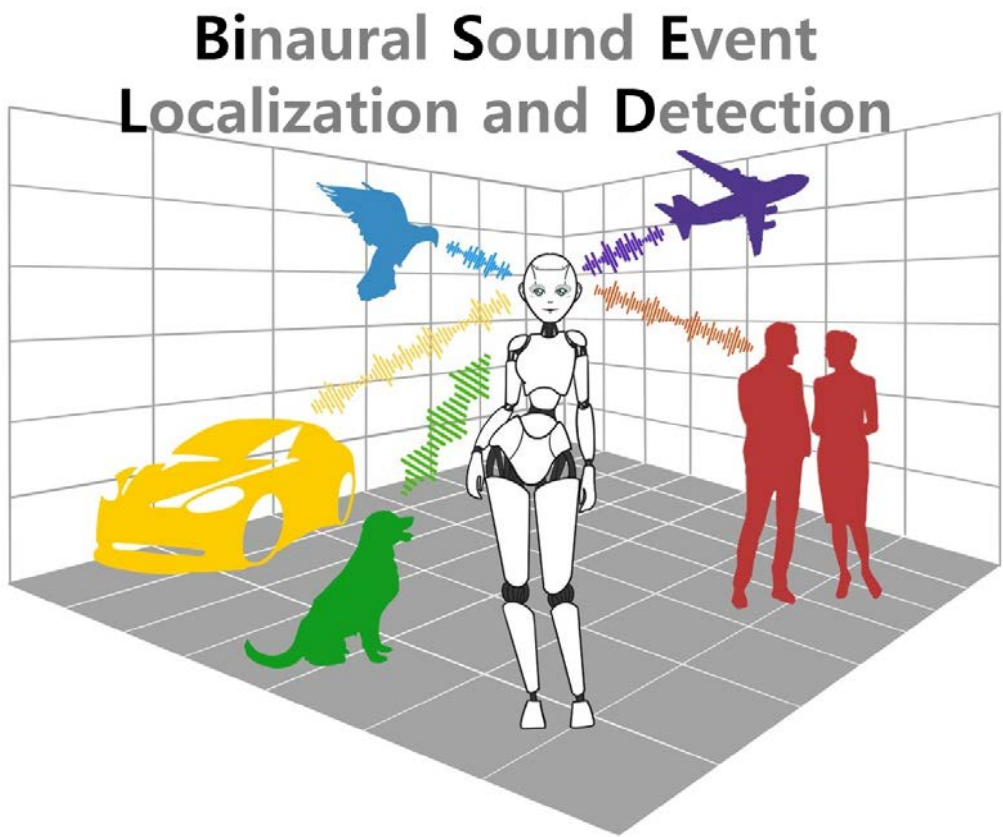

**Figure 1.3.** Abstract concept of binaural sound event localization and detection (BiSELD).

The BiSELD research includes definition of task requirements; understanding the human auditory process; HRTF measurement of an artificial head; analysis of binaural localization cues from the measured HRTFs; derivation of binaural input feature based on the HRTF analysis; construction of binaural sound event datasets using the HRTF database; implementation of the BiSELD task as artificial neural networks; evaluation of the proposed binaural input feature and BiSELD network; and state-of-the-art (SOTA) comparison in the background noise condition. The main requirements of the BiSELD task are defined as follows.

1. The task should be able to detect a selected subset of sound event classes among myriad potential sound classes.

2. Using only two-channel audio input, the task should be able to estimate the direction of various sound events in 3D spherical space when each sound event is activated.

3. The task should maintain its performance under realistic background noise conditions with a range of SNR from 30 dB to 0 dB.

## 1.3 Dissertation Overview

The remainder of this dissertation is organized as follows. Chapter 2 briefly reviews the structure and function of the human ear and the auditory pathway to the brain to understand how humans perceive sounds in three-dimensional space through both ears. Chapter 3 measures the HRTFs of an artificial head for later localization cue analysis, binaural dataset construction, and performance evaluation. To this end, a speaker arc array is designed based on electro-acoustic modeling and a HRTF measurement system is established. The HRTFs are measured by providing clear standards for origin transfer function measurement and non-causality compensation. The HRTF database and related source codes are available at GitHub (https://github.com/han-saram/HRTF-HATS-KAIST). Chapter 4 derives patterns of ITD, ILD, and spectral cue (SC) from the measured HRTF database to analyze binaural sound localization cues of HRTF. In addition, horizontal plane directivity (HPD) of the left and right HRTF pairs are investigated in the context of azimuth estimation. Then, BTFF, an input feature specialized for BiSELD task, is proposed based on the HRTF localization cue analysis. The BTFF consists of respective sub-features for sound event detection, azimuth estimation, and elevation estimation. Lastly, spatial sound event dataset is generated by synthesizing the measured HRTFs with collected sound event databases for later evaluation. Chapter 5 first provides background knowledge on the unit neural networks of BiSELD network (BiSELDnet): DNN, RNN, and CNN. Then, various versions of BiSELDnet architecture based on CRNN, hierarchical CRNN, Xception module, and Trinity module are presented. Each BiSELDnet is presented in the process of finding an effective architecture that can accelerate learning and efficiently learn the proposed input feature, BTFF. In this process, depthwise separable convolution suitable for BTFF, which has low correlation between channels, is introduced, and ultimately, Trinity module that learns input features of various sizes with a small number of parameters is proposed. Lastly, the training procedure of BiSELDnet is presented. Chapter 6 evaluates the proposed BTFF and BiSELDnet architectures. First, the evaluation metrics to measure SELD performance are presented. Next, the effectiveness of each sub-feature of BTFF is verified in terms of sound event detection, horizontal localization, and vertical localization, respectively. Then, the SELD performance of each BiSELDnet is compared and the architecture with the best performance is selected as the final model. In addition, for the analysis of the final BiSELDnet, the output of its internal layer is visualized, and a new visualization method is proposed to check the BTFF contribution to the output DOA vector. Lastly, the evaluation and visualization results of the BiSELD model are discussed in terms of comparison with human auditory processes. In Chapter 7, the detection and localization performances of the BiSELD model, SOTA SELD model, and baseline SELD model are compared for sound events on the horizontal or median plane under realistic background noise conditions with various SNRs. Finally, its feasibility of real environment application is discussed. Chapter 8 presents the conclusions of this dissertation and further discusses the future work for BiSELD research.

# Chapter 2. Human Auditory System

## 2.1 Human Ear

Before conducting full-scale BiSELD research, it is necessary to briefly review how humans can recognize the type and direction of sound sources located in three-dimensional space. The human ear is divided into the outer, middle, and inner ear as shown in Fig. 2.1.

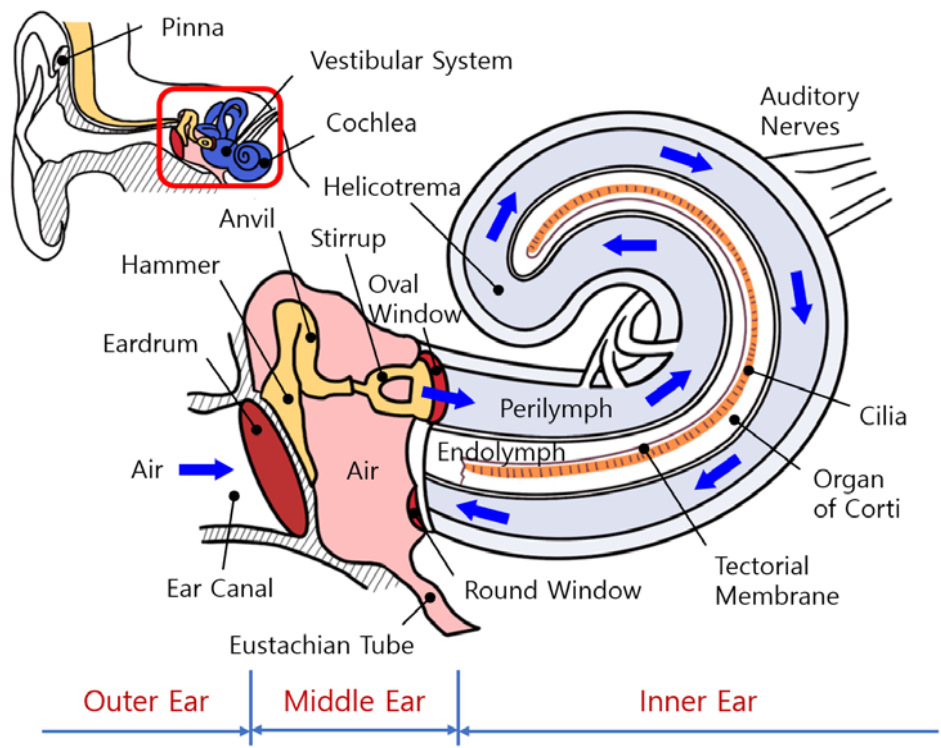

**Figure 2.1.** Anatomy of the human ear.

The outer ear is at the outermost level of the auditory system visible to our eyes and includes the pinna, the ear canal, and the eardrum. The pinna collects sound and guides it to the ear canal, and its shape helps us estimate the direction of sound source. The ear canal is an open-closed duct about 2.5 cm long, and it resonates at the frequency whose quarter wavelength is equal to the length of the ear canal. Hence, the frequency range of the maximum sensitivity of human hearing is around 3,400 Hz. Sound passing through the ear canal is converted from acoustic energy into vibrational energy through the eardrum and transmitted to the middle ear [100].

The middle ear is between the eardrum and the cochlea, and consists of the auditory ossicles: the hammer, the anvil, and the stirrup. In addition, it communicates with the outside air because it is connected to the pharynx through the Eustachian tube. The main function of the ossicles is to match the impedance between the eardrum of the outer ear and the oval window of the inner ear. Since the medium outside the eardrum is gas (air) while the medium inside the oval window is liquid (lymph), the density difference between the two media is large, resulting in a large impedance difference. The force transmitted from the eardrum to the oval window is increased by about 1.3 times due to the lever action of the ossicles [101]. On the other hand, the area of the oval window (3.2 mm$^2$) is about 17 times smaller than that of the eardrum (55 mm$^2$). Hence, the pressure transferred from the eardrum to the oval window increases by about 22 times. Consequently, the impedance matching is achieved, and energy transfer is maximized in the voice range of 300 to 3,000 Hz [102].

The inner ear is the most complex organ in the auditory system, consisting of the cochlea as the auditory organ, and the vestibular system as the equilibrium sensory organ. The anatomical structure of the cochlea is shown in Fig. 2.2.

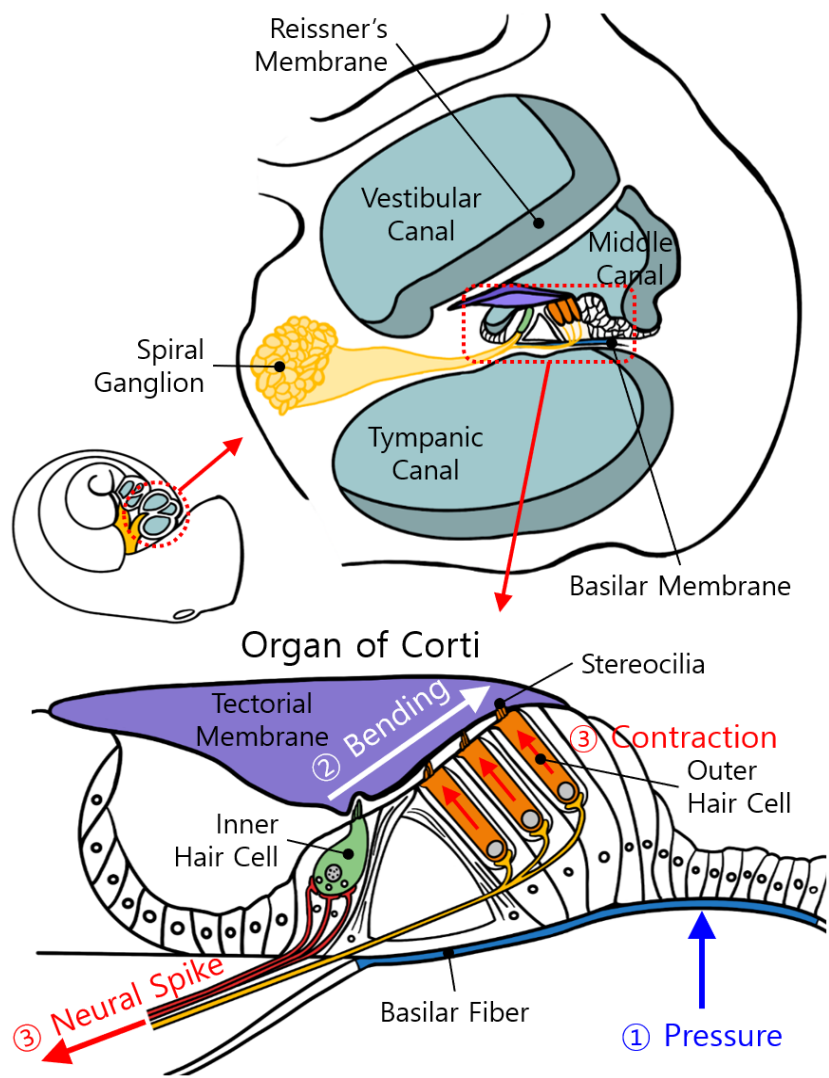

**Figure 2.2.** Cross section of the human cochlea.

The cochlea is twisted 2.75 turns in a spiral, and its periphery is surrounded by bone. The cochlea consists of three tubes: the vestibular canal (or scala vestibuli, SV), the middle canal (or scala media, SM), and the tympanic canal (or scala tympani, ST). The SM is filled with endolymph. The SV and ST are connected by the helicotrema at the top of the cochlea and are filled with the perilymph inside. The SV and SM are divided by the Reissner's membrane, and the SM and ST are separated by the basilar membrane. The basilar membrane consists of approximately 20,000 to 30,000 basilar fibers, which increase in length and decrease in thickness toward the top of the cochlea. To be specific, the width of the basilar membrane is the narrowest near the oval window (0.08–0.16 mm) and the widest at the top of the cochlea (0.42–0.65 mm), and the stiffness decreases by more than 100 times as the basilar fiber progressively becomes thinner. Thus, the basilar membrane resonates at high frequencies near the oval window and resonates at low frequencies near the top of the cochlea. The organ of Corti is located in the SM. It consists of the outer hair cell (OHC) and the inner hair cell (IHC). The OHC actively amplifies the vibration of the basilar membrane and the IHC generates an auditory signal in response to the amplified basilar membrane vibration. The IHCs are arranged in a row along the length of the cochlea, and the number is about 3,500. The OHCs are arranged in three to four rows and the number is about 12,000. Each of the IHC and OHC has stereocilia at the top and nerve fibers at the bottom, and the stereocilia are covered by the tectorial membrane [102].

The process in which the pressure transmitted to the oval window is converted into an auditory signal, and then transmitted to the auditory nerve is as follows. When the oval window's vibration forms a pressure gradient, the pressure wave propagates through the SV in the form of a traveling wave into the ST. Consequently, when the basilar membrane is lifted up by pressure as shown in Fig. 2.2, the stereocilia are bent by the tectorial membrane and actuate the OHCs and IHCs. The OHC contracts by the electro-mobility transduction (EMT) process, further lifts the basilar membrane, and therefore amplifies its amplitude. The IHC converts the amplified mechanical amplitude into an electrical signal and then fires it into the auditory nerve. Conversely, when the basilar membrane descends, the stereocilia do not bend, and nerve firing does not occur. As a result, each auditory signal is generated by the corresponding OHC and IHC rows on the vibrating basilar membrane in the form of a sequence of half-wave rectified neural spikes. Each auditory signal has a different frequency band according to the resonant frequency of the corresponding basilar membrane location. Therefore, humans can perceive sound for each frequency band. This is why two sounds of different frequencies are heard as two sounds, whereas two lights of different wavelengths appear as one combined color. The residual pressure wave after passing through the ST excites the round window like a passive radiator, discharging its energy to the middle ear [102].

## 2.2 Auditory Pathway to the Brain

Each of the left and right auditory signals arrives at the cochlear nucleus (CN) of the brainstem through the auditory nerve. The brainstem consists of the medulla, pons, and midbrain, with the superior olivary complex (SOC) between the medulla and pons. The SOC transmits auditory information to the inferior colliculus of the midbrain after sharing the signals of the left and right CNs. Finally, left and right auditory information from the inferior colliculus is transmitted through the medial geniculate nucleus of the thalamus to the auditory cortex of the cerebrum as shown in Fig. 2.3.

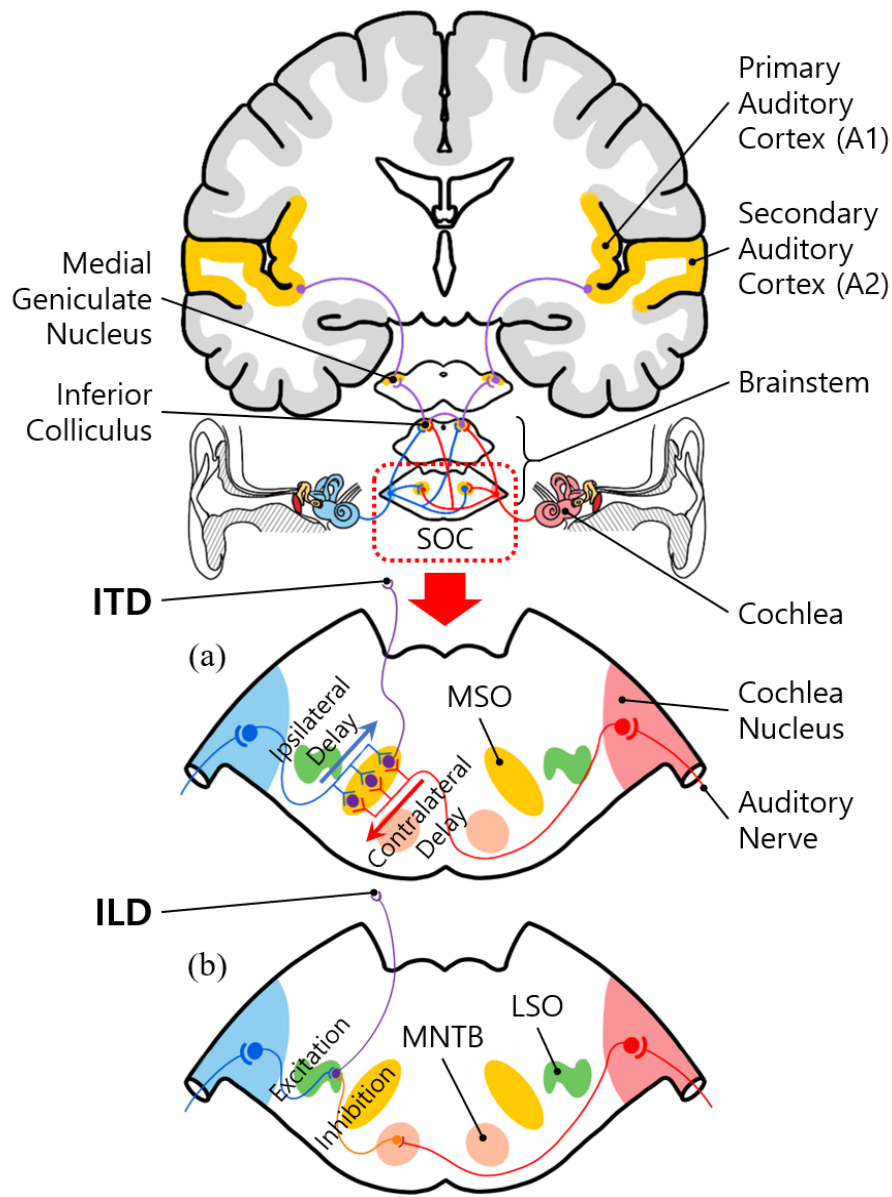

**Figure 2.3.** Auditory pathways to the brain and the superior olivary complex (SOC) in the brainstem: (a) interaural time difference (ITD) sensitivity of the medial superior olive (MSO) neurons and (b) interaural level difference (ILD) sensitivity of the lateral superior olive (LSO) neurons.

The SOC is a collection of the brainstem nuclei, and includes the medial superior olive (MSO) and lateral superior olive (LSO). The MSO and LSO are known to be responsible for the initial encoding of ITDs and ILDs, respectively [103,104]. The auditory system evaluates ITD as a low-frequency phase shift for wavelengths exceeding the head diameter and a high-frequency envelope shift for wavelengths shorter than the head diameter [105]. On the other hand, ILD is perceived as a localization cue for wavelengths shorter than the head diameter. The accuracy of human azimuthal sound localization by ITD and ILD according to frequency is summarized in Table 2.1.

As auditory signals travel from the cochlea towards the auditory cortex, additional features are extracted and represented in overlapping maps. These features include onset, offset, periodicity, amplitude modulation (AM), frequency modulations (FM), ITD, and ILD [18]. Together these features form the basis for human perception of the surrounding acoustic situation. Finally, the perceived pattern is interpreted to judge and evaluate the corresponding sound. In the final stage, a sound image is formed by learning, knowledge, and vision that have been accumulated so far [106]. Notably, the perception cues of sound direction appear in the spectral pattern of input sound. The torso, head, and pinna act as filters, interfering with incident sound waves through reflection, diffraction, and absorption. These interferences modify the sound spectrum by strengthening (spectral peaks) or weakening (spectral notches) of specific frequency bands, providing localization cues of sound source. These spectral cues generated by the filtering effects are HRTFs, a set of spectral deviations exerted on the sound transmitted to the eardrum. Since the head shape of each person is different, HRTFs are unique characteristics of each individual, like fingerprints. They are memorized by the brain throughout life in a learning process of different sound source directions [105]. HRTF contains various information about the direction of sound, including ITD and ILD. In addition, it contains information about the pinna, which has a decisive influence on vertical sound source localization. However, the pinna is difficult to be modelled due to its complex shape, so HRTFs are usually obtained by measurement.

**Table 2.1.** Accuracy of human azimuthal sound localization by interaural time difference (ITD) and interaural level difference (ILD) according to frequency [105].

| Binaural localization cue | Localization accuracy | | |
|---|---|---|---|
| | < 1000 Hz | 1000–3000 Hz | > 3000 Hz |
| ITD | Good | Mediocre | Impossible |
| ILD | Impossible | Mediocre | Good |

# Chapter 3. Head-Related Transfer Function (HRTF)

## 3.1 Related Works

Humans analyze an auditory scene by localizing and interpreting surrounding sounds. The human brain can localize sound sources by taking advantage of the way sound is modified on its way to the ears. When a sound wave in a certain direction reaches both ears, it interacts with the torso, head, and pinna, causing temporal and spectral transformations. The resulting effects provide meaningful clues about the location of the sound source. In duplex theory [107], Lord Rayleigh suggested that the human ability to localize sounds depended on the ITD and ILD, which actually influence azimuth localization. In addition, the pinna generates direction-dependent SCs that can be used by the brain to estimate source elevation. Although there is no simple relationship between direction and sound localization cues, the human brain can use these cues to accurately estimate the location of a sound source in space. Therefore, to simulate an acoustic scene with sound sources in different directions, the sound sources must be modified according to their directions. In binaural audio, such simulations are implemented using direction-dependent acoustic filters, referred to as head-related impulse responses (HRIRs) in the time domain, or as HRTFs in the frequency domain. An HRTF is a frequency response describing a sound transmission from a source position to the ear canal [108–112]. HRTFs can be measured in the form of linear time-invariant filters and synthesized by various models for real-time applications [113].

Since HRTFs contain all binaural and spectral cues for sound source localization, they play an essential role in binaural rendering for virtual auditory display (VAD) and 3D audio reproduction over headphones or loudspeakers, especially in immersive listening experiences for virtual reality (VR) and augmented reality (AR) applications [114,115]. Recently, HRTFs have been used as development datasets for deep learning-based binaural sound source localization (BSSL), which aims to localize sound sources using two microphones by mimicking the principle of binaural hearing [86–97]. In contrast to VAD, which implements binaural rendering by convolving an input signal with left and right HRIRs of a certain direction, deep learning-based BSSL encodes the information related to sound localization cues of HRTF from binaural input signals to estimate the direction of sound source. Therefore, this HRTF-learning-based BSSL is a suitable sound source localization method for humanoid robots with two ears.

In recent decades, various laboratories have constructed HRTF databases. Some databases are publicly available for scientific or commercial purposes [116–123]. Although HRTFs have been measured for decades, there appears to be no commonly used standard measurement method, rather a variety of methods remain unclear or inaccurate. In particular, loudspeaker design, measurement of origin transfer function at the head center, selection of time window interval, and compensation for non-causality of ipsilateral HRTFs, all commonly encountered issues in the HRTF measurement process, are unclear or inaccurate.

Most of previous studies have measured HRTFs using commercial loudspeakers [116,118–120,124–132]. Since a commercial loudspeaker is designed for high sound quality, a vent or passive radiator is installed to expand the low-frequency band, or a tweeter is added to a mid-range speaker to enhance the high-frequency response. Therefore, commercial loudspeakers emit sound from multiple sources rather than a single source. Since accurate measurement is possible when there is a single sound source in each direction, commercial loudspeakers are not suitable for accurate HRTF measurement. Although other studies have made sealed speaker modules using a single speaker driver, those modules did not sufficiently reproduce the frequency band of interest [133–136]. This is because individual speaker modules were not designed in consideration of the speaker driver's electro-acoustic characteristics for a given volume. For a circular baffle with a speaker driver in the center [137], there will be many dips and peaks in the frequency response due to the interference caused by the baffle edge diffraction. Because circular baffles have equal source-to-edge distances, their on-axis responses exhibit worst-case edge diffraction [138]. In an unbaffled speaker driver [139], the sounds from the front and rear sides of the speaker diaphragm cause destructive interference, reducing the low to mid-range frequency level.

According to Blauert [140], a free-field HRTF can be calculated by dividing a binaural transfer function (BTF) by an origin transfer function (OTF) at the head center position in the absence of the head to cancel out the influence of the measurement system characteristics. In previous works [116,126–128,141,142], only BTFs were measured and incorrectly called HRTFs. BTFs are significantly different from actual HRTFs unless the frequency bandwidth of the measurement system, especially the speaker module, is sufficiently wide and its tonal balance is almost perfect. In previous works [116,130,133,143], OTF was measured with a microphone different from that used for BTF measurement. In those cases, even if BTF is normalized to OTF, the difference between the frequency responses of microphones affects the resulting HRTF. Also, since the microphone used for HRTF measurement have directivity, OTF must be measured by pointing the microphone toward the acoustic on-axis of each speaker module to obtain the correct frequency response of the total measurement system. Some previous studies related to OTF measurement have pointed the microphone toward the acoustic on-axis of a sound source [135,136,143,144], while others have tilted the microphone 90° off-axis [125,133,145]. When a microphone is tilted at right angles to the speaker on-axis, there is no directivity effect in the circumferential direction of the microphone diaphragm, but its high frequency response degrades significantly relative to the 0° on-axis measurement [146]. Therefore, normalizing BTFs with OTF of 90° off-axis measurement results in inaccurate HRTFs with overemphasized high frequencies and large errors.

For reference, the time domain response corresponding to BTF is binaural impulse response (BIR) and that to OTF is origin impulse response (OIR). In BIR and OIR, the essential information is contained in a time interval of just a few milliseconds. Therefore, a window function is needed to extract the essential time intervals from BIR and OIR respectively. In OIR, the maximum peak position is fixed because the distance from the speaker module to the microphone is constant, whereas in BIR, the maximum peak position changes because the distance varies according to the direction of the head. Several studies applied a window function considering propagation delay of sound sources [114,116,133], whereas others applied it to remove reflection [118,120,126,135,136,141,143,145,147]. However, those studies did not clearly provide the measurement-setting criteria for both the start and end points of the most essential time interval.

A pair of left and right HRTFs can be computed by complex division of the corresponding pair of BTFs by OTF. Here, it is important to note that ipsilateral HRTFs are non-causal from the mathematical definition of HRTF because the ipsilateral ear is closer to the sound source than the head center. On the other hand, contralateral HRTFs are causal because the contralateral ear is farther from the sound source than the head center. In general, derived HRTFs are converted into HRIRs in the time domain through inverse Fourier transform (IFT) when stored in a database, and they are used for data synthesis through convolution with sound samples. Since an ipsilateral HRIR is non-causal, its early maximum peak should precede 0 seconds. However, the maximum peak appears in the later part of the ipsilateral HRIR. This is because the Fourier transform treats a time series as a repeating periodic signal. Therefore, post-processing is needed to prevent the discontinuity of ipsilateral HRIRs. Møller [109] noted that some HRTFs are non-causal, but did not suggest a post-processing procedure to secure causality. Xie [111] suggested a method to ensure causality only when the denominator of HRTF is a non-minimum phase function. However, he did not clearly address how to compensate for the non-causality of ipsilateral HRTFs. Iida [112] systematically summarized a procedure for obtaining HRTFs, but he did not mention the non-causality issue of ipsilateral HRIRs or how to compensate for it. In other HRTF measurements [116–123,125–129,134,135,142,143,145], non-causality issues and methods to guarantee causality were also ignored.

In this chapter, accurate and practical HRTF measurement methods, and the resulting HRTF database are presented by tackling all above-mentioned issues of previous HRTF measurement procedures such as wideband speaker module design, OTF measurement, selection of time window interval, and compensation for non-causality of ipsilateral HRTFs. The remainder of this chapter is organized as follows. Section 3.2 defines HRTFs. Section 3.3 describes the configuration of the HRTF measurement system, and then presents the electro-acoustic based speaker module design. In Section 3.4, OTF measurement using a 0° on-axis microphone is described in detail, and a procedure for time window setting is presented. Section 3.5 deals with compensation for the non-causality of ipsilateral HRTFs. Then the characteristics of the derived HRTFs are presented and discussed. The BTF and OTF measurement results, source codes for building HRTF database, and data files about derived HRTFs are available on GitHub (https://github.com/han-saram/HRTF-HATS-KAIST).

## 3.2 Definition of HRTF

The spherical coordinate system and head transverse planes for specifying the location of a sound source are shown in Fig. 3.1. In Fig. 3.1(a), the origin of the coordinate system is the center of the head, between the entrances to the two ear canals. From the origin, the *x*, *y*, and *z*-axes point to the right ear, front, and top of the head, respectively. In Fig. 3.1(b), the horizontal, median, and lateral planes are defined by these three axes. The position of a sound source is defined in the spherical coordinate system as $(r, \theta, \phi)$. The *r* is the distance from the sound source to the origin. The azimuth $\theta$ is the angle between the *y*-axis and the horizontal projection of the position vector, defined as $-180° < \theta \leq +180°$, where $-90°$, $0°$, $+90°$, and $+180°$ indicate the left, front, right, and backward directions, respectively, on the horizontal plane. The elevation $\phi$ is the angle between the horizontal plane and the position vector of the sound source, defined as $-90° \leq \phi \leq +90°$, where $-90°$, $0°$, and $+90°$ represent the bottom, front, and top directions, respectively, in the median plane.

The sound emitted from a sound source is diffracted and reflected from the torso, head, and pinna, and then reaches both ears as shown in Fig. 3.2. HRTFs are acoustic transfer functions due to the sound transmission process that account for the overall acoustic filtering effect by human anatomy. A far-field HRTF of the left or right ear for a sound source of $P_S(r, \theta, \phi)$ is defined as follows:

$$H_{L,R}(\theta, \phi, f, s) = \frac{P_{L,R}(r, \theta, \phi, f, s)}{P_0(r, f)}, \tag{3.1}$$

where $P_{L,R}$ is a complex-valued sound pressure in the frequency domain at the entrance of the left or right ear canal of a subject; $P_0$ is a complex-valued sound pressure in the frequency domain at the center of the subject's head in the absence of the subject; subscripts *L* and *R* denote the left and right ears; *f* refers to frequency; and *s* refers to a set of parameters related to the dimensions of the subject's anatomical structures. Although $P_{L,R}$ and $P_0$ are functions of distance *r*, the effects of *r* on $P_{L,R}$ and $P_0$ can be regarded as identical under the far-field assumption, so that the effects of *r* can be canceled out in $H_{L,R}$. Even though Eq. (3.1) is expressed in terms of ideal sound pressures, when actually measuring HRTFs, the transfer function between the measured sound pressure and the input signal from the measurement system is used. Therefore, it is useful to express HRTFs based on the measured transfer functions. Regarding this point, Eq. (3.1) can be re-written as follows:

$$H_{L,R}(\theta,\phi,f,s) = \frac{\tilde{P}_{L,R}(r,\theta,\phi,f,s)/X(f)}{\tilde{P}_0(r,\phi,f)/X(f)} = \frac{P_{L,R}(r,\theta,\phi,f,s)\cdot H_s(\phi,f)/X(f)}{P_0(r,f)\cdot H_s(\phi,f)/X(f)}, \qquad (3.2)$$

where $\tilde{P}_{L,R}$ is a measured sound pressure at the entrance of the left or right ear, $\tilde{P}_0$ is a measured sound pressure at the head center, and $X$ is an input signal. The measured values $\tilde{P}_{L,R}$ and $\tilde{P}_0$ are composed of $H_s(\phi,f)$ as well as $P_{L,R}$ and $P_0$, respectively. The $H_s(\phi,f)$ is the transfer function of the measurement system consisting of digital-to-analog converter (DAC), speaker amplifier, speaker module at elevation $\phi$ in a vertical speaker arc array, microphone, microphone conditioner, and analog-to-digital converter (ADC). The numerator of Eq. (3.2) is the BTF, which denotes the transfer function between the measured sound pressure at the left or right ear and the input signal, and the denominator is the OTF, which denotes the transfer function between the measured sound pressure at the head center and the input signal. The BTF and OTF are respectively defined as follows:

$$G_{L,R}(r,\theta,\phi,f,s) = \frac{\tilde{P}_{L,R}(r,\theta,\phi,f,s)}{X(f)}, \qquad (3.3)$$

$$G_0(r,\phi,f) = \frac{\tilde{P}_0(r,\phi,f)}{X(f)}. \qquad (3.4)$$

In general, when measuring far-field HRTFs, both $s$ and $r$ are constant because the measurement subject is predetermined and the distance from a speaker module to the head center is also fixed for a specific measurement setup. Therefore, based on the BTF and OTF to be measured, the far-field HRTF is defined as follows:

$$H_{L,R}(\theta,\phi,f) = \frac{G_{L,R}(\theta,\phi,f)}{G_0(\phi,f)}. \qquad (3.5)$$

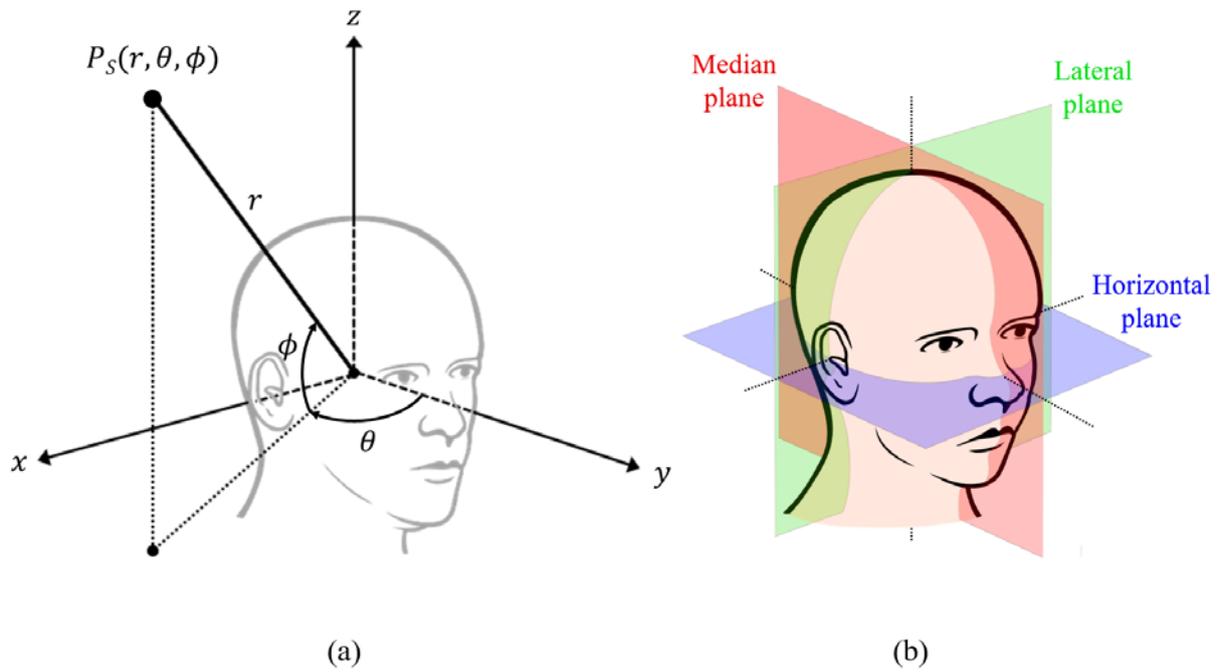

**Figure 3.1.** Illustrations of (a) spherical coordinate system and (b) head transverse planes.

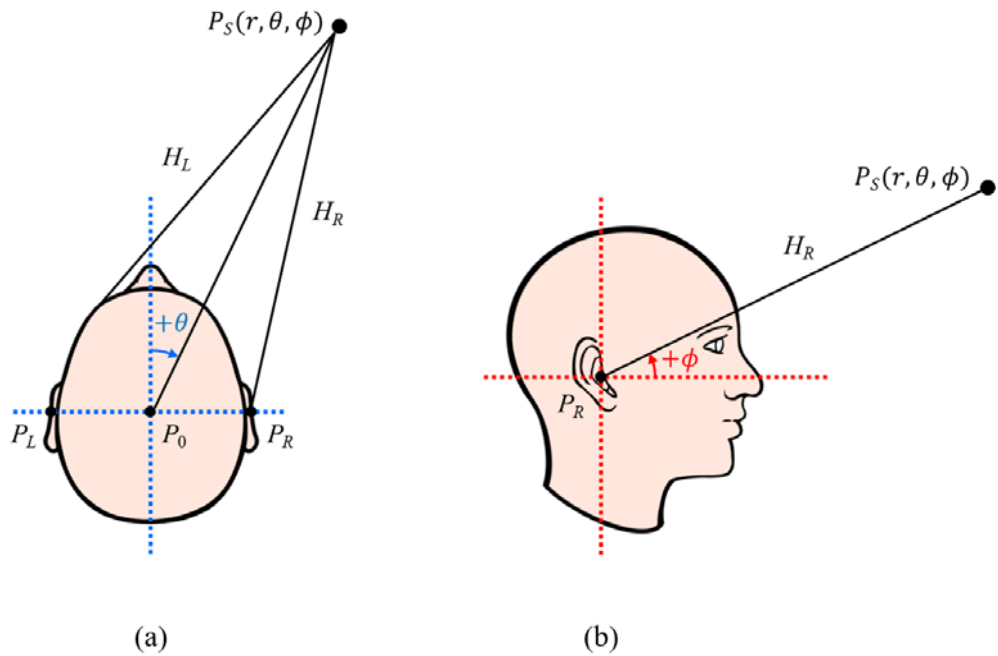

**Figure 3.2.** Illustrations of sound transmission from a sound source to both ears: (a) top view and (b) side view.

## 3.3 HRTF Measurement System

### 3.3.1 Configuration

The HRTF measurement system used in this study is designed to measure HRTFs not only on artificial heads but also on humans. Since this study is for humanoid robot applications, a standard dummy head, Brüel & Kjær (B&K) HATS Type 4100, was selected as a measurement subject. The HRTF measurement system was installed in the anechoic chamber at the Korea Advanced Institute of Science and Technology (KAIST). The size of the KAIST anechoic chamber is 3.6 m in width, 3.6 m in length, and 2.4 m in height, and the cut-off frequency is about 100 Hz. Thus, the frequency band of interest of the HRTF measurement system is bounded from 120 Hz to around 20 kHz. In addition, the distance $r$ from the diaphragm center of the speaker module to the head center of the dummy head is set to 1.1 m in consideration of the height of the anechoic chamber. When the distance $r$ exceeds 1 m, the frequency characteristics of HRTFs are not sufficiently affected by the distance change [112], such that the HRTFs become distance-independent and are called far-field HRTFs.

The range of the sound source azimuth $\theta$ is set from −180° to +180°. The range of the sound source elevation $\phi$ is set from −40° to +90° in consideration of the height of the anechoic chamber. According to several studies [148,149], the angular resolution of an HRTF database should be less than 5° in the horizontal plane and less than 10° in the vertical plane. Therefore, both the azimuth and elevation resolutions of the HRTF measurement system are set to 5°. A turntable driven by a servomotor was designed to precisely realize the azimuth resolution of 5°. As shown in Fig. 3.3, the measurement subject is mounted on the turntable and rotated so that the speaker arc array faces the subject from the azimuth angle designated by the DOA controller. To realize the 5° elevation resolution, a semicircular speaker array was designed with speaker modules spaced at 5° intervals from −40° to +90° in elevation. The elevation angle is set by the DOA controller, so only the switch connected to the corresponding speaker module is turned on in the speaker selector, enabling accurate elevation control of sound source. The total number of sound source directions is 1,944 (72 points in azimuth × 27 points in elevation).

The raw transfer functions, BTFs and OTFs, were measured with an audio interface (Audiomatica CLIO FW-02) connected to a host computer via USB 2.0. The sampling rate of the audio interface is 48 kHz, and the sample size of measured impulse responses is 4,096. In addition, the output signal of the audio interface for full frequency excitation is set as a maximum length sequence (MLS). The MLS from the audio interface is amplified through the speaker amplifier (YBA Heritage A200) and then reproduced as a sound source through the speaker module selected by the speaker selector. The acoustic signal input to the microphone of the dummy head is amplified through the microphone conditioner (B&K NEXUS) and then input to the audio interface. The transfer function is calculated using the ensemble average of acoustic signals measured eight times in the audio interface software (Audiomatica CLIO 12 Standard) installed on the host computer.

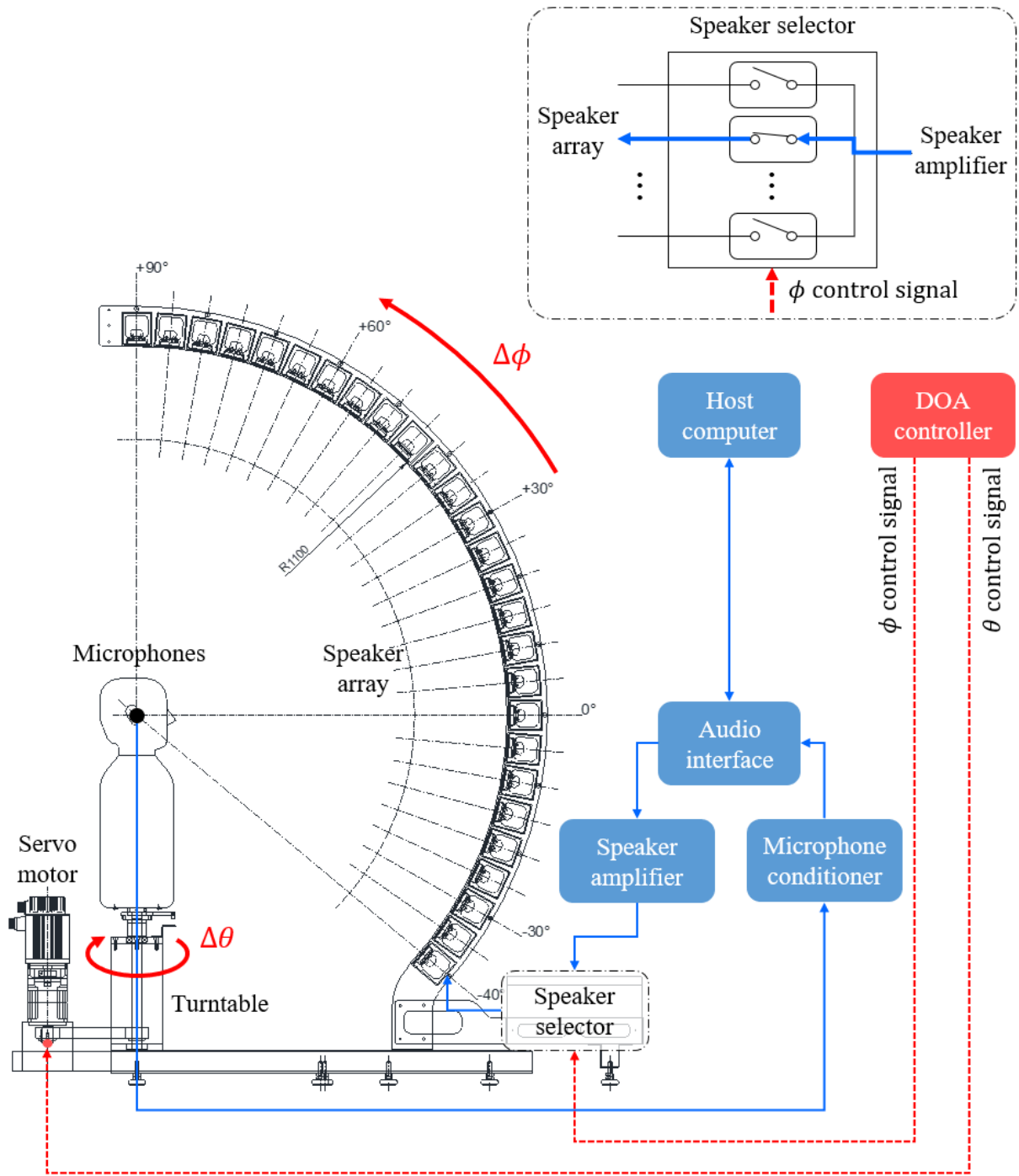

**Figure 3.3.** Block diagram of HRTF measurement system.

Among the measurement modules, the microphone conditioner is an acoustic measurement device, while the speaker amplifier is a commercial audio product. Thus, to verify the speaker amplifier, frequency response and total harmonic distortion (THD) were measured using an audio analyzer (Audio Precision), as follows. The input voltages were 2.828 $V_{rms}$, 4.000 $V_{rms}$, and 4.899 $V_{rms}$ (power inputs of 1 W, 2 W, and 3 W based on 8 Ω speaker driver), respectively. As shown in Fig. 3.4, the magnitude response of the speaker amplifier was guaranteed to be flat from 20 Hz to 20 kHz, covering the frequency band of interest. In addition, the phase response was almost 0° from 10 Hz to 80 kHz, so was sufficiently small phase modulation for the input signal. Moreover, there was no noticeable harmonic distortion because the measured THD was almost 0% in the same frequency range. Therefore, the speaker amplifier was found to be suitable for HRTF measurement.

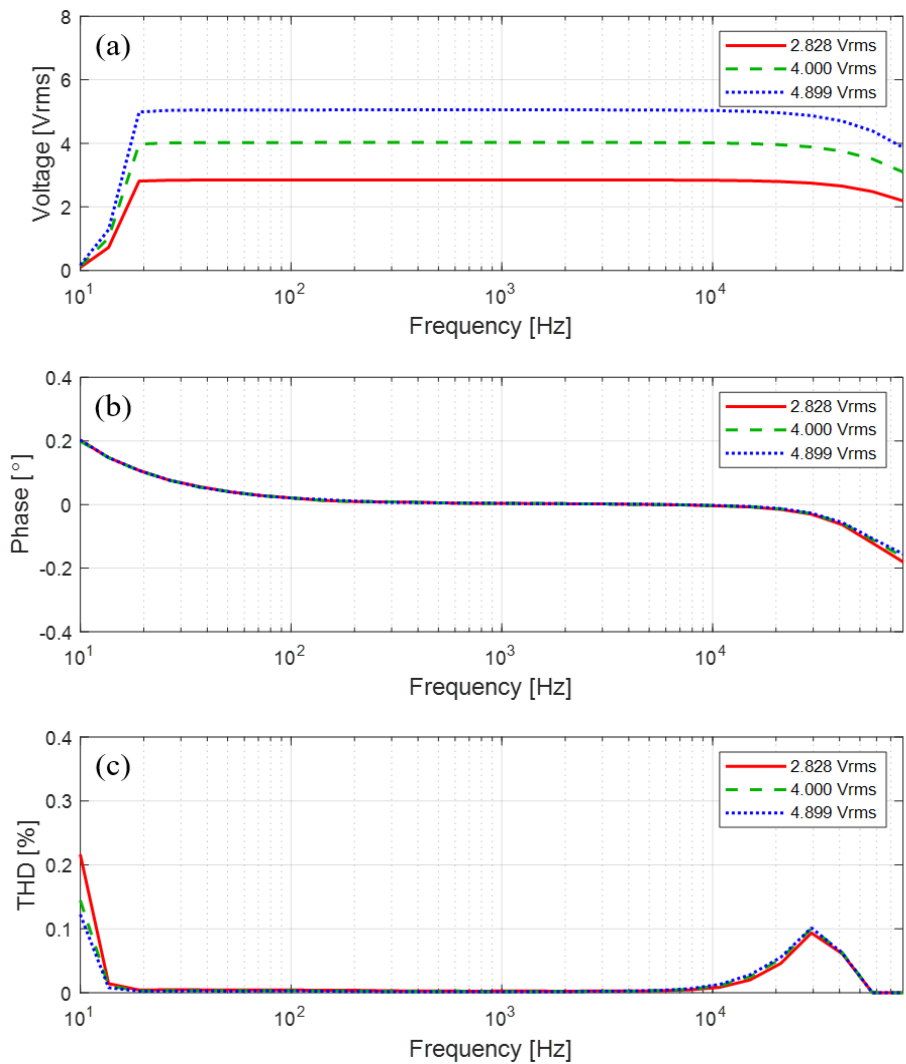

**Figure 3.4.** Frequency responses of speaker amplifier (YBA Heritage A200): (a) magnitude response, (b) phase response, and (c) THD.

- 28 -

### 3.3.2 Speaker Module Design

A full-range speaker driver (Peerless by Tymphany PLS-P830986) was selected to reproduce the frequency band of interest (120 Hz to 20 kHz) with a single sound source. Since the diameter of the speaker driver is 3 inches, it can be mounted on speaker modules arranged at 5° intervals in the speaker arc array. To simulate the frequency response of a speaker module using an equivalent acoustic circuit, the Thiele-Small parameters (TSPs) of the speaker driver were obtained using the delta mass method [150], which determines the electroacoustic parameters of a speaker driver through model curve fitting by referring to the electrical impedance curves of the speaker driver according to the presence or absence of additional mass on the diaphragm. As shown in Fig. 3.5, when 1.0 g of mass was attached to the diaphragm, the resonant frequency decreased. Table 3.1 shows the obtained TSPs of the speaker driver.

Resonance frequency $F_0$ appears at 101.221 Hz, enabling reproduction of the frequency band of interest under an infinite baffle condition. In addition, the measure of total loss $Q_{ts}$ was 0.708, indicating that the damping of the speaker driver was appropriate for low-frequency reproduction. Finally, $SPL_0$ was 83.778 dB, suggesting that the sensitivity of the speaker driver was sufficiently high for acoustic measurement. A sealed speaker enclosure shifts the resonant frequency of a loudspeaker to a higher frequency because the air inside acts as an elastic support. As a result, the sealed speaker enclosure reduces the bandwidth of the loudspeaker, and this tendency is exacerbated as the internal air volume gets smaller. Therefore, to design a speaker module that can reproduce the band of interest in a limited space, the frequency response of sealed type speaker module according to the internal air volume was simulated using the equivalent acoustic circuit shown in Fig. 3.6.

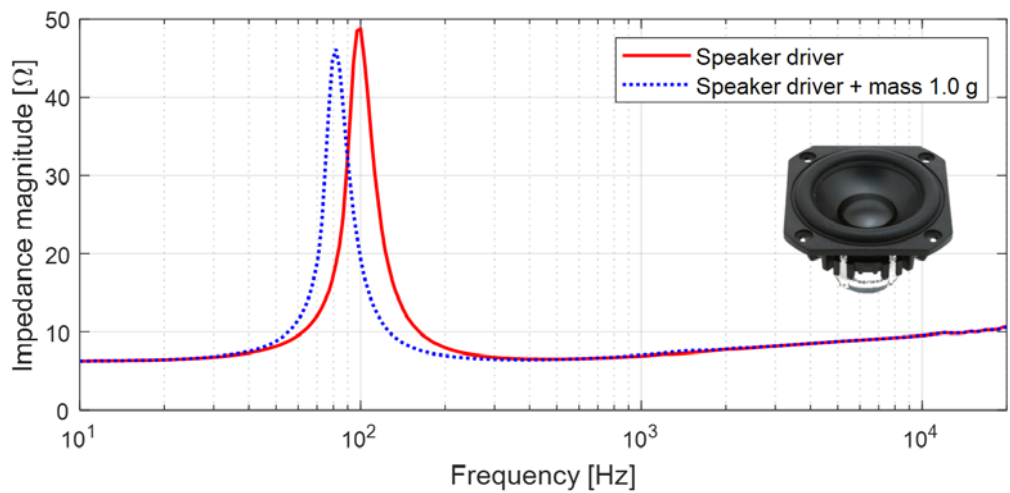

**Figure 3.5.** Electrical impedance curves of speaker driver (Peerless by Tymphany PLS-P830986) used to calculate Thiele-Small parameters (TSPs) by delta mass method.

**Table 3.1.** TSPs of a speaker driver used in the speaker module of the HRTF measurement system.

| Parameter | Value | Description |
|---|---|---|
| $R_{evc}$ | 6.291 Ω | Voice coil resistance. |
| $F_0$ | 101.221 Hz | Resonance frequency. |
| $S_d$ | 0.002827 m² | Equivalent diaphragm area. |
| $K_{rm}$ | 0.010251 Ω | Resistance constant of the motor impedance. |
| $E_{rm}$ | 0.503 | High frequency slop of the motor resistance. |
| $K_{xm}$ | 0.040639 H | Reactance constant of the motor impedance. |
| $E_{xm}$ | 0.392 | High frequency slop of the motor reactance. |
| $V_{as}$ | 1.255 ℓ | Equivalent acoustic volume. |
| $C_{ms}$ | 0.001106 m/N | Equivalent mechanical compliance. |
| $M_{md}$ | 2.150 g | Mechanical mass of the diaphragm without air load. |
| $M_{ms}$ | 2.236 g | Equivalent mechanical mass of the diaphragm with air load. |
| $BL$ | 3.265 Tm | Product of magnetic flux density and length of wire in the flux. |
| $Q_{ms}$ | 4.531 | Measure of mechanical loss in the suspension. |
| $Q_{es}$ | 0.839 | Measure of electrical loss in the voice coil. |
| $Q_{ts}$ | 0.708 | Measure of total loss. |
| $N_0$ | 0.150% | Conversion efficiency from electrical to acoustical energy. |
| $SPL_0$ | 83.778 dB | Sensitivity for half space radiation with 1 W input at 1 m. |

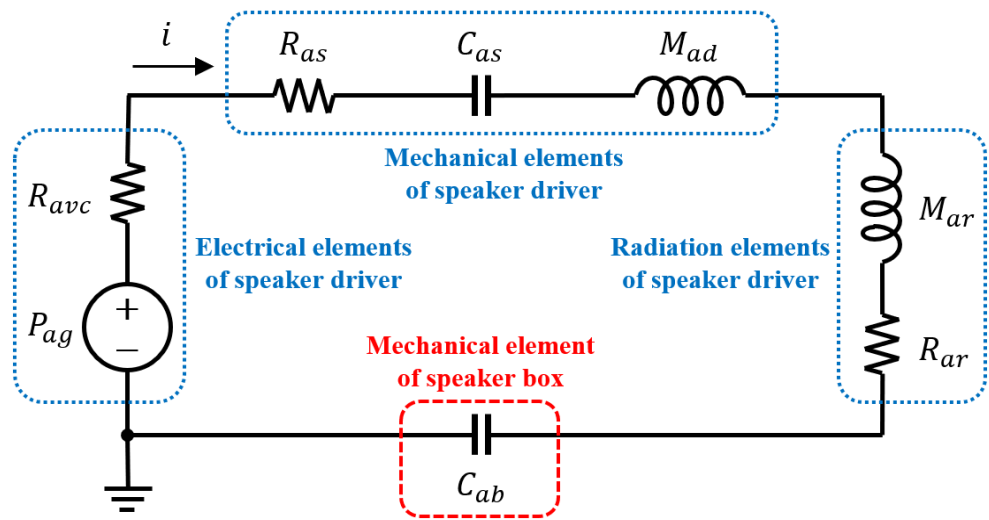

**Figure 3.6.** Equivalent acoustic circuit of sealed speaker module composed of speaker driver and speaker box.

This circuit consists of two parts: the speaker driver and the speaker box. For the electrical elements of the speaker driver, voice coil current $i$, pressure generator $P_{ag}$, and voice coil resistance $R_{avc}$ are expressed as in [150]:

$$i = V_{eg}/R_{evc}, \tag{3.6}$$

$$P_{ag} = BL \cdot i/S_d, \tag{3.7}$$

$$R_{avc} = (BL/S_d)^2/R_{evc}, \tag{3.8}$$

where $V_{eg}$ is the input voltage of the speaker driver. The mechanical elements of the speaker driver consist of suspension resistance $R_{as}$, suspension compliance $C_{as}$, and diaphragm mass $M_{ad}$, and are defined as in [150]:

$$R_{as} = R_{ms}/S_d^2, \tag{3.9}$$

$$C_{as} = C_{ms} \cdot S_d^2, \tag{3.10}$$

$$M_{ad} = M_{md}/S_d^2, \tag{3.11}$$

where $R_{ms}$ denotes the mechanical suspension resistance of the speaker driver. The radiation elements of the speaker driver are composed of radiation mass $M_{ar}$ and radiation resistance $R_{ar}$, and are expressed as in [146]:

$$M_{ar}(\omega) = X_{ar}/\omega, \tag{3.12}$$

$$X_{ar}(ka) = (\rho_0 c/S_d)[H_1(2ka)/ka], \tag{3.13}$$

$$R_{ar}(ka) = (\rho_0 c/S_d)[1 - J_1(2ka)/ka], \tag{3.14}$$

where $X_{ar}$ is the radiation reactance of the diaphragm, $\omega$ the angular frequency, $k$ the wave number, $a$ the radius of the diaphragm, and $\rho_0 c$ the characteristic impedance of air. In addition, $H_1()$ and $J_1()$ denote the first order Struve function and the first order Bessel function of the first kind, respectively. In Fig. 3.6, the speaker box consists of only the mechanical element, $C_{ab}$, and is defined as in [146]:

$$C_{ab} = S_d^2/k_{box}, \tag{3.15}$$

$$k_{box} = \rho_0 c^2 S_d^2/V_{box}, \tag{3.16}$$

where $k_{box}$ is the effective stiffness of the air inside the speaker box, $\rho_0$ the density of air, $c$ the speed of sound, and $V_{box}$ the internal air volume of the speaker box. Therefore, the total acoustic circuit impedance, $Z_{as}(\omega)$, and the volume velocity, $U_a(\omega)$, generated by the diaphragm are expressed as follows:

$$Z_{as}(\omega) = [R_{avc} + R_{as} + R_{ar}(\omega)] + j\omega[M_{ad} + M_{ar}(\omega)] + [1/j\omega C_{as} + 1/j\omega C_{ab}], \tag{3.17}$$

$$U_a(\omega) = P_{ag}/Z_{as}(\omega). \tag{3.18}$$

Finally, from Eq. (3.18), the speaker diaphragm excursion, $X(\omega)$, and the sound pressure along the acoustic axis, $p(\omega, r)$, can be obtained as in [146]:

$$X(\omega) = U_a(\omega)/j\omega S_d, \tag{3.19}$$

$$p(\omega, r) = U_a(\omega)\frac{2\rho_0 c}{S_d}\left|\sin\left[\frac{\omega}{2c}\left(\sqrt{r^2 + \frac{S_d}{\pi}} - r\right)\right]\right|, \tag{3.20}$$

where $r$ is the microphone distance from the speaker diaphragm on the acoustic axis. Fig. 3.7 shows simulation results for the frequency responses of the speaker module. Simulation conditions were an input voltage of 2.828 V$_{rms}$, a microphone distance of 1 m from the speaker, and a speaker air volume of 800 cc. From Fig. 3.7(a), the maximum excursion of the speaker diaphragm occurs at 127 Hz and is expected to be less than 1.0 mm. As shown in Fig. 3.7(b), the volume velocity generated by the speaker diaphragm is expected to exceed 0.002 m³/s at 162 Hz. Fig. 3.7(c) shows the sound pressure in dB SPL at 1 m in front of the speaker diaphragm. The −6 dB roll-off frequency appears at 116 Hz, which is expected to sufficiently cover the frequency band of interest. Finally, a speaker module prototype was made based on the internal air volume of 800 cc as shown in Fig. 3.8(a).

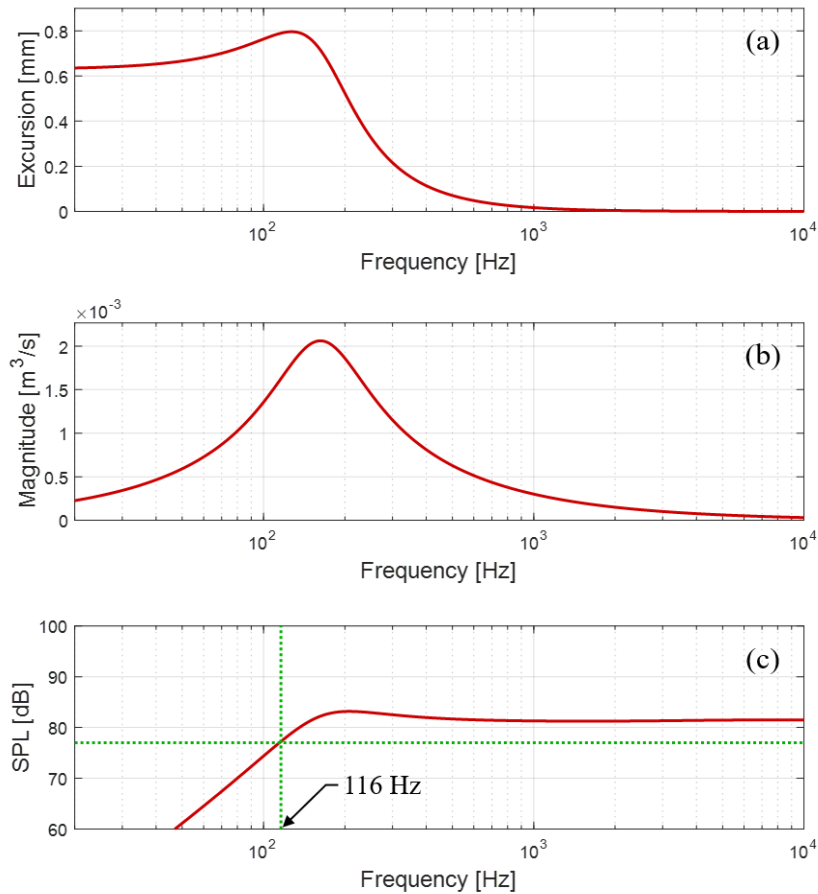

**Figure 3.7.** Computed frequency responses of sealed speaker module under conditions of $V_{eg}$ = 2.828 V$_{rms}$, $r$ = 1 m, and $V_{box}$ = 800 cc: (a) speaker diaphragm excursion; (b) volume velocity generated by the diaphragm; and (c) sound pressure along acoustic axis in dB SPL.

To verify the performance of the speaker module prototype, the frequency response was measured 1 m in front of the speaker when the input voltage was 2.828 $V_{rms}$. As shown in Fig. 3.8(c), the −6 dB roll-off frequency was 119 Hz, which was almost identical to the simulation results, and it was also confirmed that the frequency band of interest (120 Hz ~ 20 kHz) was sufficiently reproduced. The deviation of the frequency response is a unique characteristic of the speaker driver and is mitigated when HRTFs are obtained by normalizing BTFs to OTFs. Fig. 3.8(d) is the phase response of the speaker module prototype, showing a typical linear phase. Therefore, it can be expected that a sound wave reproduced by the speaker module will radiate without the distortion of waveform.

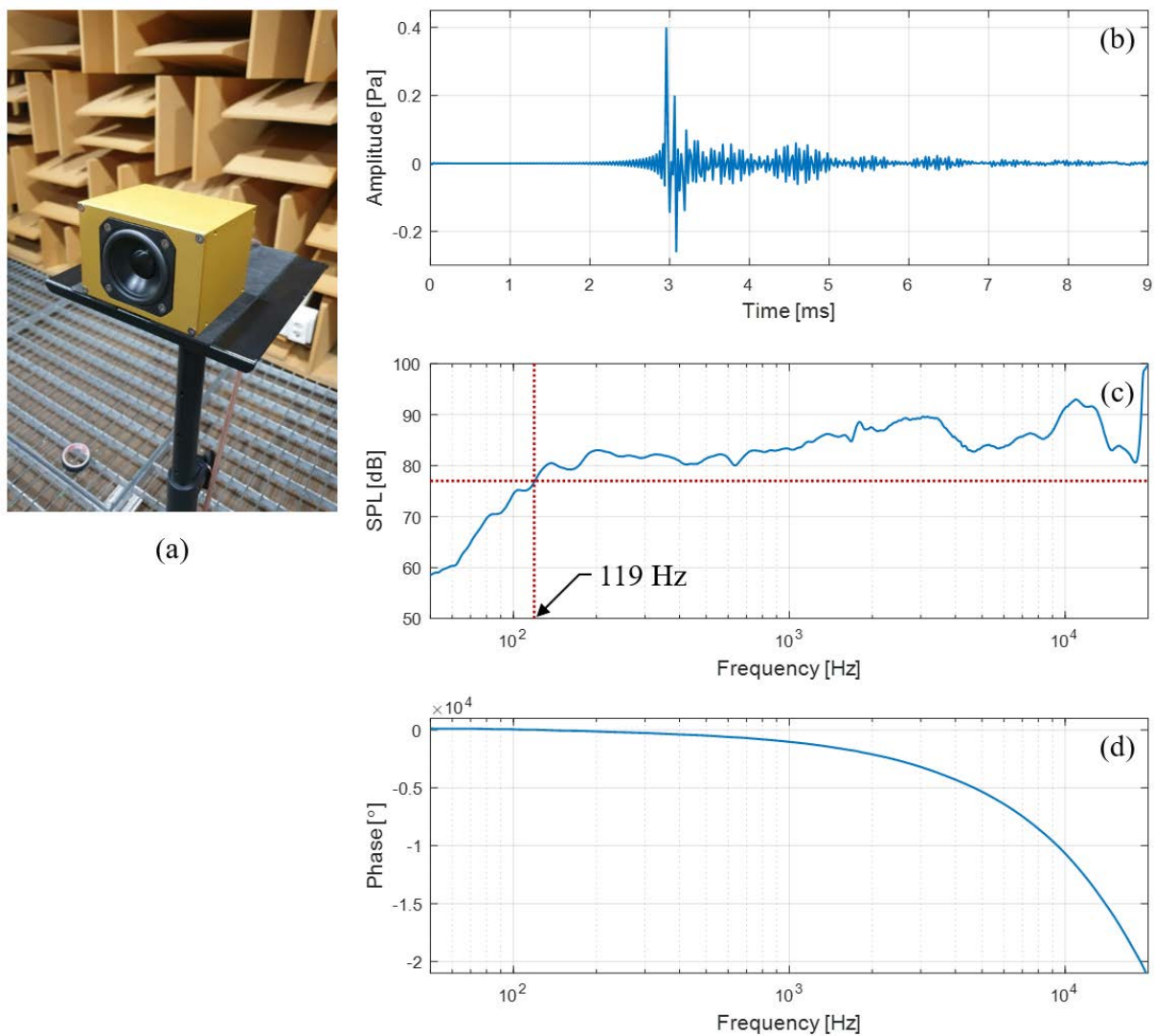

**Figure 3.8.** Responses of the speaker module measured at a microphone placed 1 m in front of the speaker diaphragm when input voltage is 2.828 $V_{rms}$ (1 W for the 8 Ω speaker driver): (a) speaker module prototype; (b) impulse response; (c) magnitude response (1/12 octave bands); and (d) unwrapped phase response.

## 3.4 Measurement of Raw Transfer Functions

### 3.4.1 Binaural Transfer Function (BTF)

As shown in Fig. 3.9, a speaker arc array and measurement system were built and installed based on the confirmed speaker module specifications. To measure BTFs, the positions of the dummy head and the speaker arc array were aligned so that the microphone of the dummy head faced the center of the speaker diaphragm at 0° in elevation. After the installation in the anechoic chamber, sound-absorbing material was covered on the support of the measurement system to minimize the reflection effect due to the support, as shown in Fig. 3.9(f). All measurements including BTFs and OTFs were conducted with input voltage of 2.828 $V_{rms}$, corresponding to 1 W for the 8 Ω speaker driver.

Measurements of BTFs were performed for a total of 1,944 points, including 72 points of azimuth (from −180° to +180° with 5° resolution) and 27 points of elevation (from −40° to +90° with 5° resolution). Fig. 3.10 shows the results of BTFs for the horizontal plane ($\phi = 0°$). Figs. 3.10(a1) and (a2) are BIRs, which are the time domain impulse responses of BTFs. The impulse response started first when the sound source was in the ipsilateral 90° direction ($\theta = -90°$ for the left or $\theta = +90°$ for the right), and last when it was in the contralateral 90° direction ($\theta = +90°$ for the left or $\theta = -90°$ for the right). Figs. 3.10(b1) and (b2) show the magnitude responses of BTFs. It can be seen that the levels of the mid and high frequency bands of the contralateral BTFs dropped significantly compared to the ipsilateral BTFs due to the head shadow effect. The peaks appearing around 20 kHz were due to the resonance frequency of the aluminum diaphragm of the speaker driver. Figs. 3.10(c1) and (c2) show the phase responses of BTFs, in which it can be seen that the contralateral BTF phase of the mid and high frequency bands changed rapidly due to the head shadow effect. Since the BTFs had not yet been post-processed, the reflection effect was noticeable in the time domain, and the speaker characteristics were revealed in the frequency domain. Therefore, additional post-processing is required to estimate HRTFs from BTFs. In particular, OTF including the transfer function of the entire measurement system is essential in the post-processing of BTFs. Detailed analysis is conducted after obtaining HRTFs through post-processing including OTF normalization.

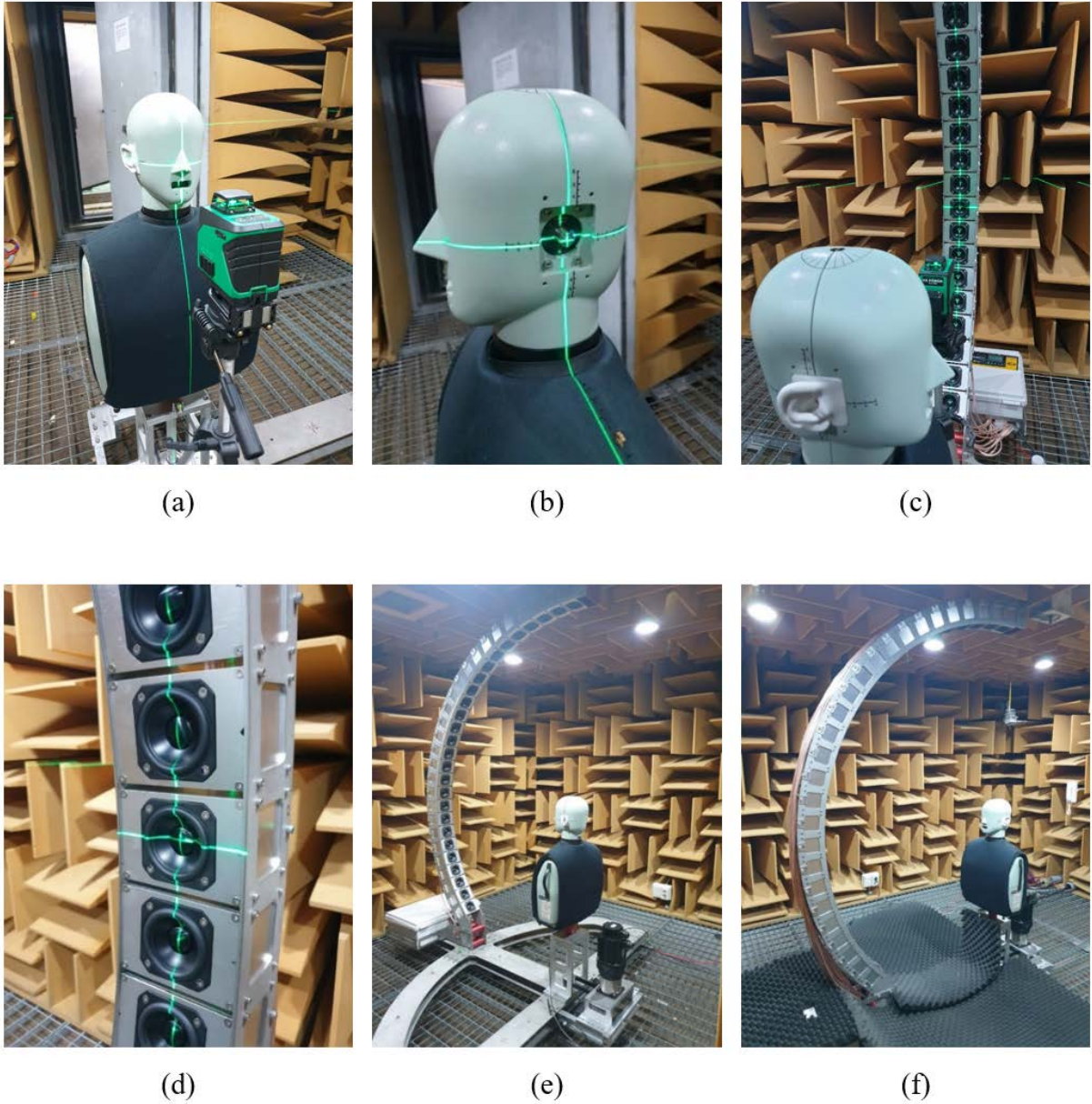

**Figure 3.9.** BTF, $G_{L,R}$, measurement setup with the speaker arc array and dummy head on a rotating turntable in the anechoic chamber at KAIST: (a) dummy head positioning in front of the head; (b) dummy head positioning on the left side of the head; (c) dummy head positioning with the speaker arc array; (d) speaker arc array positioning; (e) installation completed in the anechoic chamber; and (f) floor finished with sound-absorbing materials.

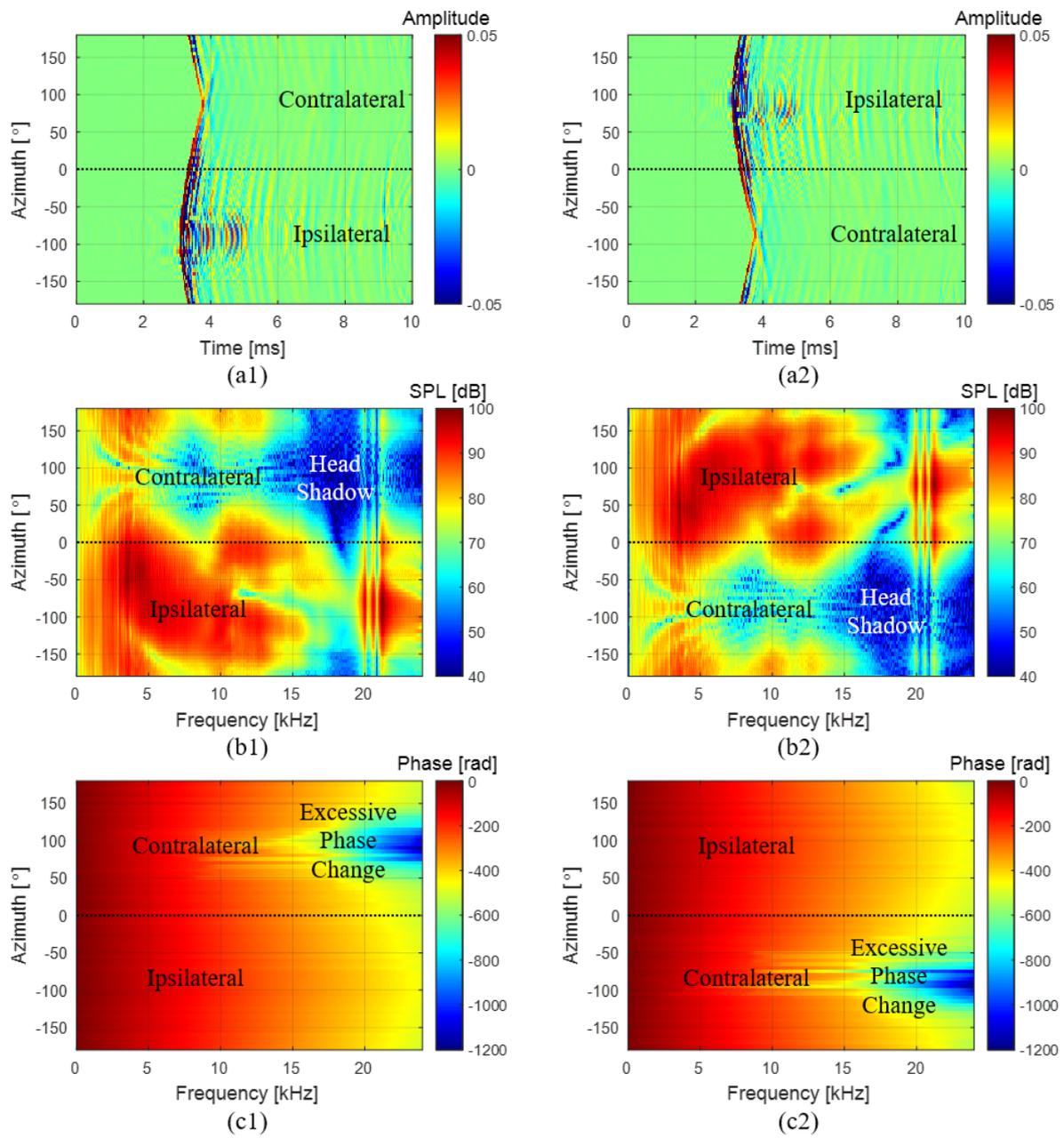

**Figure 3.10.** BTF, $G_{L,R}$, measurement results at $\phi = 0°$: (a1) left BIRs; (a2) right BIRs; (b1) magnitude of left BTFs; (b2) magnitude of right BTFs; (c1) phase of left BTFs; and (c2) phase of right BTFs.

3.4.2 Origin Transfer Function (OTF)

To obtain HRTFs, the transfer function of the measurement system included in the BTFs should be canceled out. For this, it is necessary to measure the OTFs, $G_0(\phi, f)$. Since OTF measurement is performed in the absence of the measurement target, only the transfer function of the measurement system is included. When measuring the OTFs, the measurement devices used for BTFs should be used as they are, so that the transfer function of the measurement system can later be completely excluded from HRTF data. Therefore, as shown in Figs. 3.11(a) and (b), OTF measurement was performed using one of the dummy head microphones. As shown in Fig. 3.11(c), the level difference between the left and right microphones was within ± 0.5 dB. Therefore, the right microphone was used for the OTF measurements.

For any speaker module, OTF must be non-directional but, 1/2-inch measurement condenser microphones are directional. In some studies [125,133,145], the microphone was tilted 90° from the acoustic axis of the speaker, as shown in Fig. 3.12(b). In other studies [135,136,143,144], the microphone was directed toward the acoustic axis, as shown in Fig. 3.12(a). When a microphone is tilted 90°, it is not directional along the circumference of its diaphragm, but it is directional along the radial direction of its diaphragm, which results in decreasing and also fluctuating high frequency response, as shown in Fig. 3.12(c). Fig. 3.12(d) shows the level difference between the 0° on-axis and 90° off-axis microphones. The level difference of high frequency response is reduced by about 11 dB up to 20 kHz. Therefore, normalizing the BTFs using the OTF measured with a 90° off-axis microphone will increase the high frequencies of the HRTFs by about 11 dB. In addition, since various peaks and notches are distributed over the entire frequency range, estimating HRTFs using OTF measured with a 90° off-axis microphone may give incorrect binaural localization cues. As shown in Fig. 3.12(a), when measuring with the microphone facing the acoustic axis of each speaker module, omnidirectional OTF can be obtained but, it is cumbersome to redirect the microphone every time. Fortunately, most microphones have an angular range that is close to omnidirectional. As a result of the directivity measurement of the dummy head microphone, it is confirmed that the microphone is omnidirectional, with an error of ± 0.5 dB in an angular range within ± 15° in the radial direction. As shown in Fig. 3.13(d), the entire measurement section of the speaker array was divided into five sections in consideration of the microphone's omnidirectional range (maximum 30°). In addition, the microphone was directed to elevation angles of −30°, 0°, +30°, +60°, and +90°, and the OTF for each speaker module in the corresponding section was measured. Fig. 3.13(a) shows the impulse responses of OTFs according to the elevation angle. At about 3.2 ms, the time taken for sound wave to propagate 1.1 m, the maximum peaks of the impulse responses were equally represented. Fig. 3.13(b) shows the magnitude responses of the OTFs. The peaks around 12 kHz and 20 kHz indicate intrinsic frequency characteristics of the applied speaker drivers. Fig. 3.13(c) shows the phase responses of the BTF, showing the linear phase. In this way, segmented OTF measurement considering the omnidirectional range of the on-axis microphone made it possible to obtain the omnidirectional OTFs while minimizing the number of microphone settings.

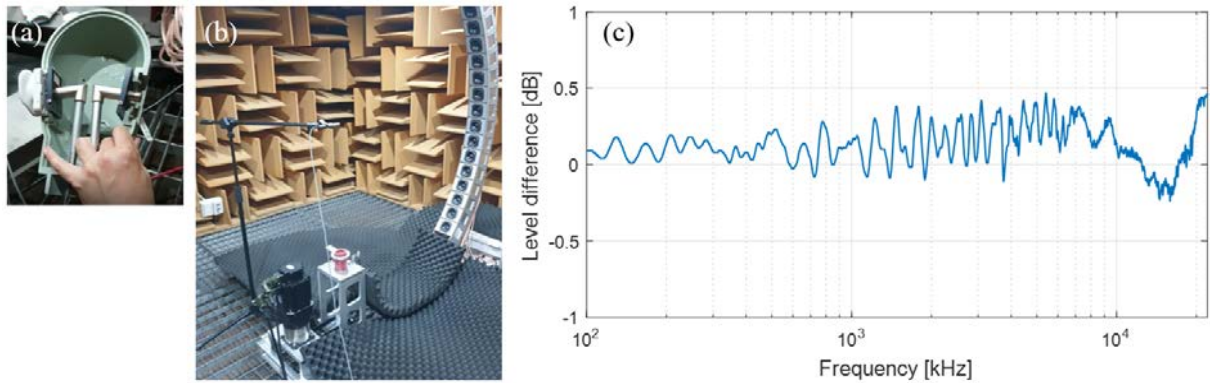

**Figure 3.11.** Difference between the left and right microphones of the dummy head: (a) microphones inside the dummy head; (b) measurement setup with the speaker module at $\phi = 0°$ turned on; and (c) level difference between the left and right microphones.

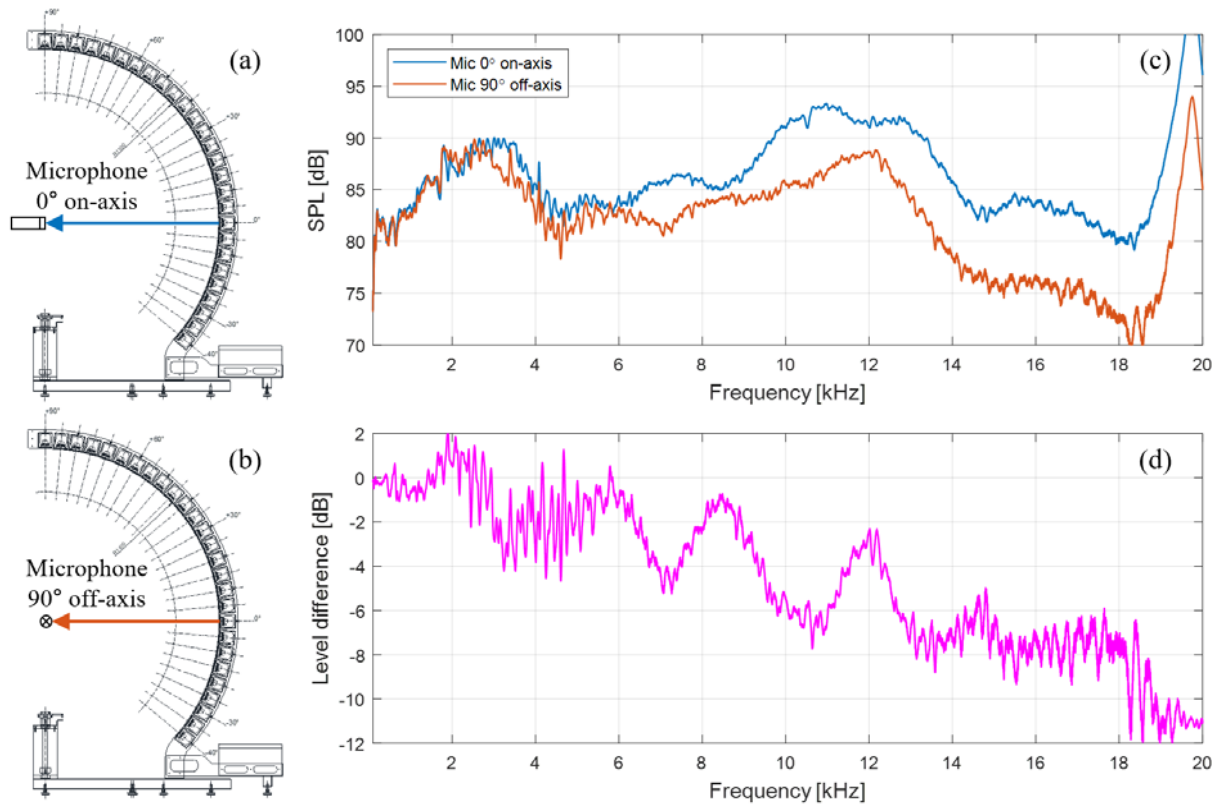

**Figure 3.12.** Difference between two representative OTF, $G_0$, measurements: (a) measurement setup with a 0° on-axis microphone; (b) measurement setup with a 90° off-axis microphone; (c) frequency responses of OTFs measured with 0° on-axis and 90° off-axis microphones; and (d) level difference between 0° on-axis and 90° off-axis OTFs.

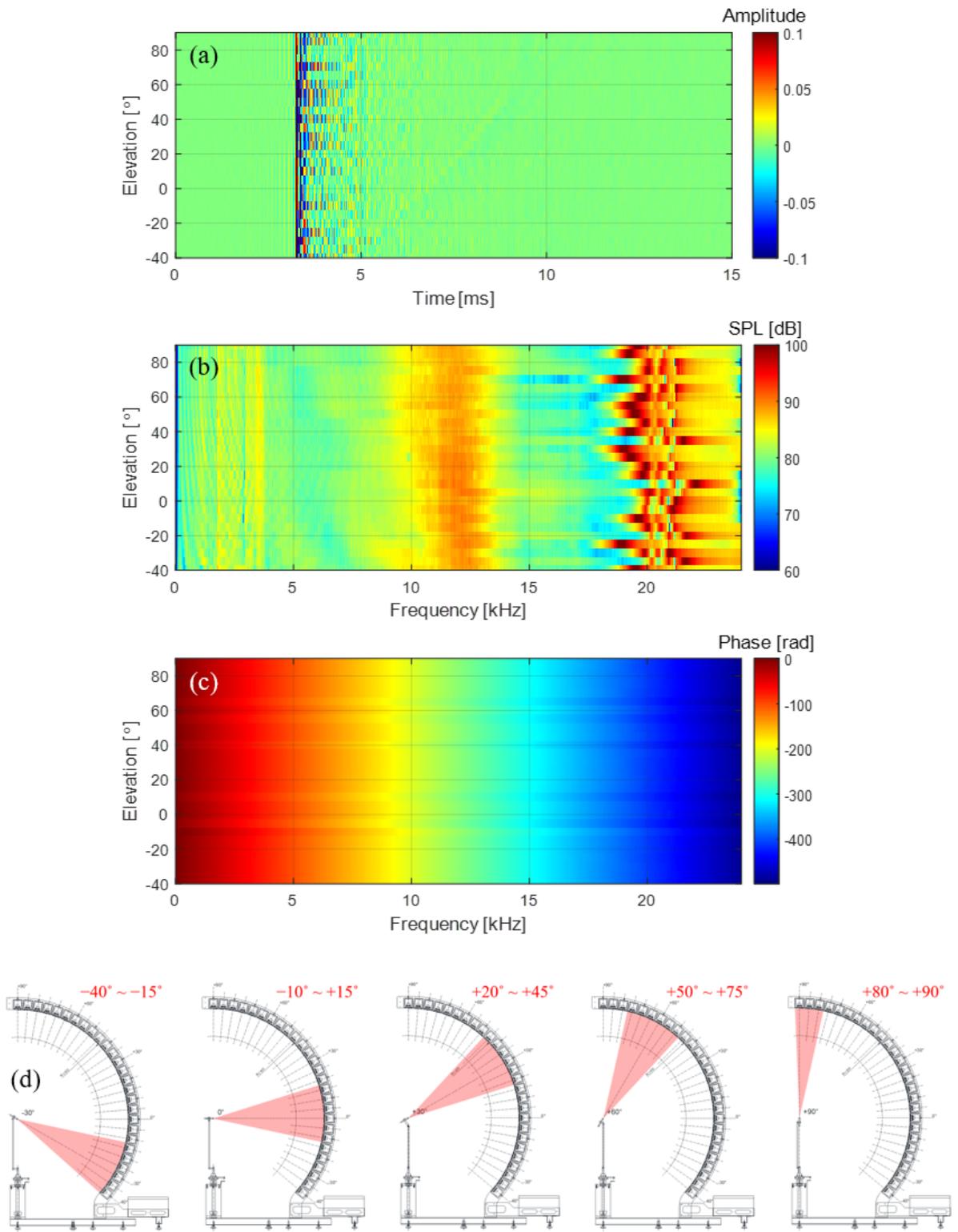

**Figure 3.13.** OTF, $G_0$, measurement setup and results: (a) OIRs; (b) magnitude of OTFs; (c) phase of OTFs; and (d) segmented OTF measurement to obtain omni-directional OTFs.

3.4.3 Time Window Setting

The sample size of the measured BIR and OIR is 4,096, which corresponds to about 85.3 ms when the sampling rate is 48 kHz. Since the essential information of the measured impulse responses is limited to several ms, it is necessary to remove unnecessary sections to minimize the storage space of HRTF database. In this measurement, the start and end points of the time window were set in consideration of the propagation delay and reflective ripples of each impulse response and then applied to each BIR and OIR. To preserve all BIR and OIR information, the start point of the time window was set based on the most preceding impulse response. Therefore, the start point was set based on the maximum sample positions of the left ipsilateral 90° BIR, $g_L(-90,0,t)$, and the right ipsilateral 90° BIR, $g_R(+90,0,t)$. As shown in Fig. 3.14, the maximum samples of both $g_L(-90,0,t)$ and $g_R(+90,0,t)$ appeared at 3.1 ms. Due to some trivial errors in signal processing, small ripples can be generated before the maximum sample of the measured impulse response [111]. Therefore, to include these ripples, the start point should be set before the maximum sample. In this measurement, a point 1 ms before the maximum sample of ipsilateral 90° BIRs was set as the start point.

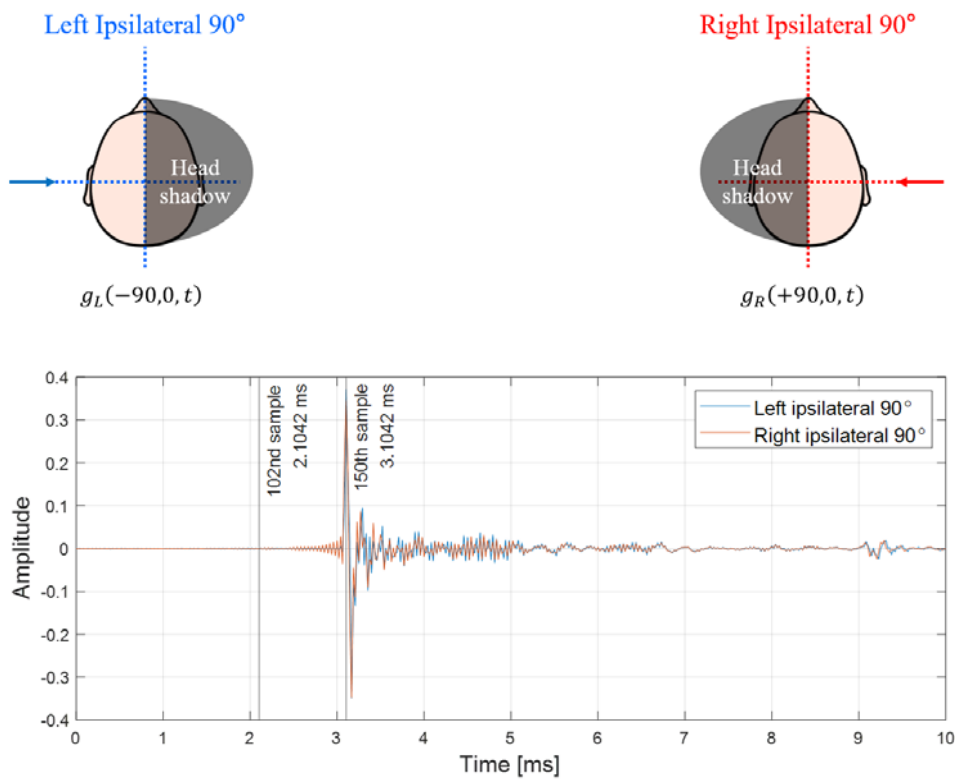

**Figure 3.14.** BIRs, $g_{L,R}$, of ipsilateral 90° directions for setting start point of time window.

The end point of the time window was set to eliminate reflective ripples in the measured impulse response. Even if BIR and OIR are measured in an anechoic chamber, reflections from the measurement system are unavoidable. As shown in Fig. 3.15, the main reflections attributed to the measurement system are speaker array and floor support reflections. Since all speaker modules are arranged equidistant from the head center, the arrival time of each sound reflected from the speaker arc array is almost the same regardless of the elevation angle of speaker module. On the other hand, as the elevation angle of speaker module decreases, the arrival time of each sound reflected from the floor support becomes shorter because the reflection path is shortened. Therefore, the sound generated from the lowest speaker module and reflected from the bottom arrives first. Even if the measurement system is finished with a sound-absorbing material to mitigate reflection of the floor support, the low-frequency components of the reflected wave remain as shown in Fig. 3.15(d). Therefore, all BIRs and OIRs need to be processed to maintain the key information in a time window where there is no reflection. In this measurement, the first zero-crossing sample more than 2.5 ms (120 samples) after the max sample of each impulse response was set as the end point of the time window. Each BIR, $g_{L,R}(\theta, \phi, t)$, and OIR, $g_0(\phi, t)$, cut out by the time window, was zero-padded to 512 samples, for a resolution of 93.75 Hz.

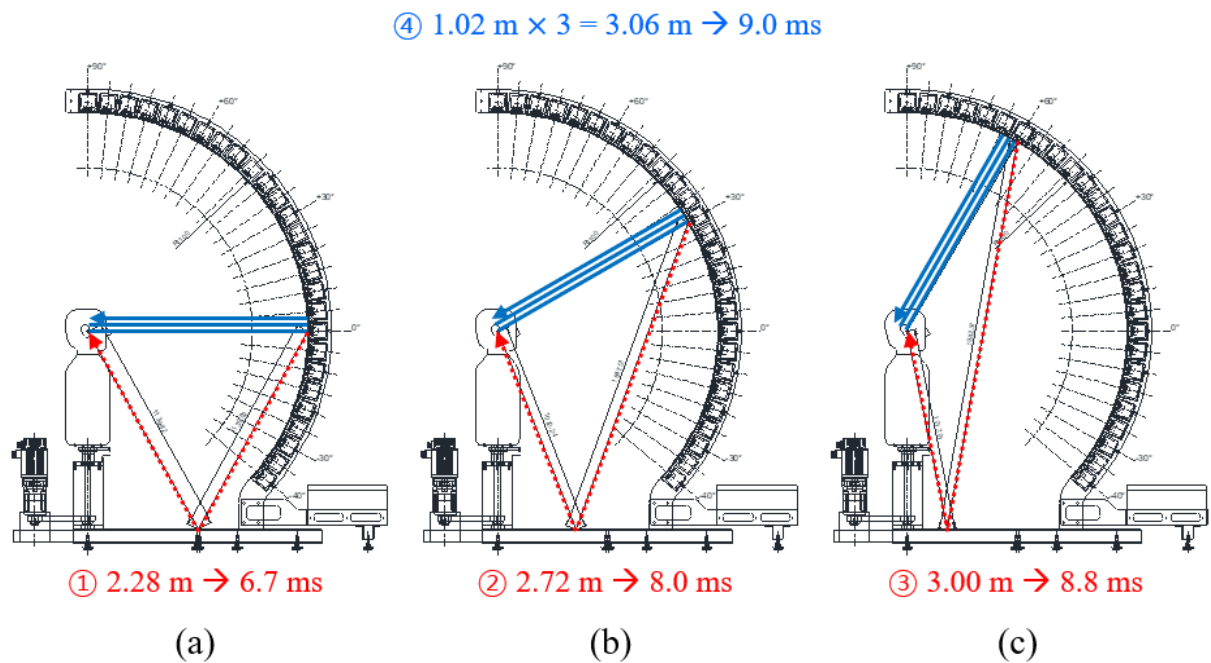

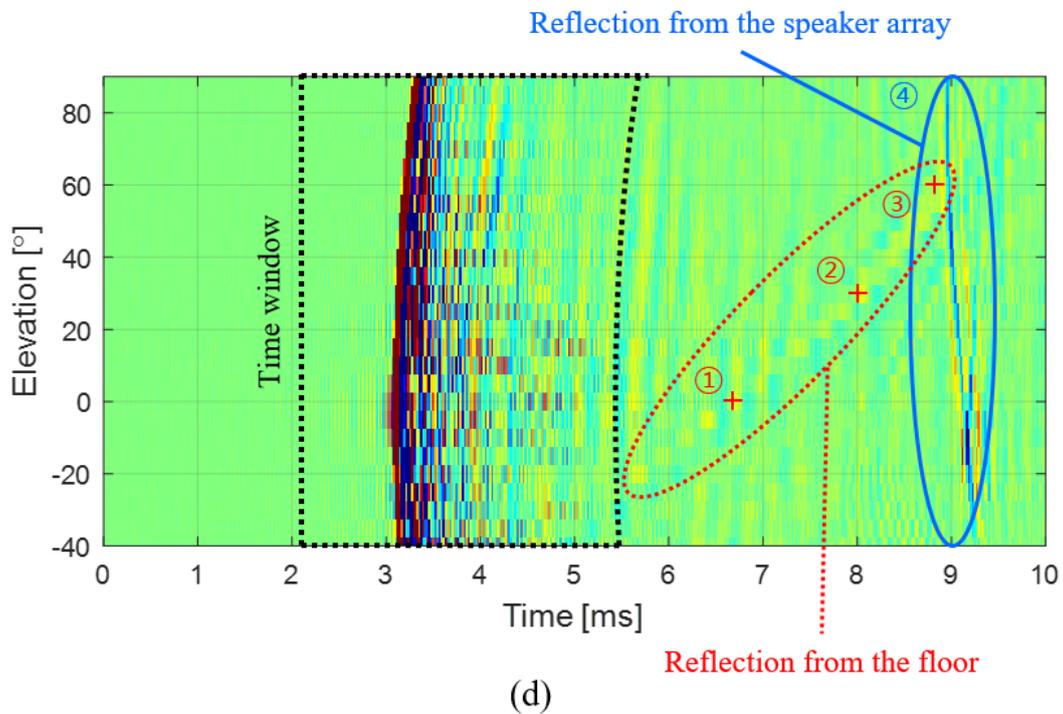

**Figure 3.15.** Illustrations for the end point setting of time window: (a) reflections when the speaker module at $\phi = 0°$ is turned on; (b) reflections when the speaker module at $\phi = 30°$ is turned on; (c) reflections when the speaker module at $\phi = 60°$ is turned on; and (d) time window applied for the right ipsilateral 90° BIRs, $g_R(+90, \phi, t)$, with reflections from the floor and speaker array.

## 3.5 Derivation of HRTF

### 3.5.1 Non-causality Compensation

The post-processed BIRs, $g_{L,R}(\theta,\phi,t)$, and OIRs, $g_0(\phi,t)$, are converted into BTFs and OTFs, respectively, through fast Fourier transform (FFT). HRTFs, $H_{L,R}(\theta,\phi,f)$, are obtained by complex division of BTFs, $G_{L,R}(\theta,\phi,f)$, and OTFs, $G_0(\phi,f)$, as shown in Eq. (3.5). Since each HRTF consists of 512 complex numbers, when saved as a data file, the 512 real numbers and 512 imaginary numbers are stored in two lines. Furthermore, when a pair of left and right HRTFs are stored together for each direction, a total of four lines are saved per direction. In general, to minimize the storage space of resulting database, HRTFs, $H_{L,R}(\theta,\phi,f)$, in the frequency domain are converted into HRIRs, $h_{L,R}(\theta,\phi,t)$, in the time domain through inverse fast Fourier transform (IFFT).

As shown in Fig. 3.16, an ipsilateral HRTF is a kind of non-causal filter because the sound pressure $P_R$ of the ipsilateral ear arrives earlier than the sound pressure $P_0$ at the head center. On the other hand, since the sound pressure $P_L$ of the contralateral ear arrives later than the sound pressure $P_0$ at the head center, the contralateral HRTF is a causal filter. Due to the circularity of the discrete Fourier transform, the maximum sample preceding 0 seconds of the non-causal filter appears later in the impulse response sequence, as shown in Fig. 3.16(a1). Therefore, additional post-processing is needed to compensate for the discontinuity of ipsilateral HRIRs. In Fig. 3.16, the difference of the arrival times between sound pressures $P_0$ and $P_R$ has maximum value when the azimuth angle is 90° on the horizontal plane, and this maximum value can be obtained as follows:

$$\tau_{max} = l/c, \tag{3.21}$$

where $l$ is the radius of the head and $c$ is the speed of sound in air. In this measurement, to compensate for the non-causality of ipsilateral HRIRs, all HRIRs were circularly shifted by at least $\tau_{max}$, as follows:

$$h_{L,R}(\theta,\phi,n) = \tilde{h}_{L,R}(\theta,\phi,\langle n-m\rangle_N)$$
$$\text{with } 0 \leq n \leq N-1, \ m > \tau_{max}F_s, \tag{3.22}$$

where $\tilde{h}_{L,R}$ is a raw HRIR, $n$ is the time index, $m$ is the time delay index, $N$ is the length of HRIR (512 samples), and $F_s$ is the sampling frequency (48 kHz). In addition, $\langle n-m\rangle_N$ denotes the circular shift operation. Through this compensation in Eq. (3.22), the maximum peak of the ipsilateral HRIR appears in the first half of the impulse response sequence as shown in Fig. 3.16(a2). Comparing Figs. 3.16(b1) and (b2), the magnitude responses do not differ before and after the non-causality compensation. However, when comparing Figs. 3.16(c1) and (c2), both the left and right time delays ($\tau_L$ and $\tau_R$) became positive after the non-causality compensation, confirming that the causality of HRIRs was secured. A pair of left and right HRIRs for each direction was saved as a text file consisting of two columns of 512 signed numbers. Finally, the HRTF database was constructed with a total of 1,944 HRIR text files (72 directions in azimuth × 27 directions in elevation).

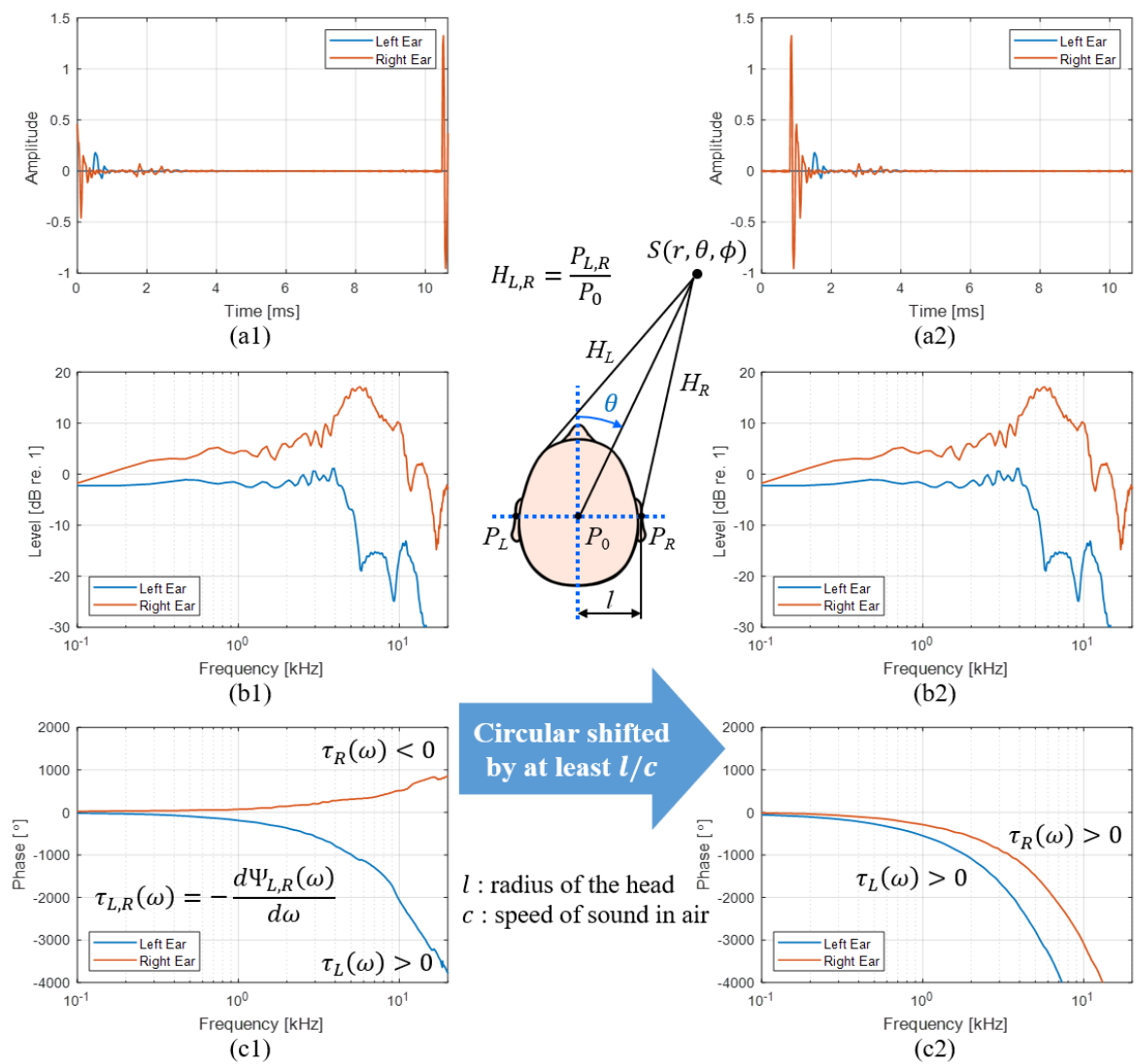

**Figure 3.16.** Non-causality issue of ipsilateral HRIRs and compensation for the non-causality: (a1) non-causal HRIRs; (a2) causal HRIRs; (b1) non-causal HRTF magnitude responses; (b2) causal HRTF magnitude responses; (c1) non-causal HRTF phase responses; and (c2) causal HRTF phase responses.

3.5.2 HRTF Pattern According to Azimuth and Elevation

The time and frequency domain characteristics of the derived HRTFs were presented as contour maps for each major azimuth and elevation angle. Figs. 3.17–3.19 show left and right HRTF pairs as impulse responses, magnitude responses in dB scale, and phase responses in radians for the entire elevation range when the azimuth angles are 0°, 45°, and 90°, respectively.

As shown in Figs. 3.17 (a1) and (a2), the essential part of the HRIRs, which reflects the complex interactions between a sound source and the torso, head, and pinna, lasts about 1.0 ms. In addition, when the azimuth angle of a sound source is 0°, it can be seen that there is almost no time difference between the left and right HRIRs, regardless of the elevation angle. On the other hand, when the azimuth of a sound source is 45°, the sound wave reaches the right ear first. Thus, the time delay of the right HRIRs is shorter than that of the left, as shown in Figs. 3.18(a1) and (a2). The arrival time difference between the left and right HRIRs is maximized in the horizontal plane ($\phi = 0°$) when the azimuth of a sound source is 90°, as shown in Figs. 3.19(a1) and (a2). This is because when the azimuth is 90°, the transmission path to the right ear is the shortest, while that to the left ear is the longest. When a sound source is on the left ($\theta < 0°$), it mirrors the case of a right sound source ($\theta > 0°$). Quantitative analysis of ITD for left and right HRIR pairs is covered in Subsection 4.1.1.

As shown in Figs. 3.17(b1) and (b2), at frequencies below 500 Hz, the level of HRTF approaches 0 dB, regardless of frequency, because the scattering effect of the head is almost negligible. As frequency increases, levels of HRTFs vary with frequency and elevation angle in a complex manner. This complexity is due to the overall filtering effects of the torso, head, and pinna. The apparent peak at 4 kHz is constant regardless of the elevation angle, while the notch at 9 kHz shifts to higher frequency as the elevation angle increases. The relative positions of the HRTF peaks and notches serve as localization cues of elevation and are discussed in detail in Subsection 4.1.3. When the azimuth angle is 0°, there is no significant difference in the left and right HRTF levels. As shown in Figs. 3.18(b1) and (b2), when the azimuth of a sound source is 45°, a level difference appears between the left and right HRTFs. Above 4 kHz, the contralateral HRTF levels are noticeably attenuated due to the low-pass filtering effect of the head shadow. The ipsilateral HRTF levels increase to some extent, and some notches appear. This is due to the reflection effect of the head on ipsilateral incidence at high frequencies, which increases the pressure on the ipsilateral sound source. This phenomenon is maximized when a sound source is in the ipsilateral 90° direction, as displayed in Figs. 3.19(b1) and (b2). Quantitative ILD analysis of left and right HRTF pairs is detailed in Subsection 4.1.2.

As shown in Figs. 3.17(c1) and (c2), when the azimuth is 0°, the phase change is close to linear, and the phase patterns of the left and right HRTFs are almost the same. However, when the azimuth angle is 45°, an excessive phase change appears at high frequencies of the contralateral HRTF, as presented in Figs. 3.18(c1) and (c2). This is another result of the head shadow effect reflected in the phase. The phase patterns at high frequencies change most rapidly when a sound source is in the contralateral 90° direction, as shown in Figs. 3.19(c1) and (c2).

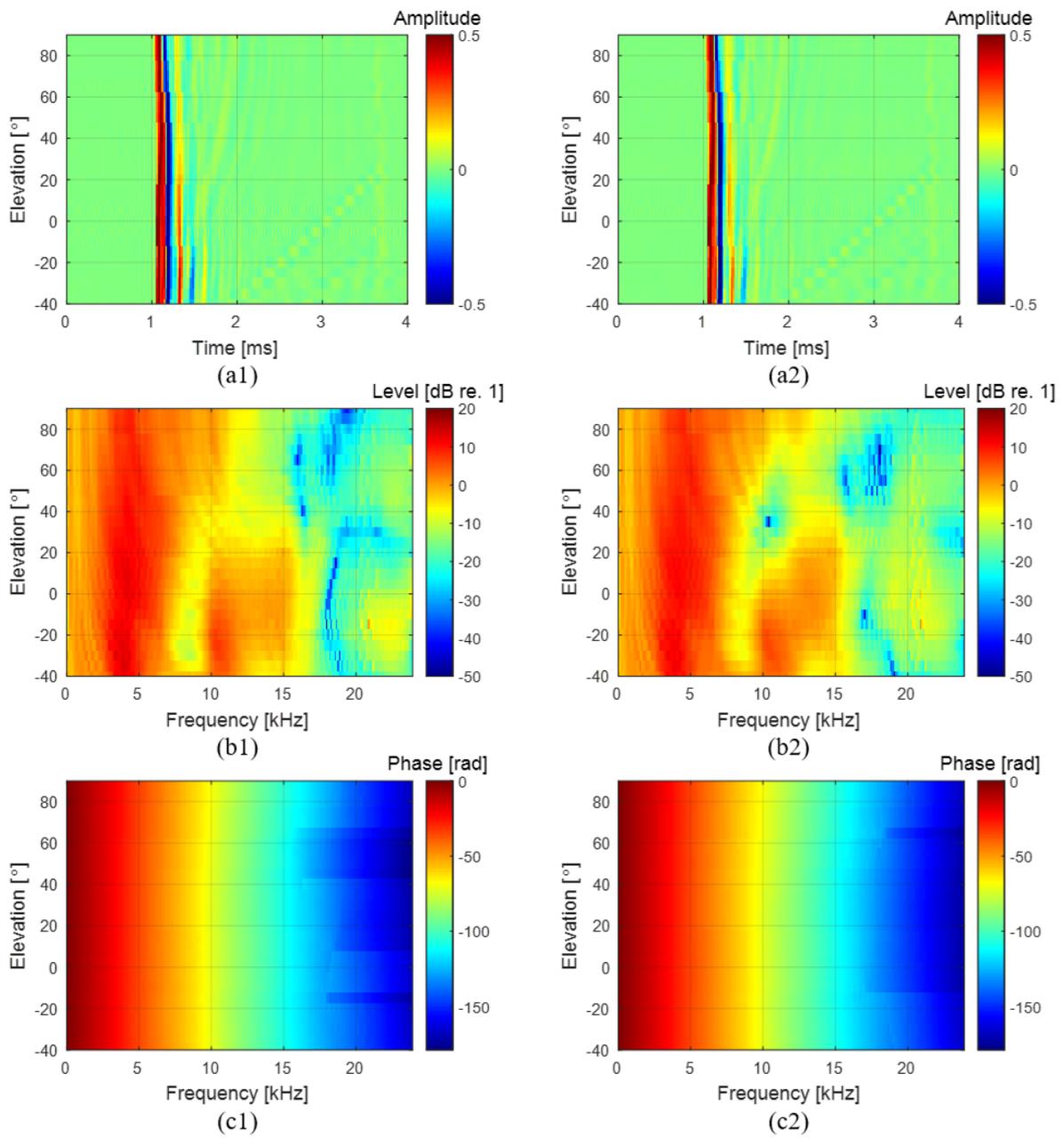

**Figure 3.17.** HRTFs at $\theta = 0°$, $H_{L,R}(0, \phi, f)$: (a1) left HRIRs; (a2) right HRIRs; (b1) magnitude of left HRTFs; (b2) magnitude of right HRTFs; (c1) phase of left HRTFs; and (c2) phase of right HRTFs.

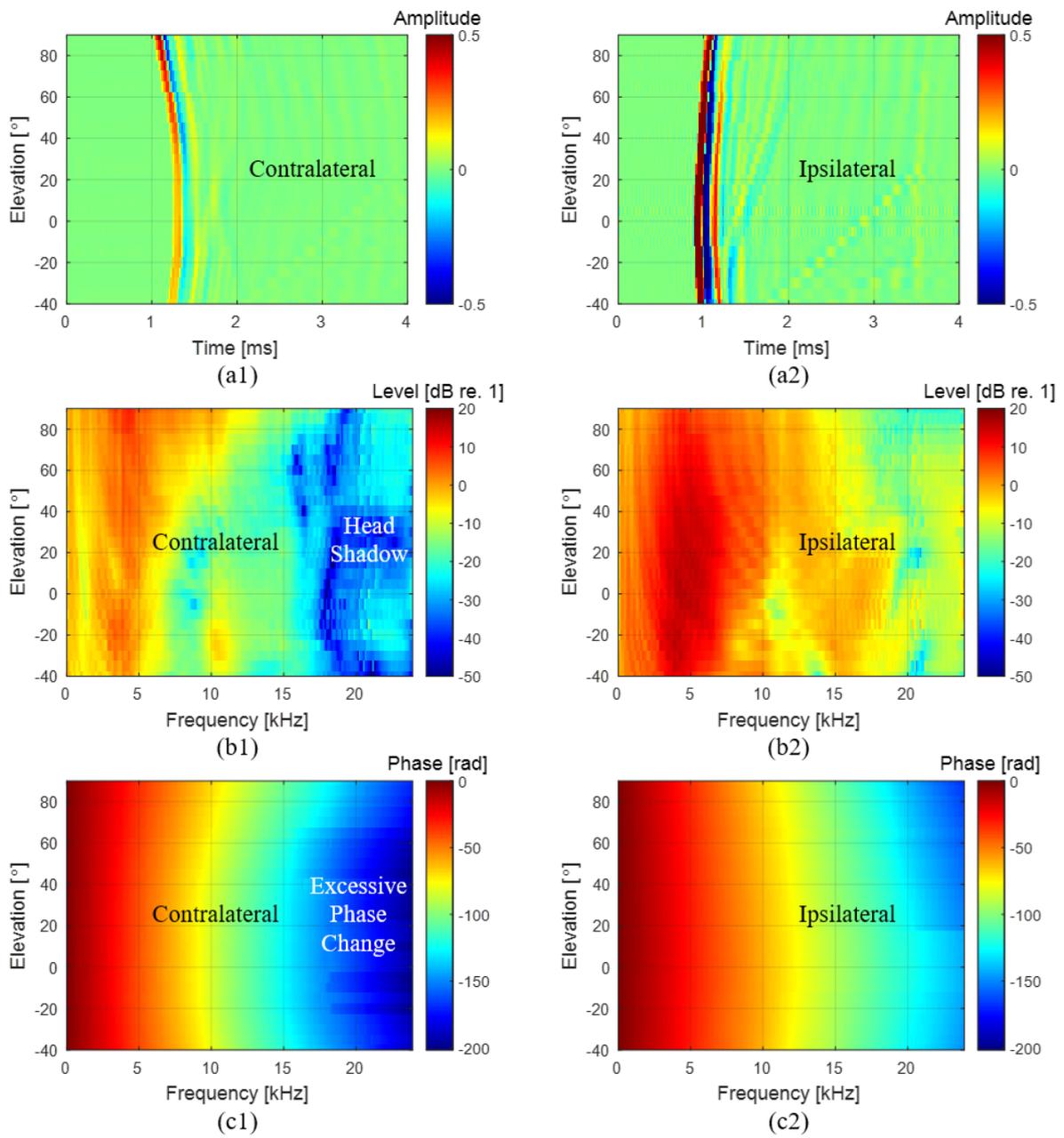

**Figure 3.18.** HRTFs at $\theta = 45°$, $H_{L,R}(45, \phi, f)$: (a1) left HRIRs; (a2) right HRIRs; (b1) magnitude of left HRTFs; (b2) magnitude of right HRTFs; (c1) phase of left HRTFs; and (c2) phase of right HRTFs.

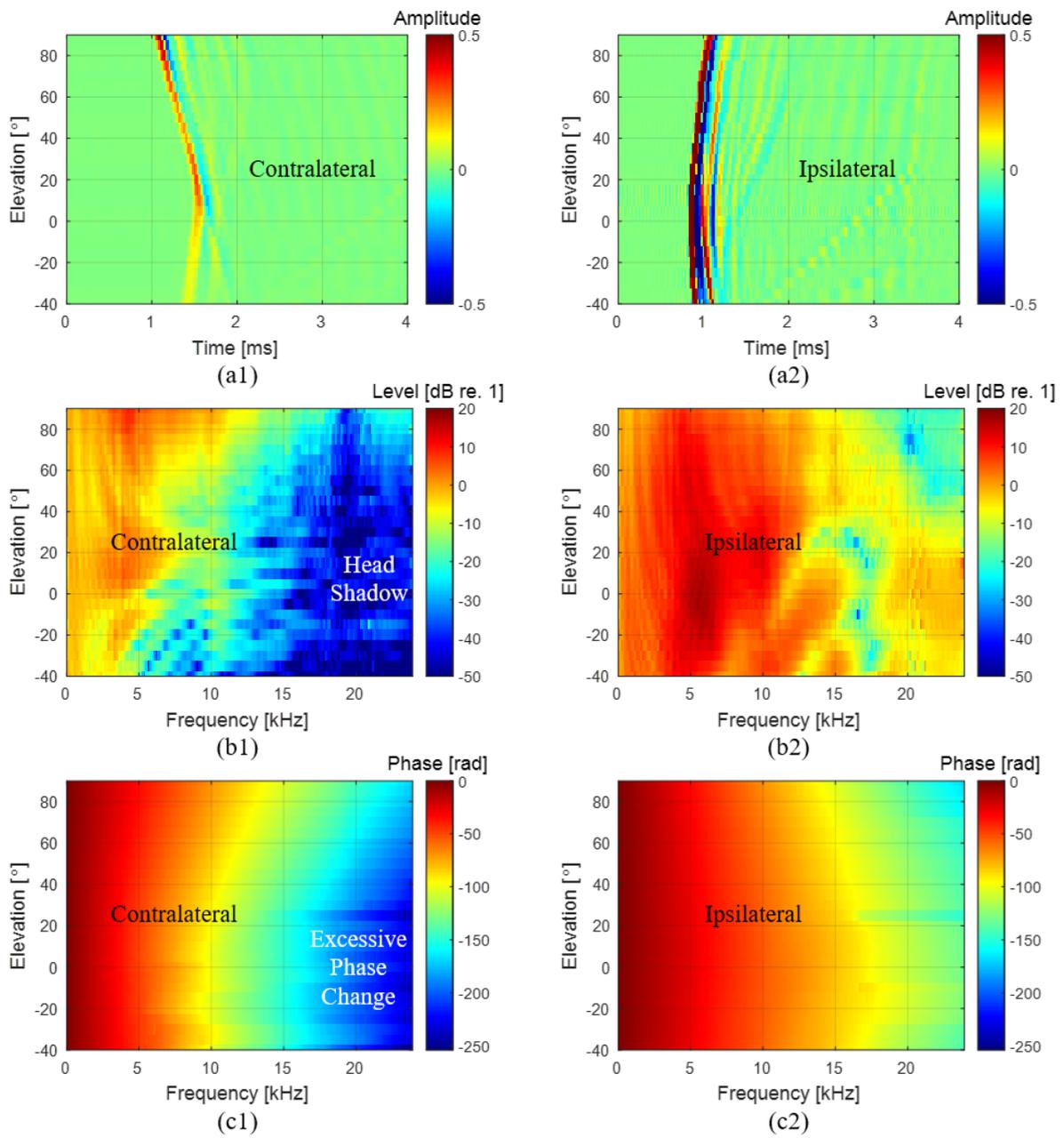

**Figure 3.19.** HRTFs at $\theta = 90°$, $H_{L,R}(90, \phi, f)$: (a1) left HRIRs; (a2) right HRIRs; (b1) magnitude of left HRTFs; (b2) magnitude of right HRTFs; (c1) phase of left HRTFs; and (c2) phase of right HRTFs.

Figs. 3.20–3.22 show left and right HRTF pairs as impulse responses, magnitude responses in dB scale, and phase responses in radians for the entire azimuth range when the elevation angles are 0°, 60°, and 90°, respectively. Looking at Figs. 3.20(a1) and (a2), when a sound source is in the ipsilateral direction in the horizontal plane ($\phi = 0°$), as for $\theta = -90°$ in Fig. 3.20(a1) or $\theta = +90°$ in Fig. 3.20(a2), the time delay of the ipsilateral HRIR is the shortest and there are many ripples after the maximum peak. On the other hand, when a sound source is in the contralateral direction, as in $\theta = +90°$ in Fig. 3.20(a1) or $\theta = -90°$ in Fig. 3.20(a2), it can be seen that the time delay of the contralateral HRIR is the longest and there are almost no ripples. The ripple change is due to the low-pass filtering effect of the head shadow. As shown in Figs. 3.20(b1) and (b2), when a sound source is located contralateral to the related ear at an azimuth of 90° for the left ear, the HRTF level decreases noticeably from 4 kHz because of the head shadow effect. As presented in Figs. 3.20(c1) and (c2), the phase change in the ipsilateral 90° direction is gentle, while the phase change in the contralateral 90° direction is steep. As shown in Figs. 3.21 and 3.22, with increasing elevation, the azimuth-dependent variations in HRIRs, magnitude responses, and phase responses decrease and become smooth.

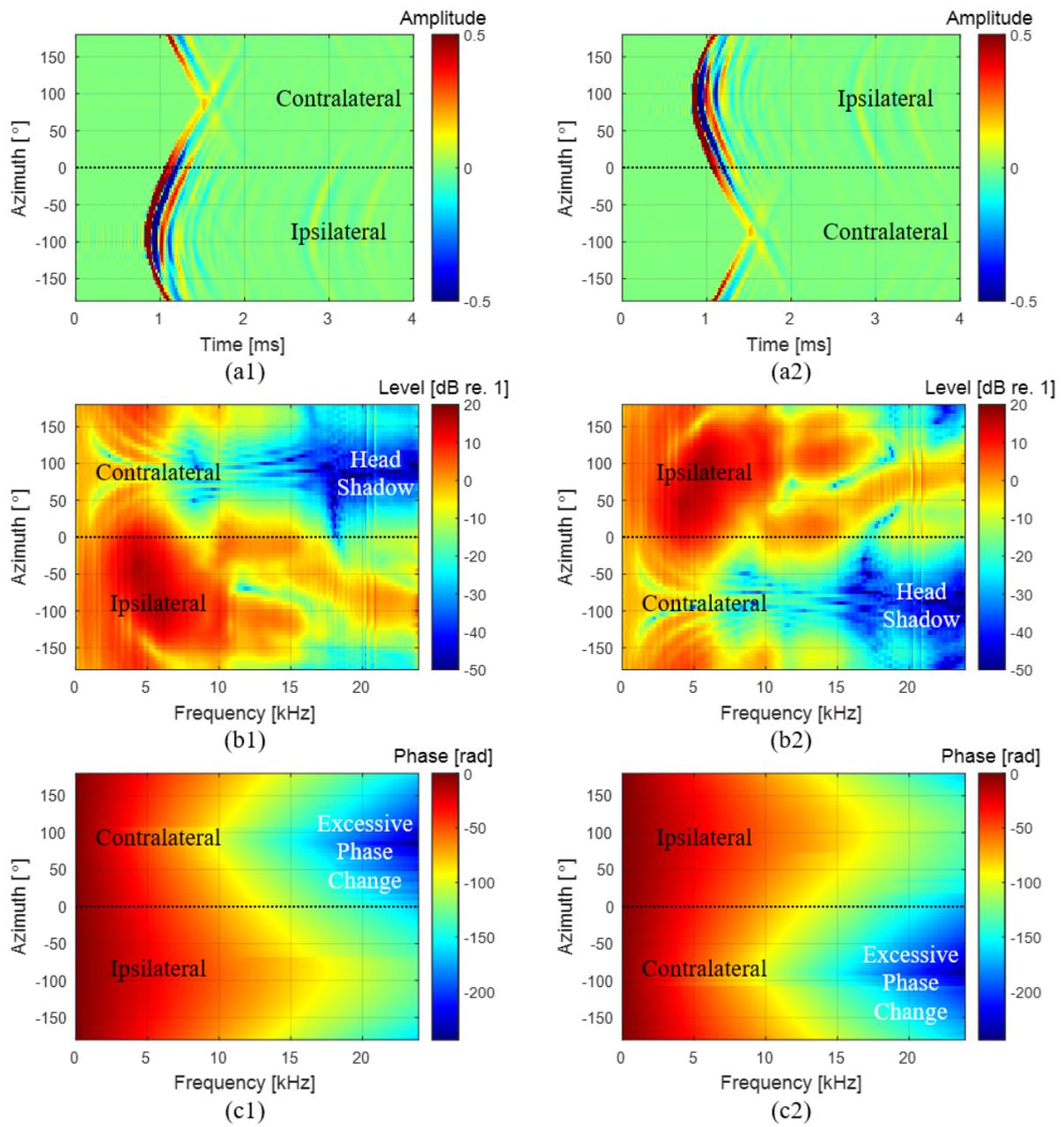

**Figure 3.20.** HRTFs at $\phi = 0°$, $H_{L,R}(\theta, 0, f)$: (a1) left HRIRs; (a2) right HRIRs; (b1) magnitude of left HRTFs; (b2) magnitude of right HRTFs; (c1) phase of left HRTFs; and (c2) phase of right HRTFs.

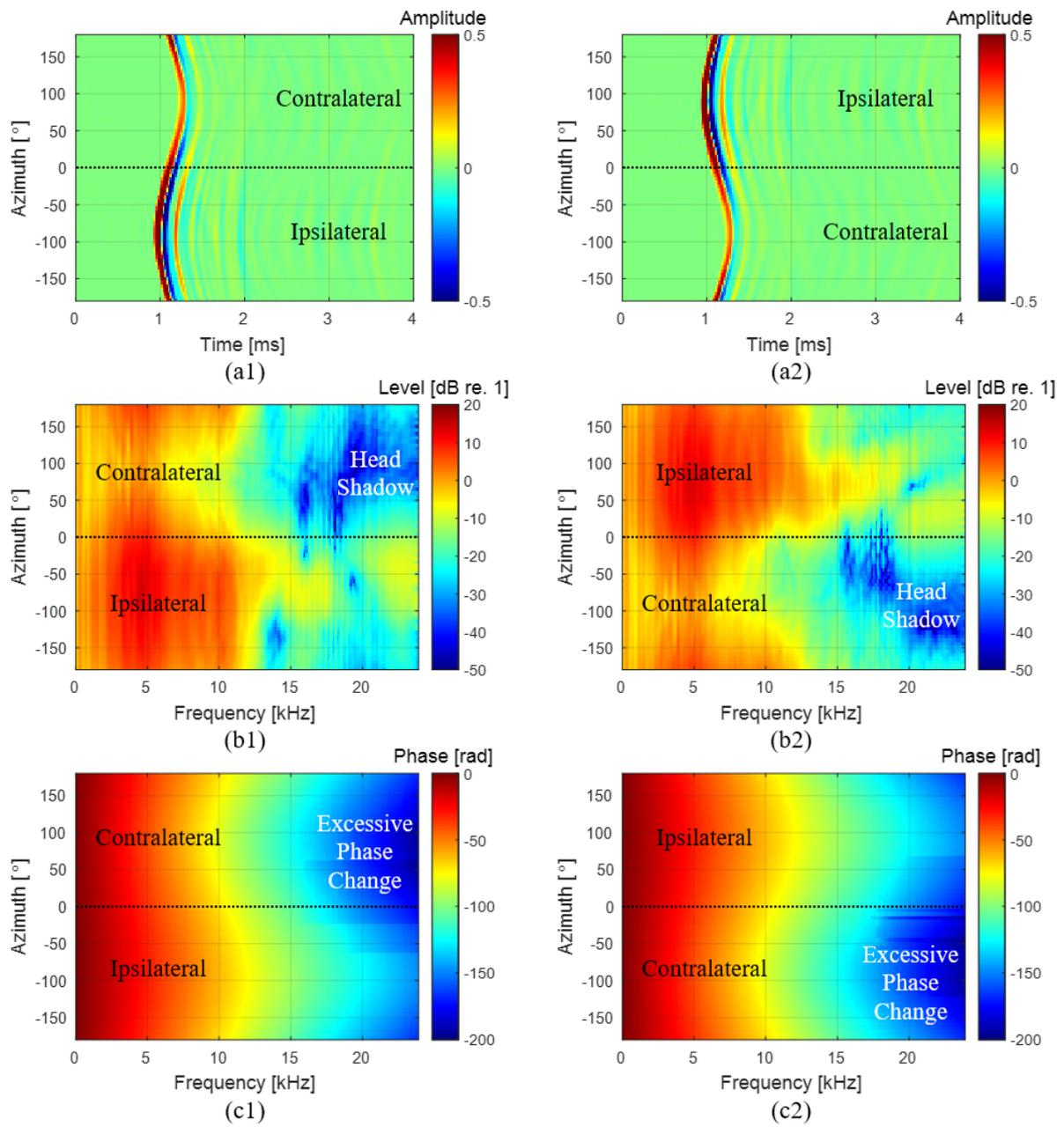

**Figure 3.21.** HRTFs at $\phi = 60°$, $H_{L,R}(\theta, 60, f)$: (a1) left HRIRs; (a2) right HRIRs; (b1) magnitude of left HRTFs; (b2) magnitude of right HRTFs; (c1) phase of left HRTFs; and (c2) phase of right HRTFs.

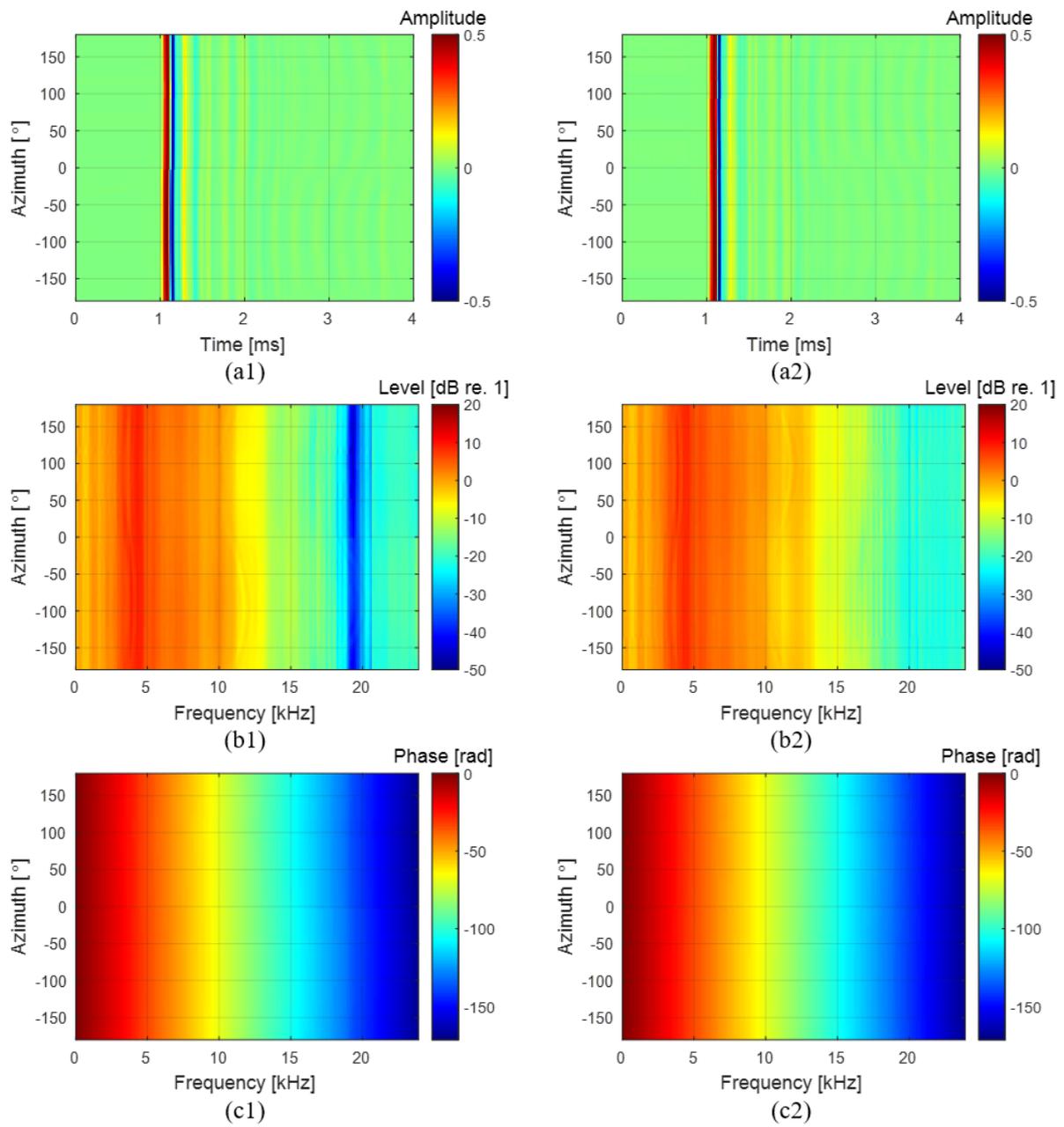

**Figure 3.22.** HRTFs at $\phi = 90°$, $H_{L,R}(\theta, 90, f)$: (a1) left HRIRs; (a2) right HRIRs; (b1) magnitude of left HRTFs; (b2) magnitude of right HRTFs; (c1) phase of left HRTFs; and (c2) phase of right HRTFs.

# Chapter 4. Input Feature and Dataset for BiSELD

## 4.1 Binaural Sound Localization Cues of HRTFs

Based on the measured HRTF database, various HRTF features can be analyzed to obtain useful information about the localization cues encoded in HRTFs. Psychoacoustic studies have shown that binaural sound localization cues for a single sound source include ITD, ILD, and SCs [140]. Hence, these cues were extracted from the measured HRTF database to analyze the binaural sound localization cues of HRTF. In addition, the directivity of the left and right HRTF pairs in the horizontal plane was obtained by defining HPD and representing it as a directional beam pattern. Through the insight into HRTF localization cues obtained in this section, acoustic input features for BiSELD model are proposed in the next section.

### 4.1.1 Interaural Time Difference (ITD)

As shown in Fig. 4.1, the time difference between the sound waves reaching the left and right ears plays an important role in BSSL. In the median plane, ITD is nearly zero because the paths from a sound source to both ears are approximately equal in length. However, if the sound source is off the median plane, the path lengths to both ears become different, so the ITD has a non-zero value.

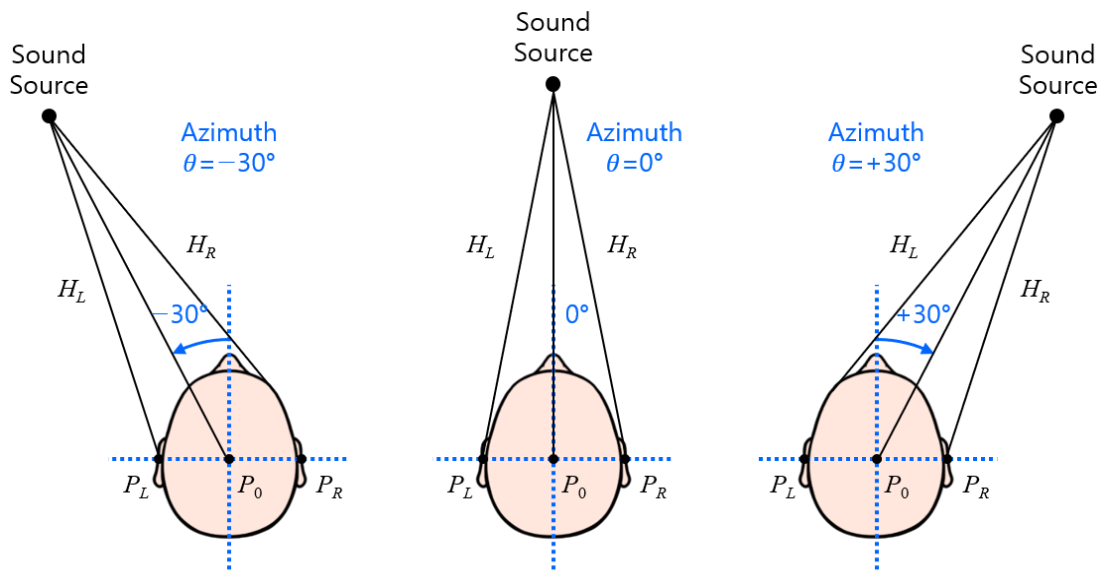

**Figure 4.1.** Illustration of the time difference between the left and right ears for a point sound source moving on the horizontal plane.

ITD can be estimated from a pair of left and right HRTFs. As stated in Ref. [111], interaural phase delay difference (ITD$_P$) is a dominant localization cue below 1.5 kHz, whereas interaural envelope delay difference (ITD$_E$) is useful for BSSL above 1.5 kHz. However, ITD$_P$ is complicated due to its frequency dependence, and ITD$_E$ cannot be directly analyzed because of its dependence on the signal type [111]. Hence, using the cross-correlation method, ITD was estimated based on the similarity between left and right HRIRs. The ITD of each direction is estimated as the time delay at which the normalized cross-correlation function of a pair of corresponding left and right HRIRs is maximized, as follows:

$$ITD(\theta,\phi) = \underset{\tau}{\mathrm{argmax}} \frac{\int_{-\infty}^{+\infty} h_L(\theta,\phi,t) h_R(\theta,\phi,t-\tau)\, dt}{\sqrt{\left[\int_{-\infty}^{+\infty} h_L^2(\theta,\phi,t)\, dt\right]\left[\int_{-\infty}^{+\infty} h_R^2(\theta,\phi,t)\, dt\right]}} \quad (4.1)$$

with $|\tau| \leq 1000$ μs,

where $\tau$ is time delay. Before the calculation of Eq. (4.1), a pair of HRIRs is subjected to low-pass filtering with a cutoff frequency of 1.5 kHz. The reason for this is that, above 1.5 kHz, when the head dimension is larger than the wavelength, the left and right phase difference exceeds $2\pi$, resulting in ambiguous ITD. To improve ITD resolution, the filtered HRIRs were upsampled four times to improve time resolution to about 5.2 μs.

Figs. 4.2(a) and (b) provide an ITD contour map and its cutting plots with azimuths from −180° to +180°, respectively, from the measured HRTF database. In Figs. 4.2(a) and (b), the ITDs are zero at 0° and 180° in azimuth, and increase gradually as sound source deviates from the median plane ($\theta = 0°$). As the sound source approaches lateral 90° directions ($\theta = -90°$ or $+90°$), the absolute value of ITD reaches its maximum. Around the lateral 90° directions, a large change in azimuth angle corresponds to a small change in ITD. Compared to the results for other elevations, the extent of ITD variation is maximized on the horizontal plane ($\phi = 0°$). As the source moves out of the horizontal plane, the range of ITD variation decreases.

Additionally, it is worth thinking about the impact on ITD of non-causality compensation mentioned in the previous chapter. Figs. 4.2(c) and (d) show an ITD contour map and its cutting plots obtained from the HRTF database without non-causality compensation. Since ipsilateral HRIR is a non-causal filter, its maximum peak appears later in the HRIR sequence. On the other hand, since contralateral HRIR is a causal filter, its maximum peak appears early in the HRIR sequence. Therefore, when the ITD is obtained from the non-causal HRIR pair, a sharp discontinuity occurs compared to that obtained from the causal HRIR pair. If a sound source is located in the front (near 0° azimuth), rear (near 180° azimuth), and top (elevation over 70°) of the head, the sound wave reaches the head center before it reaches both ears, so ITD discontinuity due to non-causality disappears.

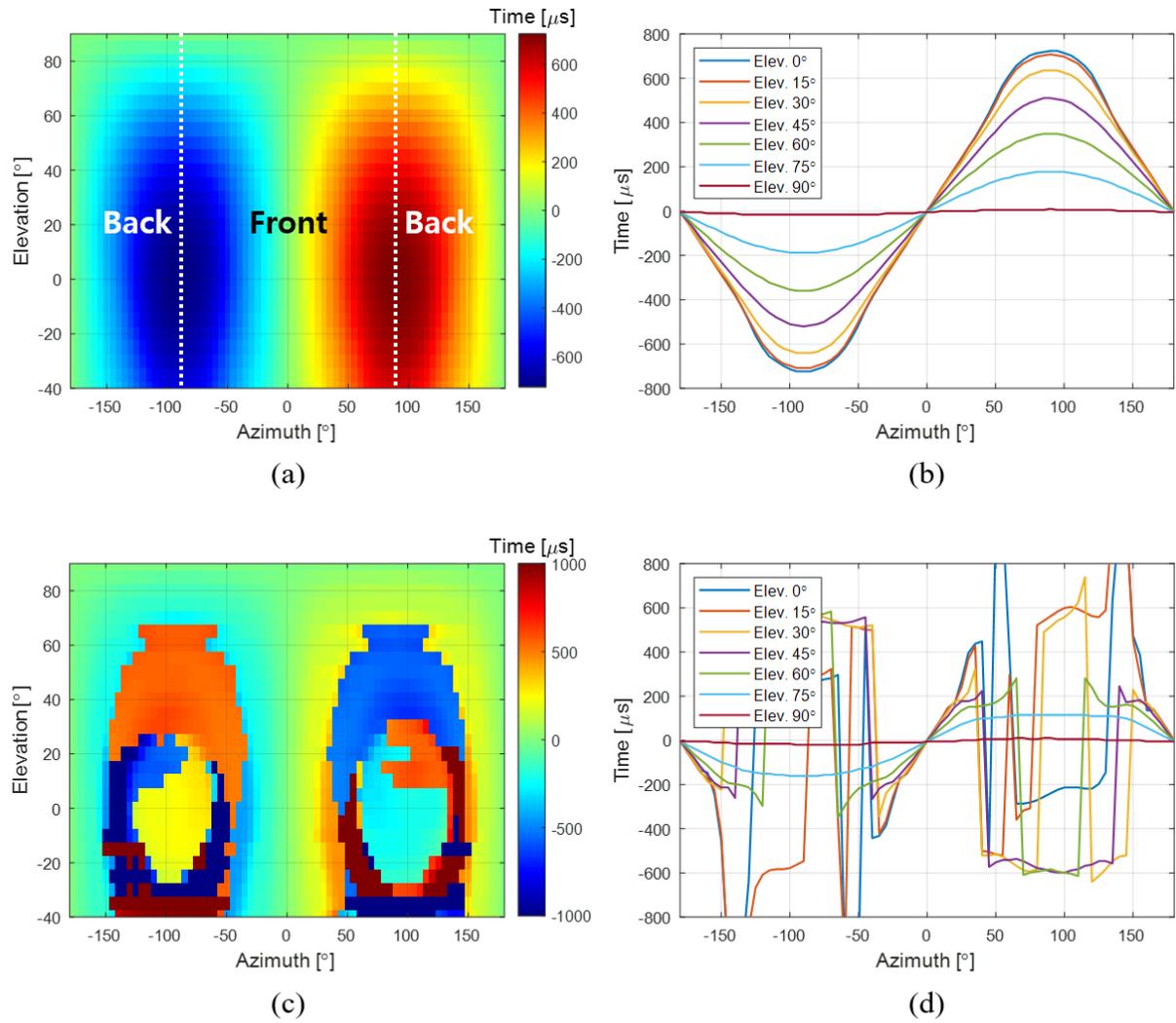

**Figure 4.2.** ITDs that change with azimuth and elevation: (a) ITD contour map of causal HRTF; (b) ITD plots of causal HRTF for seven elevation angles; (c) ITD contour map of non-causal HRTF; and (d) ITD plots of non-causal HRTF for seven elevation angles.

### 4.1.2 Interaural Level Difference (ILD)

ILD is another important localization cue above 1.5 kHz. As shown in Fig. 4.3, when a sound source deviates from the median plane, the sound pressure at the contralateral ear is attenuated, especially at high frequencies, due to the head shadow effect, while the sound pressure at the ipsilateral ear is amplified to some extent.

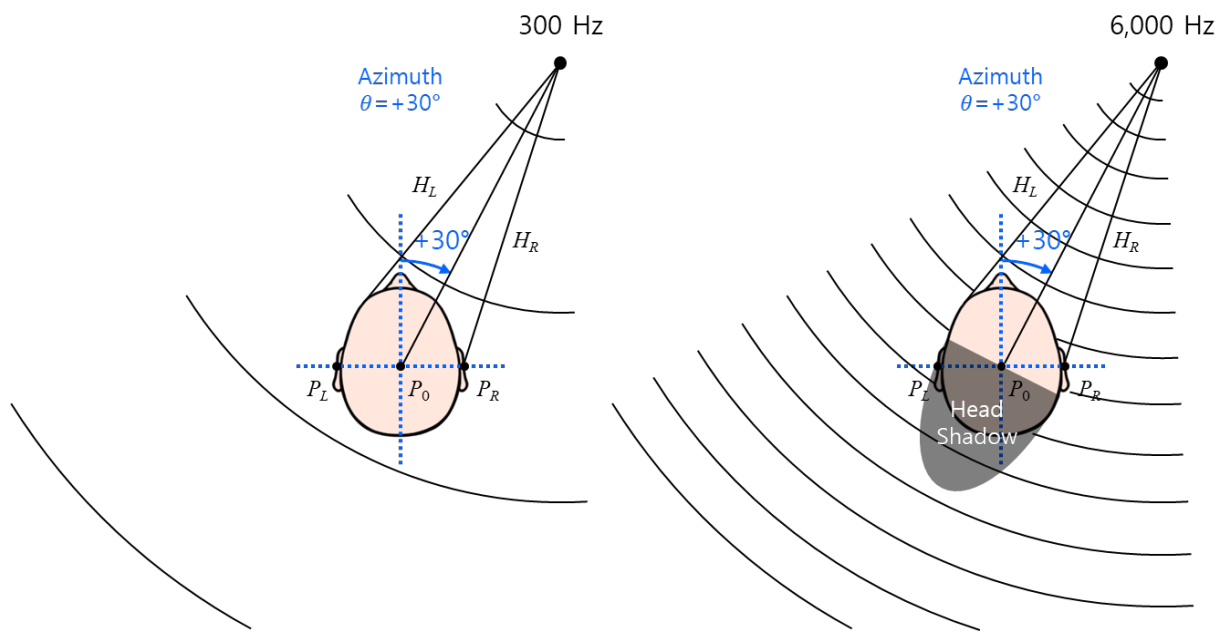

**Figure 4.3.** Illustration of the level difference between the left and right ears for 300 Hz and 6,000 Hz point sound sources propagating from a direction of $+30°$ azimuth.

In the far-field with distance $r$ far larger than the head radius $l$, narrowband ILD is defined as follows:

$$ILD_{narr}(\theta, \phi, f) = 20 \log_{10} \left| \frac{H_R(\theta, \phi, f)}{H_L(\theta, \phi, f)} \right|. \tag{4.2}$$

Here, each HRIR of 512 samples is zero-padded to 4,800 samples, and the frequency resolution is thus 10 Hz. ILD is a multivariate function of the distance, direction, and frequency of sound source, but far-field ILD is almost independent of the distance of sound source. Figs. 4.4(a) and (b) show a narrowband ILD contour map and its cutting plots with azimuth angles from $-180°$ to $+180°$ at several different frequencies. Comparing Figs. 4.2 and 4.4, unlike ITD, narrowband ILD seems to be an ambiguous localization cue because ILD does not change monotonically with respect to azimuth. The absolute value of ILD is nearly zero at the front and rear directions, and reaches maximum values around the lateral 90° directions. Fig. 4.4(b) shows that the ILD at 0.8 kHz has a small level and changes smoothly with respect to the azimuth, which indicates that the head shadow effect is negligible at low frequencies in the far-field. On the other hand, at high frequencies, especially above 5 kHz, the absolute value of ILD tends to increase and varies in a complex manner with respect to azimuth. At frequencies of 1.6 kHz and 3.2 kHz, the maximum ILD does not appear at the azimuth of 90°, where the sound source is located exactly opposite to the contralateral ear. This is because the sound pressure level at the contralateral ear is enhanced by the in-phase interference of multi-path diffracted sound waves around the head. The sound pressure enhancement at the contralateral ear reduces the difference in sound pressure level with the ipsilateral ear, causing ILD notches at −90° and +90° of azimuth. As frequency increases, the wavelength of sound waves decreases, and so the bandwidth of the ILD notch becomes narrower. Therefore, above 5.0 kHz, the ILD notch becomes gradually insignificant, as shown in Figs. 4.4(a) and (b). Above 3.2 kHz, the ILD curves are asymmetric with respect to ±90°. This is caused by the front-back asymmetry of the head shape and ear position, and the diffraction effect of the pinna. The front-back difference in the ILD curves is regarded as a localization clue to solve the front-back confusion in BSSL.

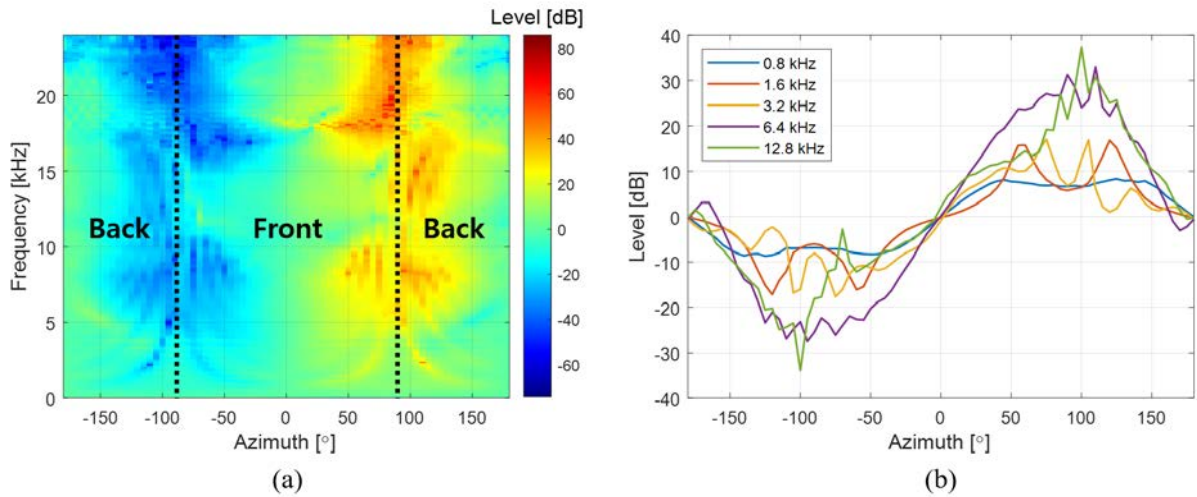

**Figure 4.4.** Narrowband ILDs on the horizontal plane: (a) ILD contour map for all frequencies and (b) ILD plots for five frequencies.

Although wideband ILD is not directly used as a binaural sound localization cue, it can help characterize the features of the HRTF database. Wideband ILD can be obtained by integrating the entire audio frequency range, as in [151]:

$$ILD_{wide}(\theta, \phi) = 10 \log_{10} \left[ \frac{\int_{f_L}^{f_H} |H_R(\theta, \phi, f)|^2 \, df}{\int_{f_L}^{f_H} |H_L(\theta, \phi, f)|^2 \, df} \right] \quad (4.3)$$

with $f_L = 20$ Hz, $f_H = 20$ $k$Hz.

Figs. 4.5(a) and (b) show wideband ILD contour map and its cutting plots with azimuths from −180° to +180° and elevations from −40° to +90°. Compared with Fig. 4.4, the variation in the wideband ILD according to azimuth is smaller than that in the high frequency narrowband ILD. The variation range of wideband ILD is maximum in the horizontal plane ($\phi = 0°$) and decreases as the elevation angle of the sound source increases. At azimuths of −90° and +90°, the ILD notch appears from −40° to +30° of elevation. At elevations outside this range, variations of wideband ILD with azimuth become smooth, as shown in the ITD curves. In addition, at elevation angles below 45°, the broadband ILD curves exhibit front-back asymmetry, which can be used as a localization cue to solve the front-back confusion in BSSL.

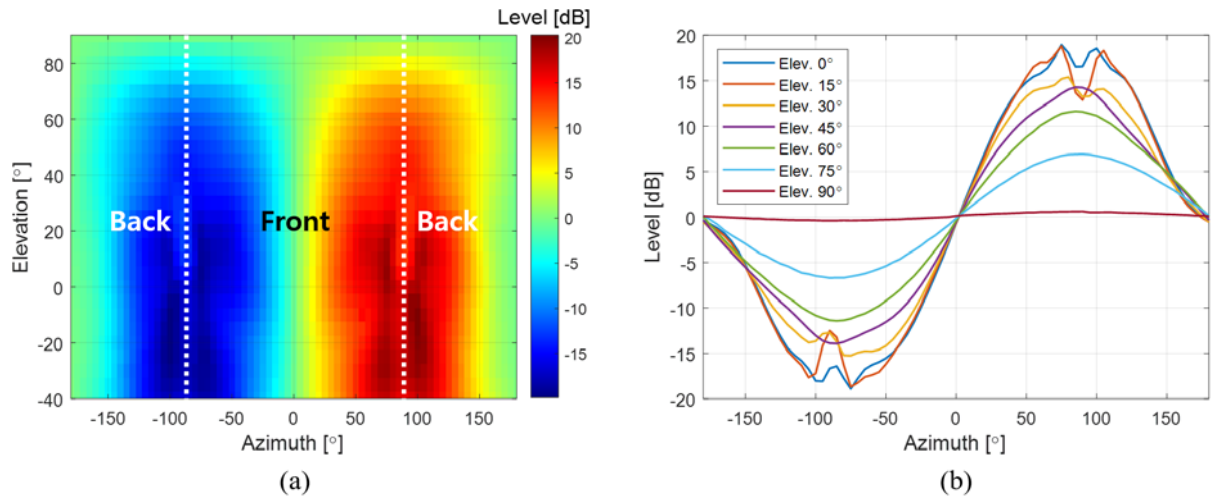

**Figure 4.5.** Wideband ILDs that change with azimuth and elevation: (a) ILD contour map and (b) ILD plots for seven frequencies.

4.1.3 Spectral Cue (SC)

Many researches have suggested that spectral cues above 5 kHz resulting from reflection and diffraction from the pinna provide useful information for vertical localization [111]. Unlike binaural cues such as ITD and ILD, SC is a kind of monaural cue as shown in Fig. 4.6.

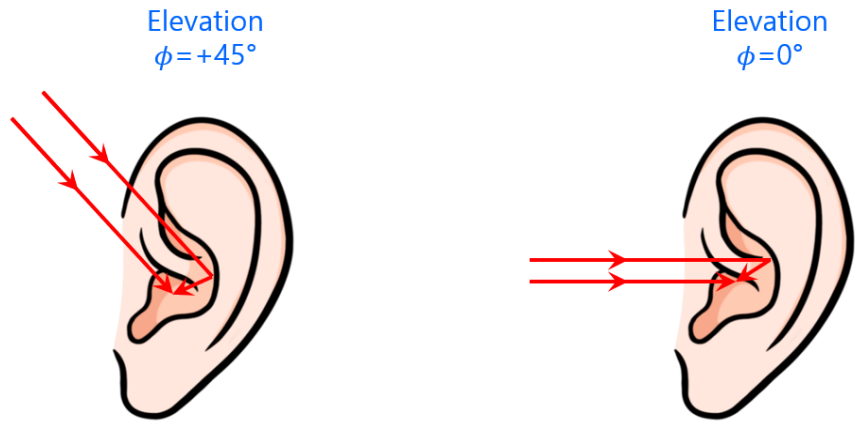

**Figure 4.6.** Illustration of the pinna interacting with a point sound source propagating from a direction of +45° or 0° elevation.

The spectral features of the pinna can be analyzed via the measured HRTFs. The high-frequency peaks and notches of the HRTFs are reported as being generated in the pinna [112]. Specifically, it is considered that notches are generated in the cavity of the concha [152], whereas peaks are generated by the resonance of the pinna [153]. Since first reported in Ref. [152], the elevation dependence of the pinna notch frequency has been considered an important vertical localization cue. To analyze SC in the measured HRTF database, the patterns of peaks and notches according to elevation angles in the median plane were explored. For this purpose, pinna-related transfer functions (PRTFs), i.e., acoustic transfer functions related only to the pinna, were extracted from the measured HRTFs. Since the pinna response reaches the entrance of the ear canal before the torso response, the pinna influence is considered to be involved in the early part of HRIR. Thus, it is assumed that information on spectral peaks and notches of the pinna is included in the early part of HRIR. Since the first 1 ms of HRIR contains information about the spectral peaks and notches for the pinna [112], HRIRs in the median plane were clipped by a 2 ms long Hanning window centered on the maximum sample of each HRIR, leaving only the pinna effect. Then, the windowed HRIRs were Fourier transformed to obtain PRTFs [154]. To analyze SCs, the local maxima and minima of PRTFs were searched according to frequency and elevation angle to obtain the distribution of peaks and notches.

Fig. 4.7(a) shows the SC distribution of the right PRTFs on the median plane, with elevations from −40° to 220°. The peak at about 4 kHz is almost constant and is thus almost independent of the elevation angle of the sound source. This peak is called the first peak, P1, and results from the modal response of the ear canal in the depth direction. The second peak, P2, is formed around 10 kHz, and there is no significant change with elevation angle. Since the P1 and P2 frequencies are almost independent of elevation angle, vertical localization cues are not included in P1 and P2. On the other hand, the first notch, N1, and the second notch, N2, are highly dependent on the elevation angle of sound source. In addition, notches are deep for sound sources close to the horizontal plane and shallow for sources far from it. This elevation angle dependency of the N1 and N2 frequencies is thought to be a vital cue for vertical localization. The N1 frequency is observed at about 8 kHz and changes from 8 kHz to 10 kHz when elevation varies from −40° to 90°. The N2 frequency varies greatly as a sound source moves from −40° to 90° in elevation. This phenomenon explains why two notches are needed for vertical localization. If the N1 and N2 frequencies were to change monotonically with the elevation angle of a sound source, two notches would not be required to determine the elevation angle. However, since the relationship between the notch frequencies and the elevation angle is not simple, at least two notches are needed to determine the elevation angle of a sound source. Notably, the front-back asymmetry in the PRTF pattern provides an important localization clue to solve the front-back confusion of BSSL.

In addition, it is worth considering about the impact on SC of the microphone direction of OTF measurement, as mentioned in the previous chapter. Fig. 4.7(b) shows the SC distribution of right PRTFs in the median plane, which is based on OTF measured by a 90° off-axis microphone. As shown in Fig. 3.12(c), when OTF is measured with a 90° off-axis microphone, the high-frequency level decreases compared to when OTF is measured with a 0° on-axis microphone. Hence, the 90°-OTF-based HRTF has a higher level in the high frequency region than the 0°-OTF-based HRTF. The high-frequency level increase is confirmed in the PRTF shown in Fig. 4.7. Moreover, as shown in Fig. 3.12(d), various peaks and notches appear in the level difference between the 0° on-axis and the 90° off-axis OTFs. These relative peaks and notches mean that the 90° off-axis OTF may affect the positions of the HRTF peaks and notches. In particular, the dominant peaks of 8 kHz and 12 kHz in the level difference affect the notch positions of HRTF. Compared to N1 and N2 in Fig. 4.7(a), N1 and N2 in Fig. 4.7(b) are concentrated around 8 kHz and 12 kHz, respectively. Since the elevation dependency of the notch frequency is regarded as an important vertical localization cue, a change in notch frequency of HRTF may confuse the elevation estimation.

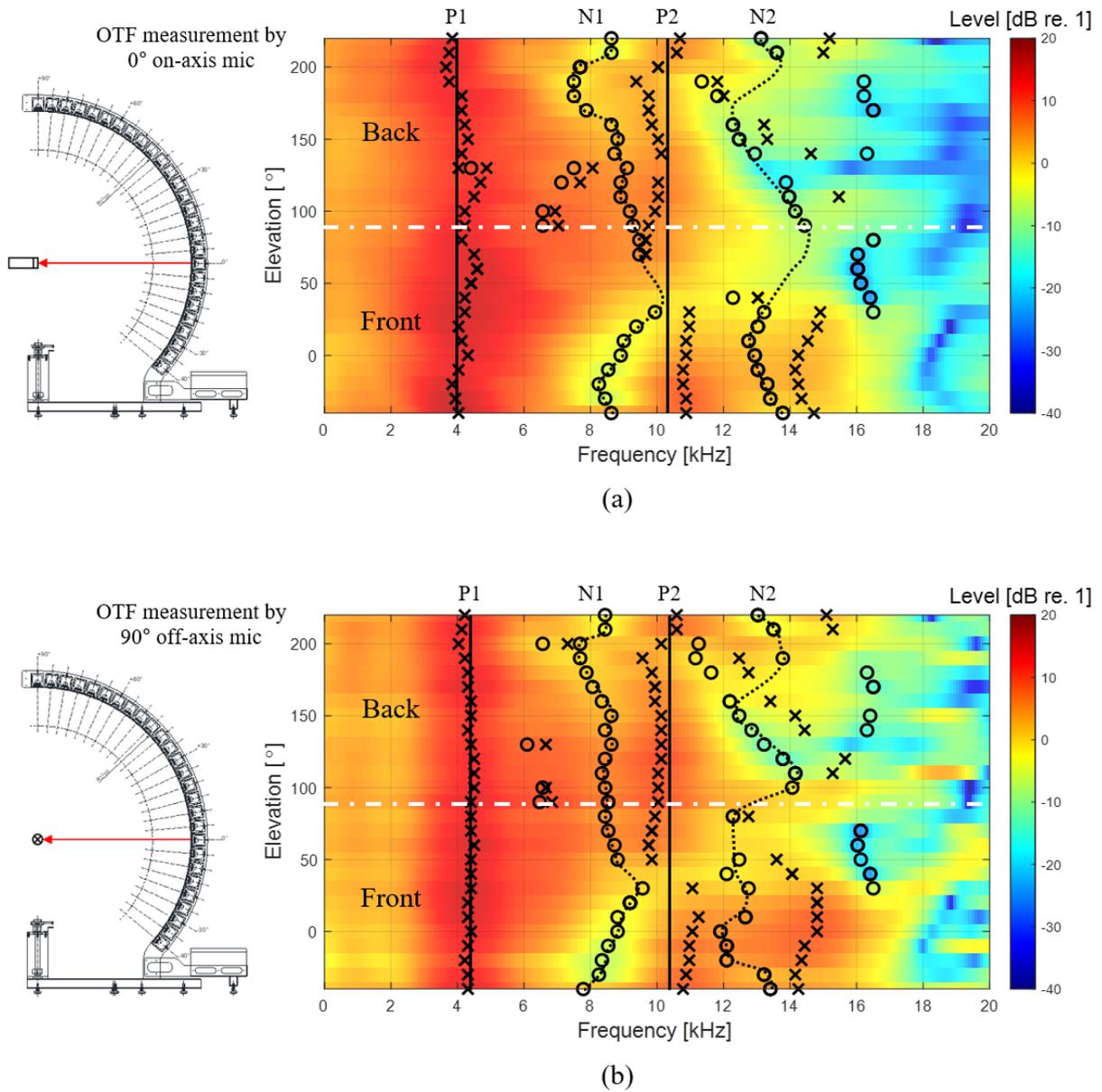

**Figure 4.7.** SC distribution of right PRTFs on the median plane with elevation angles from −40° to 220° (**X**: peaks, **O**: notches): (a) SC distribution based on the OTF measured by a 0° on-axis microphone and (b) SC distribution based on the OTF measured by a 90° off-axis microphone.

### 4.1.4 Horizontal Plane Directivity (HPD)

As mentioned above, variations of ITD and ILD are maximized in the horizontal plane compared to other elevations, and spectral notches become more pronounced for sound sources close to the horizontal plane. In general, human ability of localization is maximized in front of the horizontal plane [111]. Therefore, on the horizontal plane, the sensitivity patterns of the subject's left and right ears, according to azimuth, were investigated by extracting HPD from the measured HRTFs for each frequency. The beam pattern of HPD was defined by normalizing all HRTFs on the horizontal plane by the front HRTF at $\theta = 0°$ and $\phi = 0°$, as follows:

$$HPD_{L,R}(\theta, f) = 20 \log_{10} \left| \frac{H_{L,R}(\theta, 0, f)}{H_{L,R}(0, 0, f)} \right|. \qquad (4.4)$$

Fig. 4.8 shows the left and right HPD beam patterns as a function of the azimuth of a sound source at different frequencies ($f$ = 0.75, 1.5, 3.0, 6.0, and 12.0 kHz). Due to the front HRTF normalization, the HPD beam pattern at 0° is always 0 dB over the entire frequency range. Since the shape of the head is close to left-right symmetry, this symmetry is also observed in the HPD beam patterns. As in the narrowband ILD curves, it can be seen that level difference between the left and right becomes larger as frequency increases for a sound source located on the side. The beam pattern at 0.75 kHz shows a smooth change between –5 dB and +5 dB. At 1.5 kHz and 3 kHz, directivity forms as the main lobe gradually narrows in the ipsilateral direction, while the beam pattern level drops to –15 dB due to the head shadow effect in the contralateral direction. The beam pattern level at 6 kHz increases by about 10 dB in the ipsilateral direction, but drops by about –20 dB in the contralateral direction. At 12 kHz, a large dip appears at 70° in the ipsilateral direction, and the beam pattern level drops below –20 dB in the contralateral direction.

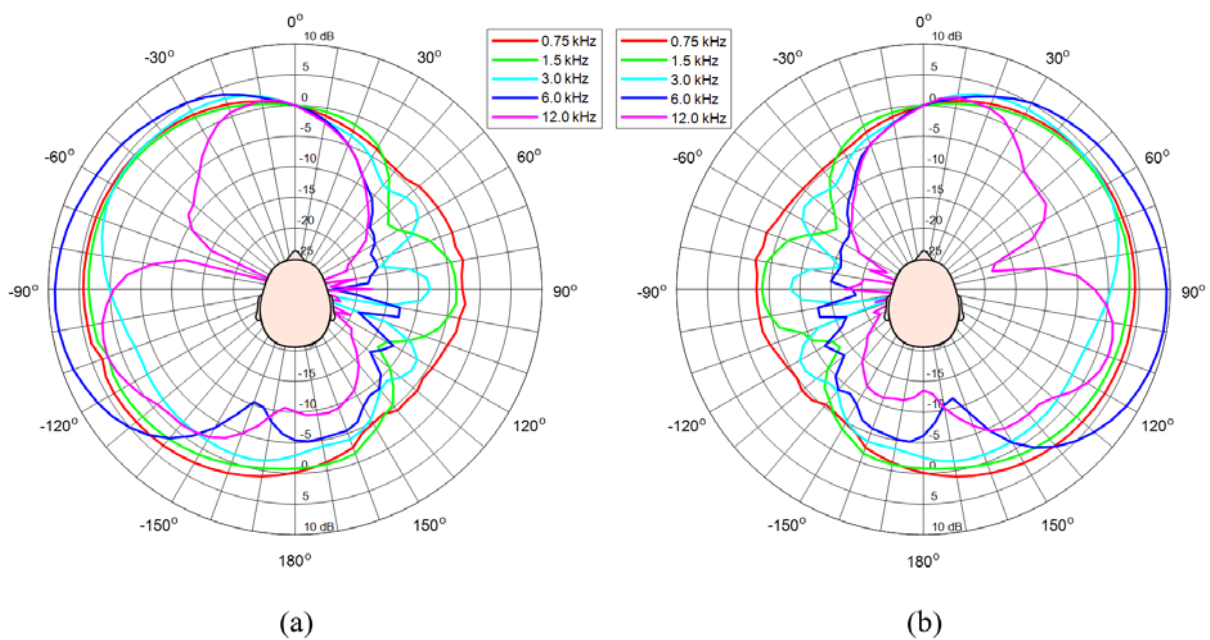

**Figure 4.8.** Beam patterns of HPD: (a) left ear HPD beam pattern and (b) right ear HPD beam pattern.

## 4.2 Binaural Time-Frequency Feature (BTFF)

In deep learning on audio signals, the key to feature engineering is to represent the input in a simpler way, making the latent manifold of data smoother, simpler, and more structured. Good features can make the model more robust to variations in audio data. In the BiSELD task, domain knowledge can be used to create features that are tailored to the task, which can significantly improve performance. Considering the human auditory process, as auditory signals travel from the cochlea to the brain, perceptual features are extracted and finally encoded at the top level into categorical representations. In the process of extraction, the features include onset, offset, periodicity, AM, FM, ITD, and ILD [18]. Together with the binaural localization cues discussed above, these time-frequency features form the auditory information for the detection and localization of surrounding sound sources. Based on the analysis of HRTF localization cues, BTFF, a binaural input feature for BiSELDnet, is proposed in this section. This audio input feature consists of mel-spectrogram (MS) and velocity-map (V-map) for sound event detection, interaural-time-difference-map (ITD-map) and interaural-level-difference-map (ILD-map) for azimuth estimation, and spectral-cue-map (SC-map) for elevation estimation.

The mel scale (derived from the word *melody*) is a perceptual scale of pitches judged by humans to be equal in distance from one another. It is designed to better match human auditory perception of sound, and is used in various fields, especially in audio processing, music, and speech recognition, to account for the way our ears perceive differences in pitch and frequency. The reference point between mel scale and linear scale is defined by assigning a perceptual pitch of 1,000 mels to a 1,000 Hz, 40 dB above the listener's threshold. Above 500 Hz, increasingly wide intervals are judged by humans to produce equal pitch increments. A formula to convert $f$ Hz into $m$ mels is as follows [155]:

$$m = 1127 \ln\left(1 + \frac{f}{700}\right). \quad (4.5)$$

To match human perception, all sub-features of BTFF below are finally converted to mel scale using Eq. (4.5). Moreover, this mel scale mapping reduces the size of the frequency dimension of the input feature, thereby reducing the number of model parameters.

### 4.2.1 Mel-spectrogram (MS) and Velocity-map (V-map) for Detection

- Mel-spectrogram (MS)

MS is suitable for sound event detection because it expresses the periodicity, AM, and FM of sound event, and it is useful for detecting the start and end points of sound event because it includes onset and offset information. The mel scale of MS is closer to the perceptual scale of human than the linear scale of spectrogram. So, the gap between the harmonic components of MS remains the same even when a pitch is shifted to a neighboring frequency band. Therefore, MS is suitable for applying convolutional layer, which learns local shift-invariant patterns from input feature. In addition, MS can be used as an aid to avoid front-back confusion when estimating azimuth. The time frame of a left or right input signal is as follows:

$$p_{L,R}(n:m) = p_{L,R}(n) \cdot w(m-n), \tag{4.6}$$

where $p_{L,R}(n)$ is a left or right input signal; $w(n)$ is a time window; $n$ is a time index; and $m$ is a time frame index. The Fourier transform of $p_{L,R}(n:m)$ and the dB-scaled magnitude of the Fourier transform are as follows:

$$P_{L,R}(m,k) = \frac{1}{N} \sum_{n=0}^{N-1} p_{L,R}(n:m) \, e^{-i\frac{2\pi}{N}kn}, \tag{4.7}$$

$$S_{L,R}(m,k) = 20 \log_{10} |P_{L,R}(m,k)|, \tag{4.8}$$

where $k$ is a frequency index and $N$ is the length of $p_{L,R}(n:m)$. The left or right MS is defined as follows:

$$MS_{L,R}(m,b) = Mel \begin{bmatrix} S_{L,R}(1,1) & S_{L,R}(1,2) & \cdots & S_{L,R}(1,K) \\ S_{L,R}(2,1) & S_{L,R}(2,2) & \cdots & S_{L,R}(2,K) \\ \vdots & \vdots & \ddots & \vdots \\ S_{L,R}(M,1) & S_{L,R}(M,2) & \cdots & S_{L,R}(M,K) \end{bmatrix}, \tag{4.9}$$

where $M$ is the max time frame index; $K$ is the max frequency index; $Mel[\,]$ refers to the mel scale mapping; and $b$ is the index of mel bins.

- Velocity-map (V-map)

The human brain tends to respond sensitively to the rate of change over time. Also, since most sound events are transient, it is effective to reflect the rate of change in binaural feature to distinguish it from the other similar sound events. V-map helps to detect sound events because it represents the amount of change in a spectrogram-based feature over time [156,157]. The left or right V-map is composed of forward difference (at $m = 1$), central difference (at $2 \leq m \leq M - 1$), and backward difference (at $m = M$) as follows:

$$V\text{-}map_{L,R}(m,b) = Mel \begin{bmatrix} V_{L,R}(1,1) & V_{L,R}(1,2) & \cdots & V_{L,R}(1,K) \\ V_{L,R}(2,1) & V_{L,R}(2,2) & \cdots & V_{L,R}(2,K) \\ \vdots & \vdots & \ddots & \vdots \\ V_{L,R}(M,1) & V_{L,R}(M,2) & \cdots & V_{L,R}(M,K) \end{bmatrix}, \tag{4.10}$$

$$V_{L,R}(m,k) = \begin{cases} S_{L,R}(m+1,k) - S_{L,R}(m,k); & (m=1), \\ \{S_{L,R}(m+1,k) - S_{L,R}(m-1,k)\}/2; & (2 \leq m \leq M-1), \\ S_{L,R}(m,k) - S_{L,R}(m-1,k); & (m=M), \end{cases} \tag{4.11}$$

where $V_{L,R}(m,k)$ is a V-map element at $m$ and $k$.

### 4.2.2 ITD-map and ILD-map for Azimuth Estimation

- Interaural-Time-Difference-map (ITD-map)

ITD-map is proposed inspired by the ITD process of the MSO in the brainstem. Humans detect the azimuth of sound based on time-of-arrival differences by comparing neuron firing times in the MSO [100,102]. As discussed in Subsection 4.1.1, ITD is the dominant localization cue below 1.5 kHz. Above 1.5 kHz, the absolute value of interaural phase difference exceeds $2\pi$ because the head size becomes larger than the wavelength. Thus, only the interaural phase delay difference below 1.5 kHz can be a useful binaural feature for azimuth estimation. For phase delay to be used, short-time phase spectrum has to be unwrapped to produce a continuous estimate. However, short-time phase spectrum is not commonly used due to computational issues in phase unwrapping, caused by the effects of window truncation and spectral nulls [158,159]. In this subsection, a new phase-related feature, ITD-map, is derived from the magnitude components of the binaural complex spectra, without the need for phase unwrapping. The phase delay difference between the left and right input signals can be expressed as follows:

$$\Delta\tau(m,k) = \tau_L(m,k) - \tau_R(m,k) = \left\{-\frac{\psi_L(m,k)}{\omega}\right\} - \left\{-\frac{\psi_R(m,k)}{\omega}\right\}$$
$$= \frac{1}{\omega}\{\psi_R(m,k) - \psi_L(m,k)\}, \tag{4.12}$$

where $\tau_L(m,k)$ and $\tau_R(m,k)$ are the left and right phase delays at $m$ and $k$, respectively; $\psi_L(m,k)$ and $\psi_R(m,k)$ are the left and right unwrapped phases at $m$ and $k$, respectively; and $\omega$ is the angular frequency at $k$. Also, $P_L(m,k)$ and $P_R(m,k)$ can be represented as follows:

$$P_L(m,k) = |P_L(m,k)|e^{i\psi_L(m,k)} \text{ with } \psi_L(m,k) = \angle P_L(m,k), \tag{4.13}$$

$$P_R(m,k) = |P_R(m,k)|e^{i\psi_R(m,k)} \text{ with } \psi_R(m,k) = \angle P_R(m,k), \tag{4.14}$$

where $|\ |$ and $\angle$ are magnitude and phase of spectrum, respectively. When Eq. (4.14) is divided by Eq. (4.13), it is represented as follows:

$$\frac{P_R(m,k)}{P_L(m,k)} = \frac{|P_R(m,k)|}{|P_L(m,k)|}e^{i\{\psi_R(m,k)-\psi_L(m,k)\}}. \tag{4.15}$$

Applying log to both sides of Eq. (4.15) gives:

$$ln\frac{P_R(m,k)}{P_L(m,k)} = ln\frac{|P_R(m,k)|}{|P_L(m,k)|} + i\{\psi_R(m,k) - \psi_L(m,k)\}. \tag{4.16}$$

The right side of Eq. (4.16) can be expressed as follows:

$$\psi_R(m,k) - \psi_L(m,k) = Im\left[ln\frac{P_R(m,k)}{P_L(m,k)}\right], \tag{4.17}$$

where $Im[\ ]$ is the imaginary part of a complex number. Substituting Eq. (4.17) into Eq. (4.12) gives:

$$\Delta\tau(m,k) = \frac{1}{\omega} Im\left[\ln \frac{P_R(m,k)}{P_L(m,k)}\right]. \tag{4.18}$$

In addition, Eq. (4.15) can be expressed as follows:

$$\frac{P_R(m,k)}{P_L(m,k)} = \frac{\{P_{R.re}(m,k) + iP_{R.im}(m,k)\} \cdot \{P_{L.re}(m,k) - iP_{L.im}(m,k)\}}{|P_L(m,k)|^2}, \tag{4.19}$$

where $P_{L.re}(m,k)$ and $P_{L.im}(m,k)$ are the real and imaginary parts of $P_L(m,k)$, respectively; and $P_{R.re}(m,k)$ and $P_{R.im}(m,k)$ are the real and imaginary parts of $P_R(m,k)$, respectively. Then, Eq. (4.19) can be simply expressed as:

$$\frac{P_R(m,k)}{P_L(m,k)} = \frac{Z_{re}(m,k) + iZ_{im}(m,k)}{|P_L(m,k)|^2} = \frac{|Z(m,k)|e^{i\psi_Z(m,k)}}{|P_L(m,k)|^2}, \tag{4.20}$$

where

$$Z_{re}(m,k) = P_{L.re}(m,k) \cdot P_{R.re}(m,k) + P_{L.im}(m,k) \cdot P_{R.im}(m,k), \tag{4.21}$$

$$Z_{im}(m,k) = P_{L.re}(m,k) \cdot P_{R.im}(m,k) - P_{L.im}(m,k) \cdot P_{R.re}(m,k), \tag{4.22}$$

$$Z(m,k) = Z_{re}(m,k) + iZ_{im}(m,k) = |Z(m,k)|e^{i\psi_Z(m,k)}, \tag{4.23}$$

$$\psi_Z(m,k) = \angle Z(m,k). \tag{4.24}$$

Applying log to Eq. (4.20) and taking its imaginary part gives:

$$Im\left[\ln \frac{P_R(m,k)}{P_L(m,k)}\right] = Im[\ln|Z(m,k)| + i\psi_Z(m,k) - 2\ln|P_L(m,k)|] = \psi_Z(m,k). \tag{4.25}$$

Substituting Eq. (4.25) into Eq. (4.18) gives:

$$\begin{aligned}\Delta\tau(m,k) &= \frac{1}{\omega}\psi_Z(m,k) = \frac{1}{\omega}tan^{-1}\left\{\frac{Z_{im}(m,k)}{Z_{re}(m,k)}\right\} \\ &= \frac{1}{\omega}tan^{-1}\left\{\frac{P_{L.re}(m,k) \cdot P_{R.im}(m,k) - P_{L.im}(m,k) \cdot P_{R.re}(m,k)}{P_{L.re}(m,k) \cdot P_{R.re}(m,k) + P_{L.im}(m,k) \cdot P_{R.im}(m,k)}\right\}.\end{aligned} \tag{4.26}$$

Thus, the ITD-map below 1.5 kHz is defined as follows:

$$ITD\text{-}map(m,b) = Mel\begin{bmatrix} \Delta\tau(1,1) & \Delta\tau(1,2) & \cdots & \Delta\tau(1,k_{1500}) \\ \Delta\tau(2,1) & \Delta\tau(2,2) & \cdots & \Delta\tau(2,k_{1500}) \\ \vdots & \vdots & \ddots & \vdots \\ \Delta\tau(M,1) & \Delta\tau(M,2) & \cdots & \Delta\tau(M,k_{1500}) \end{bmatrix}, \tag{4.27}$$

where $k_{1500}$ is the frequency index corresponding to 1.5 kHz. Therefore, the feature including ITD information can be directly extracted without phase unwrapping, using Eqs. (4.26) and (4.27).

- Interaural-Level-Difference-map (ILD-map)

ILD-map is proposed, inspired by the ILD encoding of the LSO in the brainstem. Comparison of relative intensities in the LSO involves excitatory–inhibitory interactions of inputs from the both ears [160]. As mentioned in Subsection 4.1.2, ILD contributes to localization above 1.5 kHz. Below 1.5 kHz, there is almost no difference in the left and right sound pressure levels due to the effect of sound diffraction. As frequency increases (especially above 5 kHz), the ILD gradually becomes dominant due to the head shadow effect. Hence, the ILD above 5 kHz is another effective binaural feature for azimuth estimation. The ILD-map is defined as follows:

$$ILD\text{-}map(m,b) = Mel \begin{bmatrix} \Delta S(1,k_{5000}) & \Delta S(1,k_{5000}+1) & \cdots & \Delta S(1,K) \\ \Delta S(2,k_{5000}) & \Delta S(2,k_{5000}+1) & \cdots & \Delta S(2,K) \\ \vdots & \vdots & \ddots & \vdots \\ \Delta S(M,k_{5000}) & \Delta S(M,k_{5000}+1) & \cdots & \Delta S(M,K) \end{bmatrix}, \quad (4.28)$$

$$\Delta S(m,k) = 10\log_{10}\left|\frac{P_R(m,k)}{P_L(m,k)}\right|^2 = S_R(m,k) - S_L(m,k), \quad (4.29)$$

where $\Delta S(m,k)$ is the ILD at $m$ and $k$; $k_{5000}$ is the frequency index corresponding to 5 kHz.

### 4.2.3 SC-map for Elevation Estimation

SC-map is proposed based on the fact that the spectral pattern, especially notch frequencies, caused by the pinna contribute to vertical localization. The high-frequency notches of HRTF depend on pinna shape as well as vertical angle [112]. As discussed in Subsection 4.1.3, the notches above 5 kHz are highly dependent on the elevation of sound source, so the spectral notch pattern above 5 kHz can be the distinct monaural feature for elevation estimation. Therefore, I extended the MS-map above 5 kHz and used it as additional input feature. The left or right SC-map is defined as follows:

$$SC\text{-}map_{L,R}(m,b) = Mel \begin{bmatrix} S_{L,R}(1,k_{5000}) & S_{L,R}(1,k_{5000}+1) & \cdots & S_{L,R}(1,K) \\ S_{L,R}(2,k_{5000}) & S_{L,R}(2,k_{5000}+1) & \cdots & S_{L,R}(2,K) \\ \vdots & \vdots & \ddots & \vdots \\ S_{L,R}(M,k_{5000}) & S_{L,R}(M,k_{5000}+1) & \cdots & S_{L,R}(M,K) \end{bmatrix}. \quad (4.30)$$

### 4.2.4 BTFF Pattern According to Azimuth and Elevation

To confirm that the proposed BTFF clearly expresses the time-frequency characteristics and HRTF localization cues of sound event, I examined the pattern change of BTFF according to the direction change of sound event.

Figs. 4.9–4.14 show the BTFF patterns of a baby crying sound event for various azimuth angles on the horizontal plane ($\phi = 0°$). In Figs. 4.9(a1) and (a2), the MS expresses the periodicity, AM, and FM of the sound event as well as its onset and offset information. A similar spectral pattern is repeated three or four times, and distinct vertical lines appear at the beginning and end of each sound event over the entire frequency range. Also, tonal components related to the vocal tract stand out between the onset and offset lines. All of these represent the unique time-frequency patterns of the sound event and thus provide important clues for BiSELDnet to classify the sound event into the corresponding class. Also, the onset and offset boundaries of the MS provide useful information for labeling the start and end times of each sound event. In Figs. 4.9(b1) and (b2), the V-map shows the rate of change of frequency components over time inherent in the sound event. Since most sound events are transient sounds whose frequency characteristics change in a short period of time, it is often difficult to distinguish them using MS alone. On the other hand, since V-map represents the time change rate of MS, the shorter the transient time of MS, the higher the value, providing unique time-frequency information differentiated from MS. The MS of the baby crying may be similar to that of other vocalizations because their spectral patterns of the vocal cords are similar, but the V-maps of the two will be different due to their differences in transient characteristics. Therefore, MS and V-map are considered suitable for sound event detection. As defined in Eqs. (4.12) and (4.29), when the sound event is on the right ($\theta > 0°$), positive number dominates in ITD-map and ILD-map, respectively. On the left ($\theta < 0°$), negative number dominates. As shown in Figs. 4.11(c1) and (c2), both ITD-map and ILD-map are almost red (positive number) because the sound event is on the right ($\theta = 90°$). If the sound event is on the left ($\theta = 270°$), they turn blue (negative number) as shown in Figs. 4.14(c1) and (c2). Hence, it can be seen that both ITD-map and ILD-map are specialized for azimuth estimation. Looking at Fig. 4.10(c1) and Fig. 4.12(c1), their ITD-map patterns are almost the same because their sound sources are symmetric to each other with the lateral plane in between. However, comparing Fig. 4.10(c2) and Fig. 4.12(c2), their ILD-map patterns are different, which confirms the fact that the front-back asymmetry of ILD pattern provides a clue to solve the front-back confusion. Therefore, ILD-map complements the azimuth estimation of ITD-map. Looking at Figs. 4.9(c1) and (c2), and Figs. 4.13(c1) and (c2), their ITD-map and ILD-map patterns are almost zero because their sound events are on the median plane, which means there is no time difference and no level difference between the left and right input signals. However, comparing Figs. 4.9(d1) and (d2); and Figs. 4.13(d1) and (d2), their SC-map patterns are different, which means that SC-map also provides a clue to solve the front-back confusion on the median plane.

Figs. 4.15 and 4.16 show the BTFF patterns of the baby crying sound event for two elevation angles on the median plane ($\theta = 0°$). Comparing SC-maps in Fig. 4.15 and Fig. 4.16, several thin and thick notch lines appear between 8 kHz and 16 kHz, and their high frequency patterns change according to the elevation angle of the sound event. Notably, it can be seen that the N1 notch near 8.5 kHz shifts to high frequency as the sound event rises from −30° to +30° in elevation. This was already predicted from the SC analysis of the measured HRTFs in Subsection 4.1.3. Therefore, it is expected that the elevation angle of sound event can be estimated using deep learning of SC-map pattern. As described above, it could be confirmed with the naked eye that both the inherent time-frequency patterns and HRTF localization cues of sound event clearly appear on the proposed BTFF.

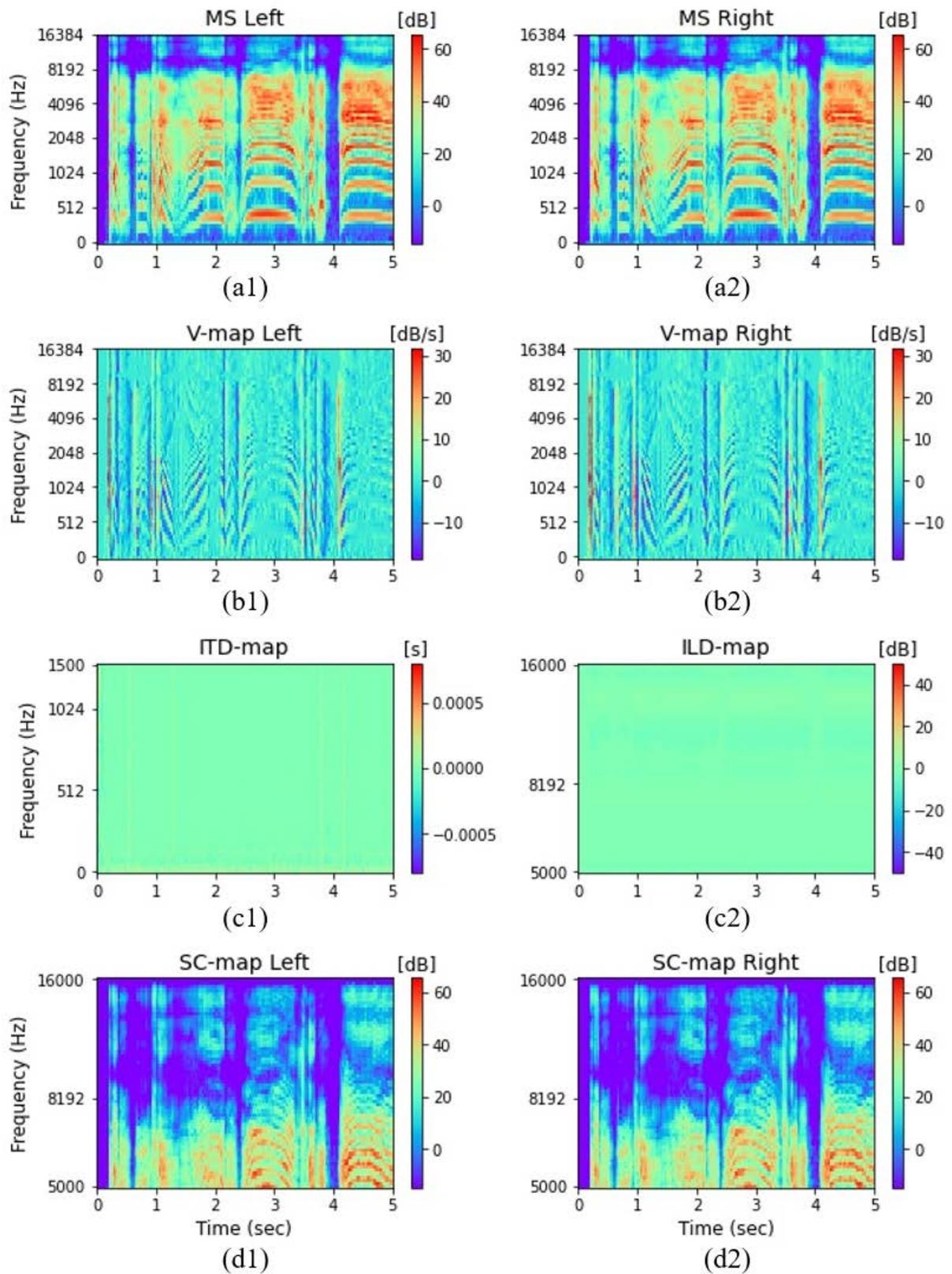

**Figure 4.9.** Binaural time-frequency feature (BTFF) of a baby crying sound event from $\theta = 0°$ and $\phi = 0°$: (a1) left MS, (a2) right MS, (b1) left V-map, (b2) right V-map, (c1) ITD-map, (c2) ILD-map, (d1) left SC-map, and (d2) right SC-map.

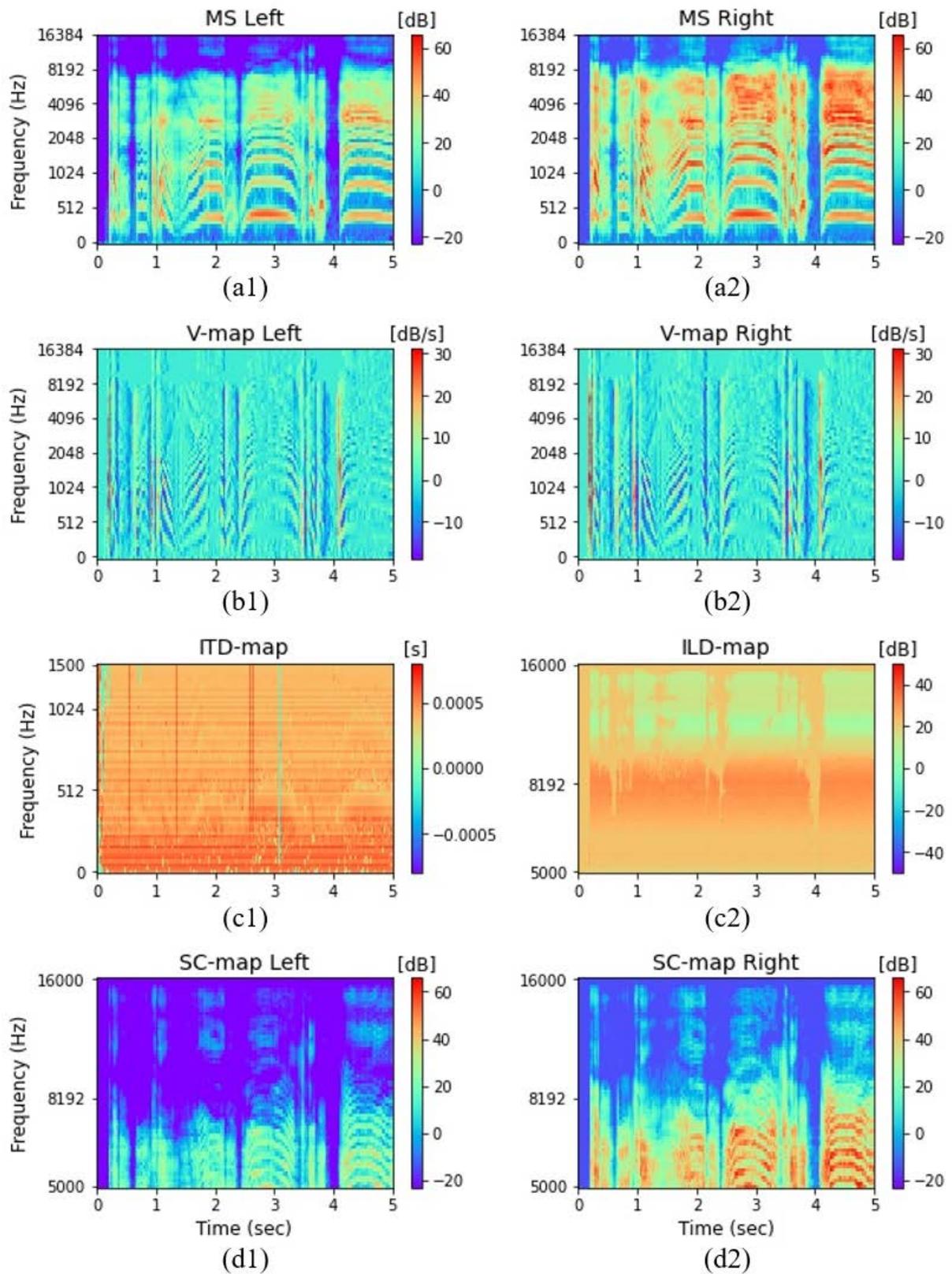

**Figure 4.10.** Binaural time-frequency feature (BTFF) of a baby crying sound event from $\theta = 60°$ and $\phi = 0°$: (a1) left MS, (a2) right MS, (b1) left V-map, (b2) right V-map, (c1) ITD-map, (c2) ILD-map, (d1) left SC-map, and (d2) right SC-map.

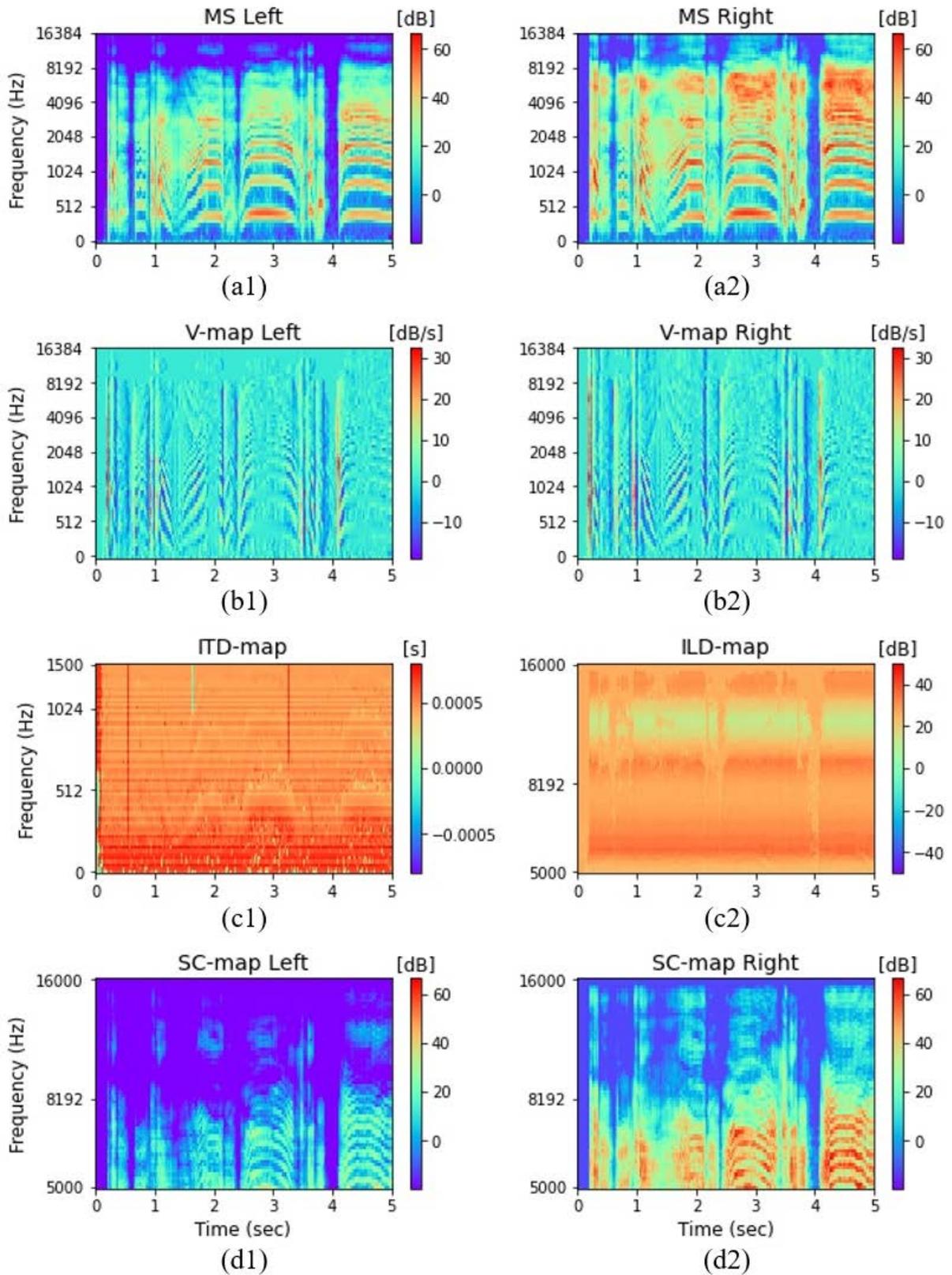

**Figure 4.11.** Binaural time-frequency feature (BTFF) of a baby crying sound event from $\theta = 90°$ and $\phi = 0°$: (a1) left MS, (a2) right MS, (b1) left V-map, (b2) right V-map, (c1) ITD-map, (c2) ILD-map, (d1) left SC-map, and (d2) right SC-map.

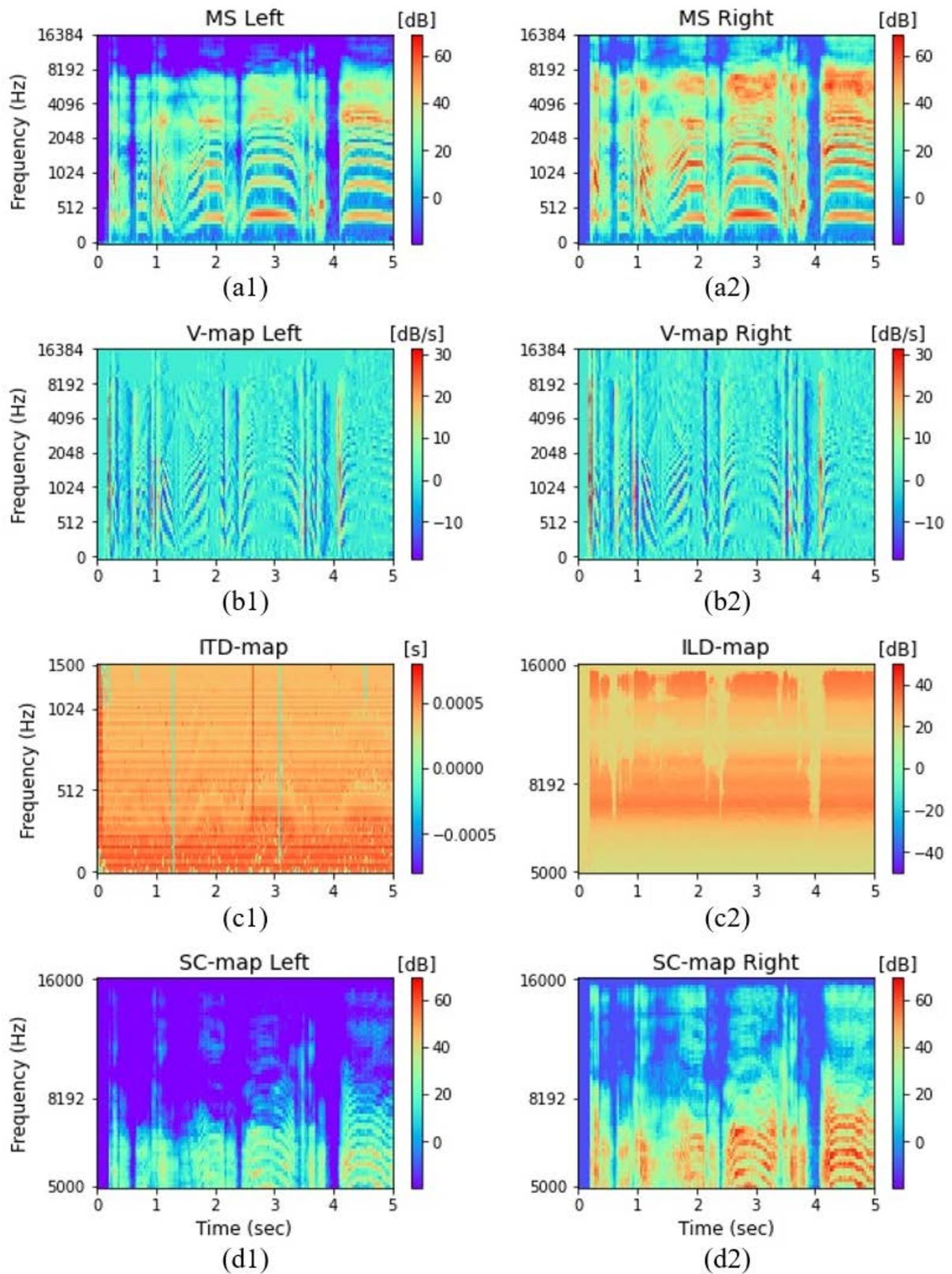

**Figure 4.12.** Binaural time-frequency feature (BTFF) of a baby crying sound event from $\theta = 120°$ and $\phi = 0°$: (a1) left MS, (a2) right MS, (b1) left V-map, (b2) right V-map, (c1) ITD-map, (c2) ILD-map, (d1) left SC-map, and (d2) right SC-map.

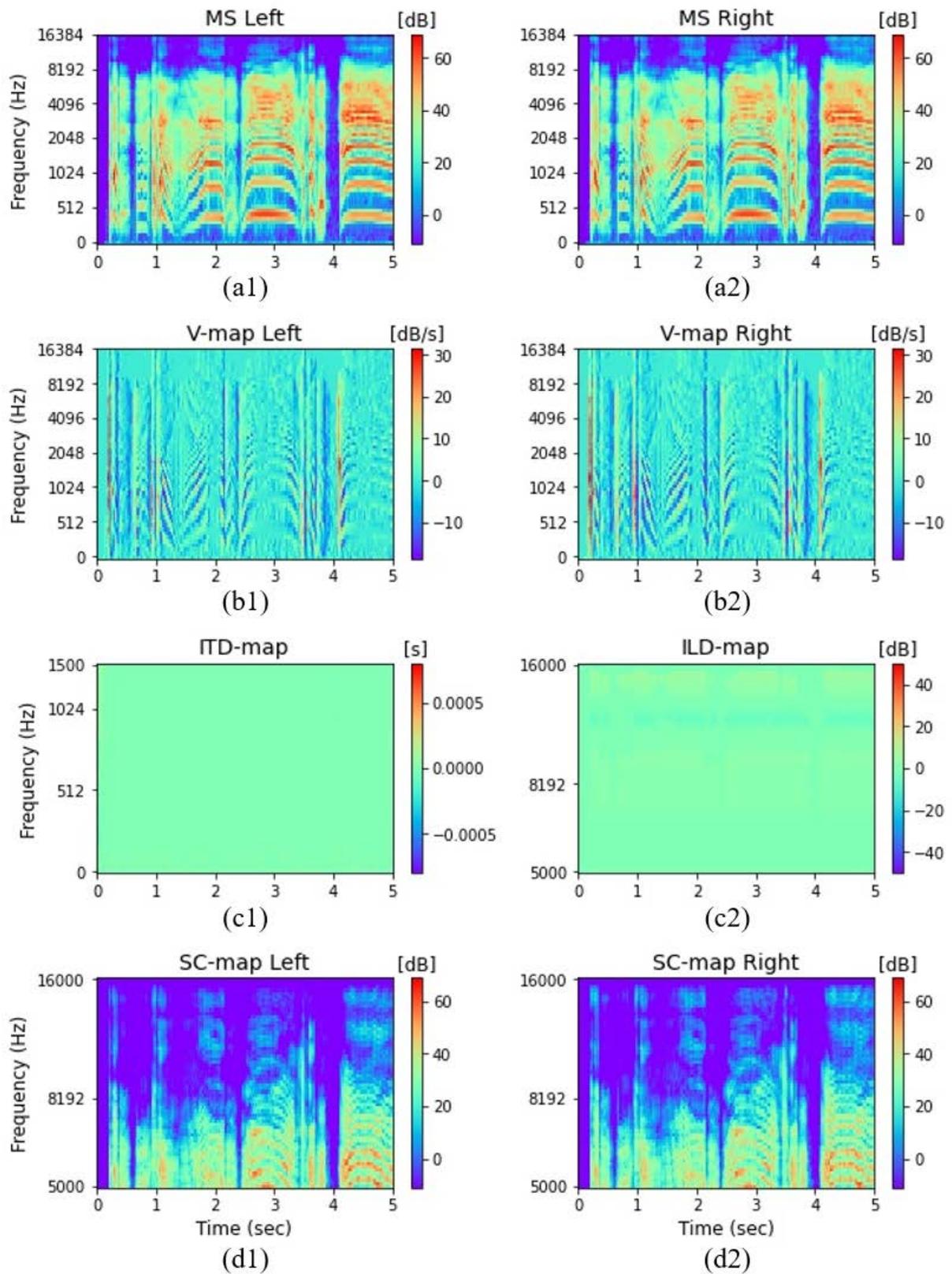

**Figure 4.13.** Binaural time-frequency feature (BTFF) of a baby crying sound event from $\theta = 180°$ and $\phi = 0°$: (a1) left MS, (a2) right MS, (b1) left V-map, (b2) right V-map, (c1) ITD-map, (c2) ILD-map, (d1) left SC-map, and (d2) right SC-map.

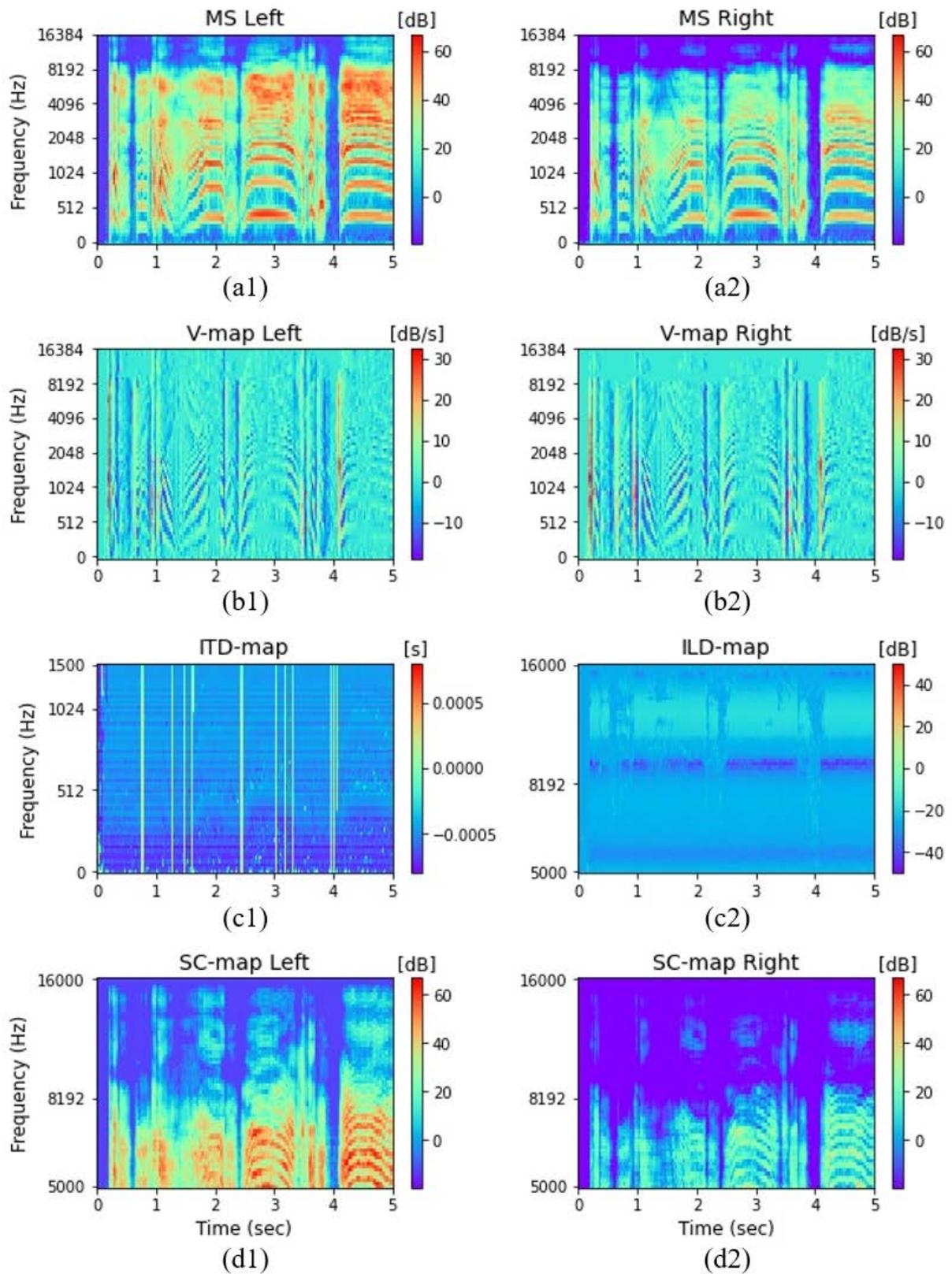

**Figure 4.14.** Binaural time-frequency feature (BTFF) of a baby crying sound event from $\theta = 270°$ and $\phi = 0°$: (a1) left MS, (a2) right MS, (b1) left V-map, (b2) right V-map, (c1) ITD-map, (c2) ILD-map, (d1) left SC-map, and (d2) right SC-map.

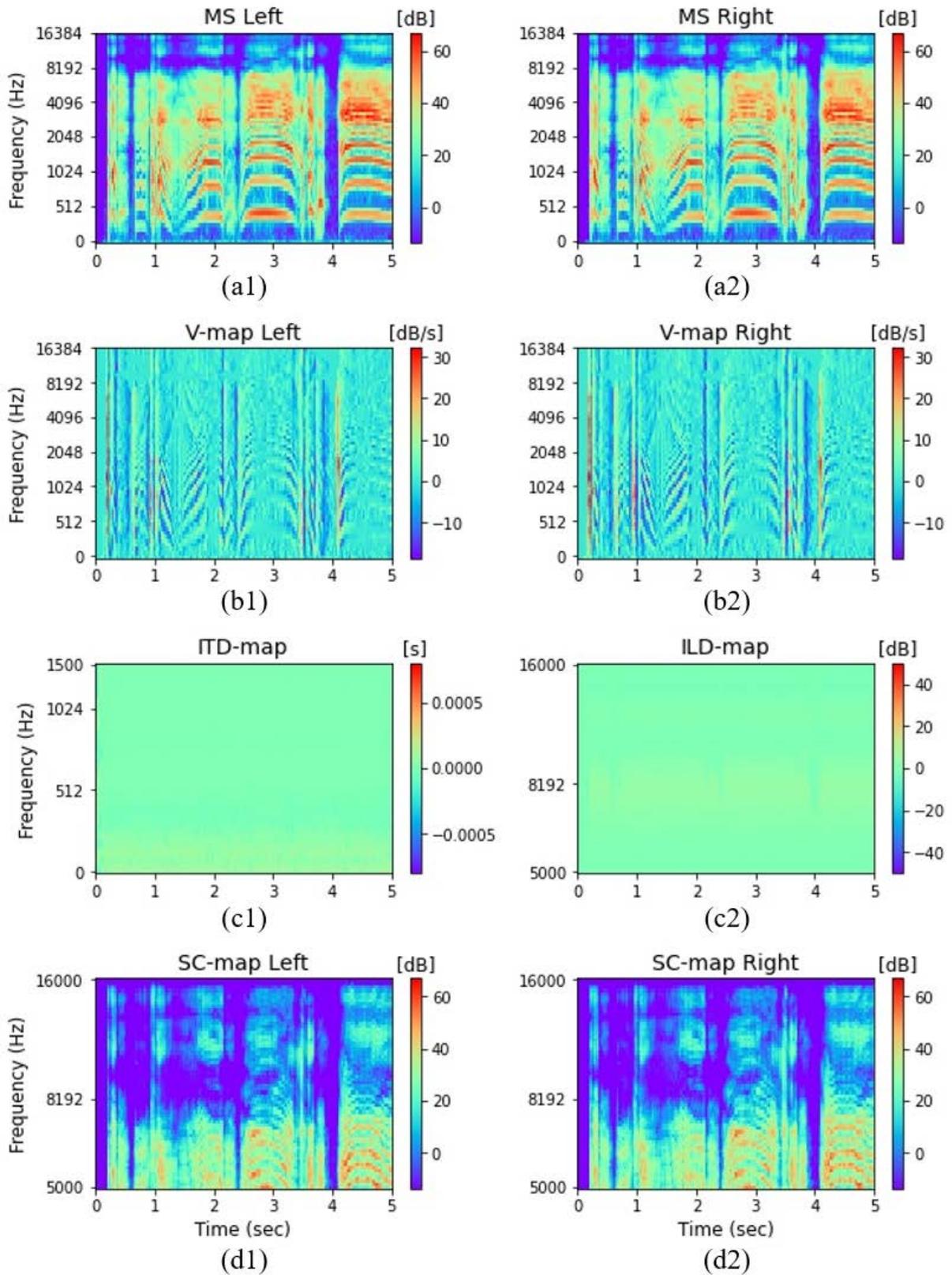

**Figure 4.15.** Binaural time-frequency feature (BTFF) of a baby crying sound event from $\theta = 0°$ and $\phi = -30°$: (a1) left MS, (a2) right MS, (b1) left V-map, (b2) right V-map, (c1) ITD-map, (c2) ILD-map, (d1) left SC-map, and (d2) right SC-map.

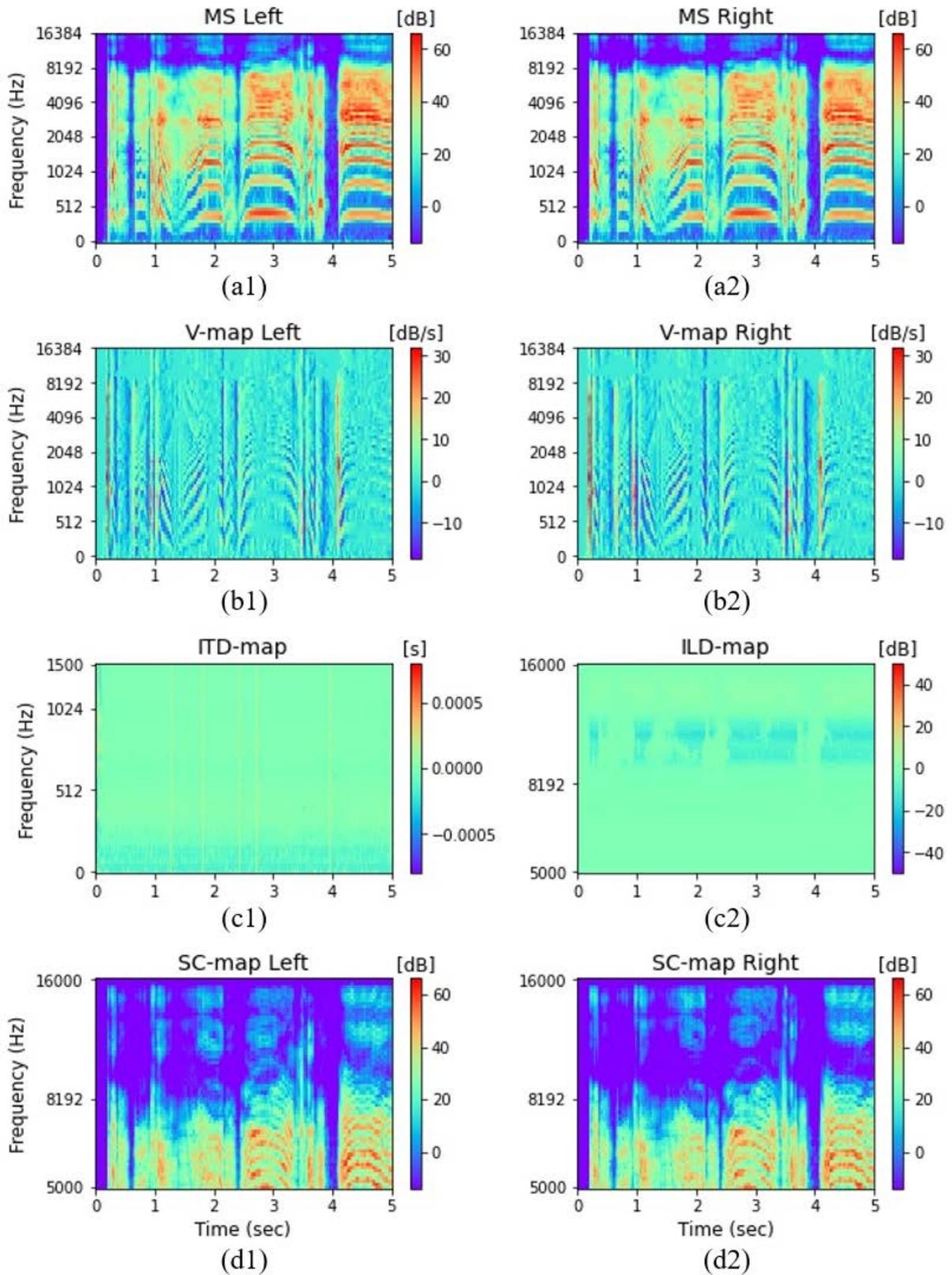

**Figure 4.16.** Binaural time-frequency feature (BTFF) of a baby crying sound event from $\theta = 0°$ and $\phi = +30°$: (a1) left MS, (a2) right MS, (b1) left V-map, (b2) right V-map, (c1) ITD-map, (c2) ILD-map, (d1) left SC-map, and (d2) right SC-map.

## 4.3 Binaural Sound Event Dataset (Binaural Set)

Real-world sound event datasets are limited in terms of diversity and availability. Creating large and diverse datasets from actual recordings can be challenging, time-consuming, and costly. Synthesizing datasets allows us to generate data representing a wide range of acoustic scenarios, sound event types, and spatial locations, which is crucial for training and evaluating a BiSELDnet effectively. Also, synthesizing data offers the advantage of controlled experiments. We can define specific scenarios, control the number and characteristics of sound events, and adjust acoustic conditions like background noise. This control allows systematic exploration of different conditions and provides insights into how BiSELD models perform in various contexts. In some cases, using real-world audio data might raise privacy or ethical concerns, especially when recording in private spaces or capturing conversations without consent. Synthesizing data allows us to avoid such issues while still developing and evaluating BiSELD models.

### 4.3.1 Database Collection

We can easily generate a large volume of training and testing data, which is essential for training complex deep learning models used in BiSELD. Large databases help in improving the robustness and generalization of BiSELD models. Each binaural sound event data can be generated by synthesizing a sound event waveform and the HRIR of a certain direction, and then adding background noise. Therefore, several databases of HRTFs, sound events and background noise were collected to build a binaural sound event dataset (Binaural Set).

First, an HRTF subset was curated from the measured KAIST HRTF database [161] presented in Chapter 3. Considering the size of the final dataset, 12 azimuth angles were selected at 30° intervals from −180° to +180°, and four elevation angles were selected at 30° intervals from −30° to +60°. In total, the HRIRs for 48 directions were used to create the Binaural Set.

Then, sound event databases were collected as shown in Table 4.1. Database 1 (NIGENS) is provided for sound-related modeling in the field of CASA, particularly for sound event detection [162]. The NIGENS database provides 714 wav-files containing 14 classes of isolated high-quality sound events and additional wav-files excluding these 14 classes. All sound events are strongly labeled with onset and offset times, and its quality of annotations distinguishes it from other databases. Database 2 (DCASE2016-2) is the training dataset available on the website of Task 2 of the DCASE challenge in 2016 [163]. All data were recorded in Finland, and to achieve high acoustic variability, each recording was made in a different location: a different street, a different park, and a different house. All recorded sound event data were cut into segments of 30 seconds length.

Lastly, as shown in Table 4.1, Database 3 (DCASE2019-1) is the background noise collection from the Task 1 of the DCASE 2019 challenge [164]. This database consists of ten different acoustic scenes, recorded in six large European cities: Barcelona, Helsinki, London, Paris, Stockholm, and Vienna, and therefore it has high acoustic variability. Since a two-channel microphone was used, each recorded acoustic scene contains multi-path directional interferences as well as reflections similar to real-life situations, mimicking the ambient sounds around the person who carried the microphone.

Databases 1 and 2 were used for building Binaural Set under clean condition, and Databases 1, 2 and 3 were used for building Binaural Sets under background noise conditions, which are datasets for the later SOTA performance comparison. Among the sound event classes of the database collection, I selected five background noise classes (airport, metro station, park, shopping mall, and street pedestrian), and 12 sound event classes related to emergency situations (0: Alarm, 1: Baby, 2: Cough, 3: Crash, 4: Dog, 5: Female Scream, 6: Female Speech, 7: Fire, 8: Knock, 9: Male Scream, 10: Male Speech, and 11: Phone) [165]. The selected classes of sound event and background noise are bolded in Table 4.1.

**Table 4.1.** Database collection of sound event and background noise for Binaural Set construction.

| Database | Type | Class | Sampling Freq. (kHz) | Total (wav-files) | Length (seconds) |
|---|---|---|---|---|---|
| NIGENS [162] | Sound event | ***Alarm**, **baby**, **crash**, **dog**, engine, **female scream**, **female speech**, **fire**, footsteps, general, **knock**, **male scream**, **male speech**, **phone**, piano.* | 44.1 | 898 | 16,759 |
| DCASE2016-2 [163] | Sound event | *Clearing throat, **cough**, door slam, drawer, keyboard, keys, **knock**, laughter, page turn, **phone**, **speech**.* | 44.1 | 220 | 265 |
| DCASE2019-1 [164] | Background noise | ***Airport**, bus, metro, **metro station**, **park**, public square, **shopping mall**, **street pedestrian**, street traffic, tram.* | 48.0 | 14,400 | 143,999 |

## 4.3.2 Data and Label Generation

Generally, synthesized datasets come with precise ground truth annotations, including the spatial coordinates (azimuth and elevation) of sound events, their temporal boundaries, and event labels. This annotation makes it easier to develop and test BiSELD models, as it provides a reference for training and evaluation. In the Binaural Set, each sound event sample consists of a pair of a data file (WAV file format) and a label file (CSV file format).

As most spectral cues are distributed below 16 kHz [161,166], the sampling frequency of the Binaural Set was set to 32 kHz, and therefore all HRIR and data samples in the database collection were resampled to 32 kHz. To create a 60-second data file, each sound event sample was prepared in 5-second increments for convenience. The 20 sound event samples placed in each sound event class are allocated to the training, validation, and test datasets in a 14:3:3 ratio, respectively. Each 5-second sound event sample is convolved with each of the HRIRs from 48 directions, resulting in 48 binaural sound event samples. For example, in Fig. 4.17, "alarm03.wav" is the 3rd sound event sample from alarm class, and "a270e+30.txt" is an HRIR file that contains HRIR pairs of 270° in azimuth and +30° in elevation. As shown in Fig. 4.17, each sound event data of the Binaural Set is generated by shuffling the binaural samples in each sound event class, selecting one from each class, and then randomly assigning the selected 12 binaural samples to background noise data at specified SNR. When mixing each binaural sample with the corresponding section of background noise data, SNR was set from 0 dB to 30 dB in 10 dB increments. In the case of dataset without background noise, each sound event data is generated by randomly arranging binaural samples from 12 classes.

An example of data and label samples of the Binaural Set is shown in Figs. 4.18(a) and (b). The first word of each data file name indicates the dataset it belongs to, the second word indicates its background noise type, and the last number indicates its mixing order. For example, the data "train_airport_mix227.wav" is the 227th mixed training data with airport background noise. Exceptionally, Data "test-a120e+00_park_mix1.wav" and "test-a000e-30_street_mix2.wav" were created to evaluate the performances of models on the horizontal and median planes, respectively, and the corresponding directions are specified in the file names. Each label is saved as a CSV format file that stores information about active sound events in four columns: time frame in deci-seconds, index of sound event class (0–11), azimuth in degrees, and elevation in degrees. In this way, each data and label pair of the Binaural Set was generated.

As a result, a total of five Binaural Sets were constructed as shown in Tables 4.2–4.6. For reference, Test-H and Test-V in each table are separate test datasets consisting only of sound events on the horizontal plane and median plane, respectively. These datasets were additionally created to evaluate the localization performance of the proposed BiSELD model separately in the horizontal and vertical directions. Table 4.2 shows the Binaural Set generated without background noise to evaluate sub-features of BTFF and various versions of BiSELDnet. Tables 4.3–4.6 represent the Binaural Sets generated with background noise with four SNRs at 10 dB intervals from 30 dB to 0 dB to compare the performance of the proposed BiSELD model with that of the SOTA SELD model.

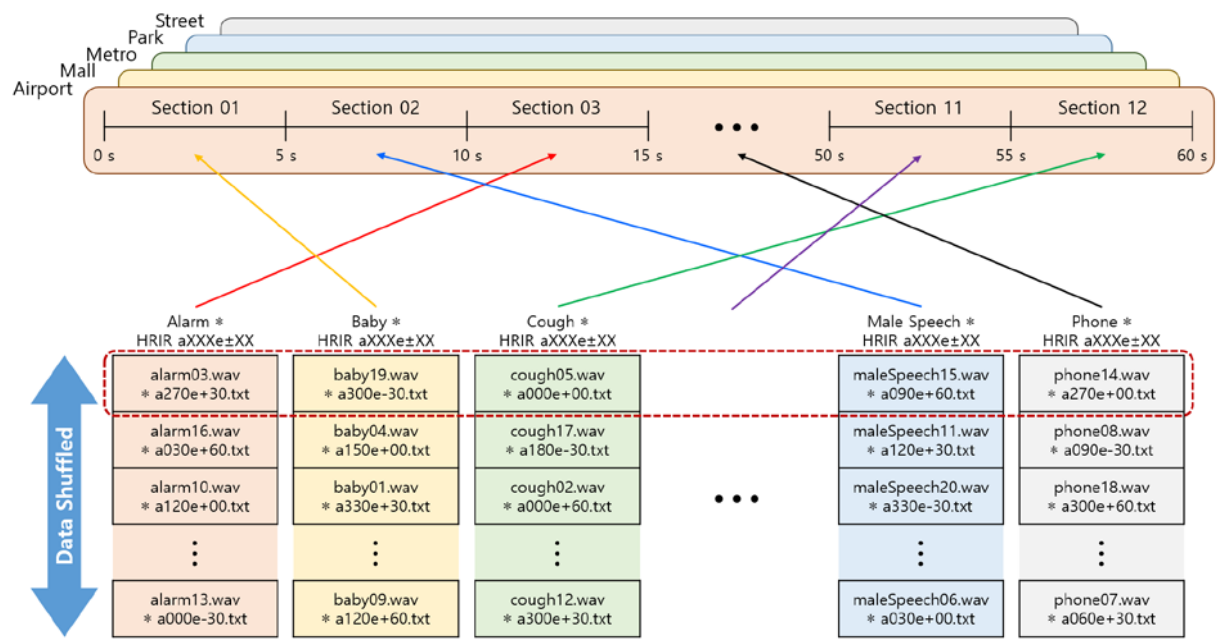

**Figure 4.17.** Illustration of data generation process for Binaural Set construction.

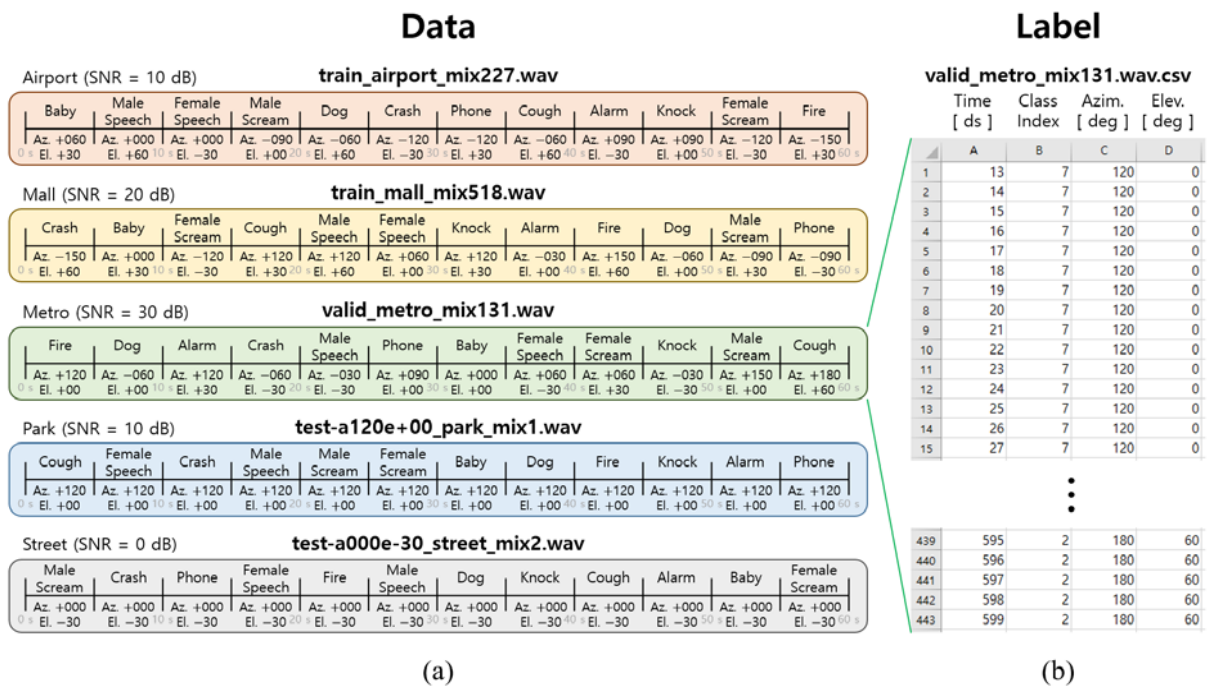

**Figure 4.18.** Data and label samples of Binaural Set: (a) data sample and (b) label sample.

**Table 4.2.** Total number of data and label pairs in the Binaural Set under clean condition.

|  | Train | Valid | Test | Test-H | Test-V | Total |
|---|---|---|---|---|---|---|
| Data | 672 | 144 | 144 | 36 | 12 | 1,008 |
| Length (seconds) | 40,320 | 8,640 | 8,640 | 2,160 | 720 | 60,480 |

**Table 4.3.** Total number of data and label pairs in the Binaural Set under background noise with SNR = 30 dB.

|  | Train | Valid | Test | Test-H | Test-V | Total |
|---|---|---|---|---|---|---|
| Data | 3,360 | 720 | 720 | 180 | 60 | 5,040 |
| Length (seconds) | 201,600 | 43,200 | 43,200 | 10,800 | 3,600 | 302,400 |

**Table 4.4.** Total number of data and label pairs in the Binaural Set under background noise with SNR = 20 dB.

|  | Train | Valid | Test | Test-H | Test-V | Total |
|---|---|---|---|---|---|---|
| Data | 3,360 | 720 | 720 | 180 | 60 | 5,040 |
| Length (seconds) | 201,600 | 43,200 | 43,200 | 10,800 | 3,600 | 302,400 |

**Table 4.5.** Total number of data and label pairs in the Binaural Set under background noise with SNR = 10 dB.

|  | Train | Valid | Test | Test-H | Test-V | Total |
|---|---|---|---|---|---|---|
| Data | 3,360 | 720 | 720 | 180 | 60 | 5,040 |
| Length (seconds) | 201,600 | 43,200 | 43,200 | 10,800 | 3,600 | 302,400 |

**Table 4.6.** Total number of data and label pairs in the Binaural Set under background noise with SNR = 0 dB.

|  | Train | Valid | Test | Test-H | Test-V | Total |
|---|---|---|---|---|---|---|
| Data | 3,360 | 720 | 720 | 180 | 60 | 5,040 |
| Length (seconds) | 201,600 | 43,200 | 43,200 | 10,800 | 3,600 | 302,400 |

# Chapter 5. Artificial Neural Network for BiSELD

## 5.1 Deep Learning and Neural Networks for BiSELD

### 5.1.1 Deep Learning for BiSELD

In the BiSELD task, the input feature **X** such as BTFF is mapped to the output **Ŷ** using a deep learning model such as BiSELDnet represented with parameters **W**. A recent approach to model **W** is using deep learning, which is a subset of machine learning, without using explicit rules to perform the task. The deep learning model is learned from a dataset such as Binaural Set.

There are three main divisions of machine learning algorithms: supervised, semi-supervised, and unsupervised algorithms. The supervised algorithm requires the target label **Y** in addition to the input feature **X** for the entire dataset. On the other hand, the unsupervised algorithm does not need the target label **Y**, while the semi-supervised algorithm only needs the target label **Y** for a subset of the dataset. In this study, the parameters **W** of BiSELDnet are learned using the supervised learning algorithm. In addition, the machine learning algorithms can be broadly categorized into classification algorithms and regression algorithms depending on the learning type. The classification algorithms learn a mapping between input feature and a discrete set of classes, whereas the regression algorithms learn a mapping between input feature and continuous-valued output.

### 5.1.2 Deep Neural Network (DNN)

A neuron, the basic functional unit of the nerve system, is connected to multiple neighboring neurons. Each neuron receives signals at the dendrites, integrates the information at the cell body, and transmits signals down the axon to communicate with other neurons through synapses using electrical action potentials and chemical neurotransmitters. An artificial neural network (ANN) is a type of supervised algorithm inspired by the human neuron system and has achieved good results in various machine learning tasks. ANN consists of input, hidden, and output layers, each of which consists of multiple artificial neurons. The output of a single artificial neuron can be defined as follows:

$$z = f\left(\sum_i w_i \cdot x_i + b\right), \quad (5.1)$$

where $f()$ is a non-linearity function; $w_i$ is a weight; $x_i$ is an input; and $b$ is a bias. Since each neuron in a layer is connected with all neurons in the previous layer, this network is usually referred to as a fully-connected (FC) neural network [167]. Initially, due to some technical limitations, ANN was constructed with a total of three layers, including only one hidden layer. ANN is sometimes used as a general term for all artificial neural networks in a broad sense, so ANNs with a single hidden layer are sometimes called shallow neural networks [168]. As shown in Fig. 5.1, DNN refers to an ANN with more than two hidden layers [169].

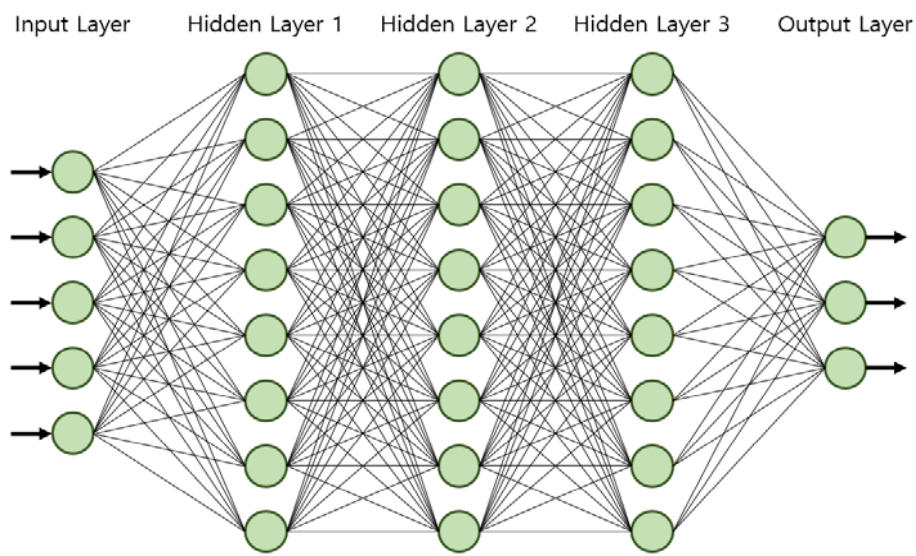

**Figure 5.1.** Deep neural network (DNN) architecture with multiple layers.

The backpropagation algorithm enables the efficient training of DNNs by propagating error gradients backward through the layers, facilitating the adjustment of model parameters and the learning of complex representations from data [170]. DNNs offer significant advantages over shallow neural networks due to their ability to learn hierarchical features and complex representations, allowing them to capture intricate patterns in data. In particular, activation functions are crucial in DNNs as they introduce non-linearity, allowing the network to learn and represent complex, non-linear relationships within data, enhancing the model's capacity to extract intricate patterns. Therefore, activation functions help DNNs learn decision boundaries of complex shapes. The sigmoid activation function takes a real value as input and produces an output between 0 and 1. It is defined as follows:

$$\sigma(x) = \frac{1}{1 + e^{-x}}. \tag{5.2}$$

Since the output range is between 0 and 1, the sigmoid output is considered a probability, and thus the sigmoid function is mainly used in classification problem. The tanh activation function has better learning efficiency than the sigmoid activation function because the slope can be both positive and negative. It is defined as follows:

$$f(x) = \tanh(x). \tag{5.3}$$

The output of the tanh function is between −1 and 1. Since its output is zero-centered, the tanh function is generally preferred over the sigmoid function. Typically, DNN is composed of multiple layers, and activation functions are repeatedly included between each layer, so when calculating error backpropagation, small gradients may accumulate and gradient loss may occur. In particular, since the sigmoid function reduces the input to a specific range, it can cause a vanishing gradient problem. On the other hand, a rectified linear unit (ReLU) can solve this problem because it outputs the same value as the input when the input value is greater than 0. Therefore, ReLU is the most popular activation function in the field of deep learning. It is defined as follows:

$$f(x) = \max(0, x). \tag{5.4}$$

In addition, compared to other activation functions, the computation of ReLU is straightforward, which greatly speeds up the learning process.

### 5.1.3 Recurrent Neural Network (RNN)

DNNs process features in a finite time frame and produce their outputs. In most cases, in sequence data such as text, audio, and video, information is spread over multiple time frames. For example, when listening to sound, each sound frame is understood based on the understanding of the previous sound frame generated from the same sound source. In order to learn sequence data with DNN, multiple time frames need to be concatenated into a single feature vector. However, this requires selecting the number of frames to connect to, so we need to create other parameters to tune in the network. Moreover, within the Binaural Set, each class event has a different duration from each other. For example, an impulsive sound event such as a knock has a shorter duration, whereas an alarm or phone sound has a much longer duration. Therefore, it is difficult to learn such class-specific information by the feature connection method in DNNs. On the other hand, RNNs overcome this shortcoming of DNNs by retaining information from previous inputs in memory [171]. An RNN is a type of ANN designed for processing sequential data. Unlike traditional feedforward DNNs, RNNs have connections that loop back on themselves, allowing them to maintain a hidden state or memory of previous inputs as shown in Fig. 5.2.

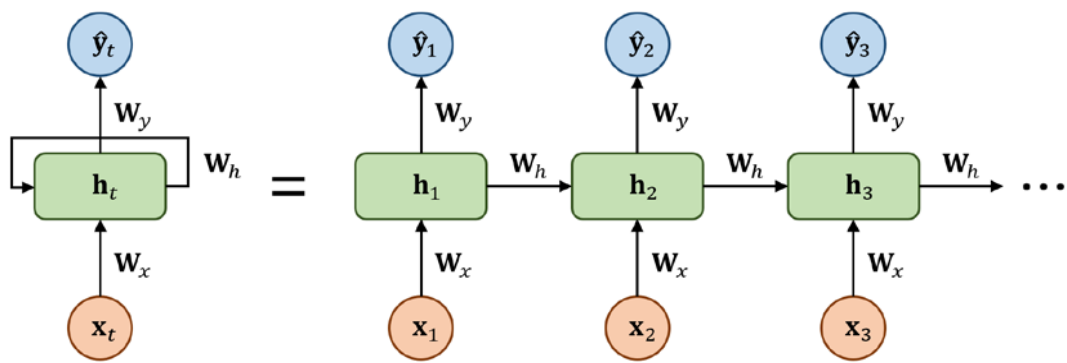

**Figure 5.2.** Recurrent neural network (RNN) architecture and its time step unfolding.

Assuming that the sequences of input feature and target label are as follows:

$$\mathbf{X} = [\mathbf{x}_1, \mathbf{x}_2, \cdots, \mathbf{x}_t, \cdots, \mathbf{x}_T], \tag{5.5}$$

$$\mathbf{Y} = [\mathbf{y}_1, \mathbf{y}_2, \cdots, \mathbf{y}_t, \cdots, \mathbf{y}_T]. \tag{5.6}$$

An RNN operation at a time frame $t$ is as follows:

$$\mathbf{h}_t = f(\mathbf{W}_h \mathbf{h}_{t-1} + \mathbf{W}_x \mathbf{x}_t), \tag{5.7}$$

$$\hat{\mathbf{y}}_t = \mathbf{W}_y \mathbf{h}_t, \tag{5.8}$$

where $\mathbf{h}_t$ is current hidden state; $\mathbf{h}_{t-1}$ is previous hidden state; $\mathbf{x}_t$ is current input feature; $\hat{\mathbf{y}}_t$ is output; $\mathbf{W}_h$ is weight for hidden state parameter; $\mathbf{W}_x$ is weight for input feature; and $\mathbf{W}_y$ is weight for output. $\mathbf{h}_t$ learns from $\mathbf{x}_t$, and $\mathbf{h}_{t-1}$ learns from all the previous feature frames. In most cases, the activation function $f()$ is tanh function. From Eqs. (5.7) and (5.8), it is confirmed that the current output $\hat{\mathbf{y}}_t$ is affected by both the current input $\mathbf{x}_t$ and the previous output state $\mathbf{h}_{t-1}$. This recurrent structure enables RNNs to capture and model temporal dependencies in data, making them particularly suitable for tasks involving sequences, such as natural language processing, speech recognition, time series prediction. For reference, backpropagation through time (BPTT) is a training algorithm used in RNNs. BPTT extends the standard backpropagation algorithm to handle sequences by unfolding the RNN through time, creating a computational graph with as many time steps as there are in the input sequence. In some sequences, the current output can be better estimated with information about future inputs as well as previous inputs. For example, a speech recognition system can perform better on both previous and future phonemes than using only the previous one. RNNs can be adapted to do forward and backward learning, and such RNNs are called bidirectional RNNs [172].

In theory, RNNs can model long-term temporal dependencies. In practice, however, RNNs suffer from the vanishing gradient problem that hinders learning of long data sequences [173]. Tanh, the activation function of RNN, becomes saturated as the input increases, so even if the input value is large, the derivative value becomes small. As the input cycles through the tanh function repeatedly, the small derivative accumulates and becomes very close to zero, making learning difficult. Therefore, to solve this problem, it is more advantageous to process past values using addition rather than multiplication. The well-known solutions are LSTM [174] and GRU [175]. RNNs of these types are composed of several gates that enable RNNs to capture and control the relevant contextual information over long sequences, address the vanishing gradient problem, and selectively use the information. The gating mechanisms make RNN powerful for handling sequential data and have been instrumental in improving its capabilities.

## 5.1.4 Convolutional Neural Network (CNN)

DNNs do not consider spatial relationships and treat each input neuron independently, which is less suitable for tasks where local patterns matter. On the contrary, CNNs are particularly suited for data with local patterns or spatial hierarchies. For example, in image processing, they recognize features like edges, corners, and textures in a spatially local manner. This localized processing is more biologically inspired and effective for tasks like object recognition and image analysis [170]. In addition, the FC layers of DNN have a large number of parameters, which can lead to increased computational requirements and a higher risk of overfitting, especially when the data is spatially structured. On the other hand, the convolutional layers of CNN share their weights across different regions of input feature. The output of a convolutional layer can be defined as follows:

$$\boldsymbol{H}_{i,j,k} = f\left(\boldsymbol{b} + \sum_l \sum_m \sum_n \boldsymbol{X}_{i-l,j-m,k-n} \boldsymbol{W}_{l,m,n}\right), \tag{5.9}$$

where $\boldsymbol{X}_{i-l,j-m,k-n}$ is 3D input feature; $\boldsymbol{W}_{l,m,n}$ is weight kernel; $\boldsymbol{b}$ is bias; and $f(\ )$ is activation function. In most cases, the activation function is ReLU. As shown in Eq. (5.9), the same kernel $\boldsymbol{W}_{l,m,n}$ is shared across the input feature, and typically the kernel size of convolutional layer is much smaller than the size of input features. This parameter sharing drastically reduces the number of parameters in the model, making it more efficient and less prone to overfitting, especially when dealing with high-dimensional data like images [169]. As shown in Fig. 5.3, the convolution operation in CNN involves sliding a kernel (also known as a filter) over an input feature grid and computing the element-wise dot product at each position to generate an output feature map, which capture local patterns and spatial hierarchies within the data.

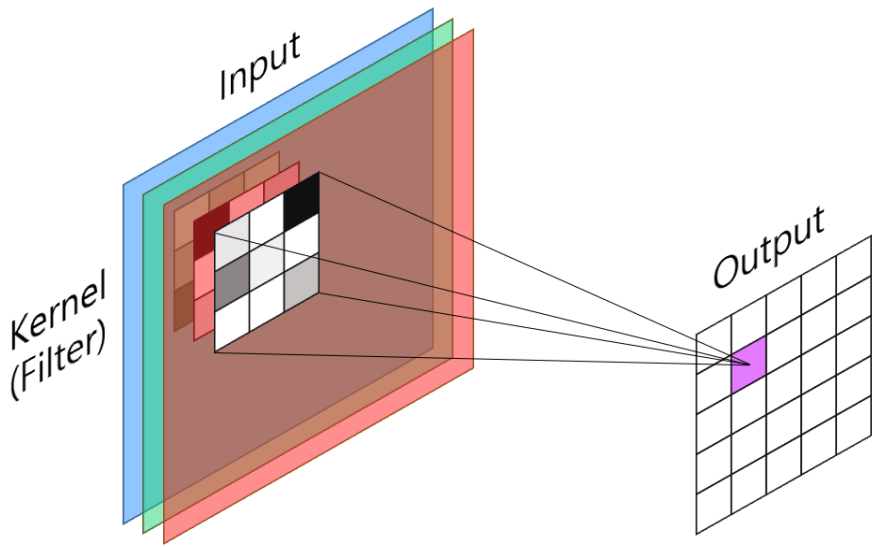

**Figure 5.3.** Convolution operation with a 3×3 kernel in convolutional neural network (CNN).

Typically, each convolutional layer consists of multiple kernels, and outputs multiple feature maps with channels corresponding to the number of kernels. The depth of input feature is one for monochrome images, whereas three for color images in RGB (red, green, and blue channels) format. Similarly, for audio input feature, if a single-channel spectrogram is used, the depth is one, whereas for multi-channel ones, the depth is equal to the number of channels. Each kernel is trained to learn one particular local pattern and detects whether this pattern is repeated within the input feature. In a mel-scaled spectrogram, if two identical sound events are separated in time and shifted on the frequency axis, both events can be detected by a single kernel that has learned the patterns of those sound events. This property of convolutional layer for detecting similar local patterns across input features is called shift-invariance. A learned feature can be detected in different parts of the input feature, making them useful for tasks where the exact location of features is not critical [171].

A CNN consists of multiple convolutional layers, each followed by a pooling layer. There are various types of pooling layers, but the most used is the max pooling layer. It is used to downsample the spatial dimensions of feature maps generated by convolutional layers. By selecting the maximum value within a local region of the feature map, max pooling retains the most salient information while reducing the computational complexity of subsequent layers. This process helps in two ways: it reduces the number of parameters in the network, which can prevent overfitting, and it increases the network's ability to be translation invariant. Max pooling also helps in capturing the most prominent features and abstracting them to higher-level representations, making the network more robust and efficient in learning hierarchical patterns [171]. In the final step, the feature map from the last convolutional layer is fed as input to the FC layer of DNN, which generates class-wise probabilities. Based on these output probabilities, the CNN's classification task is performed. In general, CNNs have been mainly used for classification tasks that do not require temporal information, such as sound event classification.

## 5.2 Learning Process

### 5.2.1 Objective Function

The objective function in deep learning serves as a crucial metric that quantifies the error between a neural network's predictions and actual target values, playing a central role in model training and optimization. By minimizing the objective function, the model's parameters are iteratively adjusted to improve prediction accuracy, making it a guiding force for learning, model selection, regularization, evaluation, and hyperparameter tuning in the pursuit of optimal model performance across various tasks, including classification, regression, and generative modeling. The choice of objective function (or loss function) in deep learning depends on the specific task and problem at hand [167]. In the SELD task, commonly used objective functions are the mean square error (MSE) and the binary cross entropy (BCE).

The MSE loss is frequently used for regression tasks. It measures the average squared difference between the predicted values and the actual target values. The MSE loss is defined as follows:

$$L_{MSE} = \frac{1}{N} \sum_{n=1}^{N} (y_n - \hat{y}_n)^2, \tag{5.10}$$

where $y_n$ is a reference value and $\hat{y}_n$ is the predicted value for $y_n$. The MSE loss produces larger values when $y_n$ and $\hat{y}_n$ are different and smaller values when they are similar.

The BCE loss is typically used for binary classification tasks. It measures the dissimilarity between the predicted class probabilities and the true binary labels. The BCE loss is defined as follows:

$$L_{BCE} = \frac{1}{NC} \sum_{n=1}^{N} \sum_{c=1}^{C} \{(1 - Y_{n,c}) \log(1 - \hat{Y}_{n,c}) - Y_{n,c} \log(\hat{Y}_{n,c})\}, \tag{5.11}$$

where $Y_{n,c}$ and $\hat{Y}_{n,c}$ are the reference and prediction for the $n$-th sample of class $c$, respectively. Similar to the MSE loss, the BCE loss is small when $Y_{n,c}$ and $\hat{Y}_{n,c}$ are similar and larger when they are not. In the SELD task, the early version of SELDnet was a multitasking neural network that processes sound event detection and its localization separately with respective FC layers [33]. In this SELDnet, BCE loss was used for the SED output, and MSE loss was used for the DOA output. Later, as the ACCDOA representation was proposed, each output for SED and DOA was integrated, and therefore the objective function of SELDnet was unified as MSE loss [79].

5.2.2 Optimization Algorithm

The goal of optimization algorithm is to find the best set of parameters that minimizes the loss function between the predicted value and the target label. One of the most famous optimization algorithms is the gradient descent algorithm. The gradient varies as the search proceeds, tending to zero as we approach the minimizer [176]. This algorithm optimizes network by multiple iterations of three steps: forward propagation, loss and gradient estimation, and backpropagation. Before forward propagation, parameters are randomly initialized to small values, usually taken from a normal distribution. During forward propagation, input features are processed to obtain prediction results using neural network parameters. Then, the similarity to the target value is evaluated using a loss function such as MSE or BCE loss, and the gradient of the loss with respect to each parameter is calculated. The loss gradient represents the slope of the loss function and helps to orient the local minima of the loss. In the final step, during backpropagation, the network parameters are updated with the loss gradient so that the new parameters produce a lower loss. The backpropagation for the update of neural network parameters was first proposed in [177]. The update process can be expressed as follows [167]:

$$\mathbf{w} \leftarrow \mathbf{w} - \alpha \frac{\partial L_\mathbf{w}(\mathbf{y}, \hat{\mathbf{y}})}{\partial \mathbf{w}}, \qquad (5.12)$$

where $\mathbf{w}$ is a set of neural network parameters; $\mathbf{y}$ are target values; $\hat{\mathbf{y}}$ are predictions; $L_\mathbf{w}(\mathbf{y}, \hat{\mathbf{y}})$ is loss between $\mathbf{y}$ and $\hat{\mathbf{y}}$; $\alpha$ is a positive number representing the learning rate controlling the update size; and $\leftarrow$ denotes an assignment operator. There are different versions of gradient descent algorithm, depending on how often the loss is calculated. In the first version, the loss is calculated after forward propagation over the entire dataset, followed by backward propagation. This is called batch gradient descent, and although this algorithm gives a better parameter $\mathbf{w}$ by giving the actual gradient, it is computationally expensive. On the other hand, it was found that calculating the loss with a small number of samples (mini-batch) from the dataset each time produces a parameter $\mathbf{w}$ similar to the batch gradient descent parameters. After the network samples the entire dataset for one epoch, the dataset is re-shuffled and resampled in mini-batches to update the parameters. This shuffling allows the network to experience various groups of samples in different combinations and learn the overall distribution of the dataset. The gradient descent with the a mini-batch size of one is called stochastic gradient descent (SGD), and that with a mini-batch size greater than one and smaller than the overall dataset size is called mini-batch gradient descent. To further accelerate gradient descent optimization, momentum updates are performed as follows [167]:

$$\mathbf{w} \leftarrow \mathbf{w} - \alpha m, \qquad (5.13)$$

$$m = \beta_1 m + \left\{ 1 - \beta_1 \frac{\partial L_\mathbf{w}(\mathbf{y}, \hat{\mathbf{y}})}{\partial \mathbf{w}} \right\}, \qquad (5.14)$$

where $m$ is a momentum; $\beta_1$ is a parameter normally set to 0.9. As shown in Eq. (5.14), momentum update is based on the weighted average of gradients across multiple iterations. By using momentum, the learning rate $\alpha$ becomes adaptive due to the weighted average of the gradient operation. Recently, Adam optimizer, which adds the second gradient moment, has been proposed and shows better results than the previous methods [178]. The update of Adam algorithm is as follows:

$$\mathbf{w} \leftarrow \mathbf{w} - \alpha \frac{m}{\sqrt{v} + \varepsilon}, \qquad (5.15)$$

$$v = \beta_2 v + (1 - \beta_2) \left\{ \frac{\partial L_\mathbf{w}(\mathbf{y}, \hat{\mathbf{y}})}{\partial \mathbf{w}} \right\}^2, \qquad (5.16)$$

where $m$ is the first moment; $v$ is the second moment; and $\varepsilon$ is a parameter used for avoiding division by zero. Commonly used parameter values for the Adam optimizer are $\beta_1=0.9$, $\beta_2 = 0.999$, and $\varepsilon = 10^{-8}$.

### 5.2.3 Training Process

When training a deep learning model, care must be taken to avoid overfitting to the given dataset. Overfitting reduces model performance on unseen data, so to prevent it, the given dataset is divided into training, validation and test datasets as in Subsection 4.3.2. The division of the dataset allows for rigorous training, hyperparameter optimization, and objective evaluation of the deep learning model. This process helps ensure that the model is both capable of capturing patterns in the training data and capable of generalizing to new, unseen data [171].

In the development phase, the deep learning model is trained using the training dataset. During training, the model is exposed to the training data, and it learns to make predictions by adjusting its parameters through backpropagation. The model's parameters are updated iteratively to minimize a loss function, which quantifies the error between its predictions and the true values. After each training epoch, the model's performance is evaluated on the validation dataset. This evaluation involves using the current model to make predictions on the validation data. The performance on the validation dataset is measured using predefined evaluation metrics (e.g., accuracy, F1 score, error rate). These metrics help assess how well the model generalizes to data it has not seen during training. If the metric begins to worsen or stagnate on the validation dataset (indicating overfitting), the training process is stopped to prevent further deterioration in generalization performance. This process is called early stopping and prevents the model from overfitting to the training data [171]. The model at the point when training is halted is typically chosen as the best model, as it has demonstrated the best generalization performance on the validation data. Moreover, the validation results are used for hyperparameter tuning, such as adjusting the learning rate, network architecture, or other hyperparameters, to improve the model's performance. Multiple model variants or configurations may be trained and evaluated on the validation dataset. Among variants, the model that performs best on the validation dataset is selected as the final model for the given task. This selection process helps ensure that the model is well-suited to the task and hyperparameters are optimized.

In the test phase, once the model and hyperparameters are fixed, the model's performance is assessed on the test dataset. The test dataset is entirely separate from the training and validation datasets and simulates unseen real-world data. The test results provide an unbiased assessment of how well the model generalizes and how it is expected to perform in a real-world application.

5.2.4 Regularization

Regularization is a method used to prevent overfitting and improve a model's generalization performance. By adding regularization, models can learn to focus on the most important features and relationships in the data while reducing sensitivity to noise or irrelevant features. This leads to improved generalization performance, making models more robust in real-world scenarios. In addition to the early stopping mentioned above, there are several common regularization methods:

• L1 and L2 Regularization (Lasso and Ridge)

L1 Regularization (Lasso) adds the absolute values of the weights to the loss function. It encourages sparsity by pushing some weights to exactly zero, effectively performing feature selection. L1 regularization is particularly useful when we suspect that only a subset of features is relevant to the task. On the other hand, L2 Regularization (Ridge) adds the square of the weights to the loss function. It encourages small weights for all features but doesn't force them to be exactly zero. This helps in reducing the impact of individual features on the model, making it more robust to noise in the data [170].

• Dropout

Dropout is a simple but effective regularization algorithm, where each neuron remains active during training with some probability *p*. The probability *p* is a hyperparameter and its value is set before training. When individual neurons are randomly turned off, connected neurons that depend on the turned off neuron learn similar information from other active neurons. This reduces the joint adaptation between the two neurons for generalization. Training by dropout is similar to training multiple neural networks and averaging the outputs to get the final result [179].

• Batch Normalization

Batch normalization is a technique in deep learning that normalizes the activations within a layer by centering and scaling them. It is typically applied to FC and convolutional layers in neural networks. By reducing internal covariate shift, batch normalization speeds up training, allows for higher learning rates, and stabilizes the training process. It not only accelerates convergence but also acts as a form of regularization, making the network less sensitive to changes in the input distribution and reducing the risk of overfitting. Batch normalization has become a fundamental component in deep learning architectures, leading to more efficient and stable training processes and improved model performance [180].

## 5.3 Architecture of BiSELDnet

As mentioned in Section 4.2, good input features make the latent manifold of data smoother, and therefore, make the model more robust to variations in data. In deep learning, the concept of a latent manifold refers to a lower-dimensional space that captures the underlying structure or representation of complex and high-dimensional data. The manifold hypothesis is a concept in machine learning that suggests that high-dimensional data, often encountered in the real world, lies on or near a much lower-dimensional manifold within the high-dimensional space. In other words, it posits that complex data can often be described or embedded in a lower-dimensional space, capturing its essential structure or patterns. This is a very strong assumption about the information structure in the universe, and this is why deep learning model works [171].

In deep learning, all data is expressed as a tensor, which is a point in geometric space. A deep learning model is a chain of simple, continuous geometric transformations to map one vector space to another. The entire process of applying complex geometric transformations to input data is similar to a human unfolding a crumpled paper ball. The crumpled paper ball is like a manifold of input data that the model has never seen before. Each movement of a person unfolding the paper ball is similar to the simple geometric transformation performed at each layer of a deep learning model. A deep learning model is a type of mathematical device that unfolds a complex manifold of high-dimensional data. Deep learning models convert data into vectors in geometric space and then gradually learn complex geometric transformations that map from one space to another. What we need is a high-dimensional space large enough to find all types of relationships in the original data, which is determined by the architecture of the deep learning model we design. **In the BiSELD task, BTFF is an input feature that makes it easy to unfold the crumpled paper ball, and BiSELDnet is a deep learning model that directly unfolds the paper ball and finds the answer written on it as shown in Fig. 5.4.** In this section, I propose a variety of BiSELDnets that can successfully perform the BiSELD task by combining basic neural networks introduced in Section 5.1.

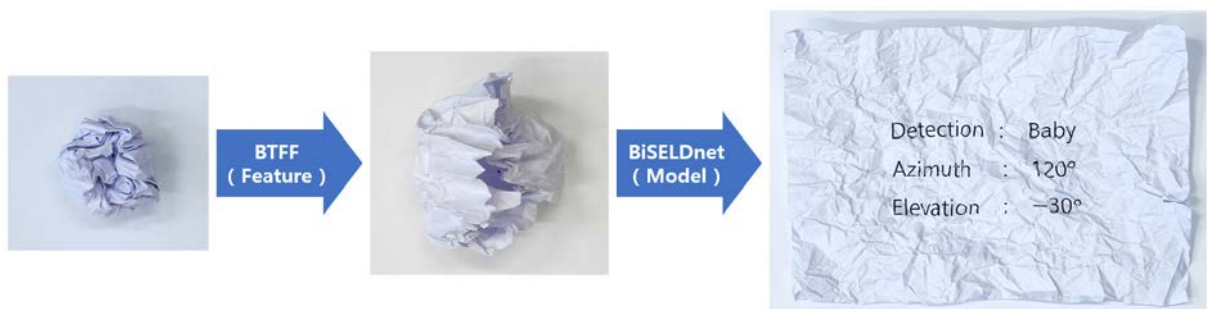

**Figure 5.4.** Illustration of unfolding a crumpled paper ball as a metaphor for unfolding a complex manifold of binaural data using BTFF and BiSELDnet.

### 5.3.1 BiSELDnet-v1 (CRNN)

BiSELDnet-v1 is a CRNN based deep neural network designed for the BiSELD task. The FC, recurrent, and convolutional layers discussed in Section 5.1 learn complementary information from each other. The FC layers are good for frame-by-frame classification, but are not good at modeling temporal structures, and the number of parameters increases with input size. On the other hand, convolutional layers are suitable for learning shift-invariant features from three-dimensional input such as BTFF. In addition, the shared kernel means that for the same input size, the number of parameters is much smaller than that of the FC layer. However, both FC and convolutional layers fail to model temporal structures. Although recurrent layers are good at learning temporal information from sequential inputs, they suffer from a similar problem as FC layers with regard to the number of parameters. Therefore, if the three layers are complemented each other, an effective model for the BiSELD task can be created. Applying RNN and DNN to the low-dimensional output of CNN helps to keep the total number of network parameters small, and this joint architecture is called CRNN. The combination of convolutional and recurrent layers of CRNN is well-suited for the BiSELD task which requires both spectro-temporal information (time-frequency characteristics of sound events) and spatial information (HRTF localization cues). Convolutional layers are suitable for extracting local time-frequency features from BTFF, while recurrent layers are specialized for modeling the sequential aspects of BTFF. Therefore, this combination is useful for the BiSELD task, where both the spectro-temporal and sequential aspects of each sound event need to be perceived simultaneously. Notably, binaural data can vary in length, making it challenging to process with conventional methods. CRNNs can handle variable-length sequences because recurrent layers can adapt to different sequence lengths. This flexibility is crucial in the BiSELD task where sound events can have varying durations.

The architecture of BiSELDnet-v1 is shown in Fig. 5.5(a). When a sound is input to the both ears of a humanoid robot, A BTFF extracted by the feature extractor is input to the BiSELDnet. The input feature has T temporal frames, 64 mel bins, and eight channels ($T \times 64 \times 8$). In the BiSELDnet, the local shift-invariant patterns of BTFF are extracted using five CNN modules. In each CNN module, two convolutional layers (64 kernels of $3 \times 3$ receptive field) is followed by a batch normalizer, a ReLU activator, and a max-pooling layer. The dimensionality of feature maps is reduced by max-pooling layers across CNN modules along the time and frequency axes, thereby reducing the time sequence length from T to T/5 and the frequency dimension from 64 to 2. The output feature map from the final CNN module is of dimension $T/5 \times 2 \times 64$. Then it is reshaped to a T/5 frame sequence of feature vectors of length 128 and fed into two bidirectional RNN modules which are used to learn the sequential context information from the feature vectors. In each RNN module, 128 nodes of GRU are used with tanh activations. This is followed by a DNN with three FC layers for regression output.

The last FC layer consists of 36 nodes and is followed by a tanh activator. The final tanh activator outputs a T/5 frame sequence of output vectors of length 36. From the output vector, each of the 12 sound event classes is represented by a 3-dimensional DOA vector (*x*, *y*, *z*). Also, from each DOA vector, its corresponding sound event is detected if the magnitude of the vector exceeds the threshold of 0.5, and then localized using the vector's azimuth and elevation as shown in Figs. 5.5(b), (c) and (d) [181]. In the localization on a unit sphere, the range of sound event location along each Cartesian axis is between −1 and 1. Hence, the tanh activator is used to keep the output vector in a similar range. The BiSELDnet is a kind of regressor that estimates a DOA vector for each active sound event class every time frame. Notably, for DOA estimation, the Cartesian coordinate system is more advantageous for learning than the spherical coordinate system because the Cartesian coordinates are continuous, while spherical coordinates are discontinuous at the wrap-around boundary ($\theta = -180°$ or $180°$) [33].

In each time frame, the target value for each active sound event class is 1 while that for inactive one is 0. Similarly, the reference location (*x*, *y*, *z*) is used as the target value for each active sound event location and the origin (0, 0, 0) is used for inactive one [182]. Since the BiSELDnet is a type of regressor, MSE loss is used for training it. Interestingly, the reduction in MSE loss means that the 3D space distance between the reference vector and the DOA vector is reduced. The BiSELDnet was trained with MSE loss for 1,000 epochs using Adam optimizer with the batch size of 128. Early stopping is used to prevent the BiSELDnet from overfitting to the training dataset. The training is set to stop if SELD error on the validation dataset does not improve for 50 epochs. Evaluation metrics, including SELD error, are introduced later in Section 6.1. The total number of parameters in the BiSELDnet-v1 is shown in Table 5.1. For reference, the deep learning server was equipped with 1 × Intel® i9-10900X CPU, 3 × NVIDIA® GeForce RTX 3090 GPUs, and 128 GB RAM. All versions of BiSELDnet were implemented on Python 3.9.5 and Keras framework with TensorFlow 2.5.0 backend under Linux Ubuntu 20.04 OS.

**Table 5.1.** Total number of parameters in the BiSELDnet-v1.

|            | Trainable | Non-trainable | Total   |
|------------|-----------|---------------|---------|
| Parameters | 762,380   | 640           | 763,020 |

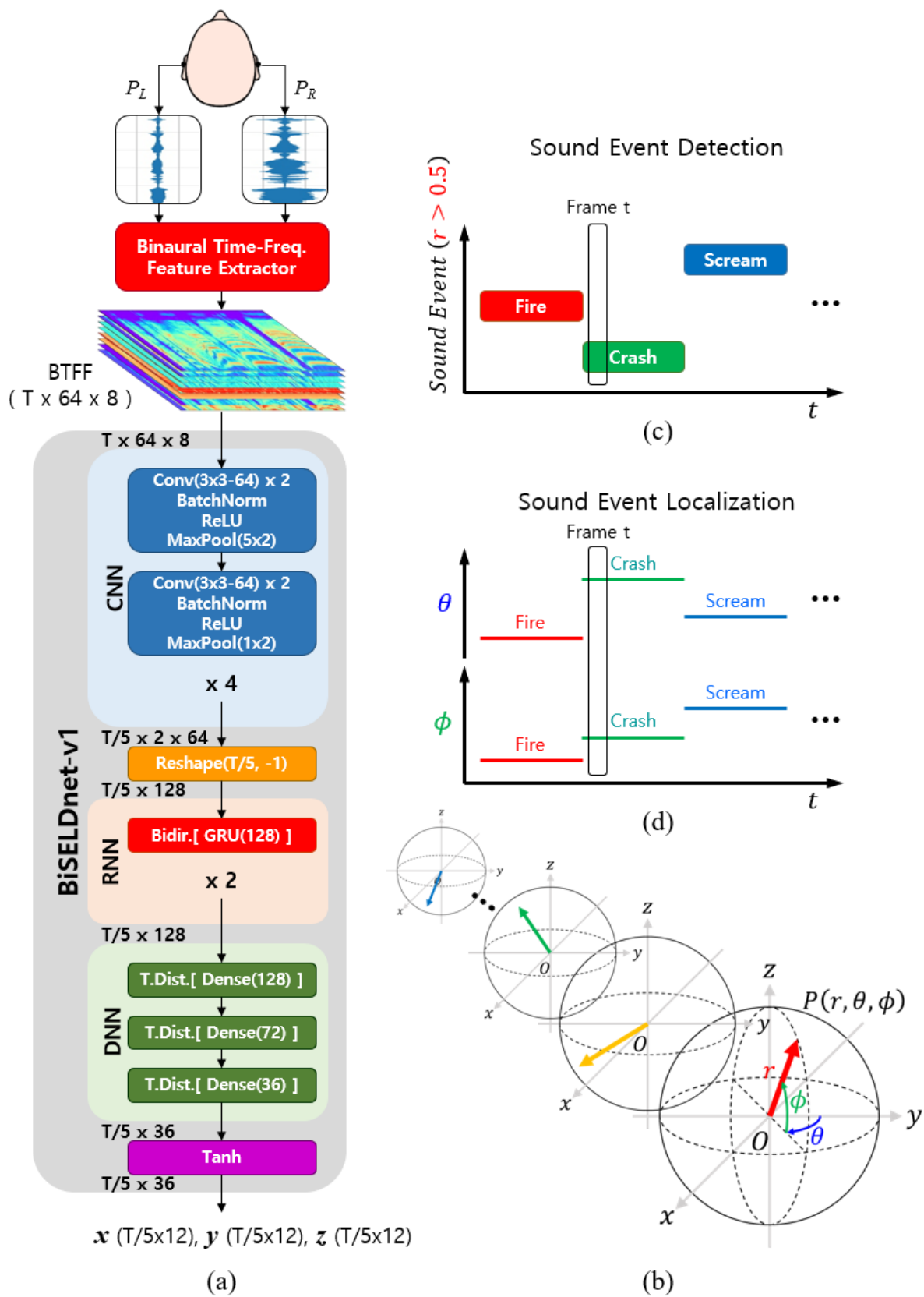

**Figure 5.5.** CRNN based BiSELD model: (a) BiSELDnet-v1 architecture with BTFF, (b) coordinate system conversion of output vectors (Cartesian → spherical), (c) result of sound event detection, and (d) result of sound event localization.

### 5.3.2 BiSELDnet-v2 (Hierarchical CRNN)

BiSELDnet-v2 is a hierarchical CRNN based deep neural network designed for the BiSELD task. In deep learning, all popular model architectures are not composed of just layers, but of repeated groups of layers (blocks or modules). For example, VGGNet, a representative model of this type, has an architecture in which 'convolution, convolution, and max pooling' blocks are repeated [183]. Designing the architecture of BiSELDnet-v2 involves key principles such as modularity and hierarchy. The modularity refers to the practice of breaking down a complex neural network into smaller, self-contained modules. Each module is responsible for a specific operation, and these modules can be interconnected to form the complete network. The hierarchy involves organizing modules in a structured manner, typically in a hierarchical fashion [171].

As shown in Fig. 5.6(a), the architecture of BiSELDnet-v2 basically inherited the CRNN structure of Version 1. However, Version 2 applied the design principles and stacked CNN and RNN modules hierarchically like a pyramid [184]. To be specific, comparing their CNN modules, in Version 1, every module uses two convolutional layers with 64 kernels, while in Version 2, as the module gets deeper, the number of convolutional layers in each module increases from one to three, and the number of the kernels increases from 32 to 1024. In Version 2, by reducing the feature map size, the network can capture increasingly abstract and high-level features. Early layers may focus on detecting low-level features like edges and textures, while deeper layers can learn more complex and higher-level patterns, such as object parts or object categories. Maintaining a hierarchy of features allows the network to understand the data at multiple scales. This ensures that the network can effectively learn and represent features in the data while avoiding information bottlenecks that could hinder its performance. Also, comparing RNN modules, in Version 1, the number of all GRU nodes is fixed at 128, while in Version 2, the number of GRU nodes decreases from 1,024 to 128 as the layer goes deeper. Reducing the number of nodes in later recurrent layers decreases the computational load. RNNs are notorious for their computational demands, and having fewer nodes in later layers helps mitigate this problem, making the network more feasible to train. For reference, the total number of parameters in the BiSELDnet-v2 is shown in Table 5.2.

**Table 5.2.** Total number of parameters in the BiSELDnet-v2.

|  | Trainable | Non-trainable | Total |
| --- | --- | --- | --- |
| Parameters | 50,086,732 | 11,584 | 50,098,316 |

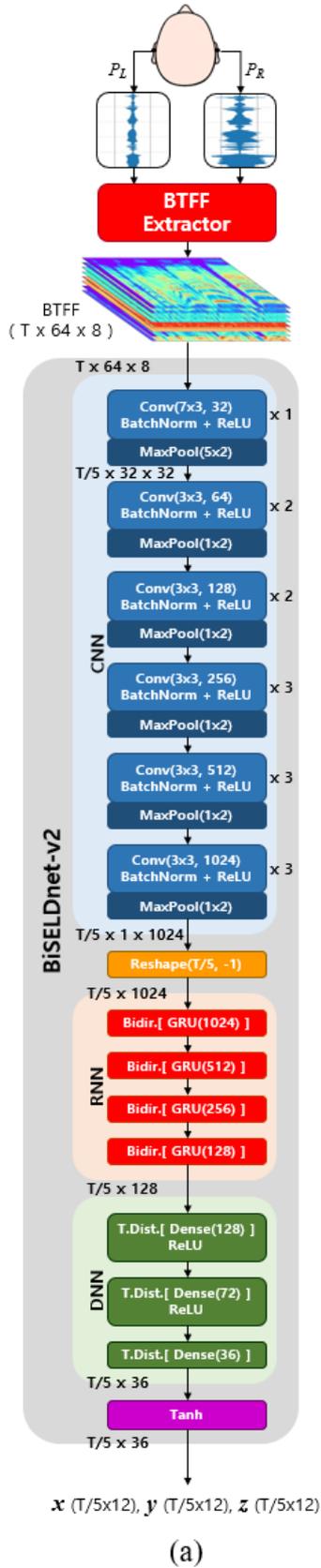
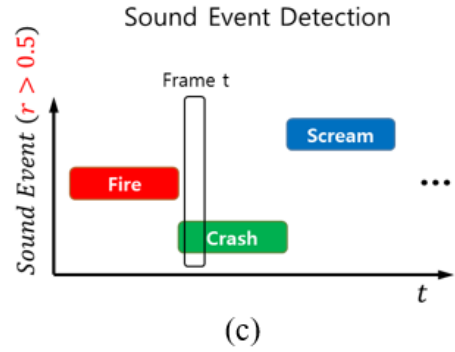
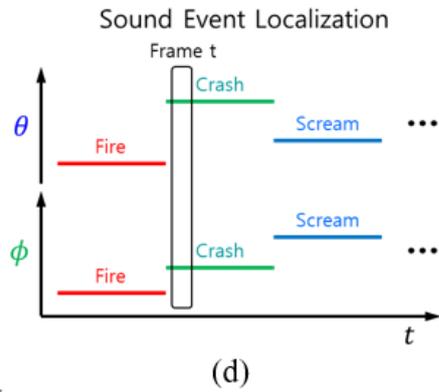
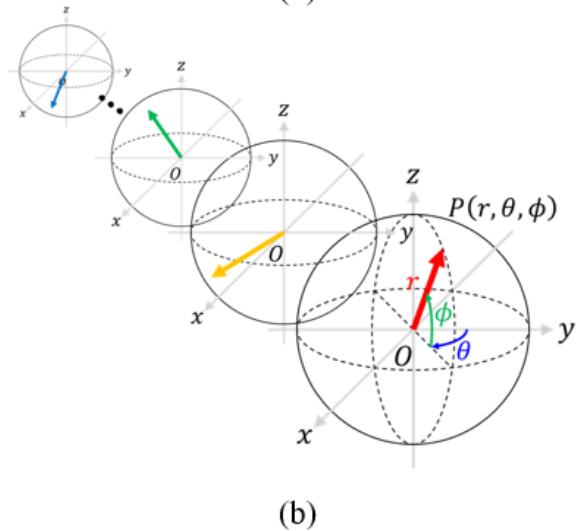

**Figure 5.6.** Hierarchical CRNN based BiSELD model: (a) BiSELDnet-v2 architecture with BTFF, (b) coordinate system conversion of output vectors (Cartesian → spherical), (c) result of sound event detection, and (d) result of sound event localization.

### 5.3.3 BiSELDnet-v3 (Xception Module)

BiSELDnet-v3 is an Xception module based deep neural network designed for the BiSELD task. Xception, short for "Extreme Inception," is a CNN architecture that was introduced by François Chollet, the creator of Keras [185]. Xception is a significant advancement in the computer vision tasks, and it is based on the Inception architecture of GoogLeNet [186] and the skip connection of ResNet [187] but introduces a novel concept called depthwise separable convolutions. The depthwise separable convolution relies on the assumption that the patterns on each plane (height and width) of input feature are highly correlated, but the patterns between channels (depth) are very independent. As shown in Fig. 5.7, a BTFF has lower cross-channel correlation than an image, so it can be considered an input feature more suitable for the assumption of depthwise separable convolution. A model with strong assumptions about the information structure of input features is a better model, as long as these assumptions are correct. Therefore, it is reasonable to design a new architecture of BiSELDnet by introducing the Xception module.

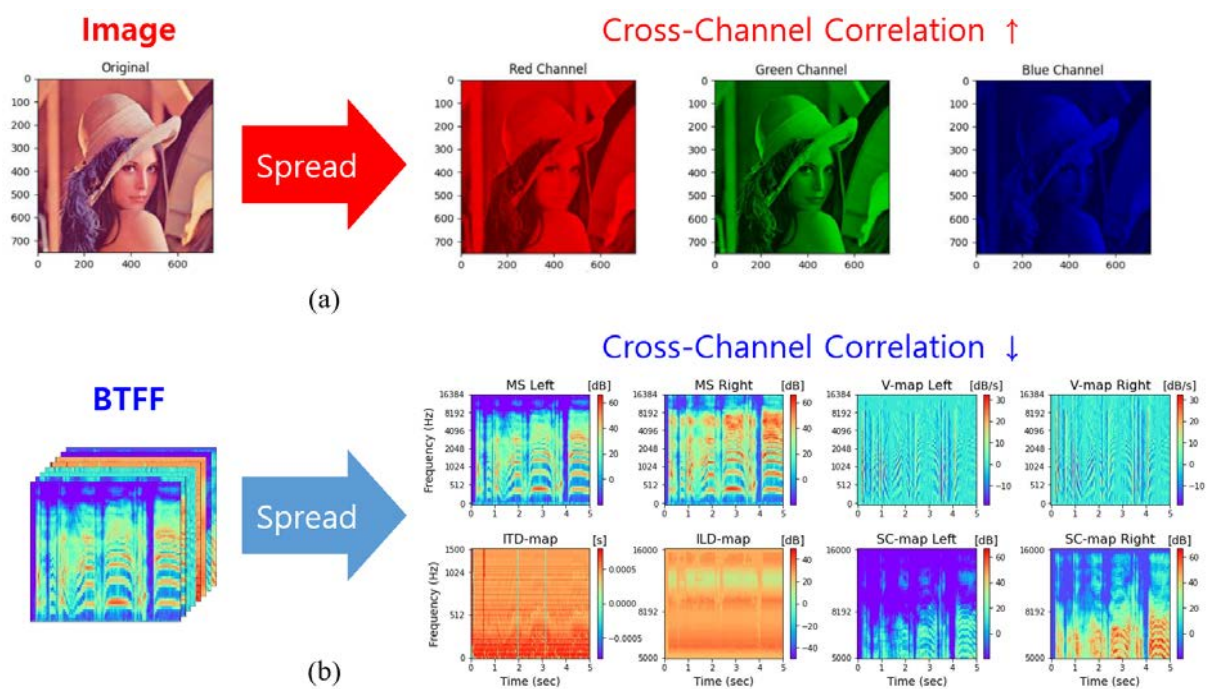

**Figure 5.7.** Difference between image and BTFF as input features: (a) high cross-channel correlation of image and (b) low cross-channel correlation of BTFF.

As shown in Fig. 5.8(a), the architecture of BiSELDnet-v3 inherited the hierarchical CRNN structure of Version 2, except for the heavy CNN modules. The CNN block of Version 3 consists of four Xception modules and one module of depthwise separable convolution at the input stage. Since the number of each kernel of the CNN block is gradually increased as in Version 2, the feature tensor that passes through each pooling layer becomes a feature vector suitable for the input of the RNN block. As shown in Fig. 5.8(b), each Xception module consists of two skip connection bundles, and each bundle consists of two depthwise separable convolution layers. In addition, since a max pooling layer is followed by the last depthwise separable convolution layer, the feature map is downsampled. The skip connection provides a shortcut for gradient flow, facilitating the training of BiSELDnet by mitigating the vanishing gradient problem and enabling the learning of both low-level and high-level features simultaneously.

As shown in Fig. 5.8(c), the input to each depthwise separable convolution is processed in two main steps: depthwise convolution and pointwise convolution. The depthwise convolution operates on each input channel independently. For a given input feature map, the depthwise convolution uses a separate 3×3 kernel for each input channel. Each kernel slides over its respective input channel to perform a spatial (time-frequency) convolution, producing a set of intermediate feature maps. These intermediate feature maps represent time-frequency information specific to each input channel. After the depthwise convolution, the pointwise convolution is applied. It involves a set of 1×1 kernels, where 1×1 means that the kernel covers a single element in the height and width dimensions but spans all input channels (depth). The pointwise convolution is responsible for combining and transforming the depthwise feature maps into a new set of feature maps. This step allows the network to capture cross-channel dependencies and create a more expressive representation. The number of 1×1 kernels determines the number of output channels of the feature map in the next layer. This step essentially controls the depth of the output feature maps.

The combination of depthwise convolution followed by pointwise convolution is expected to effectively reduce the computational complexity of conventional convolutions while preserving the network's representational power. The rest of the process is similar to previous versions. For reference, the total number of parameters in the BiSELDnet-v3 is shown in Table 5.3. Compared to Version 2, the total number of parameters in Version 3 is reduced by approximately 7.8 times.

**Table 5.3.** Total number of parameters in the BiSELDnet-v3.

|            | Trainable | Non-trainable | Total     |
|------------|-----------|---------------|-----------|
| Parameters | 6,443,380 | 7,808         | 6,451,188 |

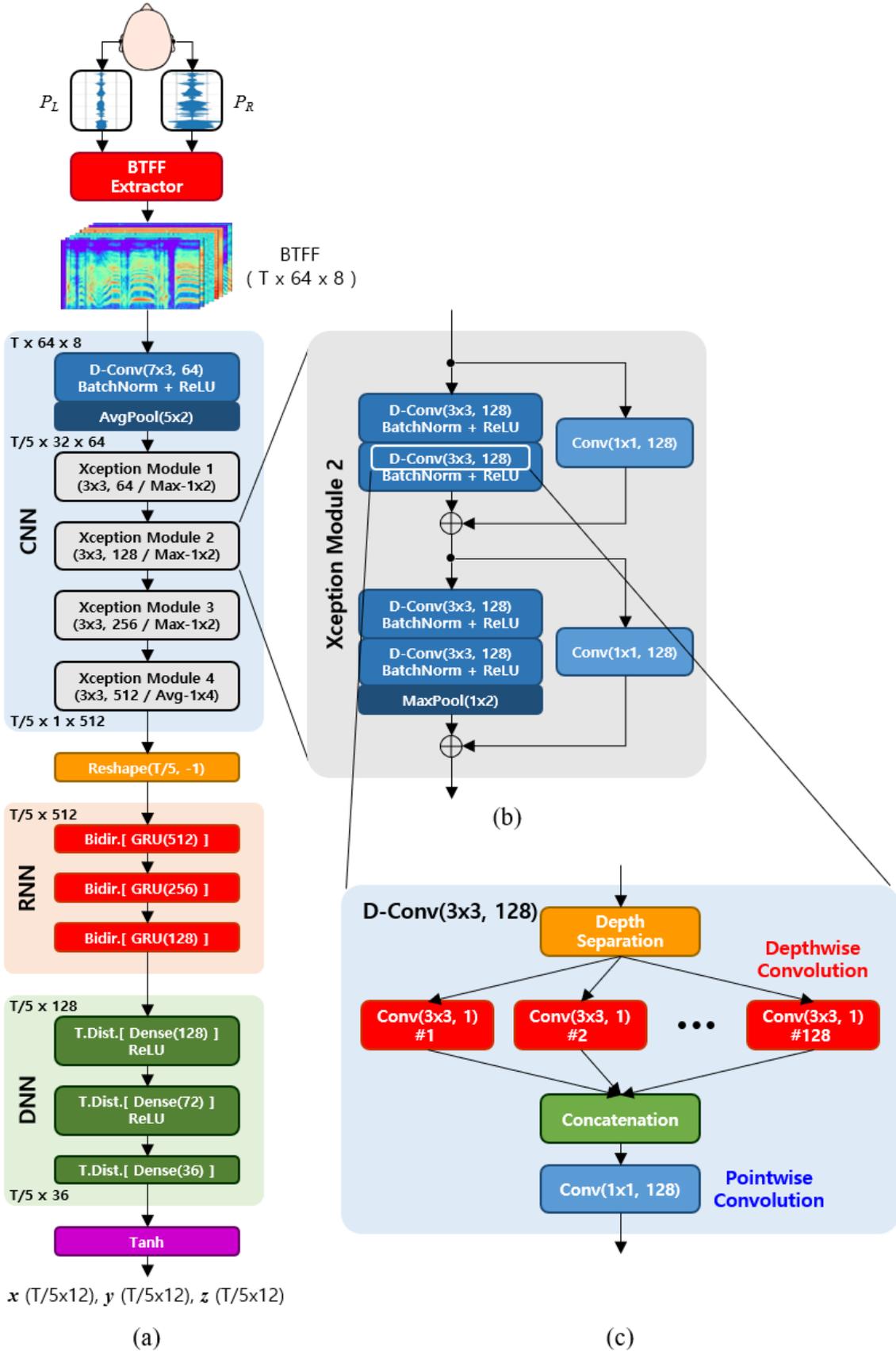

**Figure 5.8.** Xception module based BiSELD model: (a) BiSELDnet-v3 architecture with BTFF, (b) Xception module, and (c) depthwise separable convolution (D-Conv).

### 5.3.4 BiSELDnet-v4 (Trinity Module)

BiSELDnet-v4 is a Trinity module based deep neural network designed for the BiSELD task. Trinity module is a skip-connected module consisting of three concatenated depthwise separable convolutions that each output one-third of the output feature map using a receptive field of 3×3, 5×5, or 7×7. This novel architecture is proposed to simultaneously extract features of various sizes while further reducing the number of parameters compared to Xception module. In the BiSELD task, convolution with kernels of various sizes is effective because it allows the model to efficiently capture both local and global time-frequency patterns, facilitating the detection and localization of sound events at multiple resolutions, enhancing performance of BiSELDnet. To achieve this with a smaller number of parameters, each of the three depthwise separable convolutions of Trinity module with receptive fields of 3×3, 5×5, and 7×7 was implemented by factorization into 3×3 depthwise separable convolutions. The factorization of depthwise separable convolutions was inspired by the factorization of convolutions with large filter size in Inception-v3 [188], and the application of skip connection was introduced from the cases of Inception-v4 [189]. As shown in Fig. 5.9, comparing the receptive fields in the skip connection loop, the receptive field size of Xception module is 5×5, while the receptive field sizes of Trinity module are 3×3, 5×5, and 7×7. Therefore, Trinity module is more advantageous than Xception module in capturing a variety of time-frequency contexts from detailed to broader areas.

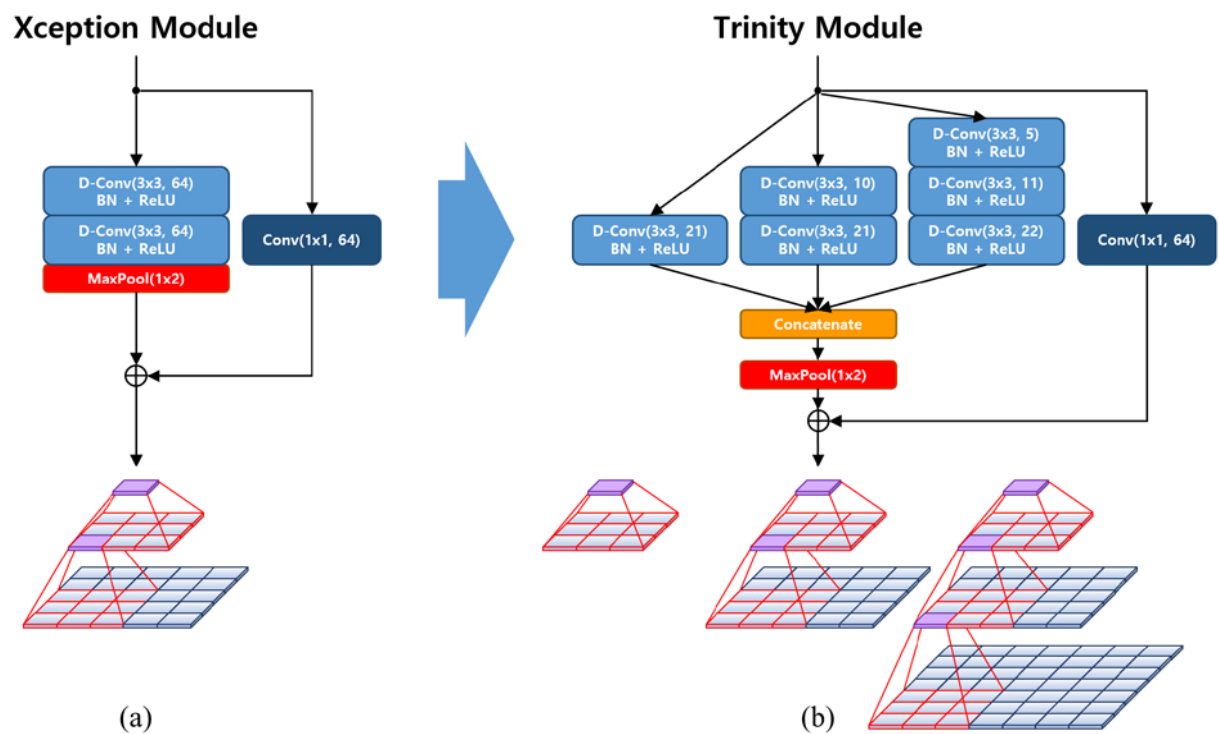

**Figure 5.9.** Differences in receptive fields between Xception module and Trinity module: (a) Xception module with a receptive field of 5×5 and (b) Trinity module with receptive fields of 3×3, 5×5, and 7×7.

In addition, Trinity module uses kernels with gradually increasing numbers, which not only reduces the number of parameters but also helps to generate hierarchical feature representations. To be specific, as shown in Fig. 5.9(b), given the number of output channels, Trinity module divides the number by three and assigns it to the number of kernels in each of the three convolution blocks, with any remainder assigned to the last block. The first convolution block consists of one depthwise separable convolution with the assigned number as the number of kernels. The second convolution block consists of two depthwise separable convolutions, where the assigned number is used as the number of kernels for the last, and half of the assigned number for the first. The third convolution block consists of three depthwise separable convolutions, where the assigned number is used in the last, half of the number in the second, and the quarter of the number in the first. The remainder that occurs when dividing the assigned number in each convolution block is discarded. For example, as shown in Fig. 5.9, when outputting a feature map with 64 channels, Xception module requires a total of 128 3×3 kernels, while Trinity module requires only a total of 90 3×3 kernels. Therefore, Trinity module can be stacked more deeply than Xception module because it is computationally more efficient, which helps in training deeper networks without overwhelming computational resources, and it is less prone to overfitting. Hierarchical feature representations of the trained Trinity modules are presented in Subsection 6.3.2.

As shown in Fig. 5.10(a), the architecture of BiSELDnet-v4 is the same as Version 3 except for the CNN module. Comparing the number of CNN modules based on the skip connection loop, Version 3 has a total of 8 Xception modules, while Version 4 has a total of 10 Trinity modules. However, comparing Tables 5.3 and 5.4, even though Version 4 consists of more CNN modules, the total number of parameters is smaller than Version 3. A deep model with a smaller number of parameters offer significant advantages in terms of resource efficiency. The reduced number of parameters results in lower memory and computational requirements, making it more accessible for training and deploying on hardware with limitations, such as edge devices or low-power processors. The rest of the process after Trinity modules is the same as Version 3.

**Table 5.4.** Total number of parameters in the BiSELDnet-v4.

|  | Trainable | Non-trainable | Total |
|---|---|---|---|
| Parameters | 6,130,852 | 13,016 | 6,143,868 |

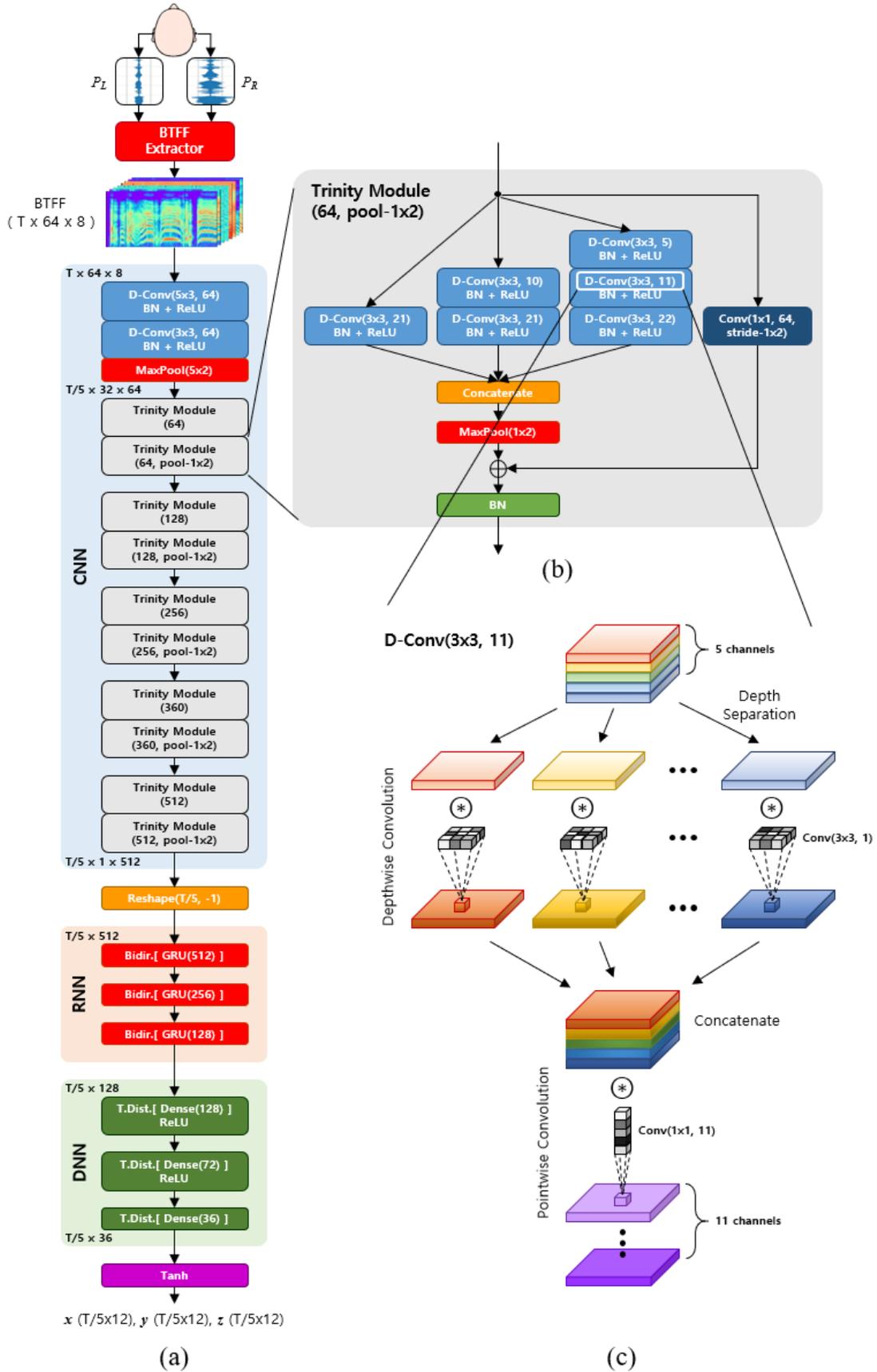

**Figure 5.10.** Trinity module based BiSELD model: (a) BiSELDnet-v4 architecture with BTFF, (b) Trinity module, and (c) depthwise separable convolution (D-Conv).

# Chapter 6. Performance Evaluation of BTFF and BiSELDnet

## 6.1 Evaluation Metrics

Evaluation metrics are essential because of the fundamental difference between the objective function and real-world performance. While the objective function, typically differentiable, guides model training by optimizing parameters, it might not directly align with the evaluation of model performance in most real-world applications. On the other hand, evaluation metrics, generally non-differentiable, provide a domain-specific perspective on model performance, offering a direct measure of how well the model accomplishes its intended task in real-world scenarios. In this section, evaluation metrics for the BiSELD task are introduced. In terms of evaluation, there is no difference between SELD task and BiSELD task, so the evaluation metrics presented in the SELD task of DCASE challenge are used to evaluate the performances of BiSELD models.

### 6.1.1 Sound Event Detection (SED) Metrics

F-score and error rate (ER) are used as SED metrics [33]. These metrics were used as the official metrics for SED tasks in the IEEE audio and acoustic signal processing (AASP) challenges [163]. A sound event is considered active in a one-second segment if it is active in at least one of the time frames within the segment. The segment-wise F-score is defined as follows:

$$F = \frac{2 \cdot \sum_{k=1}^{K} TP(k)}{2 \cdot \sum_{k=1}^{K} TP(k) + \sum_{k=1}^{K} FP(k) + \sum_{k=1}^{K} FN(k)}, \tag{6.1}$$

where true positive $TP(k)$ is the total number of active sound event classes in both reference and predicted one-second segment $k$; false positive $FP(k)$ is the total number of sound event classes that are active in prediction but inactive in reference; and false negative $FN(k)$ is the total number of sound event classes that are active in reference but inactive in prediction. In addition, ER is defined as follows:

$$ER = \frac{\sum_{k=1}^{K} S(k) + \sum_{k=1}^{K} D(k) + \sum_{k=1}^{K} I(k)}{\sum_{k=1}^{K} N(k)}, \tag{6.2}$$

$$S(k) = \min(FN(k), FP(k)), \tag{6.3}$$

$$D(k) = \max(0, FN(k) - FP(k)), \tag{6.4}$$

$$I(k) = \max(0, FP(k) - FN(k)), \tag{6.5}$$

where $S(k)$ is substitution; $D(k)$ is deletion; $I(k)$ is insertion; and $N(k)$ is the total number of active sound event classes in the reference one-second segment $k$. Substitution is obtained as the minimum value between false negatives and false positives without specifying which false positives substitute which false negatives. The remaining false negatives are counted as deletions and the remaining false positives are counted as insertions. An ideal BiSELD model will have an ER of zero and F-score of one. The detailed process of calculating F-score and ER for each time frame is shown in Fig. 6.1.

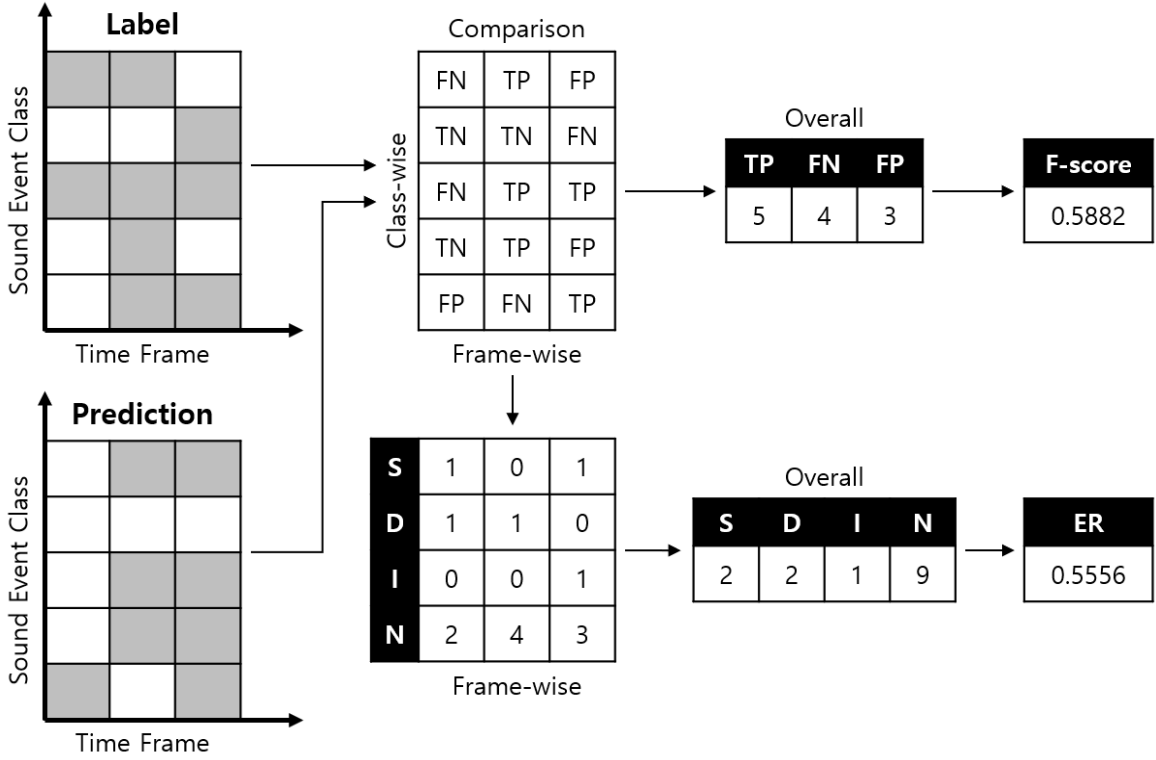

**Figure 6.1.** Illustration of the calculation process of F-score and error rate (ER) for each time frame.

### 6.1.2 Direction of Arrival (DOA) Metrics

Localization error (LE) and localization recall (LR) are used as DOA metrics [33]. LE is an averaged angular error between reference DOA and estimated DOA as shown in Fig. 6.2. LE is calculated as follows:

$$LE = \frac{1}{D}\sum_{d=1}^{D} \sigma\left(\left(x_R^d, y_R^d, z_R^d\right), \left(x_E^d, y_E^d, z_E^d\right)\right), \tag{6.6}$$

$$\sigma = 2 \cdot \sin^{-1}\left(\frac{\sqrt{\left(x_R^d - x_E^d\right)^2 + \left(y_R^d - y_E^d\right)^2 + \left(z_R^d - z_E^d\right)^2}}{2}\right) \cdot \frac{180}{\pi}, \tag{6.7}$$

where $D$ is the total number of DOA estimates across the entire dataset; $\left(x_R^d, y_R^d, z_R^d\right)$ is the $d$-th reference DOA; $\left(x_E^d, y_E^d, z_E^d\right)$ is the $d$-th estimated DOA; and $\sigma$ is the central angle between $(x_R^d, y_R^d, z_R^d)$ and $(x_E^d, y_E^d, z_E^d)$. The range of $\sigma$ is between 0° and 180°. In addition, LR is proposed to account for time frames where the number of estimated DOAs is not equal to the number of reference DOAs [33]. LR is defined as follows:

$$LR = \frac{TP_{DOA}}{TP_{DOA} + FN_{DOA}}, \qquad (6.8)$$

where true positive $TP_{DOA}$ is the total number of time frames where the number of estimated DOAs is equal to the number of reference DOAs; and false negative $FN_{DOA}$ is the total number of time frames where the number of estimated DOAs is not equal to the number of reference DOAs. An ideal BiSELD model will have a LE of zero and LR of one.

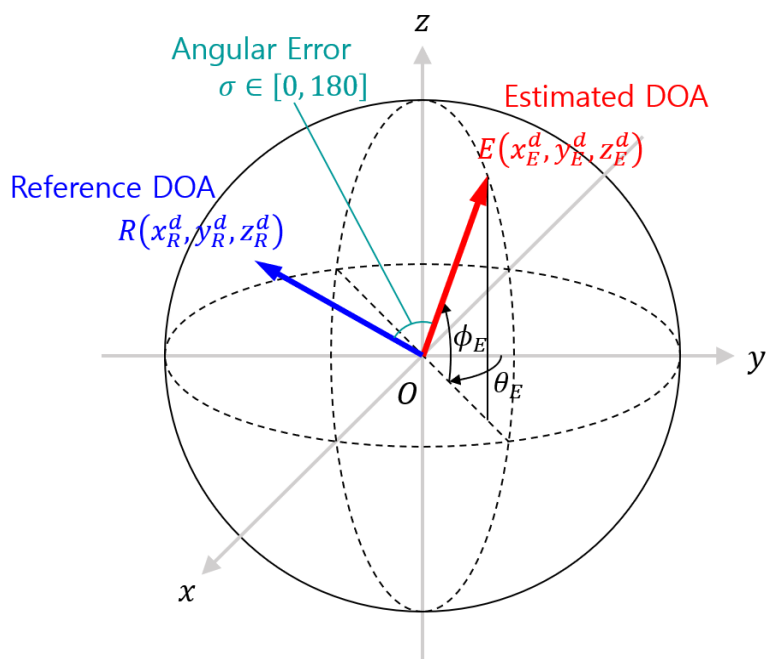

**Figure 6.2.** Illustration of angular error between reference DOA and estimated DOA.

### 6.1.3 Sound Event Localization and Detection (SELD) Metrics

Following the conventional SELD evaluation, BiSELD models are evaluated using the four joint localization and detection metrics [190]. The first two metrics are location-aware detection metrics, such as the error rate ($ER_{20°}$) and F-score ($F_{20°}$). The prediction is considered to be true positive only when the prediction and reference classes are the same, and the angle between them is less than 20°. The next two metrics are class-aware localization metrics, such as the localization error ($LE_{CD}$) in degrees and localization recall ($LR_{CD}$), where the subscript $CD$ refers to classification-dependent. Unlike the location-aware detection metrics, there is no angle threshold. However, the localization error represents the averaged angular distance between the predictions and references of the same class. Lastly, the localization recall expresses the true positive rate of the number of localization estimates detected in a class out of the total number of class events. As mentioned in Subsection 5.3.1, based on SELD error, training is stopped to prevent overfitting. The SELD error can be obtained by averaging SED and DOA errors. SED error integrates $ER_{20°}$ and $F_{20°}$ to evaluate the sound event detection performance of BiSELD model. The SED error is defined as follows:

$$SED\ error = \frac{ER_{20°} + (1 - F_{20°})}{2}, \tag{6.9}$$

An ideal BiSELD model will have a SED error of zero. In addition, DOA error integrates $LE_{CD}$ and $LR_{CD}$ to evaluate the sound event localization performance of BiSELD model. The DOA error is defined as follows:

$$DOA\ error = \frac{LE_{CD}/180 + (1 - LR_{CD})}{2}. \tag{6.10}$$

An ideal BiSELD model will have a DOA error of zero. Finally, SELD error integrates SED error and DOA error to jointly evaluate the sound event localization and detection performance of BiSELD model. The SELD error is defined as follows:

$$SELD\ error = \frac{SED\ error + DOA\ error}{2}. \tag{6.11}$$

An ideal BiSELD model will have a SELD error of zero.

## 6.2 Performance Evaluation of BTFF

In Chapter 4, BTFF, an input feature for BiSELD models, was proposed based on the analysis of human auditory process and HRTF localization cues. In addition, by examining the pattern change according to the direction of a sound event, it was visually confirmed that BTFF clearly expressed the time-frequency pattern of the sound event and HRTF localization clues. In this section, in order to quantitatively analyze the effect of BTFF, the performance of various combinations of BTFF sub-features is evaluated based on the BiSELDnet proposed in Chapter 5. Since the main purpose is to verify the effectiveness of BTFF, the dataset used for evaluation is the 'Binaural Set under clean condition' presented in Table 4.2. MS was used as the baseline sub-feature, and each other sub-feature was concatenated to MS for the corresponding evaluation. To be specific, MS and V-map were combined to evaluate the detection performance; MS, ITD-map, and ILD-map were combined to evaluate the horizontal localization performance; and MS and SC-map were combined to evaluate the vertical localization performance. For faster training and evaluation, CRNN based BiSELDnet-v1, which has the smallest number of parameters among BiSELDnets, was used for the BTFF evaluation. A total of ten training and evaluation sessions were conducted for each evaluation scenario.

### 6.2.1 Combination of MS and V-map

To examine the effect of V-map on binaural sound event detection, the evaluation results of the BiSELDnet-v1 trained using MS are compared to that trained using MS + V. For convenience, the concatenation of MS and V-map is referred to MS + V. The evaluation results are shown in Fig. 6.3, and the median values for the results are summarized in Table 6.1, with the best results bolded and underlined. When MS + V was used as the input feature, ER was decreased from 0.392 to 0.350 and F-score was improved from 75.0% to 77.9%, resulting in a decrease in SED error from 0.321 to 0.286. Therefore, it is verified that adding V-map to MS to create input features is effective in detecting binaural sound events. In addition, in the case of MS + V, LE was decreased from 19.4° to 18.9° and LR was increased from 89.9% to 91.5%, resulting in a slight decrease in DOA error from 0.106 to 0.093. This is because, as mentioned in Subsection 6.1.3, the detection and localization metrics in the SELD task are interdependent on each other. Therefore, the localization performance was slightly improved due to the improvement in detection performance.

Using V-map as an additional input feature for deep learning models provides an important advantage by enabling the capture of temporal dynamics and transitions in sound events [156]. V-map represents the rate of change in audio characteristics over time, facilitating the detection of variations in pitch, timbre, loudness, and tempo. This is particularly valuable in applications such as sound event detection devices, where understanding how audio evolves is essential for accurate interpretation. Moreover, V-map enhances the model's robustness against noise and background interference [156]. By modeling differences between consecutive audio frames, the model becomes less sensitive to environmental sounds and fluctuations, leading to improved noise tolerance. In real-world audio scenarios where background noises are prevalent, the inclusion of V-map is expected to ensure improved detection performance, making the model more effective in the BiSELD task.

**Table 6.1.** Median values for evaluation results of BiSELDnet-v1 when trained with MS and MS + V-map (MS + V) as input features, respectively.

|  | Sound Event Detection | | | Sound Event Localization | | | Total |
| --- | --- | --- | --- | --- | --- | --- | --- |
|  | $ER_{20°}\downarrow$ | $F_{20°}\uparrow$ (%) | SED Error$\downarrow$ | $LE_{CD}\downarrow$ (°) | $LR_{CD}\uparrow$ (%) | DOA Error$\downarrow$ | SELD Error$\downarrow$ |
| MS | 0.392 | 75.0 | 0.321 | 19.4 | 89.9 | 0.106 | 0.214 |
| MS + V | **<u>0.350</u>** | **<u>77.9</u>** | **<u>0.286</u>** | **<u>18.9</u>** | **<u>91.5</u>** | **<u>0.093</u>** | **<u>0.189</u>** |

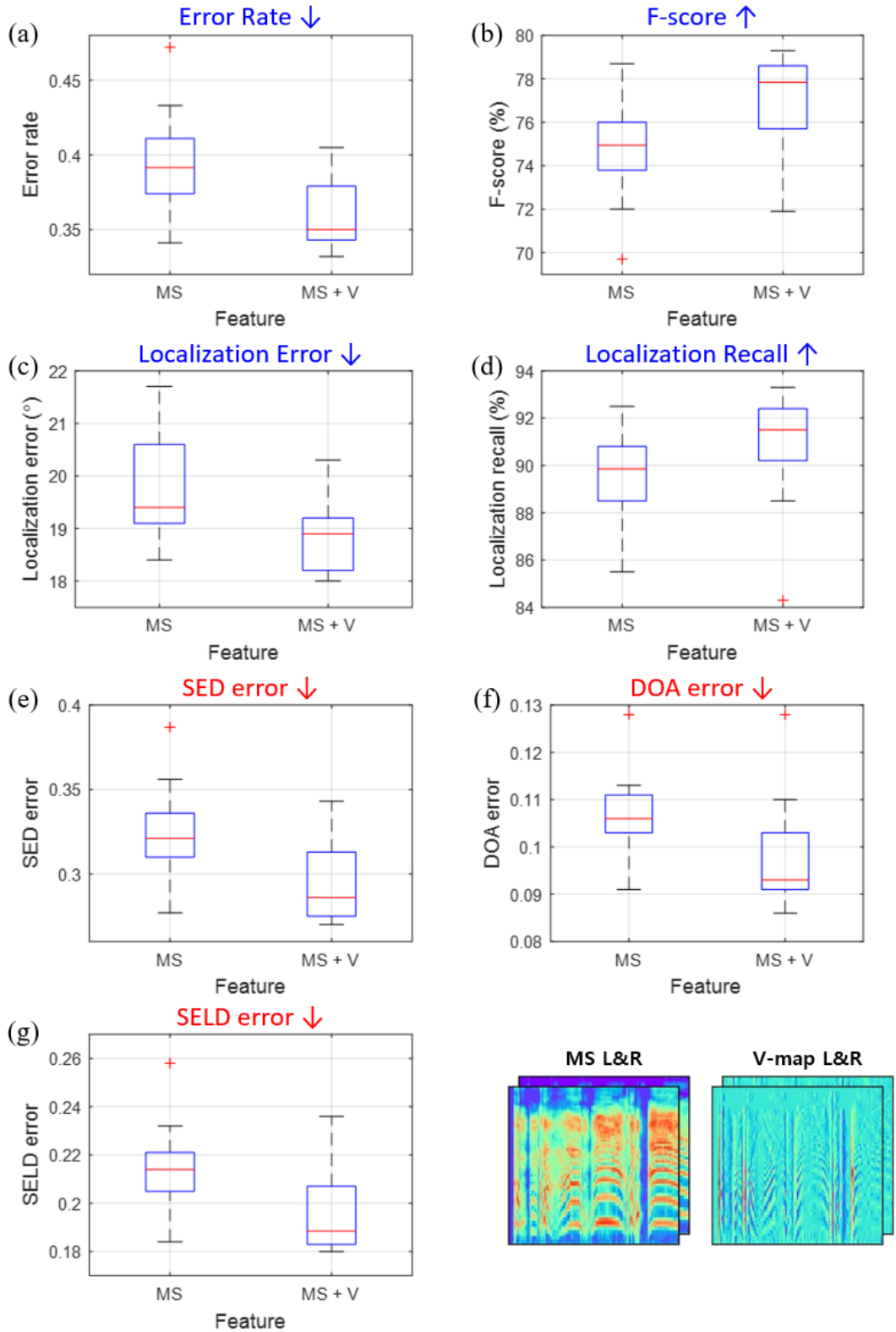

**Figure 6.3.** Evaluation results of BiSELDnet-v1 when trained with MS and MS + V-map (MS + V) as input features, respectively: (a) error rate, (b) F-score (%), (c) localization error (°), (d) localization recall (%), (e) SED error, (f) DOA error, and (g) SELD error.

### 6.2.2 Combination of MS, ITD-map, and ILD-map on the Horizontal Plane

To examine the effect of ITD-map and ILD-map on binaural sound event localization, the performance of the BiSELDnet-v1 trained using MS is compared on the horizontal plane with that trained adding ITD-map and ILD-map. In this case, the Test-H dataset, which consists only of sound event data on the horizontal plane as specified in Table 4.2, was used as a test dataset for the horizontal localization evaluation. For convenience, the concatenation of MS and ITD-map is referred to MS + ITD; and the concatenation of MS, ITD-map, and ILD-map is referred to MS + ITD + ILD. The evaluation results are shown in Fig. 6.4, and the median values for the results are summarized in Table 6.2, with the best results bolded and underlined. When MS + ITD was used as the input feature, LE was decreased from 17.3° to 11.9° and LR was increased from 87.4% to 90.1%, resulting in a decrease in DOA error from 0.112 to 0.081. Moreover, when MS + ITD + ILD was used, LE was reduced to 4.2° and LR was improved to 90.8%, resulting in further decrease in DOA error to 0.060. Therefore, it is verified that adding ITD-map and ILD-map to MS to create input features is effective in localizing binaural sound events on the horizontal plane. In particular, the qualitative observation in Subsection 4.2.4 that the pattern of ILD-map complements the azimuth estimation of ITD-map is proven quantitatively. In addition, in the case of MS + ITD, ER was decreased from 0.387 to 0.278 and F-score was enhanced from 75.0% to 81.6%, resulting in a decrease in SED error from 0.319 to 0.236. Furthermore, when MS + ITD + ILD was used, ER was reduced to 0.235 and F-score was improved to 85.8%, resulting in further decrease in SED error to 0.187. Therefore, it is seen that the improvement in localization performance on the horizontal plane by ITD-map and ILD-map leads to the improvement in detection performance in the BiSELD task.

**Table 6.2.** Median values for evaluation results of BiSELDnet-v1 when trained with MS, MS + ITD-map (MS + ITD), and MS + ITD-map + ILD-map (MS + ITD + ILD) as input features, respectively, for sound events on the horizontal plane.

|  | Sound Event Detection | | | Sound Event Localization | | | Total |
| --- | --- | --- | --- | --- | --- | --- | --- |
|  | $ER_{20°}\downarrow$ | $F_{20°}\uparrow$ (%) | SED Error$\downarrow$ | $LE_{CD}\downarrow$ (°) | $LR_{CD}\uparrow$ (%) | DOA Error$\downarrow$ | SELD Error$\downarrow$ |
| MS | 0.387 | 75.0 | 0.319 | 17.3 | 87.4 | 0.112 | 0.217 |
| MS + ITD | 0.278 | 81.6 | 0.236 | 11.9 | 90.1 | 0.081 | 0.162 |
| MS + ITD + ILD | **<u>0.235</u>** | **<u>85.8</u>** | **<u>0.187</u>** | **<u>4.2</u>** | **<u>90.8</u>** | **<u>0.060</u>** | **<u>0.124</u>** |

When estimating the location of a sound event on the horizontal plane, BiSELDnet relies on several input features, with two of the most important ones being ITD-map and ILD-map. These sub-features play complementary roles in helping BiSELDnet determine the direction of a sound event.

ITD-map is based on the time delay with which a sound event arrives at each ear. When a sound event is on one side of the head (to the left or right), the ear closest to the source will receive the sound slightly earlier than the ear on the opposite side. ITD-map is most effective for low-frequency sounds below 1.5 kHz, as the wavelength of low-frequency sounds is long enough to create noticeable time differences between the ears. In contrast, for high-frequency sounds above 1.5 kHz, the head size becomes larger than the wavelength, so the interaural phase difference exceeds $2\pi$, causing confusion in localization. BiSELDnet processes these time differences and uses them to estimate the horizontal direction of the sound source.

ILD-map is based on the level difference of a sound event at each ear. When a sound event is on one side of the head, the ear closer to the sound event receives a higher-intensity sound compared to the ear on the other side. ILD-map is most effective for high-frequency sounds, especially above 5 kHz, as their shorter wavelengths create noticeable intensity differences between the ears, due to the head shadow effect. For low-frequency sounds, the intensity differences are generally too small to be reliable cues. BiSELDnet processes these intensity differences and uses them to estimate the direction of the sound event. If the right ear receives a louder sound, the source is perceived as coming from the right, and vice versa for the left.

ITD-map and ILD-map are complementary because they provide spatial localization information for different frequency ranges. ITD-map is more effective for low-frequency sounds below 1.5 kHz, while ILD-map is more effective for high-frequency sounds above 5 kHz. This division of role allows BiSELDnet to estimate the direction of sound event on the horizontal plane. In addition, ILD-map plays a complementary role in resolving the front-back confusion problem when using ITD-map to estimate the azimuth of a sound event, as described in Subsection 4.2.4. ITD-map is the dominant localization cue at low frequencies, but there is no difference in front and back patterns, causing front-back confusion problems. However, ILD-map, unlike this, changes into a complex pattern depending on the azimuth at high frequencies, so it can be a clue to resolving ambiguous localization. This is caused by the front-back asymmetry of the head shape and ear position, and the diffraction effect of the pinna. Therefore, the resulting front-back asymmetry of ILD-map provides a clue to resolve front-back confusion when estimating the azimuth of sound event.

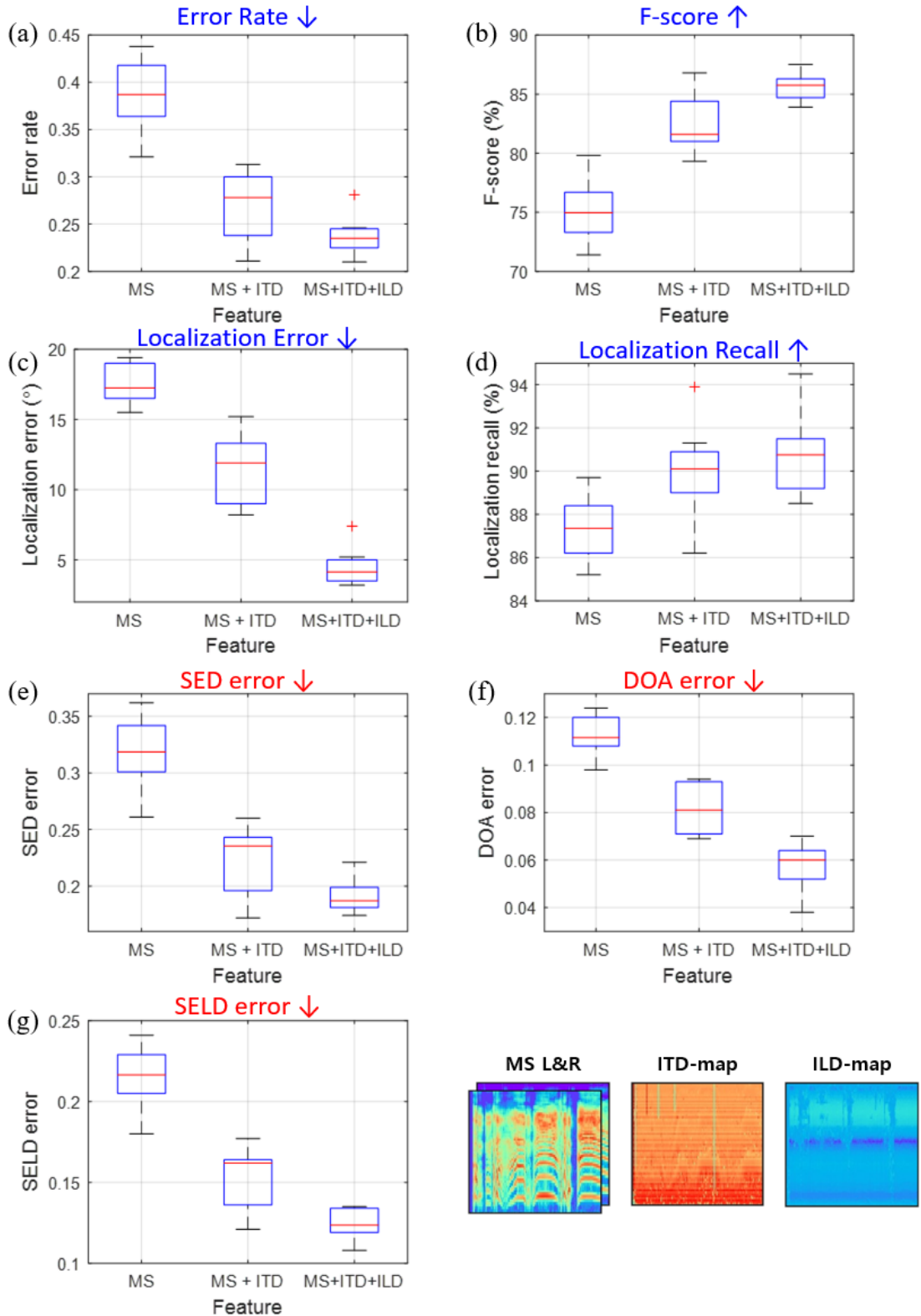

**Figure 6.4.** Evaluation results of BiSELDnet-v1 when trained with MS, MS + ITD-map (MS + ITD), and MS + ITD-map + ILD-map (MS + ITD + ILD) as input features, respectively, for sound events on the horizontal plane: (a) error rate, (b) F-score (%), (c) localization error (°), (d) localization recall (%), (e) SED error, (f) DOA error, and (g) SELD error.

### 6.2.3 Combination of MS and SC-map on the Median Plane

To examine the effect of SC-map on binaural sound event localization, the performance of the BiSELDnet-v1 trained using MS is compared on the median plane with that trained adding SC-map. In this case, the Test-V dataset, which consists only of sound events on the median plane as specified in Table 4.2, was used as a test dataset for the vertical localization evaluation. For convenience, the concatenation of MS and SC-map is referred to MS + SC. The evaluation results are shown in Fig. 6.5, and the median values for the results are summarized in Table 6.3, with the best results bolded and underlined. When MS + SC was used as the input feature, LE was decreased from 25.2° to 12.2° and LR was increased from 90.1% to 91.2%, resulting in a decrease in DOA error from 0.124 to 0.080. Therefore, it is verified that adding SC-map to MS to create input features is effective in localizing binaural sound events on the median plane. Notably, it is reconfirmed through SC-map that spectral cues or patterns above 5 kHz, generated by reflection and diffraction of the pinna, provide useful information for vertical localization. In addition, in the case of MS + SC, ER was reduced from 0.369 to 0.259 and F-score was improved from 76.6% to 84.2%, resulting in a decrease in SED error from 0.300 to 0.208. Therefore, it is also seen that the improvement in localization performance on the median plane by SC-map leads to the improvement in detection performance in the BiSELD task.

Spectral cues above 5 kHz provide crucial information for vertical localization by enhancing the perception of sound elevation. These high-frequency spectral cues arise from the complex filtering effects of the pinna and head on an incoming sound wave and manifest as distinctive spectral patterns that are specific to the elevation angle of the sound event. As confirmed in the spectral cue analysis of HRTFs, the N1 notch frequency is observed at approximately 8 kHz and moves to higher frequencies as the elevation of sound source increases. Therefore, inferring from these results, BiSELDnet might have learned the movement patterns of the N1 frequency according to the elevation of sound events shown in the SC-map. Details are discussed in the BiSELDnet visualization in Subsection 6.3.3.

**Table 6.3.** Median values for evaluation results of BiSELDnet-v1 when trained with MS and MS + SC-map (MS + SC) as input features, respectively, for sound events on the median plane.

|  | Sound Event Detection | | | Sound Event Localization | | | Total |
|---|---|---|---|---|---|---|---|
|  | $ER_{20°}\downarrow$ | $F_{20°}\uparrow$ (%) | SED Error$\downarrow$ | $LE_{CD}\downarrow$ (°) | $LR_{CD}\uparrow$ (%) | DOA Error$\downarrow$ | SELD Error$\downarrow$ |
| MS | 0.369 | 76.6 | 0.300 | 25.2 | 90.1 | 0.124 | 0.211 |
| MS + SC | **<u>0.259</u>** | **<u>84.2</u>** | **<u>0.208</u>** | **<u>12.2</u>** | **<u>91.2</u>** | **<u>0.080</u>** | **<u>0.145</u>** |

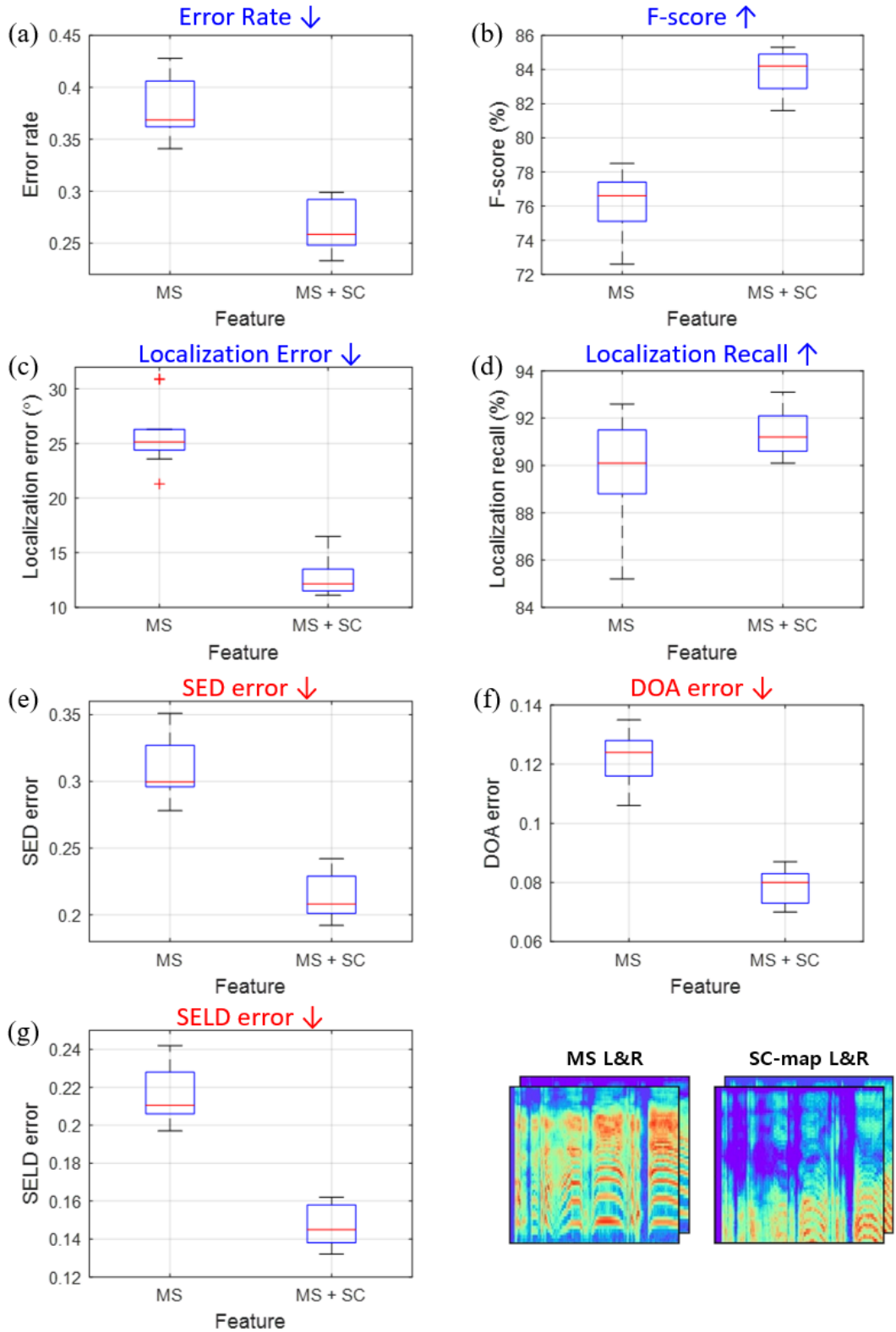

**Figure 6.5.** Evaluation results of BiSELDnet-v1 when trained with MS and MS + SC-map (MS + SC) as input features, respectively, for sound events on the median plane: (a) error rate, (b) F-score (%), (c) localization error (°), (d) localization recall (%), (e) SED error, (f) DOA error, and (g) SELD error.

## 6.3 Performance Evaluation and Visualization of BiSELDnet

In Chapter 5, various versions of BiSELD networks were proposed, including CRNN based BiSELDnet-v1, hierarchical CRNN based BiSELDnet-v2, Xception module based BiSELDnet-v3, and Trinity module based BiSELDnet-v4, which are expected to effectively learn the informative patterns of BTFF and successfully perform the BiSELD task. In this section, the performance of each model is quantitatively evaluated, and the model with the best performance is selected as the final BiSELDnet. Additionally, in order to understand what BiSELDnet learned and the decisions it made, the internal modules of BiSELDnet are analyzed using visualization techniques such as layer output visualization and vector activation map. To evaluate performance in the basic mode, the 'Binaural Set under clean condition' presented in Table 4.2 was used as the dataset for training and evaluation of each BiSELDnet. Since the effect of all sub-features was verified, an eight-channel BTFF with all sub-features was used as the input feature for BiSELDnets. A total of 10 training and evaluation sessions were conducted for each BiSELDnet.

### 6.3.1 Performance Comparison of BiSELDnets

To verify the architecture and select the best model, the detection and localization performance of each BiSELDnet was quantitatively evaluated and compared with each other. In this case, the dataset used for evaluation is the 'Test' dataset of the Binaural Set specified in the Table 4.2. The evaluation results are shown in Fig. 6.6, and the best performance of each BiSELDnet is summarized in Table 6.4, with the best results bolded and underlined.

As a result of the evaluation, Version 4 was found to have the best performance in all evaluation metrics of detection and localization, even though it has the smallest number of parameters except Version 1. To be specific, the best evaluation results of Version 4 are as follows: ER of 0.114, F-score of 91.8%, LE of 2.5°, LR of 93.3%, SED error of 0.098, DOA error of 0.040, and SELD error of 0.069. These results show that Trinity architecture, designed to simultaneously extract features of various sizes while reducing the number of parameters based on factorized depthwise separable convolutions, is most suitable and effective for the BiSELD task. Since sound events have various sizes in time and frequency dimensions, Trinity architecture, which captures these patterns of various sizes simultaneously, seems to have an advantage over other architectures.

The next best model was the Xception module based Version 3, and the evaluation results are as follows: ER of 0.140, F-score of 90.0%, LE of 2.7°, LR of 92.2%, SED error of 0.120, DOA error of 0.047, and SELD error of 0.083. Compared to Version 4, the localization performance of Version 3 is not much different, but the detection performance is relatively low, as shown by the higher ER. The reason Version 3's ER is higher than that of Version 4 is because Trinity module uses three kernels of different sizes, while Xception module uses only a fixed-size kernel, making it disadvantageous to detect sound events of various time-frequency scales. However, Version 3 shows high performance compared to other lower versions because it is based on depthwise separable convolution. Depthwise separable convolution relies on the assumption that the patterns in each plane (height and width) of input features are highly correlated, but those between channels (depth) are highly independent, making it applicable to the input features with low cross-channel correlation, such as BTFF. Therefore, when designing a new model for BiSELD, it is recommended to use depthwise separable convolution.

The evaluation results of Version 2 based on hierarchical CRNN are as follows: ER of 0.158, F-score of 88.6%, LE of 5.3°, LR of 91.3%, SED error of 0.136, DOA error of 0.058, and SELD error of 0.097. Even though Version 2 has the largest number of parameters compared to other versions, its localization performance was inferior to Version 1. A large number of model parameters increases the risk of overfitting, as the model may become overly complex and fit the training data too closely, leading to poor generalization to unseen data. Interestingly, this is the background from which GoogLeNet [186] and ResNet [187] emerged, leading to Xception architecture [185] later.

CRNN based Version 1 is the baseline model for the BiSELD task and it showed competitive performance relative to the number of parameters. Finally, Trinity module based Version 4, which has the best detection and localization performance, was selected as the final model for the BiSELD task. Hereafter, BiSELDnet or BiSELD model refers to BiSELDnet-v4 based on Trinity module.

**Table 6.4.** Best performances of BiSELDnet-v1 (CRNN), BiSELDnet-v2 (Hierarchical CRNN), BiSELDnet-v3 (Xception module), and BiSELDnet-v4 (Trinity module).

| BiSELDnet | Total Params (M) | Sound Event Detection | | | Sound Event Localization | | | Total |
|---|---|---|---|---|---|---|---|---|
| | | $ER_{20°}\downarrow$ | $F_{20°}\uparrow$ (%) | SED Error$\downarrow$ | $LE_{CD}\downarrow$ (°) | $LR_{CD}\uparrow$ (%) | DOA Error$\downarrow$ | SELD Error$\downarrow$ |
| V1 (CRNN) | 0.8 | 0.210 | 87.1 | 0.169 | 4.4 | 92.1 | 0.052 | 0.110 |
| V2 (H-CRNN) | 50.1 | 0.158 | 88.6 | 0.136 | 5.3 | 91.3 | 0.058 | 0.097 |
| V3 (Xception) | 6.5 | 0.140 | 90.0 | 0.120 | 2.7 | 92.2 | 0.047 | 0.083 |
| V4 (Trinity) | 6.1 | **0.114** | **91.8** | **0.098** | **2.5** | **93.3** | **0.040** | **0.069** |

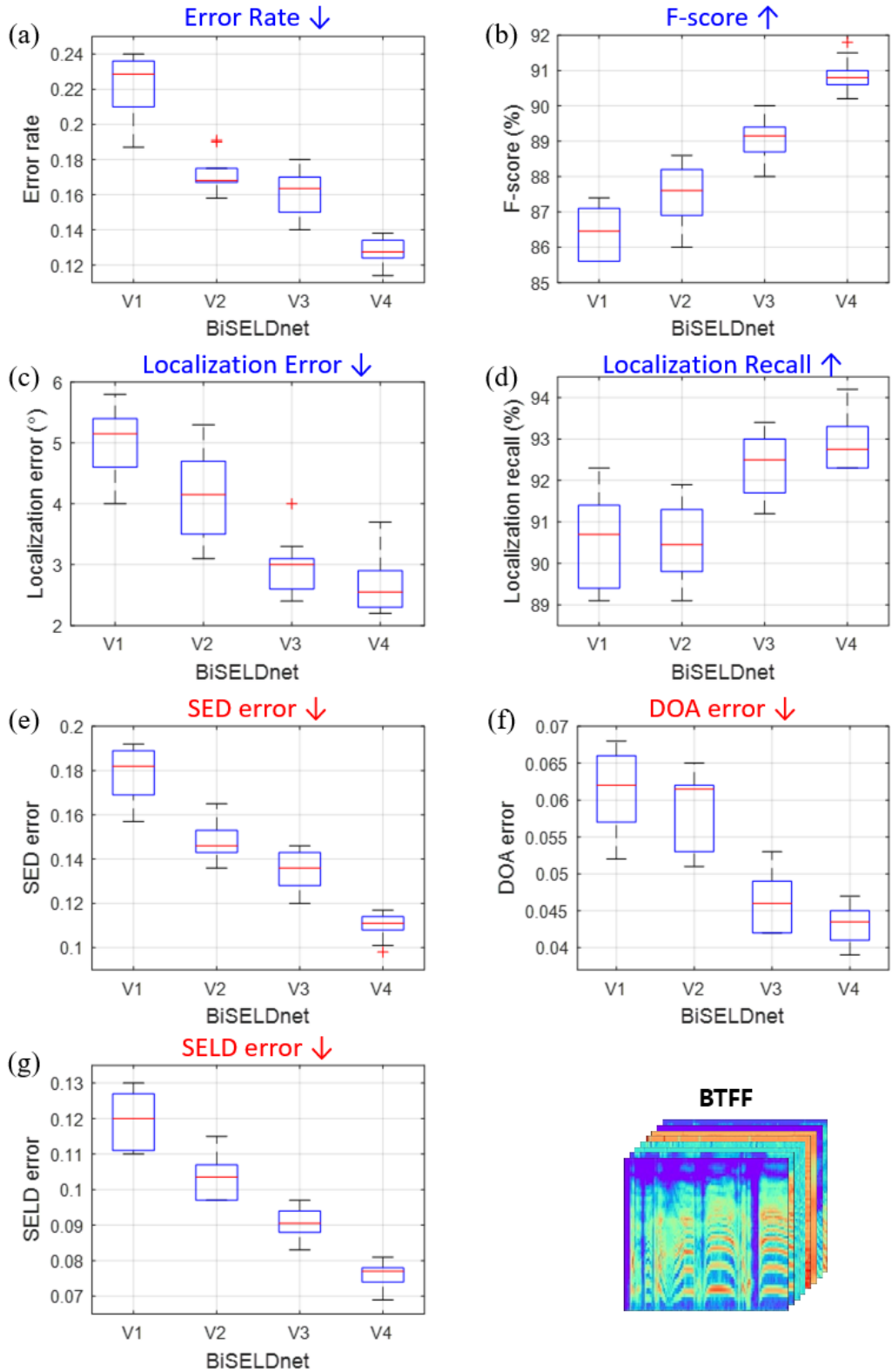

**Figure 6.6.** Evaluation results of BiSELDnet-v1 (CRNN), BiSELDnet-v2 (Hierarchical CRNN), BiSELDnet-v3 (Xception module), and BiSELDnet-v4 (Trinity module): (a) error rate, (b) F-score (%), (c) localization error (°), (d) localization recall (%), (e) SED error, (f) DOA error, and (g) SELD error.

The inference results of the BiSELDnet for a test data sample and its corresponding references are visualized in Fig. 6.7. Each *x*-axis of all sub-plots represents the same time frames. Each sound event class and its trajectory in azimuth and elevation are marked with its corresponding color. Figs. 6.7(a1) and (a2) are sub-plots for the left and right MSs of the test data sample. The unique patterns of each sound event such as onset, offset, periodicity, AM, FM, etc., and even the left and right level differences are clearly expressed.

Figs. 6.7(b1) and (b2) show the reference values and inference results of sound event detection for the test data sample, respectively. Each *y*-axis of the sub-plots represents the index of sound event class (0: Alarm, 1: Baby, 2: Cough, 3: Crash, 4: Dog, 5: Female Scream, 6: Female Speech, 7: Fire, 8: Knock, 9: Male Scream, 10: Male Speech, 11: Phone). The inference results of sound event detection are very accurate. Thus, BiSELDnet appears to have learned well the local shift-invariant features and temporal context information associated with each sound event class from the proposed BTFF. It is presumed that both MS and V-map played a major role in the sound event detection.

Figs. 6.7(c1) and (c2) show the reference values and inference results of the horizontal sound localization for the test data sample, respectively. Overall, the BiSELDnet accurately estimated the azimuth angle of each sound event. Although the azimuth estimates fluctuate slightly, their mean values are very close to their corresponding reference values. Therefore, it can be visually confirmed that the BiSELDnet learned the azimuth information from interaural cues inherent in each sound event by extracting related feature maps from the BTFF. It is assumed that both ITD-map and ILD-map played a main role in the horizontal localization of sound events.

Figs. 6.7(d1) and (d2) show the reference values and inference results of the vertical sound localization for the test data sample, respectively. Overall, the BiSELDnet accurately estimated the elevation angle of each sound event. Although the elevation estimates fluctuate slightly, their mean values are very close to their respective reference values. Therefore, it can be visually confirmed that the BiSELDnet properly learned the HRTF localization cues inherent in each sound event by extracting related time-frequency patterns from the BTFF. It is also presumed that the SC-map played a vital role in the vertical localization of sound events.

In particular, the vertical sound localization performance of the BiSELDnet is remarkable. Even the SOTA binaural SELD model [98] estimated only the azimuth of sound events. In addition, since the conventional SELD model uses four audio input channels, the physical information of the vertical direction can be used when estimating sound event elevation. Whereas, BiSELD model uses only two audio input channels arranged horizontally. Thus, in order to estimate the elevation angle of sound events, it depends entirely on the HRTF localization cues, especially spectral cues, inherent in the sound events, just like us humans.

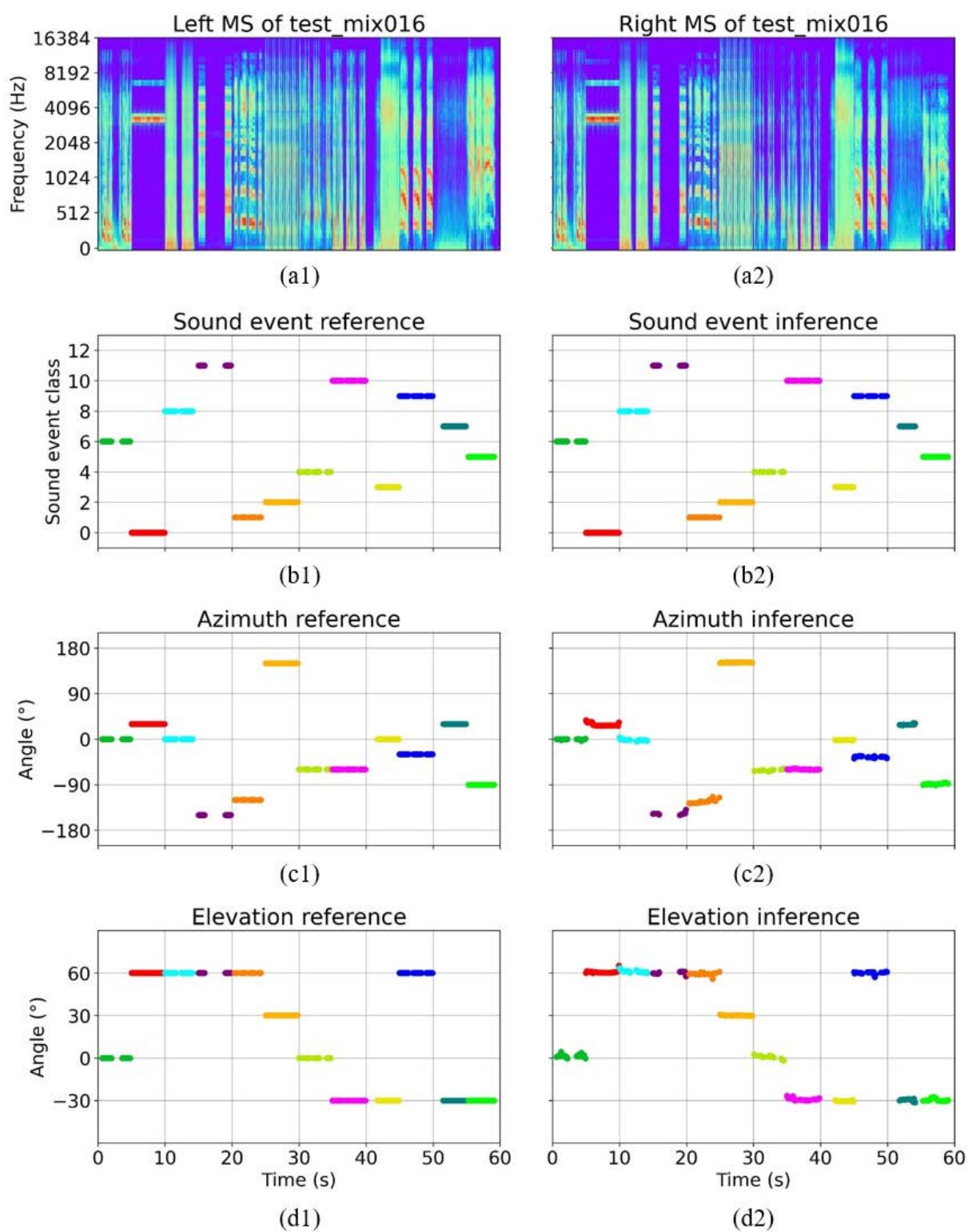

**Figure 6.7.** Inference results of BiSELDnet-v4 for the test data sample (test_mix016.wav): (a1) left MS of the test data sample, (a2) right MS of the test data sample, (b1) sound event reference (0: Alarm, 1: Baby, 2: Cough, 3: Crash, 4: Dog, 5: Female Scream, 6: Female Speech, 7: Fire, 8: Knock, 9: Male Scream, 10: Male Speech, 11: Phone), (b2) sound event inference, (c1) azimuth reference, (c2) azimuth inference, (d1) elevation reference, and (d2) elevation inference.

### 6.3.2 Layer Output Visualization of Trinity Module

Understanding how the kernels of successive convolutional layers transform the input is crucial for interpreting the learned features and hierarchical representations in CNNs. It enhances model interpretability by revealing what the network detects at various stages for decision-making. Layer output visualization is a visualization method that expresses the output of each convolutional layer as an image when an input feature is given. This method shows how the filters learned by the network decompose the input. Since each channel encodes relatively independent features, it is a good idea to draw each channel's content of feature maps as an independent 2D image.

In this subsection, the feature maps of each Trinity module were extracted to check the inside of the trained BiSELDnet. For this purpose, the trained BiSELDnet was changed to create a multi-output model that receives a BTFF batch as an input and then outputs feature maps from the concatenate layers of all Trinity modules. As shown in Fig. 6.8, since there are 10 concatenate layers, the multi-output BiSELDnet has a total of 10 outputs. The concatenate layer of each Trinity module integrates the outputs of three depthwise separable convolutions, so this layer is useful for observing the internals of the BiSELDnet. A baby crying WAV file (baby_a000e+60.wav) with azimuth of $0°$ and elevation of $60°$ was used as an input data for the multi-output BiSELDnet.

The number of feature maps for each output displayed in Fig. 6.9 was limited to 16. As shown in Fig. 6.9(a), the first layer is like a collection of several types of edge detectors. This stage of activation preserves almost all information about the input features. In Figs. 6.9(b)–(j), as the layer goes deeper, the activations become increasingly abstract and visually difficult to understand. It seems that high-level concepts such as time-frequency patterns and HRTF localization cues begin to be encoded. In addition, as the layer gets deeper, filters that become inactive begin to appear. In the first layer, as shown in Fig. 6.9(a), almost all filters are activated for the input features, but as we go through the layers, some filters become inactive. This means that the pattern encoded in the filter does not appear in the input feature maps. As a result, the feature maps of each layer become increasingly abstract depending on the depth of the layer, and the information about specific inputs gradually decreases and that about the target (class and location) increases. Typically, it is known that deep neural networks operate like an information refining pipeline for the raw input data [171]. Through repeated transformations, irrelevant information is filtered out and useful information (class and location) is emphasized and improved. This process is similar to the way humans and animals perceive objects around them. Our brain converts information received from sense organs into high-level concepts by filtering out less relevant elements [19]. The reason why it is difficult to remember the input information in detail is because learning is abstract.

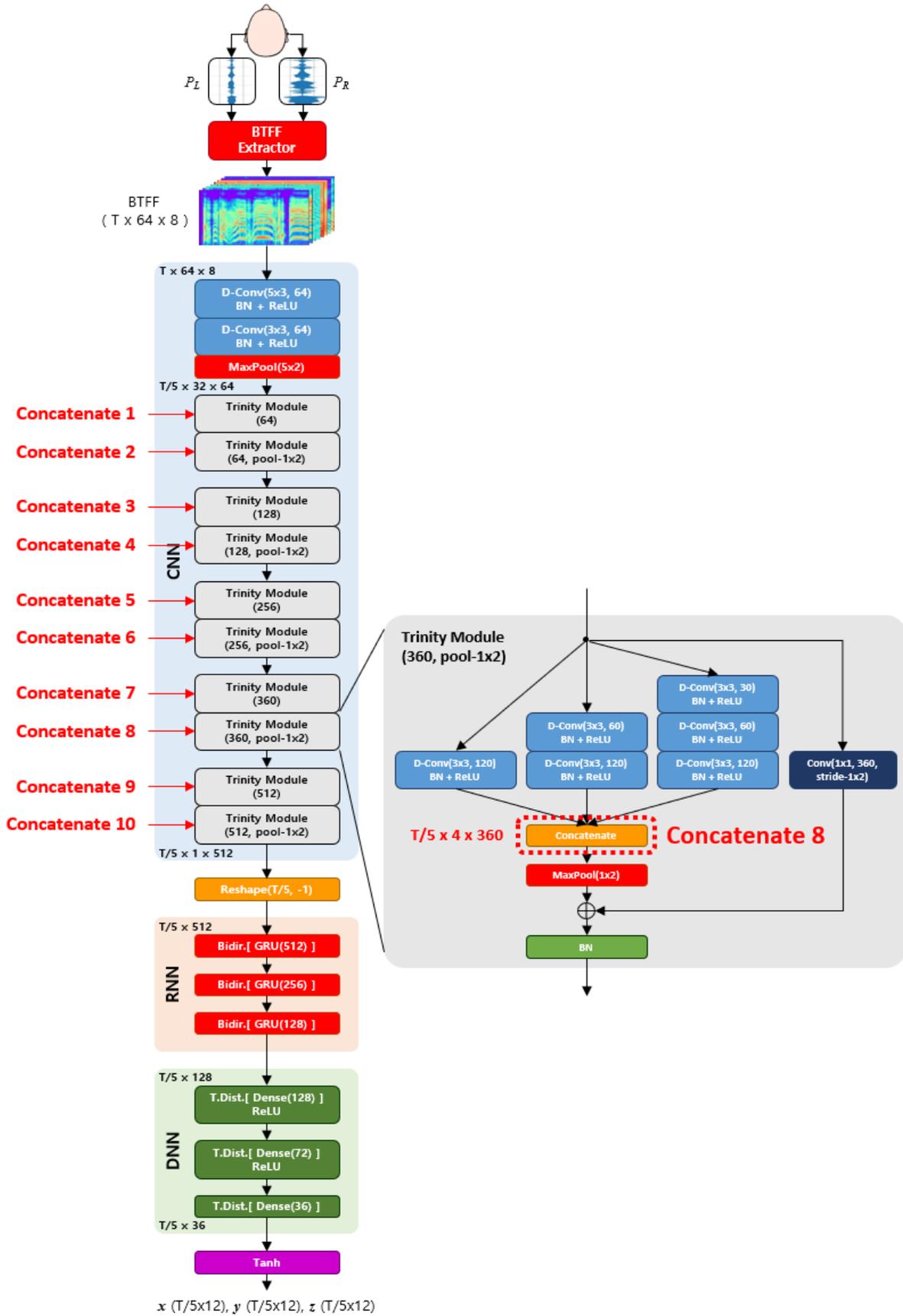

**Figure 6.8.** Concatenate layer in each Trinity module whose output feature maps are visualized through layer output visualization.

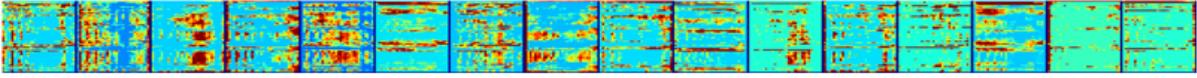

(a)

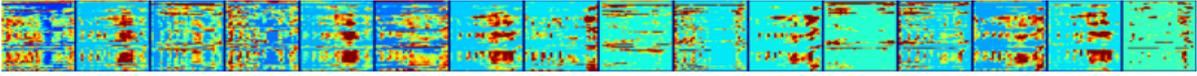

(b)

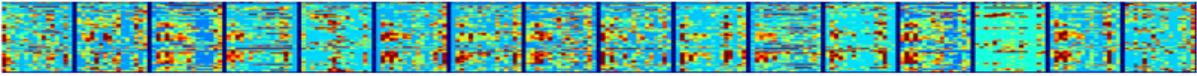

(c)

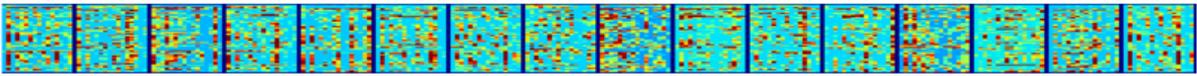

(d)

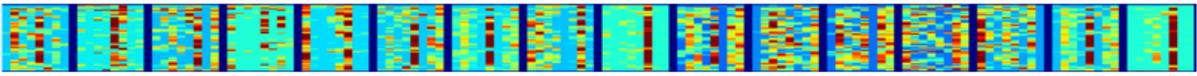

(e)

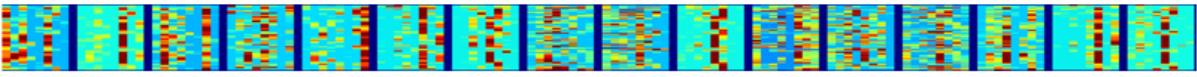

(f)

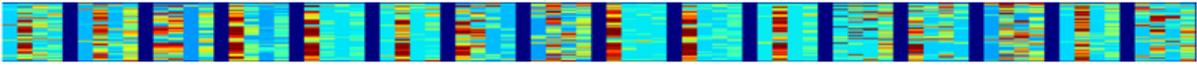

(g)

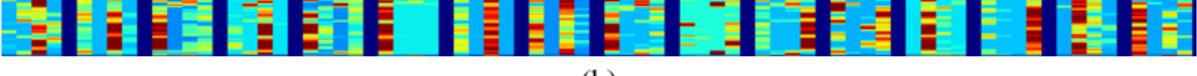

(h)

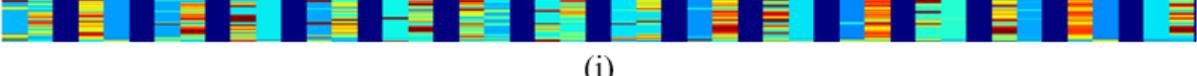

(i)

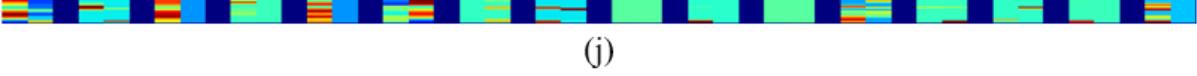

(j)

**Figure 6.9.** Output feature maps of the concatenate layer in each Trinity module for an input data sample (baby_a000e+60.wav): (a) concatenate 1, (b) concatenate 2, (c) concatenate 3, (d) concatenate 4, (e) concatenate 5, (f) concatenate 6, (g) concatenate 7, (h) concatenate 8, (i) concatenate 9, and (j) concatenate 10.

### 6.3.3 Vector Activation Map (VAM) of Trinity Module

Vector activation map (VAM) is a visualization method for SELD or BiSELD models proposed by modifying the existing visualization method to determine which part of input feature contributes to the model's final decision. Representative CNN visualization methods are class activation map (CAM) [191] and gradient-weighted class activation mapping (Grad-CAM) [192]. CAM is a method used in deep learning and computer vision to interpret and understand the decisions made by CNNs during image classification tasks. CAM provides a visual heatmap that highlights the regions of input image that are most influential in determining the network's final classification. Grad-CAM is an extension and improvement of the CAM method, as it does not require modifying the network architecture and can be applied to any CNN based model. However, since Grad-CAM is suitable for classification, it cannot be directly applied to regression-based models such as SELD or BiSELD models that need to estimate the DOA vector of each sound event. Therefore, I present a VAM visualization method that can obtain a visual heatmap for the input feature map in regression-based SELD or BiSELD models by modifying the existing Grad-CAM.

The calculation process of VAM is as follows. First, the pivot layer, which is the basis layer for VAM calculation, is selected from the network. In the case of BiSELDnet, the concatenate layer of Trinity module, which concatenates all convolution outputs, is selected as the pivot layer as shown in Fig. 6.10. In the forward pass, an input feature is fed through the model and the Cartesian coordinate values (*x*, *y*, *z*) of the DOA vector for each sound event class are calculated. Each coordinate value indicates the model's confidence in whether a sound event belongs to the corresponding class. Here, in order to simplify the backpropagation calculation, it is necessary to simplify the output. Therefore, the norm for the target DOA vector of sound event class *c* is obtained as follows:

$$|v_c| = \sqrt{x_c^2 + y_c^2 + z_c^2}, \quad (6.12)$$

where $x_c$, $y_c$, and $z_c$ are the *x*-axis, *y*-axis, and *z*-axis values of the target vector $v_c$, respectively. In the backpropagation, the gradients of the vector norm with respect to the output feature maps of the pivot layer are computed as shown in Fig. 6.11. These gradients reveal how sensitive the target vector is to changes in each part of the feature maps. In the stage of global average pooling (GAP), the gradients are globally average-pooled for each feature map to obtain a set of weights, each of which represents the importance of the corresponding feature map for the target vector. The weight of the *k*-th feature map for the target vector $v_c$ is obtained as follows:

$$w_k^c = \frac{1}{Z} \sum_i \sum_j \frac{\partial |v_c|}{\partial P_{i,j}^k}, \quad (6.13)$$

where *Z* is the total number of elements in the *k*-th feature map; and $P_{i,j}^k$ is the *i* & *j*-th element in the *k*-th feature map. In the stage of weighted sum, the feature maps of the pivot layer are linearly combined using the computed weights, creating a weighted feature map (WFM) that emphasizes the regions relevant to the target vector. The WFM is obtained as follows:

$$WFM_{i,j}^c = \sum_k w_k^c P_{i,j}^k. \tag{6.14}$$

In the final stage of VAM generation, by passing ReLU function, a VAM is generated that highlights the crucial regions in the input feature map responsible for the target vector generation. The VAM can be obtained as follows:

$$VAM_{i,j}^c = ReLU(WFM_{i,j}^c). \tag{6.15}$$

The VAM is scaled to the dimensions of the original input feature map and can be overlaid on it to visualize highlighted regions. VAM visualization allows us to understand which regions of input feature map a SELD or BiSELD model relies on when estimating the DOA vector of the sound event of interest. It offers a valuable tool for model interpretation and can be used in various SELD or BiSELD tasks to explain the reasoning behind the model's predictions.

VAM visualization was performed to verify that BiSELDnet estimates the elevation angle of a sound event based on spectral clues. For this purpose, several VAMs were obtained from the selected pivot layer for baby crying sound events with elevation angles of −30°, 0°, +30°, and +60° on the median plane. As shown in Fig. 6.10, the concatenate layer of the 8th Trinity module was selected as the pivot layer considering the frequency axis resolution of VAM. The VAMs of BiSELDnet for baby crying sound events on the median plane are shown in Fig. 6.12. For reference, on the median plane, the ITD-map and ILD-map related to horizontal localization are zero, so the SC-map related to vertical localization was used as the reference input feature map for VAM. As shown in Fig. 6.12(d1), when the elevation angle of the sound event is −30°, a thick N1 notch line appears near 8.5 kHz and a thin N2 notch line appears near 13 kHz. Then, the N1 and N2 notch lines shift to higher frequencies as the sound event rises from −30° to +60° in elevation, as shown in Figs. 6.12(c1), (b1) and (a1). This was already observed through the SC analysis of the measured HRTFs in Subsection 4.1.3, and it is known that humans perceive the elevation of sound sources through these spectral cues. Additionally, in Subsection 4.2.4, these changes were visually confirmed from the SC-map pattern changes, so it was expected that it would be possible to estimate the elevation angle of binaural sound events through deep learning. Looking at Figs. 6.12(a2), (b2), (c2) and (d2), we can see that VAMs are activated exactly on the corresponding N1 notch area and spectral pattern area of the SC-map. This means that BiSELDnet pays attention to the inherent time-frequency patterns to detect sound events, and pays attention to the N1 notch patterns to localize them vertically, just like us humans.

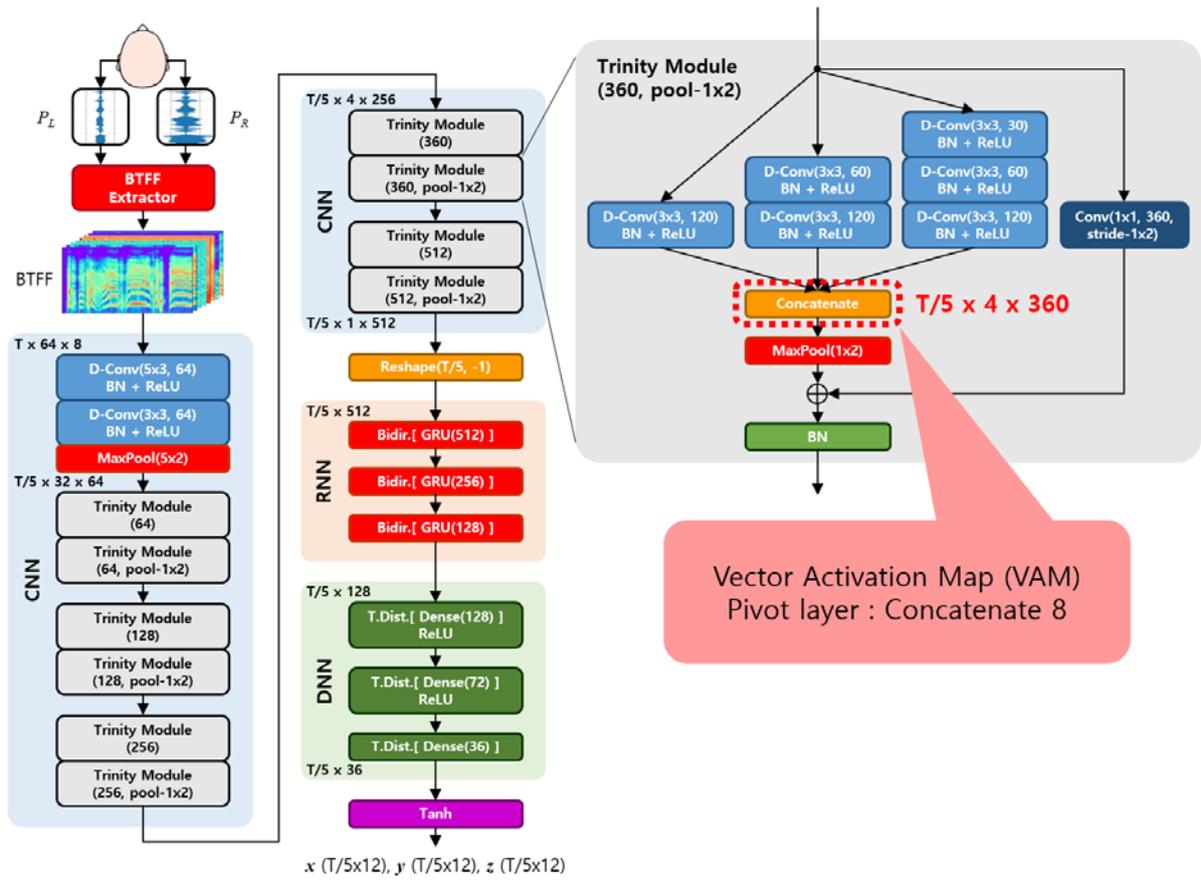

**Figure 6.10.** Concatenate layer selected as the pivot layer for vector activation map (VAM) visualization in BiSELDnet.

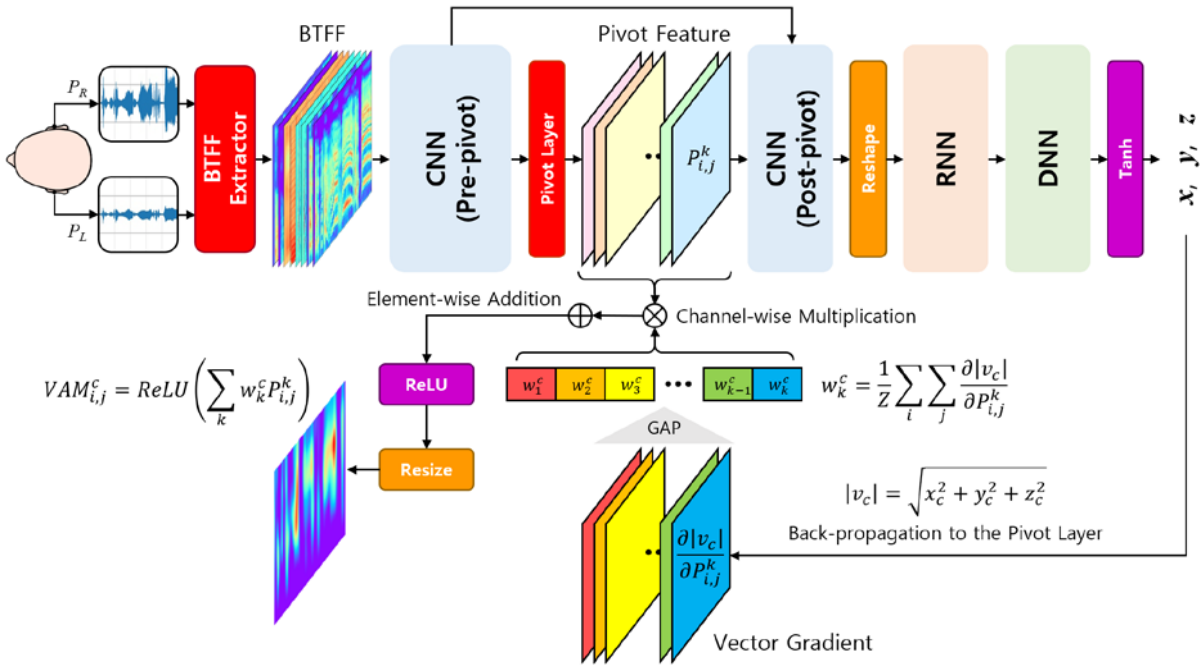

**Figure 6.11.** Overview of vector activation map (VAM) visualization for the Trinity module in BiSELDnet.

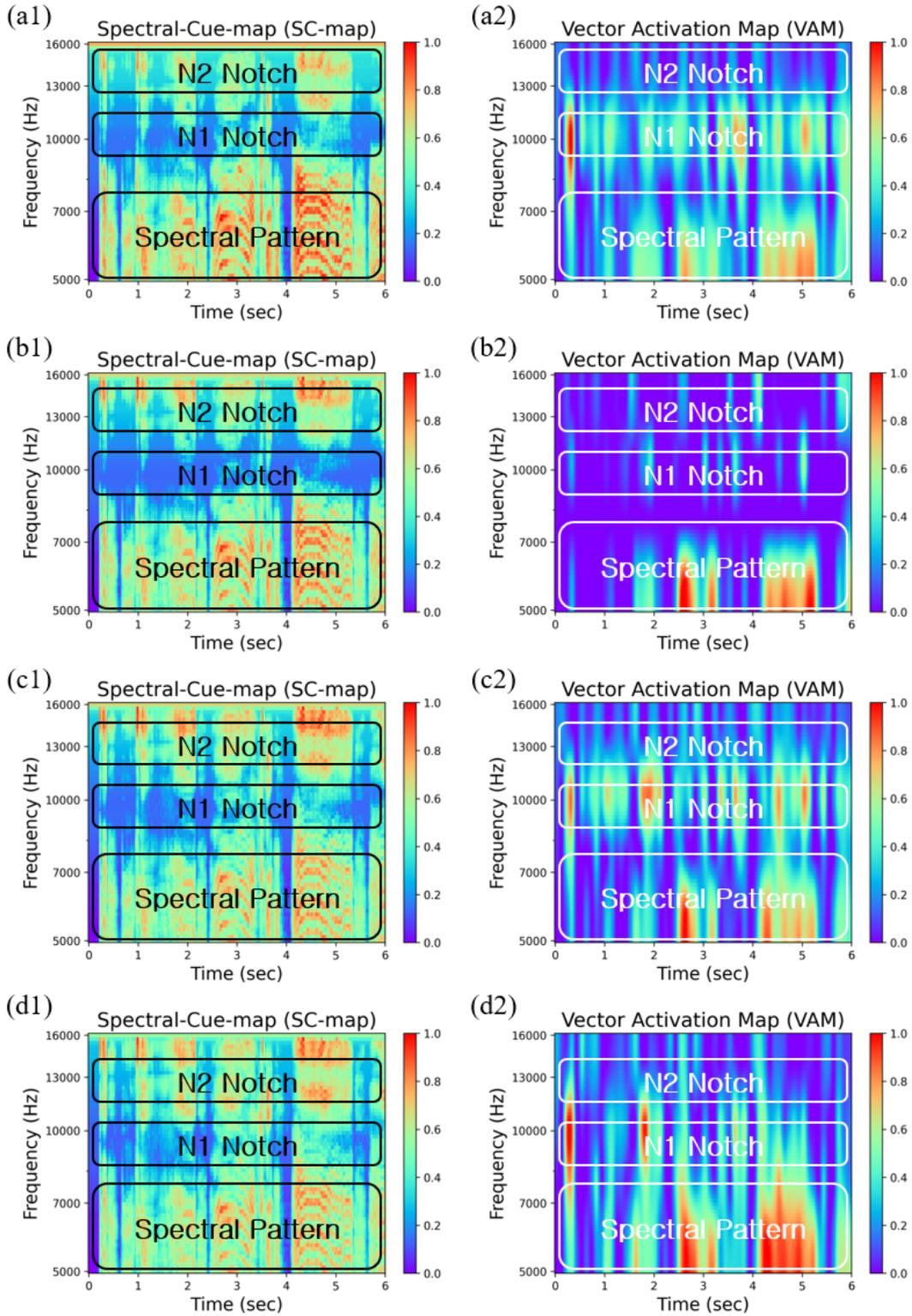

**Figure 6.12.** SC-maps and corresponding VAMs for baby crying sound events on the median plane: (a1) SC-map for $\phi = +60°$, (a2) VAM for $\phi = +60°$, (b1) SC-map for $\phi = +30°$, (b2) VAM for $\phi = +30°$, (c1) SC-map for $\phi = 0°$, (c2) VAM for $\phi = 0°$, (d1) SC-map for $\phi = -30°$, and (d2) VAM for $\phi = -30°$.

## 6.4 Discussion

In this chapter, in-depth evaluations were conducted on the sub-features of BTFF and various versions of BiSELDnet, which were proposed for the BiSELD task. The evaluation results for BTFF and BiSELDnet are discussed as follows:

As a result of verifying the effect of V-map on sound event detection, it is confirmed that adding V-map to MS to generate input features is effective in binaural sound event detection. In addition, as the detection performance improves, it is seen that its localization performance also improves slightly. Using V-map as one of the input features for BiSELD model offers important benefits by allowing it to capture the temporal dynamics and transitions of sound events.

As a result of verifying the effect of ITD and ILD-maps on horizontal localization, adding ITD and ILD-maps to MS to generate input features is found to be effective in localizing binaural sound events horizontally. Notably, the qualitative observation in Section 4.2.4 that ILD-map complements the azimuth estimation of ITD-map is quantitatively verified. Also, it is seen that the improvement in horizontal localization performance by ITD and ILD-maps leads to the improvement in detection performance. ITD-map and ILD-map are complementary because they provide spatial location information for different frequency ranges. ITD-map is effective for low-frequency sounds below 1.5 kHz, while ILD-map is effective for high-frequency sounds above 5 kHz. Unlike ITD-map, the high frequency pattern of ILD-map changes complexly with azimuth angles, providing clues for solving ambiguous horizontal localization. Through this role division, BiSELD models can more accurately estimate the horizontal direction of sound events.

As a result of verifying the effect of SC-map on vertical localization, adding SC-map to MS to generate input features is found to be effective in localizing binaural sound events vertically. In particular, it is reconfirmed through SC-map that spectral patterns above 5 kHz provide useful information for vertical localization. Also, it is seen that the improvement in vertical localization performance by SC-map leads to the improvement in detection performance. SC-map was proposed based on the fact that the spectral patterns generated by the pinna, especially the notch frequencies, contribute to vertical localization. Since the notches above 5 kHz are highly dependent on the elevation of sound source, the spectral notch pattern above 5 kHz is a distinct monaural feature for vertical localization.

As a result of evaluating and comparing the performance of each BiSELDnet version, Version 4 was found to have the best performance in all evaluation metrics of detection and localization, despite having the lowest number of parameters except Version 1. It shows that Trinity architecture, designed to simultaneously extract features of various sizes while reducing the number of parameters based on factorized depthwise separable convolutions, is best suited for BiSELD tasks. Since sound events have varying sizes in both the time and frequency dimensions, Trinity architecture, which captures these diverse patterns simultaneously, appears to have an advantage over other architectures. The next best performer was Version 3, based on Xception module. Compared to Version 4, the localization performance was not much different, but the detection performance was relatively low. The reason appears to be that Trinity module uses three kernels of different sizes, while Xception module uses only a fixed-size kernel, which is relatively disadvantageous in detecting sound events of various time-frequency scales. However, Version 3 showed higher performance than other lower versions because it is based on depthwise separable convolution, suitable for input features with low cross-channel correlation, such as BTFF. Although Version 2, based on hierarchical CRNN, has the largest number of parameters compared to other versions, its localization performance was lower than that of Version 1, the baseline model based on CRNN. A large number of model parameters increases the risk of overfitting because the model becomes overly complex and fits too closely to the training data, resulting in poor generalization.

As a result of layer output visualization for BiSELDnet, it is seen that as layers go deeper, activations become more abstract, and higher-level concepts such as time-frequency patterns and HRTF localization cues are encoded. This process is similar to how humans perceive objects around them, in which the brain transforms sensory information into higher-level concepts by filtering out less relevant elements. In addition, to identify which parts of BTFF contribute to the final decision of BiSELDnet, VAM was proposed by modifying the existing Grad-CAM visualization. As a result of VAM visualization, it is seen that VAMs are activated exactly on the corresponding N1 notch and spectral pattern areas of SC-map, which means that BiSELDnet pays attention to the time-frequency patterns to detect sound events, and pays attention to the N1 notch patterns to localize them vertically, just like us humans.

Considering the human auditory process, as auditory signals travel from the cochlea to the brain, perceptual features are extracted and finally encoded into a categorical representation at the top level. Together with the binaural localization cues described above, the time-frequency features form the auditory information for the detection and localization of surrounding sound events. Although the information processing of AI models is completely different from that of humans, the results of this chapter remind me of Alan Turing, who proposed the Turing Test as a way to evaluate the intelligence of a machine based on its ability to engage in a conversation with a human without the need for the machine's internal processes to resemble human thought.

# Chapter 7. SOTA Performance Comparison in the Background Noise Condition

## 7.1 Baseline and SOTA SELD Models

As mentioned in Section 1.2, the BiSELD task is proposed to solve problems that arise when changing the four-channel input of the existing SELD task to two-channel input for application to humanoid robots. As a solution for the two-channel application, BTFF, an input feature based on HRTF localization cues, and BiSELDnet, a model suitable for learning input features with low cross-channel correlation, were presented and verified. In this chapter, the proposed BiSELD model is verified through performance comparison with existing SOTA methods. As shown in Table 1.1, various binaural SED and SSL models have been presented recently, but these models cannot detect and localize sound events simultaneously. In addition, the SOTA binaural SELD model by Wilkins et al. [98], published in 2023, cannot estimate the elevation angle of sound events. Consequently, the well-known baseline and SOTA SELD models, whose source codes are publicly available, were selected, and their audio inputs were modified to two-channels for the performance comparison.

### 7.1.1 SELDnet with ACCDOA Output Format

In this subsection, the baseline SELD model is briefly introduced. The model is the SELDnet with ACCDOA output format (ACCDOA model) [79], which is presented as a baseline model in the SELD task of the DCASE challenge. When the SELD task first started in 2019, the original SELDnet [33] employed multi-task output, but its output format was later simplified by using the ACCDOA format, which encodes SED and DOA information into one single output, as shown in Fig. 7.1. The input of the ACCDOA model is multi-channel audio, where different acoustic features can be extracted depending on the input format. Based on the typical input format such as FOA, the model takes a sequence of consecutive feature frames, and predicts all active sound event classes and their respective spatial directions for each time frame. As a result, the model generates temporal activity and DOA trajectory for each sound event. To be specific, after feature extraction, CRNN is used to map an input feature sequence to an output ACCDOA sequence which encodes both SED and DOA estimates in the continuous 3D space as a multi-output regression task. Each sound event class of the ACCDOA format is represented by three regressors that estimate the Cartesian DOA vector ($x$, $y$, $z$) of the corresponding sound event with respect to the FOA microphone array. If the length of the DOA vector is greater than 0.5, the sound event is considered active and the coordinate values of the DOA vector are used as the predicted direction.

As noted in Subsection 1.1.3, most SELD models use four audio input channels such as FOA,

conforming to the SELD task of the DCASE challenge. To compare performance under the same input conditions, the audio input of the ACCDOA model was modified to two-channel input. The ACCDOA model has four audio input channels and seven input feature channels, which consist of four MSs and three intensity vectors (IVs). Here, the number of MS is allocated according to the number of audio input channels. Therefore, when modifying its audio input to two-channels, the input feature becomes three-channels, which consist of two MSs and one IV. Notably, this modified ACCDOA model is almost the same as the SOTA binaural SELD model [98], which is a modified ACCDOA model with MS and GCC as input features. Anyway, the total number of parameters in this two-channel based ACCDOA model is shown in Table 7.1.

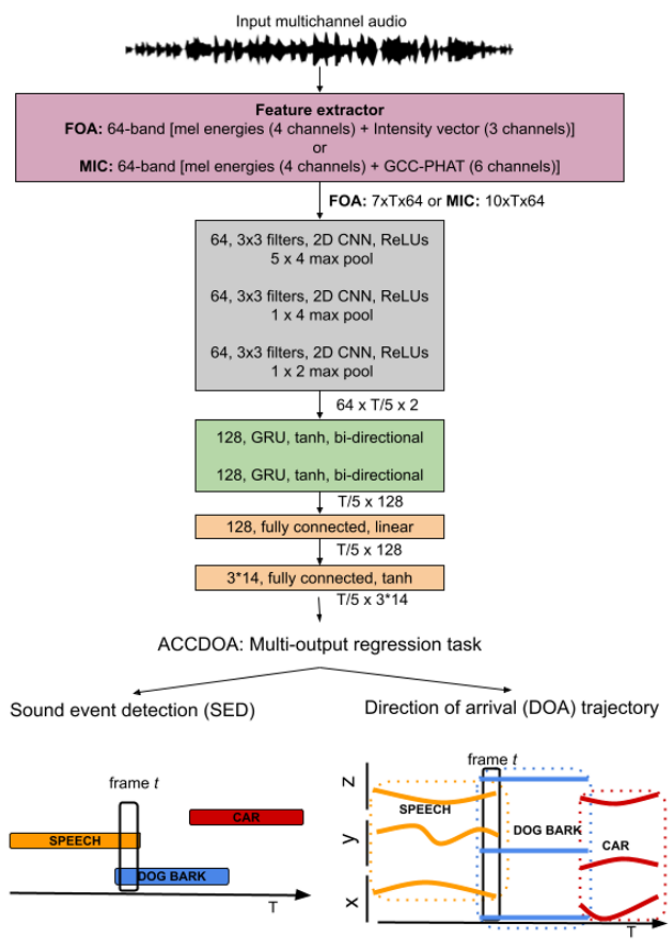

**Figure 7.1.** SELDnet architecture with ACCDOA output format used as a baseline model [79].

**Table 7.1.** Total number of parameters in the two-channel based ACCDOA model.

|            | Trainable | Non-trainable | Total   |
|------------|-----------|---------------|---------|
| Parameters | 493,476   | 384           | 493,860 |

### 7.1.2 SELDnet with SALSA Input Feature

In this subsection, the SOTA SELD model is briefly introduced. The model is the SELDnet with SALSA input feature (SALSA model) [82,83], which ranked top 2 in the SELD task of the DCASE 2021 challenge. SALSA [82] is an input feature proposed for the SELD task, and SALSA-Lite [83] is the fast version of the SALSA feature for polyphonic SELD task. The SALSA feature consists of multi-channel log-linear spectrograms stacked along with the normalized principal eigenvector of the spatial covariance matrix at each corresponding time-frequency bin [82]. In contrast to SALSA, which uses eigenvector-based spatial features, SALSA-Lite uses normalized inter-channel phase differences as spatial features, allowing a 30-fold speedup compared to the original SALSA feature [83]. Both SALSA and SALSA-Lite features maintain the exact time-frequency mapping between the signal power and the source directional cues, which is important for resolving overlapping sound sources. SALSA-Lite is designed for multi-channel microphone arrays, which are actually the most accessible and commonly used type of microphone array. Experimental results on the TAU-NIGENS Spatial Sound Events (TNSSE) 2021 dataset with directional interferences showed that SALSA-Lite achieved similar performance as SALSA for multi-channel microphone arrays, and significantly outperformed multi-channel log-mel spectrograms with generalized cross-correlation spectra (MS + GCC) feature [83]. The SELDnet architecture with SALSA input feature is shown in Fig. 7.2. This SALSA model is a CRNN-based multi-tasking model that performs SED and DOA estimation separately. To be specific, the model consists of a ResNet22-based CNN, a two-layer BiGRU, and separate FC layers for SED and DOA estimation.

This SALSA model can be adapted for different input features by setting the number of input channels in the first convolutional layer to that of the input features. To compare performance under the same input conditions as BiSELD model, the audio input of the SALSA model was modified to two-channel input. The original SALSA model has four audio input channels and seven input feature channels, which consist of four Log Power Spectrograms (LPSs) and three Frequency-Normalized IPDs (NIPDs). Here, the number of LPS is allocated according to the number of audio input channels. Therefore, when modifying its audio input to two-channels, the input feature becomes three-channels, which consist of two LPSs and one NIPD. For reference, the total number of parameters in this two-channel based SALSA model is shown in Table 7.2.

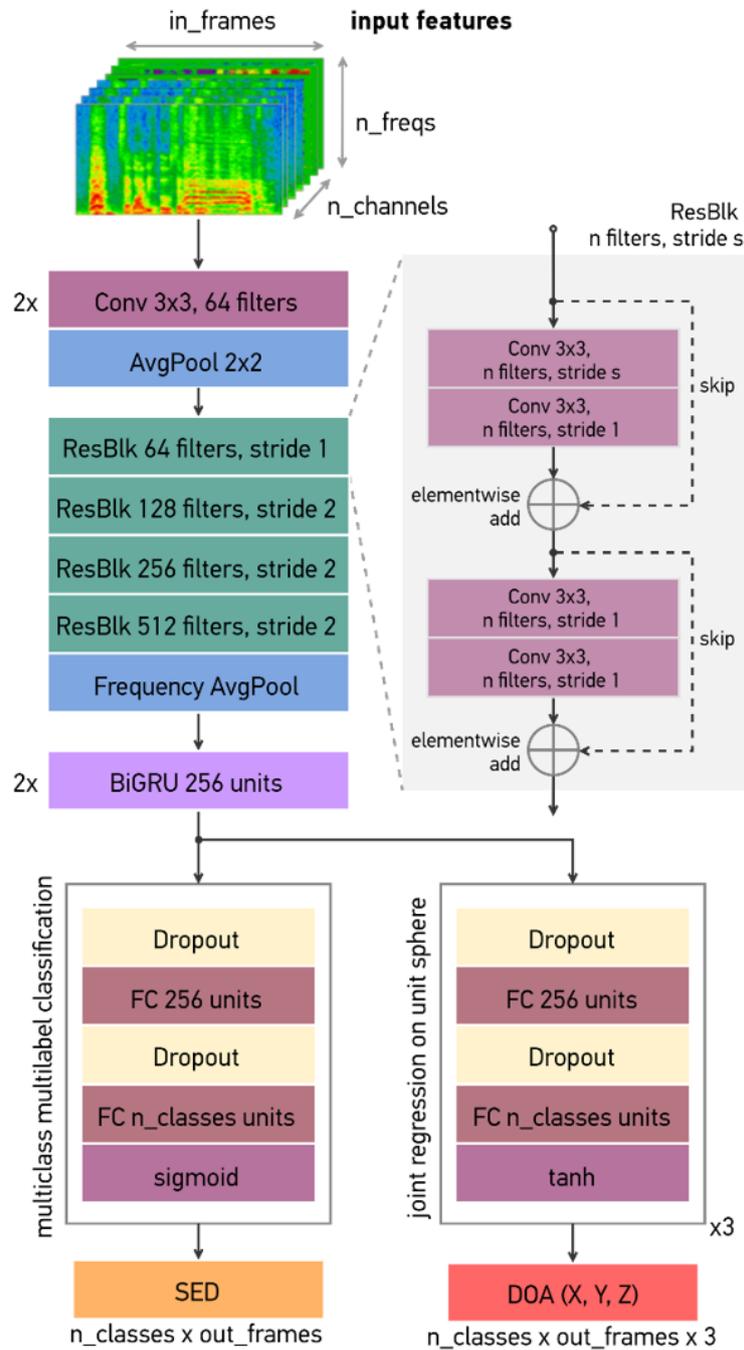

**Figure 7.2.** SELDnet architecture with SALSA input feature used as a SOTA model [83].

**Table 7.2.** Total number of parameters in the two-channel based SALSA model.

|  | Trainable | Non-trainable | Total |
|---|---|---|---|
| Parameters | 13,252,336 | 7,936 | 13,260,272 |

## 7.2 SOTA Performance Comparison

The need for SOTA performance comparison in deep learning models is driven by the relentless pursuit of pushing the boundaries of AI. In the rapidly evolving landscape of deep learning, where novel architectures and techniques continuously emerge, SOTA comparisons provide a standardized framework to evaluate and benchmark the performance of different models. These comparisons not only offer a snapshot of the current landscape but also facilitate the identification of superior models, guiding us toward the most effective solutions. In addition, these comparisons are crucial for guiding the allocation of resources, identifying promising approaches for research, and ensuring that advancements align with practical applications across diverse domains. By establishing a baseline for excellence, SOTA comparisons play a pivotal role in shaping the trajectory of deep learning, pushing the boundaries of what is achievable, and contributing to the ongoing evolution of AI.

In this section, the detection and localization performances of the BiSELD model, SOTA SELD model, and baseline (BL) SELD model are compared against each other for sound events on the horizontal or median plane under realistic background noise conditions with SNR from 0 dB to 30 dB in 10 dB steps. The datasets used for training and evaluation are 'Binaural Set under background noise with SNR = 30 dB' specified in Table 4.3; 'Binaural Set under background noise with SNR = 20 dB' specified in Table 4.4; 'Binaural Set under background noise with SNR = 10 dB' specified in Table 4.5; and 'Binaural Set under background noise with SNR = 0 dB' specified in Table 4.6. As mentioned in Subsection 4.3.1, the background noise database used to build these Binaural Sets has a high degree of acoustic variability because it was collected from a variety of locations across Europe. Moreover, since the background noise was recorded with a two-channel microphone, these Binaural Sets contain multi-path directional interferences as well as reflections similar to real-life situations. In each Binaural Set, the Test-H dataset, which consists only of horizontal sound events, was used to compare the horizontal performance of the BiSELD model, SOTA SELD model, and BL SELD model. Similarly, the Test-V dataset, which consists only of vertical sound events, was used to compare the vertical performance of the models. In preparation for the performance comparison, BTFF featureset, SALSA-Lite featureset, and MS-IV featureset were extracted from each Binaural Set for the BiSELD model, SALSA model, and ACCDOA model, respectively.

7.2.1 Performance on the Horizontal or Median Plane at SNR = 30 dB

The evaluation results on the horizontal plane is shown in Fig. 7.3 and those on the median plane is shown in Fig. 7.4. Based on SELD error, the best performance of each model on the horizontal plane is summarized in Table 7.3 and that on the median plane is summarized in Table 7.4, with the best results bolded and underlined. On the horizontal plane, the BiSELD model was found to have the best performance in all evaluation metrics of detection and localization, despite having less than half the number of parameters of the SOTA SELD model. To be specific, the best results on the horizontal plane are as follows: ER of 0.183, F-score of 85.4%, LE of 3.3°, LR of 87.4%, SED error of 0.164, DOA error of 0.072, and SELD error of 0.118. On the median plane, the SOTA SELD model performed slightly better than the BiSELD model in LR, but its LE was 12.4° higher than the BiSELD model, resulting in a higher DOA error. As a result, the BiSELD model showed the lowest SELD error on the median plane as follows: ER of 0.114, F-score of 92.5%, LE of 6.8°, LR of 95.2%, SED error of 0.094, DOA error of 0.043, and SELD error of 0.069.

**Table 7.3.** Best performances of BL-SELD (ACCDOA), SOTA-SELD (SALSA), and BiSELD (Trinity module) models for sound events on the horizontal plane in the background noise with SNR of 30 dB.

| Model | Total Params (M) | Sound Event Detection | | | Sound Event Localization | | | Total |
|---|---|---|---|---|---|---|---|---|
| | | $ER_{20°}\downarrow$ | $F_{20°}\uparrow$ (%) | SED Error$\downarrow$ | $LE_{CD}\downarrow$ (°) | $LR_{CD}\uparrow$ (%) | DOA Error$\downarrow$ | SELD Error$\downarrow$ |
| BL-SELD (ACCDOA) | 0.5 | 0.320 | 76.9 | 0.276 | 8.3 | 76.6 | 0.140 | 0.208 |
| SOTA-SELD (SALSA) | 13.3 | 0.311 | 77.5 | 0.268 | 17.9 | 86.4 | 0.118 | 0.193 |
| BiSELD (Trinity) | 6.1 | **0.183** | **85.4** | **0.164** | **3.3** | **87.4** | **0.072** | **0.118** |

**Table 7.4.** Best performances of BL-SELD (ACCDOA), SOTA-SELD (SALSA), and BiSELD (Trinity module) models for sound events on the median plane in the background noise with SNR of 30 dB.

| Model | Total Params (M) | Sound Event Detection | | | Sound Event Localization | | | Total |
|---|---|---|---|---|---|---|---|---|
| | | $ER_{20°}\downarrow$ | $F_{20°}\uparrow$ (%) | SED Error$\downarrow$ | $LE_{CD}\downarrow$ (°) | $LR_{CD}\uparrow$ (%) | DOA Error$\downarrow$ | SELD Error$\downarrow$ |
| BL-SELD (ACCDOA) | 0.5 | 0.301 | 79.3 | 0.254 | 9.7 | 78.7 | 0.133 | 0.193 |
| SOTA-SELD (SALSA) | 13.3 | 0.267 | 83.5 | 0.216 | 19.2 | **95.4** | 0.076 | 0.146 |
| BiSELD (Trinity) | 6.1 | **0.114** | **92.5** | **0.094** | **6.8** | 95.2 | **0.043** | **0.069** |

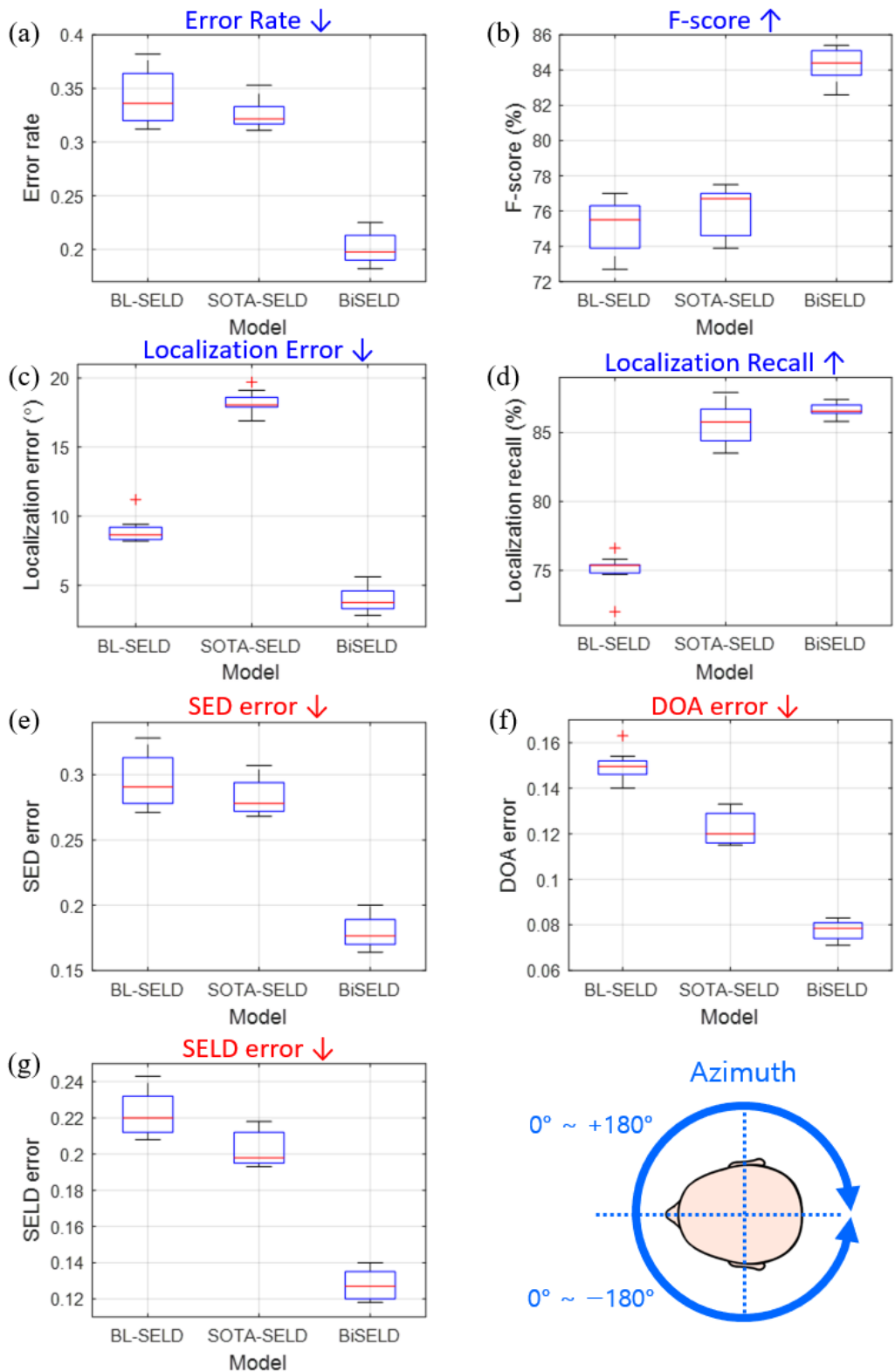

**Figure 7.3.** Evaluation results of BL-SELD (ACCDOA), SOTA-SELD (SALSA), and BiSELD (Trinity module) models for sound events on the horizontal plane in the background noise with SNR of 30 dB: (a) error rate, (b) F-score (%), (c) localization error (°), (d) localization recall (%), (e) SED error, (f) DOA error, and (g) SELD error.

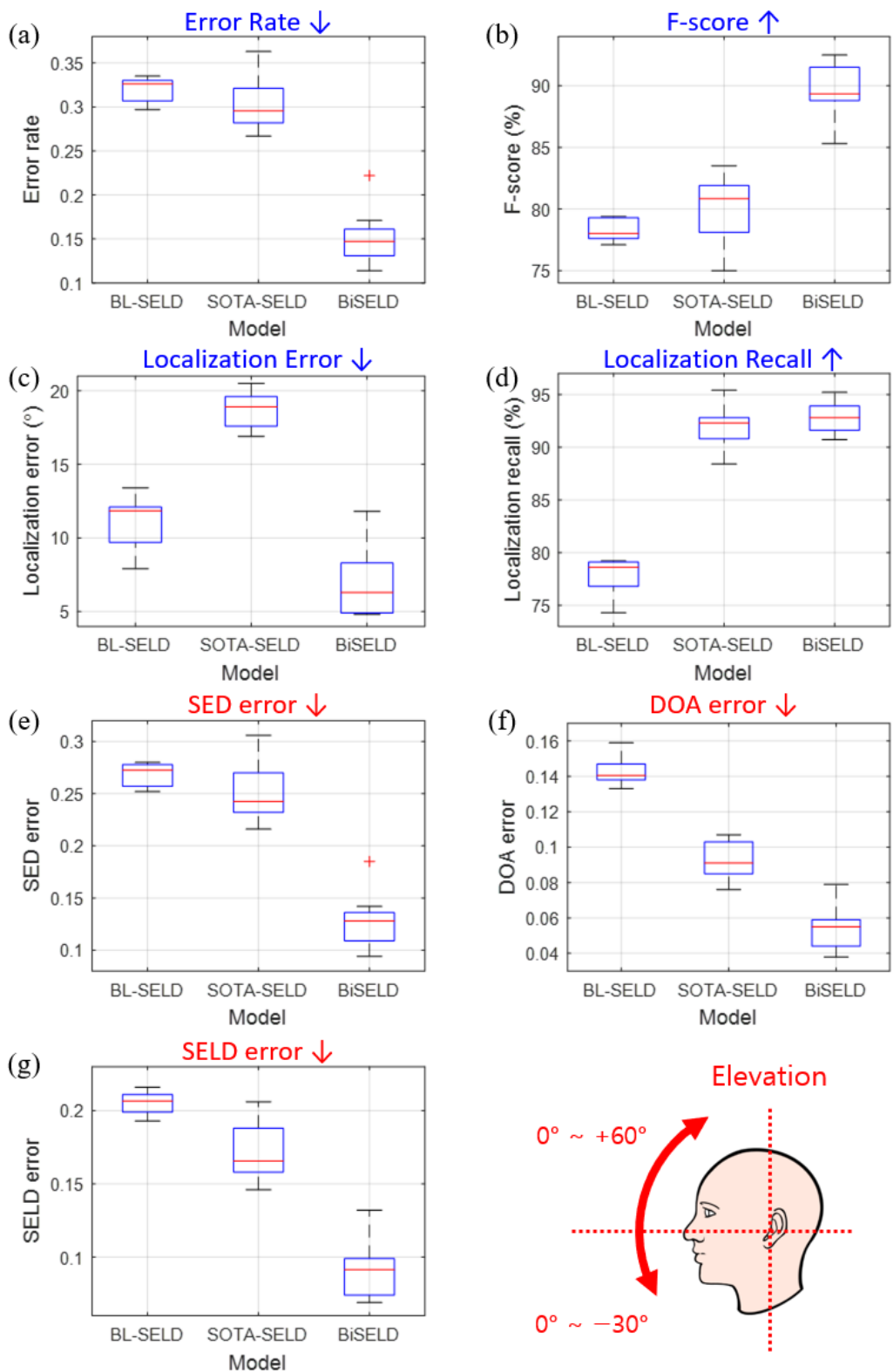

**Figure 7.4.** Evaluation results of BL-SELD (ACCDOA), SOTA-SELD (SALSA), and BiSELD (Trinity module) models for sound events on the median plane in the background noise with SNR of 30 dB: (a) error rate, (b) F-score (%), (c) localization error (°), (d) localization recall (%), (e) SED error, (f) DOA error, and (g) SELD error.

### 7.2.2 Performance on the Horizontal or Median Plane at SNR = 20 dB

The evaluation results on the horizontal plane is shown in Fig. 7.5 and those on the median plane is shown in Fig. 7.6. Based on SELD error, the best performance of each model on the horizontal plane is summarized in Table 7.5 and that on the median plane is summarized in Table 7.6, with the best results bolded and underlined. On the horizontal plane, the SOTA SELD model performed slightly better than the BiSELD model in LR, but its LE was 13.1° higher than the BiSELD model, resulting in a higher DOA error. As a result, the BiSELD model showed the lowest SELD error on the horizontal plane as follows: ER of 0.214, F-score of 83.1%, LE of 5.0°, LR of 86.4%, SED error of 0.192, DOA error of 0.082, and SELD error of 0.137. On the median plane, the BiSELD model was found to have the best performance in all evaluation metrics of detection and localization, despite having less than half the number of parameters of the SOTA SELD model. To be specific, the best results on the median plane are as follows: ER of 0.193, F-score of 86.9%, LE of 11.9°, LR of 93.0%, SED error of 0.162, DOA error of 0.068, and SELD error of 0.115.

**Table 7.5.** Best performances of BL-SELD (ACCDOA), SOTA-SELD (SALSA), and BiSELD (Trinity module) models for sound events on the horizontal plane in the background noise with SNR of 20 dB.

| Model | Total Params (M) | Sound Event Detection | | | Sound Event Localization | | | Total |
|---|---|---|---|---|---|---|---|---|
| | | $ER_{20°}\downarrow$ | $F_{20°}\uparrow$ (%) | SED Error$\downarrow$ | $LE_{CD}\downarrow$ (°) | $LR_{CD}\uparrow$ (%) | DOA Error$\downarrow$ | SELD Error$\downarrow$ |
| BL-SELD (ACCDOA) | 0.5 | 0.359 | 74.8 | 0.306 | 9.0 | 75.1 | 0.149 | 0.227 |
| SOTA-SELD (SALSA) | 13.3 | 0.324 | 76.5 | 0.280 | 18.1 | **86.7** | 0.117 | 0.198 |
| BiSELD (Trinity) | 6.1 | **0.214** | **83.1** | **0.192** | **5.0** | 86.4 | **0.082** | **0.137** |

**Table 7.6.** Best performances of BL-SELD (ACCDOA), SOTA-SELD (SALSA), and BiSELD (Trinity module) models for sound events on the median plane in the background noise with SNR of 20 dB.

| Model | Total Params (M) | Sound Event Detection | | | Sound Event Localization | | | Total |
|---|---|---|---|---|---|---|---|---|
| | | $ER_{20°}\downarrow$ | $F_{20°}\uparrow$ (%) | SED Error$\downarrow$ | $LE_{CD}\downarrow$ (°) | $LR_{CD}\uparrow$ (%) | DOA Error$\downarrow$ | SELD Error$\downarrow$ |
| BL-SELD (ACCDOA) | 0.5 | 0.354 | 76.6 | 0.294 | 12.5 | 76.6 | 0.152 | 0.223 |
| SOTA-SELD (SALSA) | 13.3 | 0.299 | 80.3 | 0.248 | 22.1 | 92.9 | 0.097 | 0.172 |
| BiSELD (Trinity) | 6.1 | **0.193** | **86.9** | **0.162** | **11.9** | **93.0** | **0.068** | **0.115** |

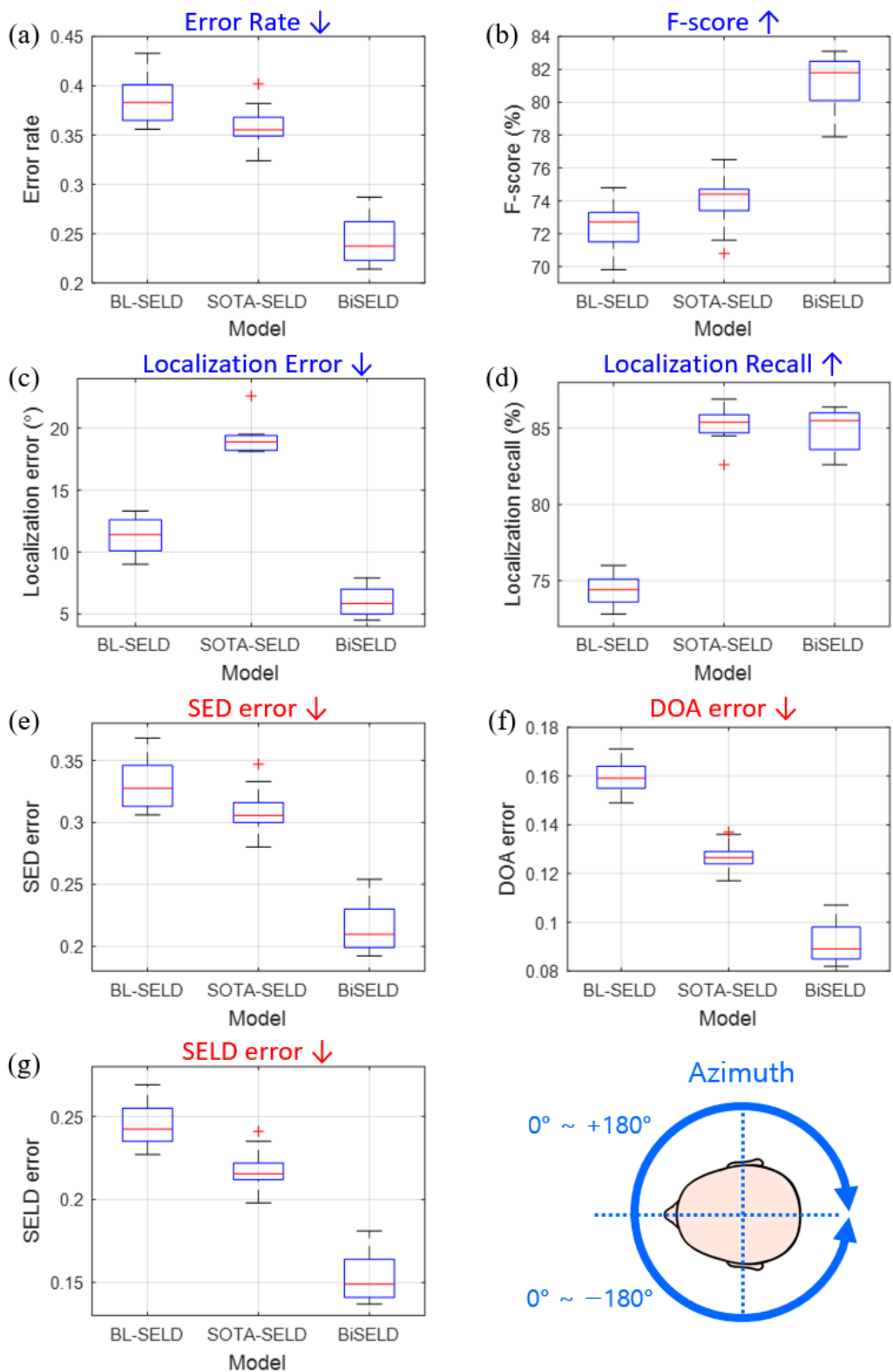

**Figure 7.5.** Evaluation results of BL-SELD (ACCDOA), SOTA-SELD (SALSA), and BiSELD (Trinity module) models for sound events on the horizontal plane in the background noise with SNR of 20 dB: (a) error rate, (b) F-score (%), (c) localization error (°), (d) localization recall (%), (e) SED error, (f) DOA error, and (g) SELD error.

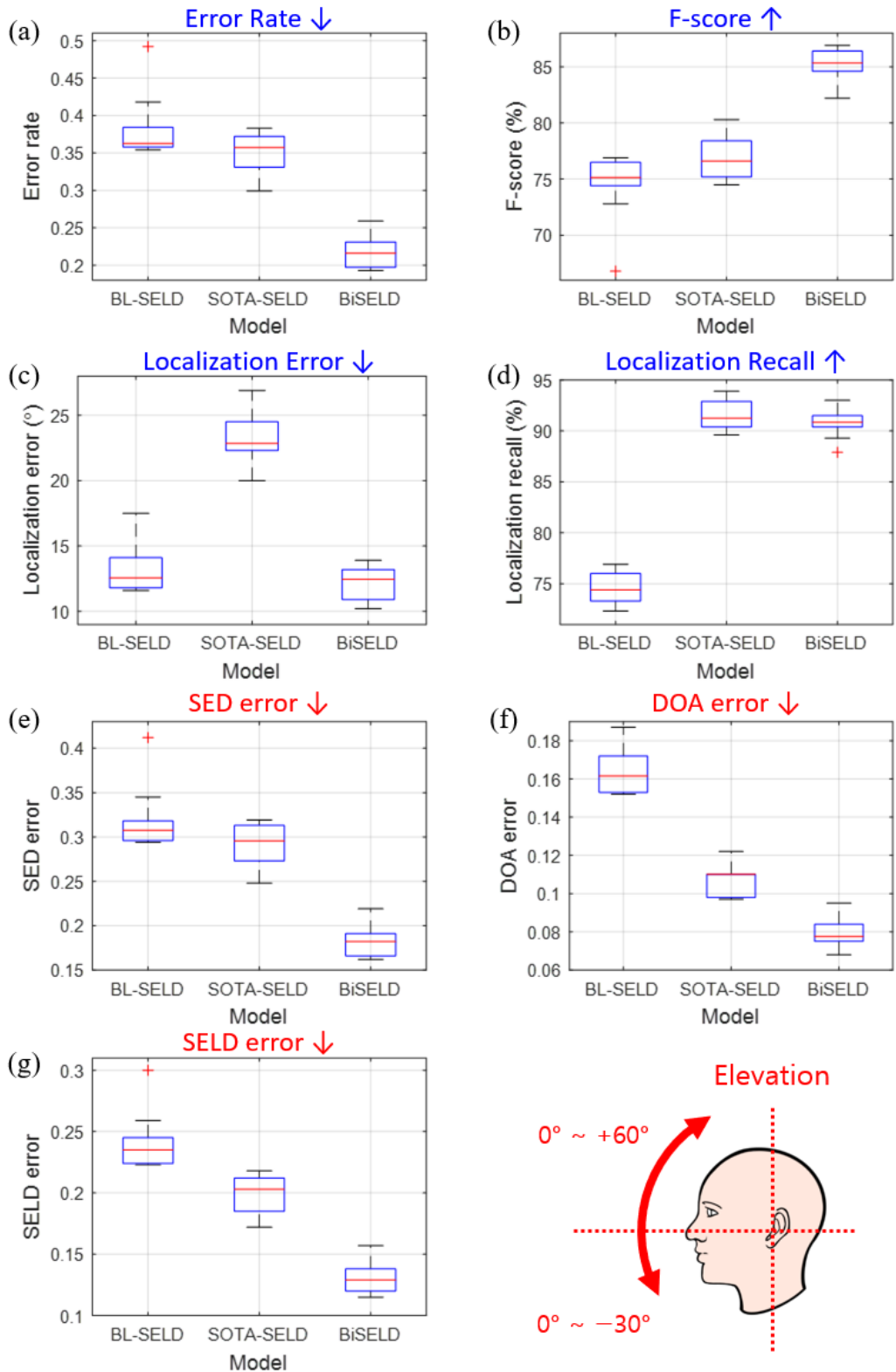

**Figure 7.6.** Evaluation results of BL-SELD (ACCDOA), SOTA-SELD (SALSA), and BiSELD (Trinity module) models for sound events on the median plane in the background noise with SNR of 20 dB: (a) error rate, (b) F-score (%), (c) localization error (°), (d) localization recall (%), (e) SED error, (f) DOA error, and (g) SELD error.

### 7.2.3 Performance on the Horizontal or Median Plane at SNR = 10 dB

The evaluation results on the horizontal plane is shown in Fig. 7.7 and those on the median plane is shown in Fig. 7.8. Based on SELD error, the best performance of each model on the horizontal plane is summarized in Table 7.7 and that on the median plane is summarized in Table 7.8, with the best results bolded and underlined. On the horizontal plane, the BiSELD model was found to have the best performance in all evaluation metrics of detection and localization, despite having less than half the number of parameters of the SOTA SELD model. To be specific, the best results on the horizontal plane are as follows: ER of 0.240, F-score of 82.0%, LE of 7.4°, LR of 86.7%, SED error of 0.210, DOA error of 0.087, and SELD error of 0.148. On the median plane, the BL SELD model performed better than the other two models in LE, but its LR was 17.2% lower than the BiSELD model, resulting in a higher DOA error. As a result, the BiSELD model showed the lowest SELD error on the median plane as follows: ER of 0.247, F-score of 83.8%, LE of 17.1°, LR of 91.8%, SED error of 0.205, DOA error of 0.088, and SELD error of 0.147.

**Table 7.7.** Best performances of BL-SELD (ACCDOA), SOTA-SELD (SALSA), and BiSELD (Trinity module) models for sound events on the horizontal plane in the background noise with SNR of 10 dB.

| Model | Total Params (M) | Sound Event Detection | | | Sound Event Localization | | | Total |
|---|---|---|---|---|---|---|---|---|
| | | $ER_{20°}\downarrow$ | $F_{20°}\uparrow$ (%) | SED Error$\downarrow$ | $LE_{CD}\downarrow$ (°) | $LR_{CD}\uparrow$ (%) | DOA Error$\downarrow$ | SELD Error$\downarrow$ |
| BL-SELD (ACCDOA) | 0.5 | 0.390 | 72.0 | 0.335 | 11.3 | 73.7 | 0.163 | 0.249 |
| SOTA-SELD (SALSA) | 13.3 | 0.382 | 72.3 | 0.330 | 19.7 | 84.7 | 0.131 | 0.230 |
| BiSELD (Trinity) | 6.1 | **<u>0.240</u>** | **<u>82.0</u>** | **<u>0.210</u>** | **<u>7.4</u>** | **<u>86.7</u>** | **<u>0.087</u>** | **<u>0.148</u>** |

**Table 7.8.** Best performances of BL-SELD (ACCDOA), SOTA-SELD (SALSA), and BiSELD (Trinity module) models for sound events on the median plane in the background noise with SNR of 10 dB.

| Model | Total Params (M) | Sound Event Detection | | | Sound Event Localization | | | Total |
|---|---|---|---|---|---|---|---|---|
| | | $ER_{20°}\downarrow$ | $F_{20°}\uparrow$ (%) | SED Error$\downarrow$ | $LE_{CD}\downarrow$ (°) | $LR_{CD}\uparrow$ (%) | DOA Error$\downarrow$ | SELD Error$\downarrow$ |
| BL-SELD (ACCDOA) | 0.5 | 0.377 | 74.9 | 0.314 | **<u>12.5</u>** | 74.6 | 0.162 | 0.238 |
| SOTA-SELD (SALSA) | 13.3 | 0.368 | 75.1 | 0.309 | 26.2 | 90.7 | 0.119 | 0.214 |
| BiSELD (Trinity) | 6.1 | **<u>0.247</u>** | **<u>83.8</u>** | **<u>0.205</u>** | 17.1 | **<u>91.8</u>** | **<u>0.088</u>** | **<u>0.147</u>** |

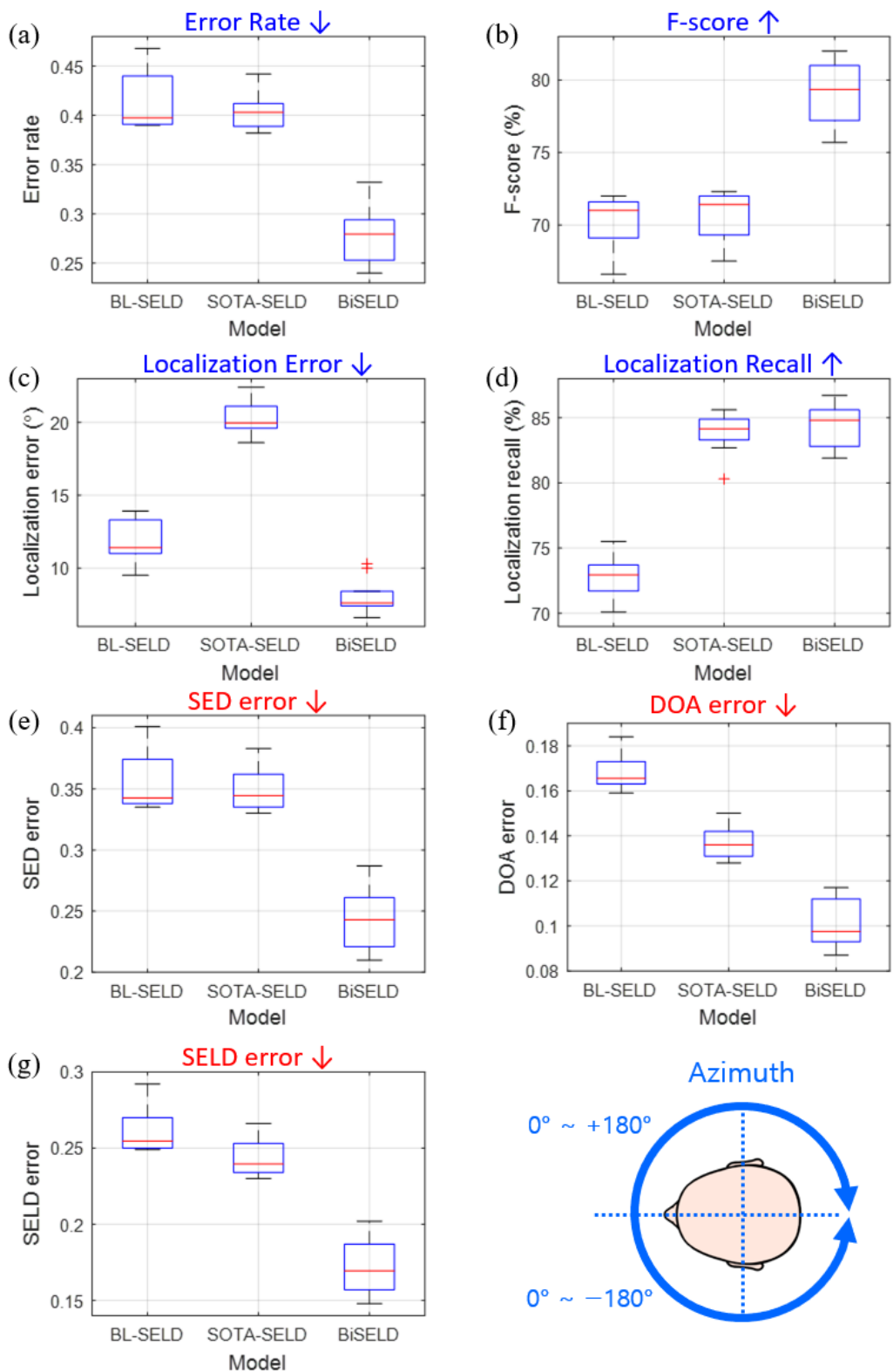

**Figure 7.7.** Evaluation results of BL-SELD (ACCDOA), SOTA-SELD (SALSA), and BiSELD (Trinity module) models for sound events on the horizontal plane in the background noise with SNR of 10 dB: (a) error rate, (b) F-score (%), (c) localization error (°), (d) localization recall (%), (e) SED error, (f) DOA error, and (g) SELD error.

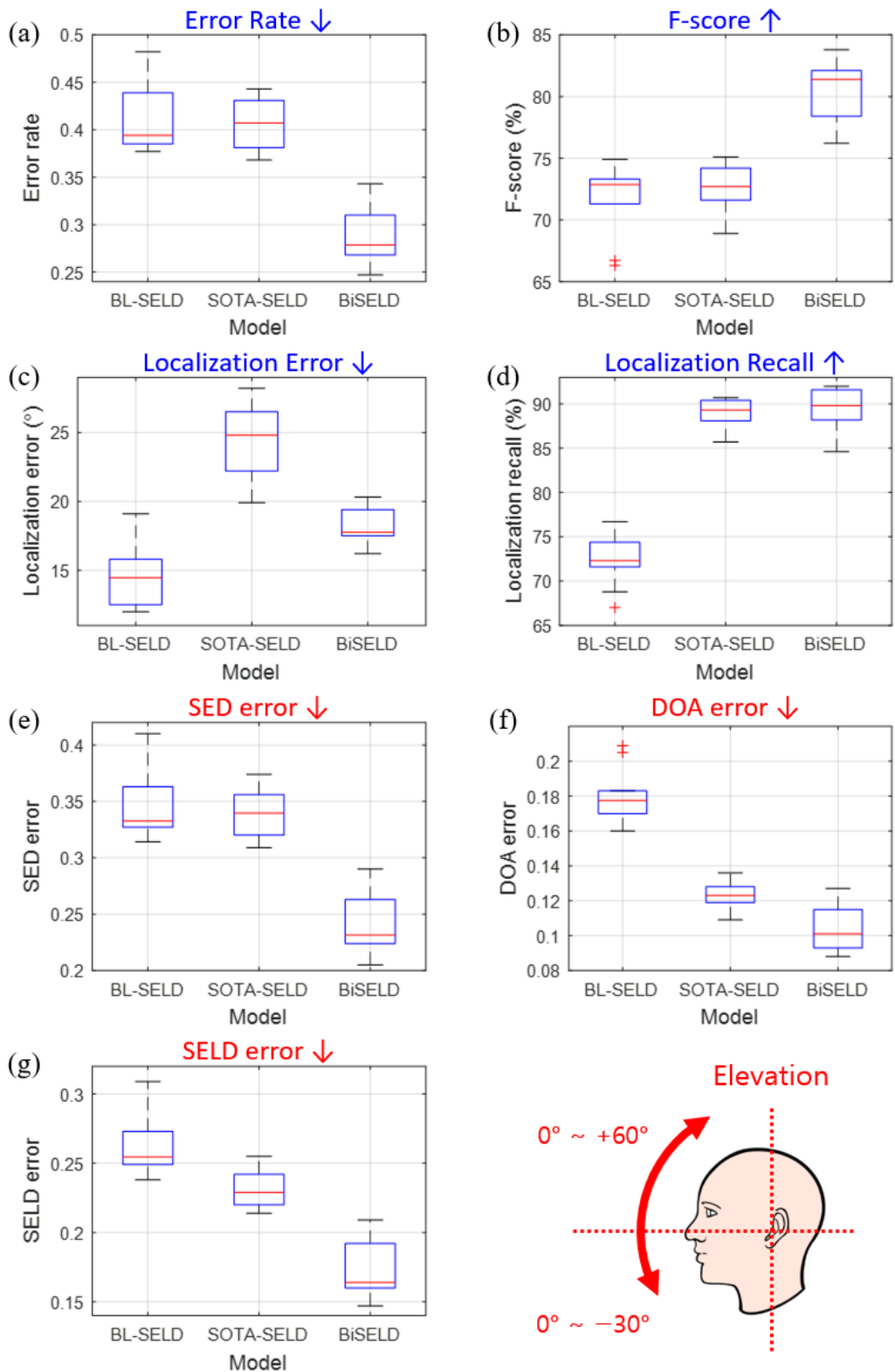

**Figure 7.8.** Evaluation results of BL-SELD (ACCDOA), SOTA-SELD (SALSA), and BiSELD (Trinity module) models for sound events on the median plane in the background noise with SNR of 10 dB: (a) error rate, (b) F-score (%), (c) localization error (°), (d) localization recall (%), (e) SED error, (f) DOA error, and (g) SELD error.

7.2.4 Performance on the Horizontal or Median Plane at SNR = 0 dB

The evaluation results on the horizontal plane is shown in Fig. 7.9 and those on the median plane is shown in Fig. 7.10. Based on SELD error, the best performance of each model on the horizontal plane is summarized in Table 7.9 and that on the median plane is summarized in Table 7.10, with the best results bolded and underlined. On the horizontal plane, the SOTA SELD model performed slightly better than the BiSELD model in LR, but its LE was 10.2° higher than the BiSELD model, resulting in a higher DOA error. As a result, the BiSELD model showed the lowest SELD error on the horizontal plane as follows: ER of 0.346, F-score of 74.0%, LE of 11.2°, LR of 81.2%, SED error of 0.303, DOA error of 0.125, and SELD error of 0.214. On the median plane, the BL SELD model performed best in LE, but its LR was 20.4% lower than the BiSELD model, resulting in a higher DOA error. In addition, the SOTA SELD model performed slightly better than the BiSELD model in LR, but its LE was 2.4° higher than the BiSELD model, resulting in a higher DOA error. As a result, the BiSELD model showed the lowest SELD error on the median plane.

**Table 7.9.** Best performances of BL-SELD (ACCDOA), SOTA-SELD (SALSA), and BiSELD (Trinity module) models for sound events on the horizontal plane in the background noise with SNR of 0 dB.

| Model | Total Params (M) | Sound Event Detection | | | Sound Event Localization | | | Total |
|---|---|---|---|---|---|---|---|---|
| | | $ER_{20°}\downarrow$ | $F_{20°}\uparrow$ (%) | SED Error$\downarrow$ | $LE_{CD}\downarrow$ (°) | $LR_{CD}\uparrow$ (%) | DOA Error$\downarrow$ | SELD Error$\downarrow$ |
| BL-SELD (ACCDOA) | 0.5 | 0.453 | 67.0 | 0.391 | 12.4 | 69.6 | 0.187 | 0.289 |
| SOTA-SELD (SALSA) | 13.3 | 0.419 | 69.5 | 0.362 | 21.4 | **82.8** | 0.145 | 0.254 |
| BiSELD (Trinity) | 6.1 | **0.346** | **74.0** | **0.303** | **11.2** | 81.2 | **0.125** | **0.214** |

**Table 7.10.** Best performances of BL-SELD (ACCDOA), SOTA-SELD (SALSA), and BiSELD (Trinity module) models for sound events on the median plane in the background noise with SNR of 0 dB.

| Model | Total Params (M) | Sound Event Detection | | | Sound Event Localization | | | Total |
|---|---|---|---|---|---|---|---|---|
| | | $ER_{20°}\downarrow$ | $F_{20°}\uparrow$ (%) | SED Error$\downarrow$ | $LE_{CD}\downarrow$ (°) | $LR_{CD}\uparrow$ (%) | DOA Error$\downarrow$ | SELD Error$\downarrow$ |
| BL-SELD (ACCDOA) | 0.5 | 0.457 | 67.0 | 0.393 | **13.4** | 66.5 | 0.205 | 0.299 |
| SOTA-SELD (SALSA) | 13.3 | 0.442 | 69.4 | 0.374 | 24.0 | **87.1** | 0.131 | 0.252 |
| BiSELD (Trinity) | 6.1 | **0.372** | **74.6** | **0.313** | 21.6 | 86.9 | **0.125** | **0.219** |

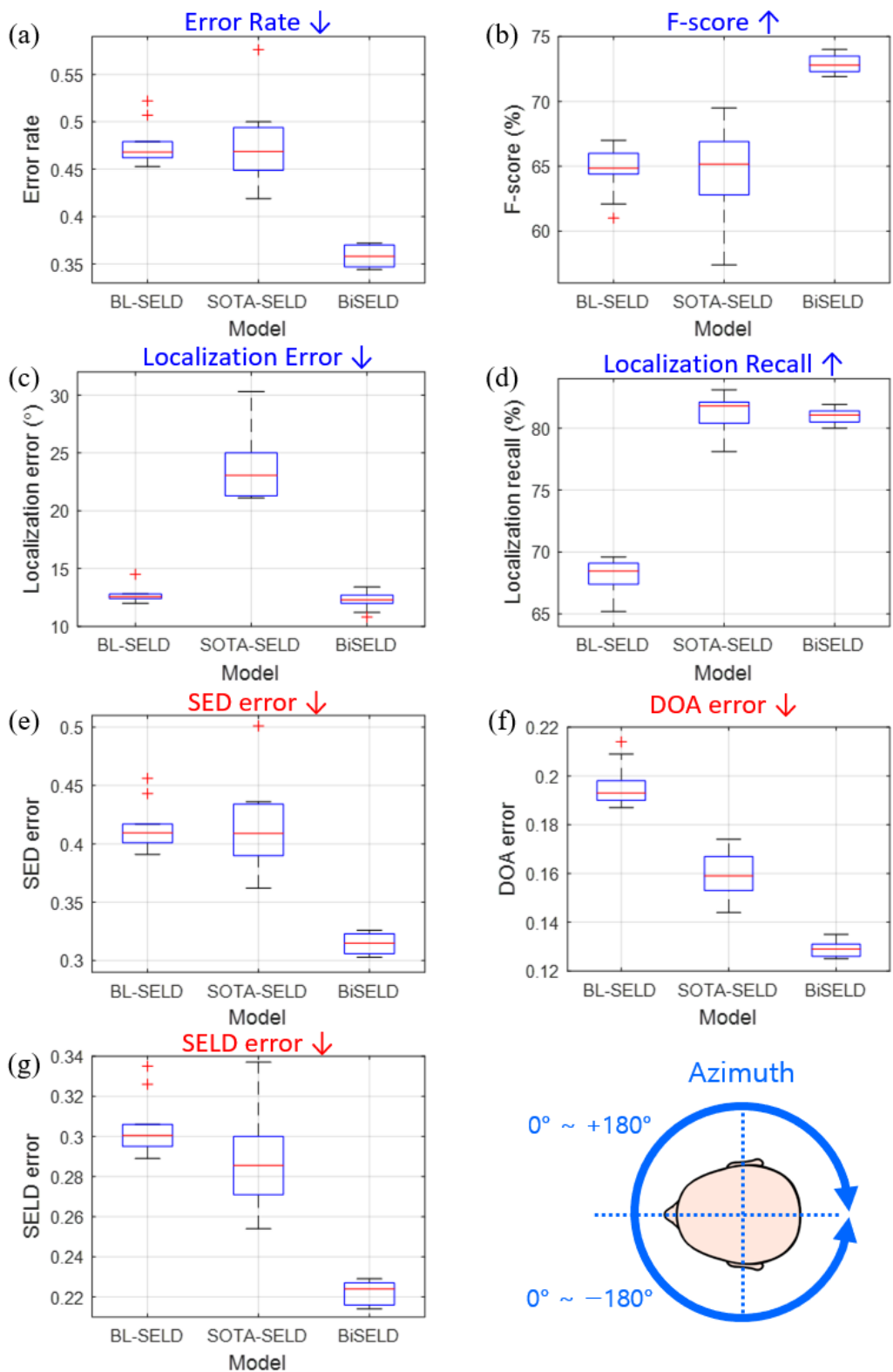

**Figure 7.9.** Evaluation results of BL-SELD (ACCDOA), SOTA-SELD (SALSA), and BiSELD (Trinity module) models for sound events on the horizontal plane in the background noise with SNR of 0 dB: (a) error rate, (b) F-score (%), (c) localization error (°), (d) localization recall (%), (e) SED error, (f) DOA error, and (g) SELD error.

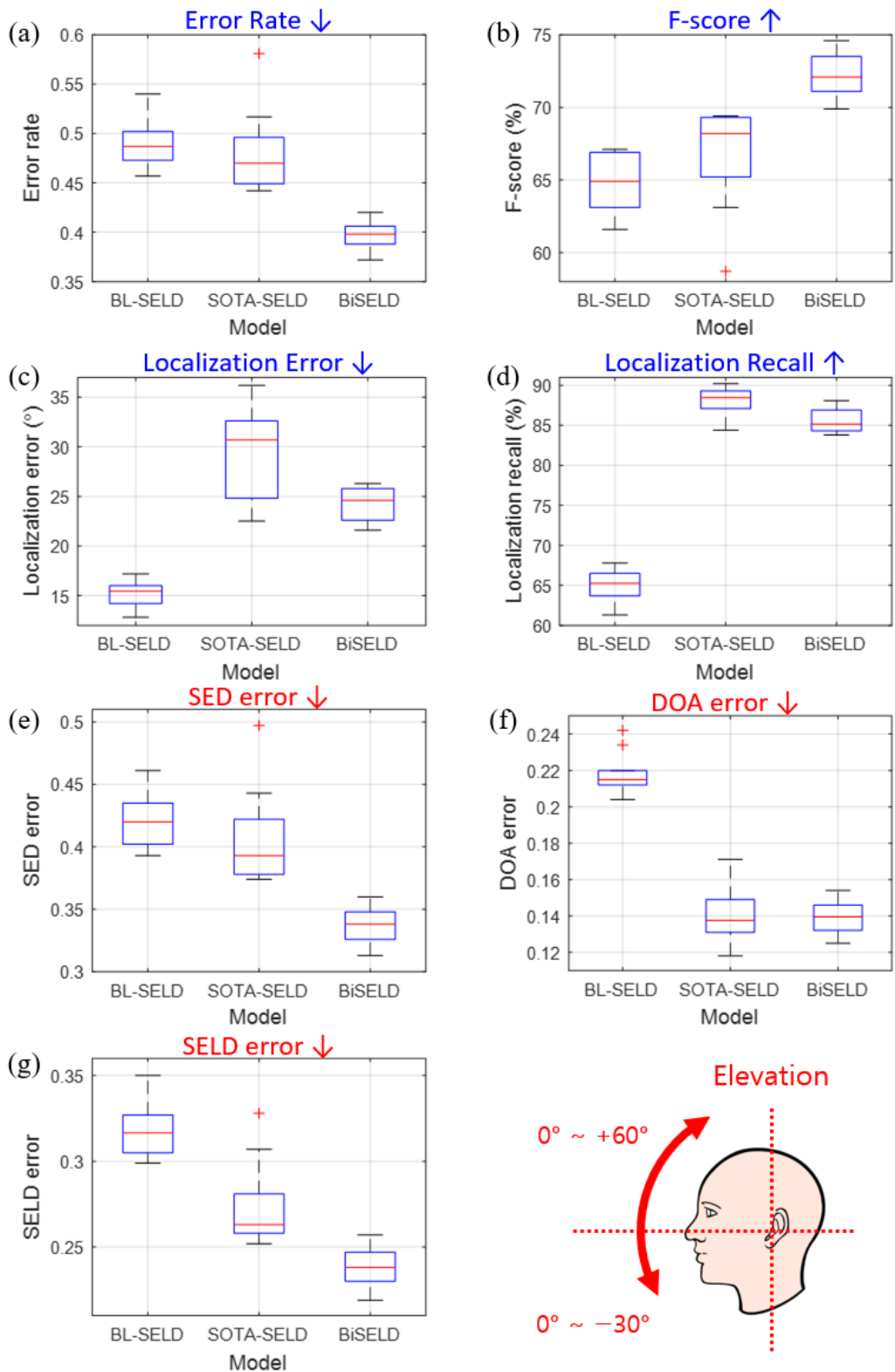

**Figure 7.10.** Evaluation results of BL-SELD (ACCDOA), SOTA-SELD (SALSA), and BiSELD (Trinity module) models for sound events on the median plane in the background noise with SNR of 0 dB: (a) error rate, (b) F-score (%), (c) localization error (°), (d) localization recall (%), (e) SED error, (f) DOA error, and (g) SELD error.

7.2.5 Change in the Best Performance According to SNR

From the above evaluation results, the change in the best performance of each model according to SNR is summarized as follows:

The changes in the best performance of BL SELD, SOTA SELD, and BiSELD models for sound events on the horizontal plane in the background noise with SNR from 0 dB to 30 dB in 10 dB steps are shown in Fig. 7.11. As shown in Figs. 7.11(a), (b) and (e), the SED error of the BiSELD model decreased steeply from 0.303 to 0.210 when SNR increased from 0 dB to 10 dB, and then gradually decreased beyond that. While, those of the SOTA SELD and BL SELD models gradually decreased in the same SNR range. As shown in Figs. 7.11(c), (d) and (f), the LEs of all three models gradually decreased as SNR increased, and the SOTA SELD model showed the largest error. In addition, the LRs of the SOTA SELD and BL SELD models gradually improved as SNR increased, but the LR of the BiSELD model showed a slight stagnation at SNRs above 10 dB. Notably, the LR of the SOTA SELD model was not significantly different from that of the BiSELD model. As a result, as SNR increased from 0 dB to 30 dB, the SELD errors of the BiSELD, SOTA SELD, and BL SELD models decreased from 0.214 to 0.118, 0.254 to 0.193, and 0.289 to 0.208, respectively.

The changes in the best performance of BL SELD, SOTA SELD, and BiSELD models for sound events on the median plane in the background noise with SNR from 0 dB to 30 dB in 10 dB steps are shown in Fig. 7.12. As shown in Figs. 7.12(a), (b) and (e), the SED errors of all three models decreased steeply as SNR increased. To be specific, as SNR increased from 0 dB to 30 dB, the SED errors of the BiSELD, SOTA SELD, and BL SELD models decreased from 0.313 to 0.094, 0.374 to 0.216, and 0.393 to 0.254, respectively. As shown in Figs. 7.12(c), (d) and (f), as SNR increased from 0 dB to 30 dB, the LE of the BiSELD model decreased from 21.6° to 6.8°, showing the greatest improvement over other models. As with the corresponding horizontal LE, the SOTA SELD model showed the largest LE compared to the other models. Additionally, the LR of the three models gradually improved with increasing SNR, and the LR of the SOTA SELD model was not significantly different from that of the BiSELD model. As a result, as SNR increased from 0 dB to 30 dB, the SELD errors of the BiSELD, SOTA SELD, and BL SELD models decreased from 0.219 to 0.069, 0.252 to 0.146, and 0.299 to 0.193, respectively.

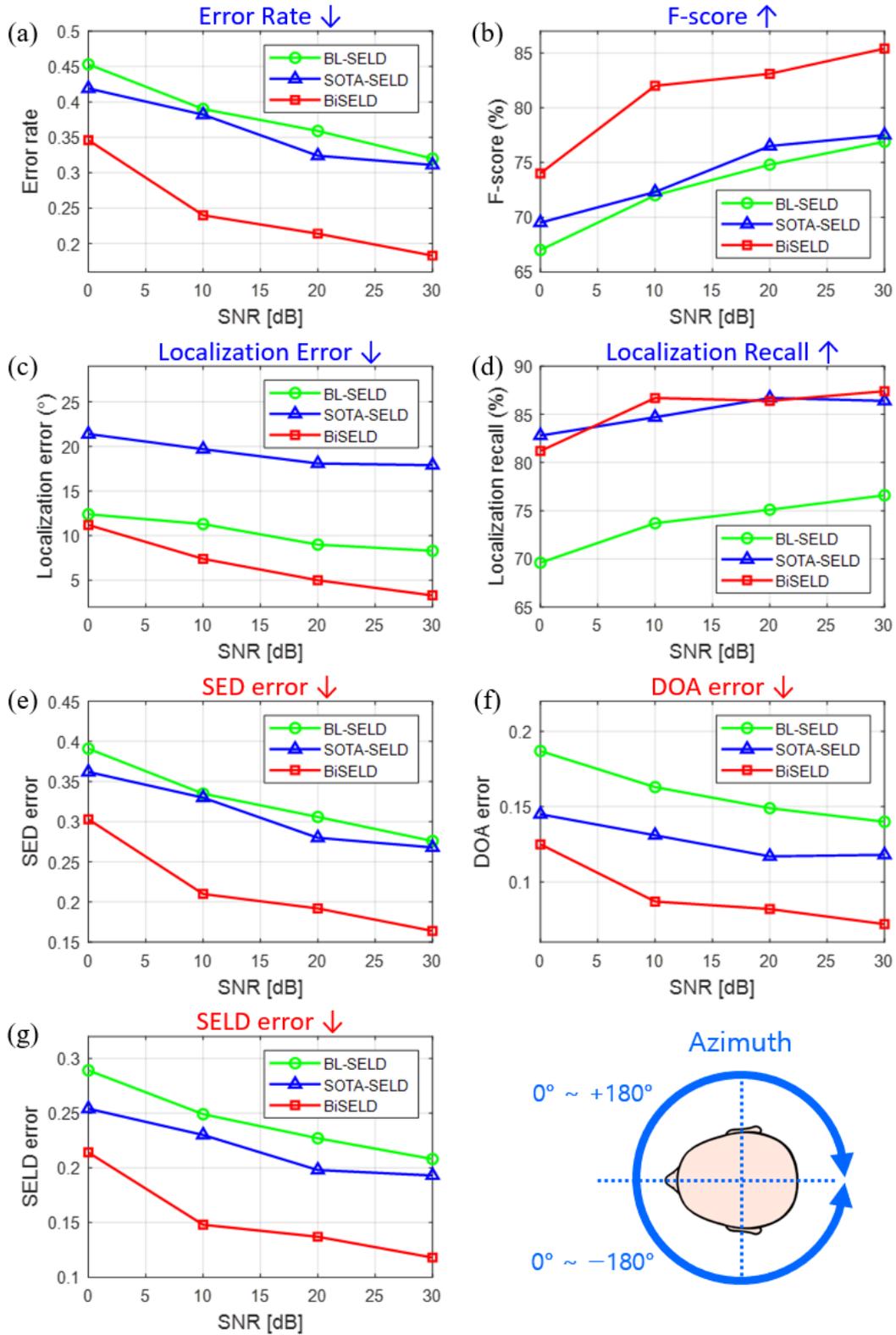

**Figure 7.11.** Changes in the best performance of BL-SELD (ACCDOA), SOTA-SELD (SALSA), and BiSELD (Trinity module) models for sound events on the horizontal plane in the background noise with SNR from 0 dB to 30 dB in 10 dB steps: (a) error rate, (b) F-score (%), (c) localization error (°), (d) localization recall (%), (e) SED error, (f) DOA error, and (g) SELD error.

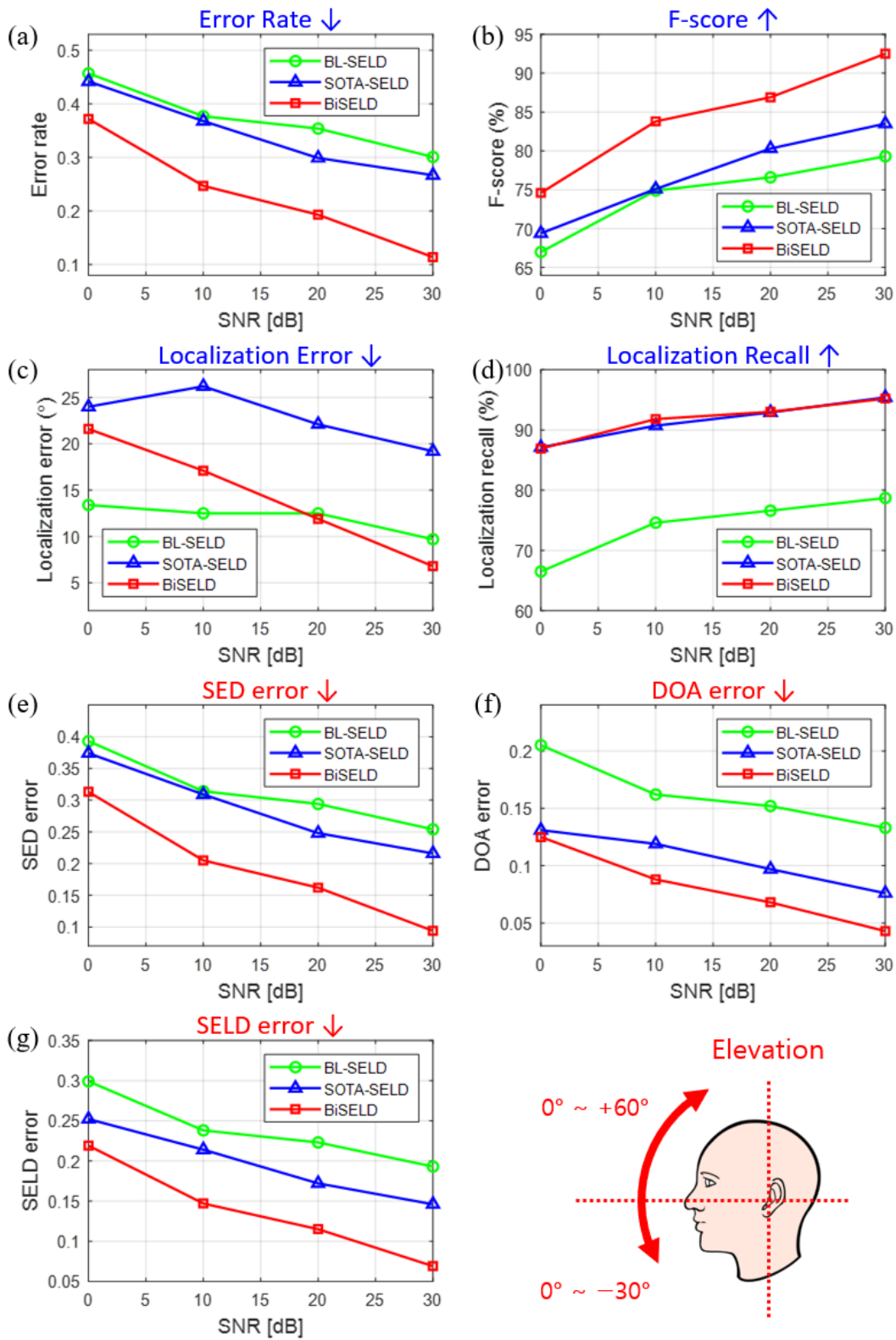

**Figure 7.12.** Changes in the best performance of BL-SELD (ACCDOA), SOTA-SELD (SALSA), and BiSELD (Trinity module) models for sound events on the median plane in the background noise with SNR from 0 dB to 30 dB in 10 dB steps: (a) error rate, (b) F-score (%), (c) localization error (°), (d) localization recall (%), (e) SED error, (f) DOA error, and (g) SELD error.

## 7.3 Discussion

As a result of the SOTA performance comparison, the detection and localization performance of the BiSELD model was found to be better than that of other models even under background noise conditions of various SNRs. This appears to be a natural consequence of the unique input feature and architecture of the BiSELD model. The BiSELD model has a dedicated input feature based on HRTF localization cues, and an architecture that can effectively learn the input feature, allowing it to localize sound events without front-back confusion even under two-channel input conditions.

BTFF, the input feature of BiSELD model, consists of MS, a basis sub-feature; V-map, a sub-feature for detection; ITD-map and ILD-map, sub-features for horizontal localization; and SC-map, a sub-feature for vertical localization. In horizontal localization, ITD-map and ILD-map complement each other, and provide spatial location information from different perspectives about sound events. On the other hand, in vertical localization, the BiSELD model has no choice but to rely on the spectral notch patterns of SC-map, which is vulnerable to background noise. From the best performance changes shown in Fig. 7.12(c), the vertical localization error of the BiSELD model appears to increase further as SNR decreases. It seems that the N1 notch of SC-map was severely affected by background noise. Therefore, additional information about the elevation of sound event is needed to compensate for this. Rather than introducing another sub-feature, there is an easy but effective way to improve SC-map. That is utilizing the N2 notch. As described in Subsection 4.1.3, we humans utilize the frequency shifting patterns of the N1 and N2 notches to estimate the elevation of sound source. Whereas, the BiSELD model only focuses on the N1 notch to estimate the elevation of sound event, as observed in Subsection 6.3.3. I think the reason the BiSELD model misses the N2 notch is because time-frequency patterns above 12 kHz in SC-map are heavily compressed by the mel scale. Therefore, the N2 notch pattern can be sufficiently learned in the Trinity module by expressing only the SC-map in a linear scale. As the vertical localization performance becomes more robust even under low SNR conditions, its applicability in real-world environments will increase.

The reason why the BiSELD model can utilize sub-features of various scales without having to worry about the exact time-frequency mapping, a limitation of existing SELD models, is because of depthwise separable convolution. This convolution is specialized for learning input features with low cross-channel correlation, and therefore it is suitable for the BiSELD model that uses BTFF as input feature. In addition, based on the factorization of depthwise separable convolutions, kernels of various sizes can be implemented with a small number of parameters, allowing the BiSELD model to be built without the risk of overfitting. This is believed to be the reason why the BiSELD model achieved the best performance despite having less than half the number of parameters of the SALSA model.

# Chapter 8. Conclusions

To realize human auditory modality in humanoid robots, BTFF, an input feature based on HRTF localization cues, and BiSELDnet, a deep learning model based on Trinity architecture for BiSELD task, were proposed. The BiSELDnet simultaneously detects and localizes sound events by learning their time-frequency patterns and HRTF localization cues represented in BTFF with three factorized kernels of depthwise separable convolutions.

An eight-channel BTFF consists of left and right MSs; left and right V-maps; ITD and ILD-maps; and left and right SC-maps. V-map helps to detect sound events because it represents the amount of change in MS over time. As a result of evaluation, it was confirmed that adding V-map to MS is effective in detecting binaural sound events. ITD and ILD-maps are complementary because they provide spatial location information for different frequency ranges. ITD-map is more effective for low-frequency sounds below 1.5 kHz, and ILD-map is more effective for high-frequency sounds above 5 kHz. Unlike ITD-map, ILD-map changes into a complex pattern depending on azimuth at high frequencies, so it is a clue to solving ambiguous localization. The evaluation results showed that adding ITD and ILD-maps to MS is effective in localizing binaural sound events horizontally. Notably, SC-map was proposed based on the fact that the spectral notch patterns generated by the pinna contribute to vertical localization. Since the notch frequencies above 5 kHz is highly dependent on the elevation of sound event, the spectral pattern above 5 kHz is a distinct monaural feature for vertical localization. The evaluation results demonstrated that adding SC-map to MS is effective in localizing binaural sound events vertically. Therefore, based on HRTF localization cues, the proposed BTFF is effective for detection and localization of binaural sound events.

As a result of evaluating various BiSELDnets, the BiSELDnet based on Trinity module showed the best performance in all evaluation metrics of detection and localization, despite its relatively small number of parameters. Trinity architecture, designed to simultaneously extract features of various sizes while reducing the number of parameters based on factorized depthwise separable convolutions, appears to be best suited for BiSELD tasks. Also, the evaluation results of the BiSELDnet based on Xception module confirmed that depthwise separable convolution is suitable for input features with low cross-channel correlation, such as BTFF. A visualization for the layers of BiSELDnet showed that as layers get deeper, activations become more abstract and higher-level concepts such as time-frequency patterns and HRTF localization cues begin to be encoded. Particularly, VAM was proposed by modifying the existing Grad-CAM visualization. VAM visualization results for sound events on the median plane confirmed that VAM is activated exactly on the N1 notch and spectral pattern areas of SC-map, which means that BiSELDnet pays attention to the unique time-frequency patterns to detect sound events, and to the N1 notch patterns to estimate the elevation angle.

As a result of SOTA performance comparison, the BiSELD model was shown to have the best detection and localization performance even under various SNR conditions. This appears to be a natural consequence of BTFF, a dedicated input feature based on HRTF localization cues, and Trinity architecture, which can utilize the sub-features of various scales without having to worry about the exact time-frequency mapping, a limitation of existing SELD models. Therefore, the BiSELD model can estimate the azimuth and elevation angles of each sound event without front-back confusion using only two-channel input, even under various SNR conditions.

In future work, SC-map needs to be modified to improve the vertical localization performance of BiSELDnet under low SNR conditions. This could be achieved by expressing only the SC-map in a linear scale so that BiSELDnet can exploit not only the N1 notch but also the N2 notch. In particular, this study was conducted on single sound events, focusing on the feasibility of BiSELDnet based on HRTF localization cues. As a next step, BiSELDnet should be evolved to maintain its performance even in polyphonic scenarios. Last but not least, a realistic acoustic environment, including indoor and outdoor reflections, should be considered.

# Acknowledgments in Korean (감사의 글)

먼저, 지난 6년 6개월 동안 저를 지도해 주고 격려해 주신 박용화 교수님께 깊은 감사를 드립니다. 돌이켜 보면, SAMSUNG Best Paper Award 2015의 인연이 결국 저를 여기까지 이끌었고, 한 사회인이 다시 대학원생으로 회귀하는 엄청난 인생의 경험이었습니다. 그동안 교수님과 토론하며 저의 부족한 점을 알게 되었고, 저와 다른 관점에서 정말 많은 것을 배울 수 있었습니다. 무엇보다 연구에 있어서 핵심적인 본질이 무엇인지 깊게 숙고하는 계기가 되었습니다. 그리고 제 학위 논문의 심사를 맡아 주신 전원주 교수님, 윤용진 교수님, 이승철 교수님, 그리고 최정우 교수님께 진심으로 감사드립니다. 제가 2017년 8월 처음 카이스트에 와서 들은 강의가 매주 월요일 아침 9시에 시작하는 전원주 교수님의 파동학이었습니다. 음파 및 진동파에서 전자파까지 파동 전반에 걸친 물리적 통찰을 얻을 수 있는 명강의 였습니다. 공학설계 조교장하며 인연을 맺게 된 윤용진 교수님, 항상 온화한 표정과 따뜻한 말로 저를 격려해 주시고 제 연구에 대해 지혜로운 조언을 해 주셔서 감사합니다. 기계공학의 인공지능 도입에 주요한 역할을 하신 이승철 교수님, 제가 처음 인공지능 공부를 시작할 때 박사과정생이라면 어디서부터 출발해야 하는지 이정표를 세워 주셔서 감사합니다. 덕분에 기초 수학부터 최적화를 거쳐 기계학습 및 딥러닝 응용까지 인공지능 전반에 걸친 지식과 경험을 쌓을 수 있었습니다. 삼성종합기술원에 계셨을 때부터 그 명성을 익히 들어온 최정우 교수님, 본 논문이 완성되기까지 노빅 선배가 후배에게 하듯 아낌없는 조언과 충고를 해 주셔서 정말 고마웠습니다. 청강으로 들었던 음향 어레이 신호처리 기법은 파동장의 관점에서 음파를 이해하는 데 도움이 된 명강의 였습니다. 또한, 미국 오디오 공학회(AES)의 20세기 주요 라우드스피커 논문에 유일한 한국인 저자로 선정되어 한국 음향학자들의 자존심을 세우신 이정권 명예교수님, 진심으로 존경하고 가끔 향긋한 커피와 함께 따뜻하게 맞이해 주셔서 고마웠습니다. 마지막으로, 다시 공부를 시작하는 나이 많은 제자를 격려하고 흔쾌히 추천서를 써 주신 한양대의 오재응 명예교수님, 제자를 위한 진심 어린 조언 및 충고에 깊이 감사드립니다.

저의 낯선 대전 생활에 도움을 준 수많은 후배님들에게 말로 다하지 못한 고마움을 여기 지면을 빌어 전합니다. 우리 휴먼랩의 첫 랩장이자 나이에 비해 성숙한 신준오, 이제는 어엿한 애아빠가 된 내 소중한 박사 동기 재덕이, 제빵학으로는 이미 박사학위를 딴 것과 다름없는 마음씨 착한 강재, FEM 전문가이자 논문 많이 쓴 길용이, 우리 음향팀의 에이스 성후, 언제나 내게 아낌없는 조언과 충고를 해주는 얼리 버드 현욱이, 나의 전속 일러스트레이터 혜원이, 창의적이고 진취적인 준혁이, 다른 랩에 갔지만 오가다 만나면 웃는 얼굴로 안부 전하는 대희, 미국으로 유학간 근육맨 인규, 타국에 와서 홀로 조용히 연구하는 Simeneh, 연구실의 큰누나 수민이, 다혈질이지만 알고 보면 속정 깊은 성현이, 재기 발랄하며 누구와도 어울릴 줄 아는 사업가 마인드의 재학이, 나와 함께 카이스트 노빅의 HRTF 측정설비를 업그레이드하고 HATS KAIST DB를 구축하여 공개

한 최상민, 그 뒤를 이어 HRTF 개인화를 위해 사람도 측정 가능하도록 측정설비를 더 업그레이드한 유쾌한 웃음 소리의 병윤이, 가끔 새벽에 4114 문을 조용히 열고 얼굴을 쑥 내밀어 나의 간담을 서늘하게 했던 천재 현준이, 세상 모든 것을 다 품을 수 있을 것 같은 넓은 마음을 가진 착하고 똑똑하고 예쁜 보미, 본인의 연구 및 후배들을 위해 열과 성을 다하고 육사 장교를 했어도 잘했을 것 같은 정민이, 연구실에서 나와 유일하게 쇼펜하우어 및 니체와 같은 철학자에 대해 담론을 나눌 수 있는 공학하는 철학자 준기, 연구실 GPU 서버를 체계적으로 구축하고 AI 세미나를 주관하여 연구실의 AI 역량을 끌어 올리는데 기여한 젊은 과학자 원호, 회사 그만두고 자신의 연구를 묵묵히 수행하는 성실한 대근이, 올해 박사과정을 시작한 차분하게 연구 잘 할 것 같은 윤섭이, 나의 박사학위 디펜스 때 늦은 밤까지 제본 및 셋팅을 기꺼이 도와준 나의 처갓집 치맥 소울메이트이자 우리 음향팀의 희망 덕기, 방이 달라 많은 얘기는 안 나누어 봤지만 왠지 잘 맞을 것 같은 윤성현, 석사 디펜스를 준비하며 늦은 밤까지 연구에 자신을 갈아 넣었던 얼핏 보면 하정우를 닮은 잘생긴 도현이, 석사 디펜스 막판에 극적으로 돌파구를 마련한 불굴의 민창희, 이미 한국 사람이 다 된 것 같은 Adelle, 나처럼 회사생활 오래하다가 박사과정을 시작한 동갑내기 애아빠 민태, LG과제 하며 그 진가를 비로소 알게 된 성실하고 똑똑한 김진, 목소리가 굵고 멋있는 키 큰 석준이, 지난 6년간 내 아내보다 더 오랜 밤을 같이 보낸 룸메이트이자 피아니스트인 넙주기 류승현, 내가 대전에 있는 동안 너무 바빠서 단 한번도 방문을 못했던 사랑하는 한양대 후배 오철이와 지명이 부부, 또 미처 언급하지 못한 한양대 및 카이스트 선후배들에게 앞으로의 인생에서 건강과 행운이 늘 함께 하시길 기원합니다.

제가 회사 나와서 박사과정을 시작할 때 여러모로 지원을 아끼지 않았고, 저의 가능성을 알아보고 새로운 계기를 마련해주신 부전전자의 서동현 사장님 및 이석순 회장님께 깊은 감사의 말씀을 전해드립니다. 항상 자기 관리에 충실하고, 임직원들과 격의 없이 소통하며, 시대의 흐름에 맞게 회사를 재편하여 성과를 내시는 서동현 사장님을 뵈면서 진정한 리더십이 무엇인지 생각해보게 되었습니다. 더욱이 그런 사장님의 비전을 믿고 묵묵히 뒤에서 지원하시는 회장님의 리더십 또한 저로 하여금 부전을 선택하게 만들었습니다. 연구소장으로서 앞으로 회사가 기존의 음향산업을 넘어 새로운 영역으로 확장할 수 있도록 열과 성을 다하겠습니다. 그리고 곧 연구소에 합류할 파리 낭테르 대학의 강현정 박사에게도 저를 믿고 저희 회사를 선택해주신데 대해서 감사의 말씀드립니다. 앞으로 재미있게 일할 수 있는 즐거운 연구소를 만들어 봅시다. 저의 대학원 선배이자, 전기음향학의 스승인 삼성전자의 김종배 수석님께도 깊은 감사의 말씀을 전해드립니다. 수석님 덕분에 기존의 소음진동을 넘어 오디오 음향의 다채로운 세계에 눈을 뜨게 되었습니다.

누구보다, 오늘의 저를 있게 해 주신 가족 여러분 모두 사랑하고 고맙습니다. 언제나 아들이 도전하기를 바라시는 아버지, 아들의 성취 보다 아들의 건강이 최우선인 어머니, 어떠한 상황에서도 사위를 믿고 격려해주시는 장인 어른, 그리고 사위가 아닌 아들로 뒷바라지를 해 주신 장모님께 깊은 사랑과 존경의 마음을 전합니다. 항상 동생이 잘되길 바라는 형과 형수님, 필리핀의 내

반쪽 언주와 형님, 처제라기 보다는 여동생에 가까운 미국 메인주의 려진이와 헐서방, 그리고 내 귀여운 조카들인 승재, 진호, 에이든, 엘리아에게도 아낌없는 사랑의 마음을 전합니다. 마지막으로, 이 영겁의 광활한 우주에서 사랑하는 내 아내 경진과 아들 윤재 그리고 딸 가원이와 같은 시간과 공간을 점유할 행운을 주신 대우주 자연 모두에게 감사드립니다.

<div style="text-align: right;">
2023년 12월 21일 오후 5시 39분<br>
학자의 성(儒城) 카이스트에서<br>
이 경 태
</div>

*다석 류영모가 풀이한 한글 천부경*

## Curriculum Vitae

### 1. Personal Information

| | |
|---|---|
| Name (in Korean) | Gyeong-Tae Lee (이 경 태) |
| Title | Ph. D. |
| Affiliation | Korea Advanced Institute of Science and Technology (KAIST) |
| Birth Date @ Place | February 27, 1980 @ Cheonan, Korea |
| Gender | Male |
| Nationality | Korea |
| Phone | +82-10-8299-2778 |
| E-mail | hansaram@kaist.ac.kr | g.t.lee333@gmail.com |
| Profile | https://www.linkedin.com/in/lgt31/ |
| | https://scholar.google.co.kr/citations?user=Ge90dWIAAAAJ&hl=ko |
| | https://github.com/han-saram |

### 2. Education

**KAIST**  Daejeon, Korea  Aug 2017 - Feb 2024

*Ph. D. in Mechanical Engineering*

- National Scholarship Student
- Academic Performance (GPA: 3.66 / 4.30)
- Award for Outstanding Teaching Assistant of Mechanical Engineering, 2019
- Supervisor: Prof. Yong-Hwa Park
- Dissertation: *Binaural Sound Event Localization and Detection Neural Network based on HRTF Localization Cues for Humanoid Robots*
- Academic Research Projects
  - *Development of Sound Camera Post-processing S/W AI Algorithm*
    (Funded by LG Electronics, Jul 2023 ~ Jen 2024)
  - *Advanced Education Track for Integrated Design of Offshore Hybrid Power Generation System and its Intelligent Operation*
    (Funded by KETEP, Apr 2018 ~ Dec 2022)
  - *Smart Protective Suit with Active Cooling/Ventilation and Antivirus Suit Changing System*

(Funded by KAIST, Aug 2020 ~ Dec 2020)
- *Development of Deep Learning based Cough Recognition Model and its Application to Acoustic Camera*
(Funded by SM Instruments, Apr 2020 ~ Jun 2020)
- *Development of Rail Inspection Car for the Monitoring System of Rail Damage Inspection and Concrete Crack Detection Module*
(Funded by KAIA, Sep 2017 ~ Dec 2019)
- *Free-Running Embedded Speech Recognition Technology for Natural Language Dialogue with Robots*
(Funded by Ministry of Trade, Industry and Energy, Apr 2018 ~ Dec 2018)
- *Optimized Novel Meta-materials and Meta-structures to Solve the Practical Acoustic and/or Multi-modal Problems in the Near Future*
(Funded by KAIST, Jul 2017 ~ Dec 2017)

**Hanyang University**          Seoul, Korea          Mar 2005 - Feb 2007

*Master of Science in Mechanical Engineering*

- Fellowship (merit-based, 100% tuition)
- Academic Performance (GPA: 4.45 / 4.50)
- Supervisor: Prof. Jae-Eung Oh
- Thesis: *A Study on the Active Noise Control of the Duct System using Co-FXLMS Algorithm*
- Academic Research Projects
    - *Development of Active Noise Control System for Washing Machines*
    (Funded by LG Electronics, Mar 2006 ~ Jan 2007)
    - *Development of Active Noise Control System for the Low Freq. Noise under Rapid Acceleration*
    (Funded by SMBA, Jul 2005 ~ Apr 2006)

**Kyonggi University**          Suwon, Korea          Mar 1998 - Feb 2005

*Bachelor of Science in Mechanical Engineering*

- Scholarship (merit-based, top 10%)
- Academic Performance (GPA: 4.12 / 4.50)
- Award for Outstanding Student of Mechanical Engineering, 2005
- Leave of Absence for Military Service (Feb 1999 ~ Apr 2001)

## 3. Professional Experience

**SAMSUNG ELECTRONICS Co., Ltd.**  Suwon, Korea  Feb 2007 - Aug 2017
Future R&D Lab, Advanced R&D Group, R&D Team, VD Business, CE  Jan 2016 - Aug 2017
Acoustic Technology Center, Multimedia R&D Team, DMC R&D Center  Jul 2007 - Oct 2015
Audio Lab, Mobile Solution Team, DM R&D Center  May 2007 - Jul 2007

- Project: *Development of New Radiation Type Speaker Solution for UHD-TV*, Task Leader as Senior Engineer  Jan 2017 - Aug 2017

- Project: *Development of High-quality Audio Solutions for the Next UHD-TV*, Task Leader as Senior Engineer  Mar 2016 - Dec 2016

- Project: *Development of Speaker Solutions for New Form Factor*, Task Leader as Senior Engineer  Jul 2014 - Dec 2015

- Project: *Development of Next Generation Acoustic Transducers*, Senior Engineer in charge of electro-acoustic research & development  Jan 2013 - Jun 2014

- Project: *Development of Super-Slim Sound-Enhanced Speaker Solutions*, Engineer in charge of electro-acoustic research & development  Jan 2012 - Dec 2012

- Project: *Development of Speaker Solutions for Differentiation in Sound Quality*, Engineer in charge of electro-acoustic research & development  Jan 2011 - Dec 2011

- Project: *Development of Technology and Infrastructure for Sound Quality Enhancement of Slim Speaker*, Engineer in charge of electro-acoustic research & development  Jan 2010 - Nov 2010

- Project: *Insight - Development of High Sound Quality Speaker Solution*, Engineer in charge of electro-acoustic research & development  Feb 2009 - Dec 2009

- Project: *Orchard*, Engineer in charge of acoustic development  Jan 2008 - Dec 2008

- Project: *Magellan*, Engineer in charge of acoustic development  May 2007 - Nov 2007

## 4. Professional Skills

**Engineering Skills**

- Binaural Sound Event Localization and Detection (BiSELD)
    - Head-Related Transfer Functions (HRTFs)
    - Binaural Sound Source Localization (BSSL)
    - Sound Event Localization and Detection (SELD)
    - Digital Signal & Speech Processing
    - Machine Learning & Deep Learning

- Acoustic Expert System Construction
    - Development of Speaker Response Simulators based on Electro-acoustics
    - Development of Sound Measurement & Tuning Systems
    - Development of Sound Quality Evaluation System (SQES)

- Electro-acoustic System Design
    - Design of New Acoustic Transducers
    - Design of New Speaker Systems for Thin Electronic Devices with Narrow Bezel
    - Experience in dealing with Many Cases of Slim & Hidden Speaker Solution

- Measurement & Tuning
    - Multi-point Measurement based on Ref. Room Target Curve
    - Sound Tuning Process using Passive (RLC) Filters
    - Sound Tuning Process using Active (IIR & FIR) Filters

- Sound Quality Evaluation
    - Sound Quality Evaluation of Speaker Systems based on SQES

- Active Noise & Vibration Control
    - Development of Active Noise Control Systems based on DSP (SISO & MIMO)
    - Development of Active Vibration Control Systems based on DSP (SISO & MIMO)

- Noise & Vibration
    - Noise Source Identification based on Sound Intensity Measurement & Analysis

- Analysis of Dynamic Characteristics of Mechanical Systems through Modal Analysis & Testing
- Experience in dealing with Many Cases of Passive Noise & Vibration Isolation

**Computer Skills**
- CAE: ANSYS, LEAP, FilterShop
- Programming Language: C/C++, Python, MATLAB, LabVIEW

**Language Skills**
- Fluent in English both oral and written

# 5. Special Activities

| | |
|---|---|
| *"Deep Learning based Cough Detection Camera" was broadcast through various news media such as YTN, MBC, and JTBC* | Aug 2020 |
| *Public Lecture on the Design of Slit-Firing Sound Plate for TV at NoViC, KAIST* | Jul 2016 |
| *Lectures on the Acoustics of Speaker Systems to Junior Engineers of VD Business* | Apr 2013 |
| *Introductory Lectures on Acoustics to New Recruits of DMC R&D Center* | Jul 2009 |
| *Training Materials on Sound & Vibration (over 400 pages in PowerPoint slides)* | Jun 2009 |

# 6. Honors

| | |
|---|---|
| *Excellent Presentation Award at the 2023 Spring Conference of the KSNVE* | Oct 2023 |
| *Excellent Presentation Award at the 2020 Fall Conference of the ASK* | Nov 2020 |
| *Best Paper Award at the 2019 Spring Conference of the KSNVE* | Dec 2019 |
| *Excellent Presentation Award at the 2019 Fall Conference of the ASK* | Nov 2019 |
| *Outstanding Teaching Assistant Award in Mechanical Engineering at KAIST* | Mar 2019 |
| *Silver Award at the "SAMSUNG Best Paper Award 2015"* | Nov 2015 |
| *DMC R&D Center Award for the Design of Co-axial Air Pumping Transducer* | Feb 2015 |
| *DMC R&D Center Award for the Development of Sound Quality Evaluation System* | Mar 2012 |
| *Special Incentive Award for the Development of Ultra-slim Acoustic Solutions* | May 2011 |
| *DMC R&D Center Award for the Development of Sound Tuning Systems* | Mar 2010 |
| *DMC R&D Center Award for the Development of Ultra-slim Woofer for LED TV* | Feb 2009 |
| *Outstanding Student Award in Mechanical Engineering at Kyonggi University* | Feb 2005 |

# 7. Patents

## Korea Patent

| | Title of Invention | Inventors | Appl. No. (Filing Date) | Patent No. (Reg. Date) |
|---|---|---|---|---|
| 1 | Method and apparatus for cough recognition based on deep learning | Y.-H. Park, **G.-T. Lee**, S. Kim, H. Nam, Y.-K. Kim, J.-S. Lee, S.-H. Park, I.-K. Kim, and K.-H. Lee | 10-2020-0163070 (2020.11.27) | 10-2453984 (2022.10.07) |
| 2 | Method and apparatus for measuring remaining time of laundry using water hose | Y.-H. Park, **G.-T. Lee**, M.-K. Kim, H.-J. Kim, and T.-Y. Woo | 10-2019-0154246 (2019.11.27) | 10-2333332 (2021.11.26) |
| 3 | Speaker apparatus | S.-H. Son, J.-B. Kim, S.-J. Kim, and **G.-T. Lee** | 10-2017-0129831 (2017.10.11) | 10-2402327 (2022.05.23) |
| 4 | Wideband slot loading loudspeaker | J.-B. Kim, S.-H. Son, **G.-T. Lee**, S.-J. Kim, and Y.-S. Lee | 10-2016-0113423 (2016.09.02) | 10-2472499 (2022.11.25) |
| 5 | Slim acoustic transducer and image display apparatus having the same | **G.-T. Lee**, J.-B. Kim, and S.-H. Son | 10-2016-0104070 (2016.08.17) | 10-2272386 (2021.06.28) |
| 6 | Loudspeaker | **G.-T. Lee**, J.-B. Kim, D.-K. Park, and S.-H. Son | 10-2015-0116105 (2015.08.18) | 10-2359269 (2022.01.28) |
| 7 | Acoustic transducer | **G.-T. Lee**, J.-B. Kim, and S.-H. Son | 10-2015-0095855 (2015.07.06) | 10-2322035 (2021.10.29) |
| 8 | Speaker | D.-K. Ahn, D.-J. Kim, S.-S. Woo, J.-B. Kim, J.-I. Jo, S.-J. Kim, H.-P. Kim, C.-Y. Ahn, and **G.-T. Lee** | 30-2014-0032230 (2014.07.01) | 30-0809965 (2015.08.07) |
| 9 | Speaker | D.-K. Ahn, D.-J. Kim, S.-S. Woo, J.-B. Kim, J.-I. Jo, S.-J. Kim, H.-P. Kim, C.-Y. Ahn, and **G.-T. Lee** | 30-2014-0032229 (2014.07.01) | 30-0809964 (2015.08.07) |
| 10 | Speaker | D.-K. Ahn, D.-J. Kim, S.-S. Woo, J.-B. Kim, J.-I. Jo, S.-J. Kim, H.-P. Kim, C.-Y. Ahn, and **G.-T. Lee** | 30-2014-0032228 (2014.07.01) | 30-0809963 (2015.08.07) |
| 11 | Audio output apparatus capable of outputting multi channel audio and display apparatus applying the same | **G.-T. Lee**, J.-B. Kim, D.-K. Park, and S.-H. Son | 10-2014-0069427 (2014.06.09) | 10-2201870 (2021.01.06) |
| 12 | Method and apparatus for outputting sound through teum speaker | J.-B. Kim, D.-J. Kim, D.-K. Park, S.-H. Son, **G.-T. Lee**, J.-Y. Lee, J.-I. Jo, S.-J. Kim, D.-H. Jung, S.-H. Hwang, D.-K. Ahn, S.-S. Woo, and S.-M. Hyun | 10-2014-0019433 (2014.02.20) | 10-2077236 (2020.02.07) |
| 13 | Audio system, method for outputting audio, and speaker apparatus thereof | **G.-T. Lee**, J.-B. Kim, J.-I. Jo, W.-O. Sung, J.-Y. Lee, and H. Shim | 10-2013-0120605 (2013.10.10) | 10-2114219 (2020.05.18) |
| 14 | Sound generation apparatus and electric apparatus comprising thereof | **G.-T. Lee**, J.-B. Kim, D.-K. Park, J.-I. Jo, S.-C. Lee, and S.-M. Hyun | 10-2013-0077905 (2013.07.03) | 10-2023189 (2019.09.11) |
| 15 | An electronic device employing a sound plate switchable between stand-type and hang-type | **G.-T. Lee**, J.-B. Kim, and J.-I. Jo | 10-2010-0009640 (2010.02.02) | 10-1631275 (2016.06.10) |
| 16 | Sound plate and electronic device employing the same | **G.-T. Lee**, J.-B. Kim, and J.-I. Jo | 10-2010-0009639 (2010.02.02) | 10-1632299 (2016.06.15) |
| 17 | Enclosure for amplifying bass sound, woofer with the enclosure, and electronic device with the woofer | **G.-T. Lee** and J.-B. Kim | 10-2009-0068411 (2009.07.27) | 10-1629822 (2016.06.07) |
| 18 | Active noise control system and method in enclosed field of 3-dimension using correlation filtered-x least mean squares algorithm | J.-E. Oh, S.-G. Park, **G.-T. Lee**, and O-C. Kwon | 10-2007-0128309 (2007.12.11) | 10-0902954 (2009.06.08) |

**United States Patent**

| | Title of Invention | Inventors | Appl. No. (Filing Date) | Patent No. (Reg. Date) |
|---|---|---|---|---|
| 1 | Deep learning-based cough recognition method and device | Y.-H. Park, **G.-T. Lee**, H. Nam, S. Kim, Y.-K. Kim, I.-K. Kim, K.-H. Lee, J.-S. Lee, and S.-H. Park | 18/051137 (2022.10.31) | US-20230078404-A1 (2023.03.16) |
| 2 | Audio output apparatus capable of outputting multi channel audio and display apparatus applying the same | **G.-T. Lee**, J.-B. Kim, D.-K. Park, and S.-H. Son | 16/693491 (2019.11.25) | US-10820092-B2 (2020.10.27) |
| 3 | Speaker apparatus | S.-H. Son, J.-B. Kim, S.-J. Kim, and **G.-T. Lee** | 16/754967 (2018.09.04) | US-11064290-B2 (2021.07.13) |
| 4 | Wideband slot-loading loudspeaker | J.-B. Kim, S.-H. Son, **G.-T. Lee**, S.-J. Kim, and Y.-S. Lee | 15/672828 (2017.08.09) | US-10499141-B2 (2019.12.03) |
| 5 | Slim acoustic transducer and image display apparatus having the same | **G.-T. Lee**, J.-B. Kim, and S.-H. Son | 15/402393 (2017.01.10) | US-10681467-B2 (2020.06.09) |
| 6 | Loudspeaker | **G.-T. Lee**, J.-B. Kim, D.-K. Park, and S.-H. Son | 15/010272 (2016.01.29) | US-9848259-B2 (2017.12.19) |
| 7 | Acoustic transducer | **G.-T. Lee**, J.-B. Kim, and S.-H. Son | 14/965969 (2015.12.11) | US-10149064-B2 (2018.12.04) |
| 8 | Speaker | D.-K. Ahn, D.-J. Kim, S.-S. Woo, J.-B. Kim, J.-I. Jo, S.-J. Kim, H.-P. Kim, C.-Y. Ahn, and **G.-T. Lee** | 29/513454 (2014.12.31) | US-D782445-S (2017.03.28) |
| 9 | Speaker | D.-K. Ahn, D.-J. Kim, S.-S. Woo, J.-B. Kim, J.-I. Jo, S.-J. Kim, H.-P. Kim, C.-Y. Ahn, and **G.-T. Lee** | 29/513451 (2014.12.31) | US-D781820-S (2017.03.21) |
| 10 | Speaker | D.-K. Ahn, D.-J. Kim, S.-S. Woo, J.-B. Kim, J.-I. Jo, S.-J. Kim, H.-P. Kim, C.-Y. Ahn, and **G.-T. Lee** | 29/513444 (2014.12.31) | US-D778260-S (2017.02.07) |
| 11 | Method and apparatus for outputting sound through speaker | J.-B. Kim, D.-J. Kim, D.-K. Park, S.-H. Son, **G.-T. Lee**, J.-Y. Lee, J.-I. Jo, S.-J. Kim, D.-H. Jung, S.-H. Hwang, D.-K. Ahn, S.-S. Woo, and S.-M. Hyun | 14/523146 (2014.10.24) | US-10038947-B2 (2018.07.31) |
| 12 | Audio system, method of outputting audio, and speaker apparatus | **G.-T. Lee**, J.-B. Kim, J.-I. Jo, W.-O. Sung, J.-Y. Lee, and H. Shim | 14/507463 (2014.10.06) | US-10009687-B2 (2018.06.26) |
| 13 | Audio output apparatus capable of outputting multi channel audio and display apparatus applying the same | **G.-T. Lee**, J.-B. Kim, D.-K. Park, and S.-H. Son | 14/902974 (2014.07.07) | US-10491985-B2 (2019.11.26) |
| 14 | Sound generating apparatus and electronic apparatus including the same | **G.-T. Lee**, J.-B. Kim, D.-K. Park, J.-I. Jo, S.-C. Lee, and S.-M. Hyun | 14/310073 (2014.06.20) | US-9485565-B2 (2016.11.01) |
| 15 | Electronic device employing sound plate switchable between stand type and hang type | **G.-T. Lee**, J.-B. Kim, and J.-I. Jo | 13/019364 (2011.02.02) | US-8447052-B2 (2013.05.21) |
| 16 | Sound plate and electronic device employing the same | **G.-T. Lee**, J.-B. Kim, and J.-I. Jo | 13/016019 (2011.01.28) | US-8494186-B2 (2013.07.23) |
| 17 | Bass sound amplifying enclosure, woofer including the same, and electronic device including the woofer | **G.-T. Lee** and J.-B. Kim | 12/683540 (2010.01.07) | US-8351631-B2 (2013.01.08) |
| 18 | Hidden speaker apparatus | **G.-T. Lee** and J.-B. Kim | 12/535857 (2009.08.05) | US-8737673-B2 (2014.05.27) |

**European Patent**

|   | Title of Invention | Inventors | Appl. No. (Filing Date) | Patent No. (Reg. Date) |
|---|---|---|---|---|
| 1 | Loudspeaker | **G.-T. Lee**, J.-B. Kim, D.-K. Park, and S.-H. Son | 16158076.6 (2016.03.01) | EP3133827 (B1) (2018.01.31) |
| 2 | Acoustic transducer | **G.-T. Lee**, J.-B. Kim, and S.-H. Son | 19168453.9 (2015.12.16) | EP3537727 (B1) (2020.06.10) |
| 3 | Audio system, method of outputting audio, and speaker apparatus | **G.-T. Lee**, J.-B. Kim, J.-I. Jo, W.-O. Sung, J.-Y. Lee, and H. Shim | 14188310.8 (2014.10.09) | EP2860992 (B1) (2018.12.19) |
| 4 | Audio output apparatus capable of outputting multi channel audio and display apparatus applying the same | **G.-T. Lee**, J.-B. Kim, D.-K. Park, and S.-H. Son | 14819325.3 (2014.07.07) | EP3017593 (B1) (2020.02.19) |
| 5 | Sound plate and electronic device employing the same | **G.-T. Lee**, J.-B. Kim, and J.-I. Jo | 11153100.0 (2011.02.02) | EP2360938 (B1) (2013.07.03) |
| 6 | Bass sound amplifying enclosure, woofer including the same, and electronic device including the woofer | **G.-T. Lee** and J.-B. Kim | 10157372.3 (2010.03.23) | EP2282553 (B1) (2020.07.01) |

## 8. Publications

**International Journal Articles**

[1]   B.-Y. Ko, **G.-T. Lee**, H. Nam, and Y.-H. Park, "PRTFNet: HRTF individualization for accurate spectral cues using a compact PRTF," *IEEE Access*, vol. 11, pp. 96119–96130, Sep. 2023.

[2]   P. R. Andersen, **G.-T. Lee**, D. G. Nielsen, J. Kook, V. C. Henriquez, N. Aage, and Y.-H. Park, "Experimental characterization of a shape optimized acoustic lens: Application to compact speakerphone design," *J. Acoust. Soc. Am.*, vol. 153, no. 4, pp. 2351–2361, Apr. 2023.

[3]   **G.-T. Lee**, H. Nam, S.-H. Kim, S.-M. Choi, Y. Kim, and Y.-H. Park, "Deep learning based cough detection camera using enhanced features," *Expert Syst. Appl.*, vol. 206, no. 117811, pp. 1–20, Nov. 2022.

[4]   **G.-T. Lee**, S.-M. Choi, B.-Y. Ko, and Y.-H. Park, "HRTF measurement for accurate sound localization cues," *arXiv:2203.03166v2*, 2022. Available: https://arxiv.org/abs/2203.03166

[5]   H.-J. Lee, **G.-T. Lee**, and J.-E. Oh, "Active control of automotive intake noise under rapid acceleration using the Co-FXLMS algorithm," *J. Syst. Des. Dyn.*, vol. 4, no. 3, pp. 429–439, May 2010.

[6]   J.-K. Hoh, Y.-S. Park, K.-J. Cha, J.-E. Oh, H.-J. Lee, **G.-T. Lee**, and M.-I. Park, "Chaotic indices and canonical ensemble of heart rate patterns in small-for-gestational age fetuses," *J. Perinat. Med.*, vol. 35, pp. 210–216, May 2007.

**Domestic Journal Articles**

**International Conference Papers**

**Domestic Conference Papers**